\documentclass[12pt]{iopart}
\usepackage{iopams}
\usepackage[square,numbers,sort&compress]{natbib}%sort&compress
\usepackage{graphicx,amssymb,bm}
\usepackage{csquotes}
\usepackage[dvipsnames]{xcolor}
\usepackage[normalem]{ulem}
\usepackage{upgreek}
\newcommand{\mb}[1]{\bm{#1}}
\newcommand{\gstwo}{\texttt{GS2}}
\newcommand{\refeq}[1]{(\ref{#1})}

\newcommand{\beqn}{\begin{eqnarray}}
\newcommand{\eeqn}{\end{eqnarray}}
\newcommand{\xp}{\times}
\newcommand{\kron}[1]{\delta_{#1}}
\newcommand{\fract}[2]{#1 / #2 }
\newcommand{\eye}{\mb{I}}
\newcommand{\expo}[1]{\exp{\left[#1\right]}}
\newcommand{\codot}{\mb{:}}
\newcommand{\trace}[1]{{\rm{Tr}}\left[#1\right]}
\newcommand{\vprod}[2]{\left\langle #1 \middle|  #2\right\rangle}
\newcommand{\imag}{{\rm{i}}}
\newcommand{\order}[1]{\Or\left( #1 \right) }
\newcommand{\ovl}[1]{\overline{#1}}

\newcommand{\ltsp}{c}
\newcommand{\bu}{\mb{b}}
\newcommand{\bvec}{\mb{B}}
\newcommand{\bmag}{B}
\newcommand{\bmagz}{B_0}
\newcommand{\bmagovl}{\ovl{B}}
\newcommand{\bmagmax}{B_{\rm max}}

\newcommand{\wfreq}{\omega}
\newcommand{\wfreqr}{\omega_{{\rm{r}}}}
\newcommand{\wfreqrz}{\omega^{(0)}_{{\rm{r}}}}
\newcommand{\wfreqrh}{\omega^{(1/2)}_{{\rm{r}}}}
\newcommand{\wfreqz}{\omega^{(0)}}
\newcommand{\wfreqh}{\omega^{(1/2)}}
\newcommand{\wfreqo}{\omega^{(1)}}
\newcommand{\wfreqn}{\omega^{(n)}}
\newcommand{\spe}{s}
\newcommand{\pbeta}{\beta}

\newcommand{\fldl}{\alpha}
\newcommand{\fldlz}{\alpha_0}
\newcommand{\flxl}{\psi}
\newcommand{\flxlz}{\psi_0}
\newcommand{\Dflxl}{\Delta \flxl}

\newcommand{\kkflxl}{k_\flxl}
\newcommand{\kkfldl}{k_\fldl}
\newcommand{\kperp}{k_\perp}
\newcommand{\kperpvec}{\mb{k}_{\perp}}
\newcommand{\kperpvecz}{\mb{k}^{(0)}_{\perp}}
\newcommand{\kperpveco}{\mb{k}^{(1)}_{\perp}}
\newcommand{\kpara}{k_\|}

\newcommand{\rr}{{\mb{r}}}

\newcommand{\gc}{{\mb{R}}}

\newcommand{\ttime}{t}

\newcommand{\nbl}{\nabla}

\newcommand{\pvel}{\mb{v}}
\newcommand{\pvelprim}{\mb{v}^\prime}
\newcommand{\gyrophase}{\gamma}
\newcommand{\energy}{\varepsilon}
\newcommand{\pitch}{\lambda}
\newcommand{\pangle}{\xi}
\newcommand{\sign}{\sigma}
\newcommand{\vpar}{v_\|}
\newcommand{\vmag}{v}
\newcommand{\vmagprim}{v^\prime}
\newcommand{\vperp}{v_\perp}
\newcommand{\wvel}{\mb{w}}
\newcommand{\wmag}{w}
\newcommand{\svel}{\mb{s}}
\newcommand{\smag}{s}
\newcommand{\Ufunc}{\mb{U}}
\newcommand{\Ufunchat}{\widehat{\mb{U}}}
\newcommand{\lpar}{\theta}
\newcommand{\lchi}{\chi}
\newcommand{\kpar}{\bu \! \cdot \! \nbl \lpar}
\newcommand{\kparprim}{\bu \!\cdot\! \nbl \lparprim}
\newcommand{\lscal}{a}
\newcommand{\toragl}{\zeta}
\newcommand{\saffac}{q}
\newcommand{\nufunc}{\nu}
\newcommand{\saffacprim}{\saffac^\prime}
\newcommand{\bcur}{I}
\newcommand{\jint}{j_\|}
\newcommand{\jintpm}{j^{\pm}_\|}
\newcommand{\jintplus}{j^{+}_\|}
\newcommand{\jintminus}{j^{-}_\|}

\newcommand{\lparav}[1]{\left\langle{#1}\right\rangle^{\lpar}}
\newcommand{\transav}[1]{\left\langle{#1}\right\rangle^{\rm{t}}}
\newcommand{\bouncav}[1]{\left\langle{#1}\right\rangle^{\rm{b}}}
\newcommand{\gav}[2]{\left\langle{#1}\right\rangle^{\gyrophase}_{#2}}
\newcommand{\intv}[1]{\int   #1 \; d^3 \mb{v} }
\newcommand{\intvprim}[1]{\int   #1 \; d^3 \mb{v}^\prime }
\newcommand{\intw}[1]{\int   #1 \; d^3 \mb{w} }
\newcommand{\ints}[1]{\int   #1 \; d^3 \mb{s} }

\newcommand{\ptlsub}[1]{\phi_{#1}} 
\newcommand{\ptl}{\phi} 

\newcommand{\ptlz}{\phi^{(0)}} 
\newcommand{\ptlo}{\phi^{(1)}} 
\newcommand{\ptlh}{\phi^{(1/2)}} 
\newcommand{\ptln}{\phi^{(n)}}

\newcommand{\hhs}{h_{\spe }}
\newcommand{\hhssub}[1]{h_{\spe, #1 }}
\newcommand{\hhsub}[1]{h_{ #1 }}
\newcommand{\hhe}{h_{{\rm{e}} }}
\newcommand{\hhi}{h_{{\rm{i}} }}

\newcommand{\fulldistf}{f}
\newcommand{\fulldistfprim}{f^\prime}
\newcommand{\fulldistg}{g}
\newcommand{\fulldistgprim}{g^{\prime}}
\newcommand{\dlf}{\delta \! f}
\newcommand{\dlfs}{\delta \! f_\spe}

\newcommand{\eqlba}{F_{0\spe}}

\newcommand{\eqlbi}{F_{0i}}
\newcommand{\eqlbe}{F_{0e}}
\newcommand{\eqlbeprim}{F_{0e}^\prime}
\newcommand{\eqlbaprim}{F_{0\spe}^\prime}

\newcommand{\hhsn}{\hhs{}^{(n)}}
\newcommand{\hhez}{\hhe{}^{(0)}}

\newcommand{\hheo}{\hhe{}^{(1)}}
\newcommand{\hhiz}{\hhi{}^{(0)}}
\newcommand{\HHs}{H_\spe}
\newcommand{\HHe}{H_{\rm{e}}}
\newcommand{\HHeSH}{H_{\rm{SH}}}
\newcommand{\HHeSHp}[1]{H^{(#1)}_{\rm{SH}}}
\newcommand{\HHesub}[1]{H_{{\rm{e}},#1}}
\newcommand{\hhisub}[1]{h_{{\rm{i}},#1}}
\newcommand{\KKeSH}{K_{\rm{SH}}}
\newcommand{\ffSH}{f_{\rm SH}}
\newcommand{\ggSH}{g_{\rm SH}}
\newcommand{\HHeSHz}{H_{\rm{SH},0}}
\newcommand{\HHi}{H_{{\rm{i}}}}
\newcommand{\HHez}{H^{(0)}_{{\rm{e}}}}
\newcommand{\HHeh}{H^{(1/2)}_{{\rm{e}}}}
\newcommand{\HHeo}{H^{(1)}_{{\rm{e}}}}
\newcommand{\HHen}{H^{(n)}_{{\rm{e}}}}
\newcommand{\HHep}[1]{H^{(#1)}_{{\rm{e}}}}
\newcommand{\HHepq}[2]{H^{(#1)}_{{\rm{e}},(#2)}}

\newcommand{\bes}[1]{ J_0(#1)}
\newcommand{\besfirst}[1]{ J_1(#1)}
\newcommand{\bessn}[1]{J_{#1\spe}}
\newcommand{\besen}[1]{J_{#1{\rm{e}}}}
\newcommand{\besenz}[1]{J^{(0)}_{#1{\rm{e}}}}
\newcommand{\besin}[1]{J_{#1{\rm{i}}}}

\newcommand{\gyrdvecs}{\bm{\rho}_\spe}
\newcommand{\gyrdvece}{\bm{\rho}_e}
\newcommand{\gyrdveci}{\bm{\rho}_i}
\newcommand{\gyrds}{\rho_{\rm{th},\spe}}
\newcommand{\gyrdi}{\rho_{\rm{th,i}}}
\newcommand{\gyrde}{\rho_{\rm{th,e}}}
\newcommand{\gyrdez}{\rho^0_{\rm{th,e}}}
\newcommand{\gyrdeovl}{\ovl{\rho}_{\rm{th,e}}}

\newcommand{\dens}{n}
\newcommand{\denss}{n_\spe}
\newcommand{\densi}{n_{\rm{i}}}
\newcommand{\dense}{n_{\rm{e}}}
\newcommand{\ddensi}{\delta \! n_{\rm{i}}}
\newcommand{\ddense}{\delta \! n_{\rm{e}}}
\newcommand{\ddenss}{\delta \! n_{\spe}}
\newcommand{\ddensns}[2]{\delta \! n^{(#1)}_{\rm{#2}}}

\newcommand{\dtempns}[2]{\delta \! T^{(#1)}_{\rm{#2}}}
\newcommand{\dtempnsq}[3]{\delta \! T^{(#1)}_{{\rm{#2}},(#3)}}
\newcommand{\ddensnsq}[3]{\delta \! n^{(#1)}_{{\rm{#2}},(#3)}}

\newcommand{\dqpare}{\delta \! q_{\|,\rm{e}}}
\newcommand{\dqpareinner}{\delta \! q_{\|,\rm{e,inner}}}
\newcommand{\dqpareouter}{\delta \! q_{\|,\rm{e,outer}}}
\newcommand{\dqpareeff}{\overline{\delta \! q}_{\|}}
\newcommand{\dqpareeffinner}{\overline{\delta \! q}_{\|,{\rm inner}}}
\newcommand{\dqpareeffboot}{\overline{\delta \! q}_{\rm B}}
\newcommand{\dqparns}[2]{\delta \! q^{(#1)}_{\|,{\rm{#2}}}}

\newcommand{\dQparns}[2]{\delta \! Q^{(#1)}_{\|,{\rm{#2}}}}

\newcommand{\dupare}{\delta \! u_{\|,\rm{e}}}
\newcommand{\dupareinner}{\delta \! u_{\|,\rm{e,inner}}}
\newcommand{\dupareouter}{\delta \! u_{\|,\rm{e,outer}}}
\newcommand{\dupareeff}{\overline{\delta \! u}_{\|}}
\newcommand{\dupareeffinner}{\overline{\delta \! u}_{\|,{\rm inner}}}
\newcommand{\dupareeffboot}{\overline{\delta \! u}_{\rm B}}
\newcommand{\duparns}[2]{\delta \! u^{(#1)}_{\|,{\rm{#2}}}}

\newcommand{\dUparns}[2]{\delta \! U^{(#1)}_{\|,{\rm{#2}}}}

\newcommand{\duveci}{\delta \! \mb{u}_{\rm{i}}}
\newcommand{\duvecns}[2]{\delta \! \mb{u}^{(#1)}_{{\rm{#2}}}}

\newcommand{\dpartc}{\overline{\delta\!\Gamma}_{\rm C}}
\newcommand{\dpartnc}{\overline{\delta\!\Gamma}_{\rm N}}
\newcommand{\dheatc}{\overline{\delta\! q}_{\rm C}}
\newcommand{\dheatnc}{\overline{\delta\! q}_{\rm N}}

\newcommand{\temps}{T_\spe}
\newcommand{\tempi}{T_{\rm{i}}}
\newcommand{\tempe}{T_{\rm{e}}}
\newcommand{\dtempe}{\delta \! T_{\rm{e}}}

\newcommand{\vtheri}{v_{\rm{th,i}}}
\newcommand{\vthere}{v_{\rm{th,e}}}
\newcommand{\vthers}{v_{\rm{th},\spe}}

\newcommand{\zeds}{Z_\spe}
\newcommand{\zedi}{Z_{\rm{i}}}
\newcommand{\charge}{e}

\newcommand{\tdrv}[2]{\frac{d #1}{d #2}}
\newcommand{\tdrvt}[2]{\fract{d #1}{d #2}}
\newcommand{\drv}[2]{\frac{\partial #1}{\partial #2}}
\newcommand{\drvt}[2]{{\partial #1}/{\partial #2}}
\newcommand{\drvsq}[2]{\frac{\partial^2 #1}{\partial #2^2}}
\newcommand{\drvsqt}[2]{{\partial^2 #1}/{\partial #2^2}}

\newcommand{\ma}{m_\spe}
\newcommand{\me}{m_{{\rm{e}}}}
\newcommand{\mi}{m_{{\rm{i}}}}
\newcommand{\md}{m_{{\rm D}}}
\newcommand{\meomi}{\left(\frac{m_{\rm{e}}}{m_{\rm{i}}}\right)}
\newcommand{\meominb}{\frac{m_{\rm{e}}}{m_{\rm{i}}}}
\newcommand{\massrt}{\left(m_{\rm{e}}/m_{\rm{i}}\right)^{1/2}}
\newcommand{\massrtn}{\left(m_{\rm{e}}/m_{\rm{i}}\right)^{n/2}}
\newcommand{\massrst}{\left(m_{\rm{e}}/m_{\rm{i}}\right)^{1/4}}
\newcommand{\massr}{\meomi^{1/2}}
\newcommand{\massru}{\left(\frac{m_{\rm{i}}}{m_{\rm{e}}}\right)^{1/2}}
\newcommand{\massrut}{\left({m_{\rm{i}}}/{m_{\rm{e}}}\right)^{1/2}}
\newcommand{\massrtp}[1]{\left({m_{\rm{e}}}/{m_{\rm{i}}}\right)^{#1}}
\newcommand{\massrutp}[1]{\left({m_{\rm{i}}}/{m_{\rm{e}}}\right)^{#1}}
\newcommand{\massrp}[1]{\left(\frac{m_{\rm{e}}}{m_{\rm{i}}}\right)^{#1}}
\newcommand{\massrup}[1]{\left(\frac{m_{\rm{i}}}{m_{\rm{e}}}\right)^{#1}}

\newcommand{\massrl}{\meominb}
\newcommand{\massrlt}{\me/\mi}

\newcommand{\vms}{\mb{v}_{{\rm{M}},{\spe}}}
\newcommand{\vme}{\mb{v}_{\rm{M,e}}}
\newcommand{\vmi}{\mb{v}_{\rm{M,i}}}

\newcommand{\cycfs}{\Omega_\spe}
\newcommand{\cycfe}{\Omega_{\rm{e}}}
\newcommand{\cycfez}{\Omega^0_{\rm{e}}}
\newcommand{\cycfeovl}{\ovl{\Omega}_{\rm{e}}}
\newcommand{\cycfi}{\Omega_{\rm{i}}}

\newcommand{\qfluxs}{Q_{\rm \spe}}
\newcommand{\qfluxgb}{Q_{\rm gb,\spe}}
\newcommand{\qfluxgbi}{Q_{\rm gb,i}}
\newcommand{\qfluxgbe}{Q_{\rm gb,e}}

\newcommand{\lparp}{\lpar^{+}_b}
\newcommand{\lparm}{\lpar^{-}_b}
\newcommand{\lparpm}{\lpar^{\pm}_b}
\newcommand{\lparprim}{\lpar^{\prime}}

\newcommand{\coplandaus}{C_{\spe\spe}}
\newcommand{\coplandaue}{C_{\rm{ee}}}
\newcommand{\coplorentze}{\mathcal{L}}
\newcommand{\sigvec}{\mb{\sigma}}
\newcommand{\tauvec}{\mb{\tau}}
\newcommand{\copbothe}{\mathcal{C}}
\newcommand{\coplandaui}{C_{\rm{ii}}}
\newcommand{\matrixlandau}[1]{C_{#1}}
\newcommand{\matrixlorentz}[1]{\mathcal{L}_{#1}}
\newcommand{\boldmatrixlandau}{\mb{C}}
\newcommand{\boldmatrixlorentz}{\mb{\mathcal{L}}}

\newcommand{\cops}{C^{\rm{GK}}_{\spe}}
\newcommand{\copsh}{\widehat{C}_{\spe}}

\newcommand{\copehz}{\widehat{C}^{(0)}_{\rm{e}}}

\newcommand{\coloumblog}{\ln \Lambda}
\newcommand{\cope}{C^{\rm{GK}}_{\rm{e}}}
\newcommand{\copi}{C^{\rm{GK}}_{\rm{i}}}

\newcommand{\pitchscatter}{\mathcal{C}_{\pitch \pitch}}

\newcommand{\cfreqii}{\nu_{\rm{ii}}}
\newcommand{\cfreqiiperp}{\nu_{\perp, {\rm i}}}
\newcommand{\cfreqiipar}{\nu_{\|, {\rm i }}}
\newcommand{\cfreqei}{\nu_{\rm{ei}}}
\newcommand{\cfreqee}{\nu_{\rm{ee}}}
\newcommand{\nustar}{\nu_{\ast}}
\newcommand{\Phifunca}{\bm{\Uppsi}}
\newcommand{\Phifuncb}{\bm{\Phi}}

\newcommand{\neophases}{\frac{\bcur \vpar}{\cycfs}}

\newcommand{\wmags}{\omega_{{\rm{M}},\spe}}
\newcommand{\wmage}{\omega_{\rm{M,e}}}
\newcommand{\wmageth}{\omega^{\rm{th}}_{\rm{M,e}}}
\newcommand{\wmageprecth}{\omega_{D}}

\newcommand{\aindex}{p}
\newcommand{\aindexmax}{N}
\newcommand{\mindex}{j}
\newcommand{\bindex}{q}
\newcommand{\acoeffn}[1]{a_{#1}}
\newcommand{\bcoeffn}[1]{c_{#1}}
\newcommand{\dmatrix}[1]{D_{#1}}
\newcommand{\dbmatrix}{\bm{D}}
\newcommand{\qbmatrix}{\bm{Q}}
\newcommand{\qmatrix}[1]{Q_{#1}}
\newcommand{\nueff}{\nu_{\rm e}}
\newcommand{\nueffhat}{\hat{\nu}}
\newcommand{\Psifunc}{\Psi}
\newcommand{\soninen}[1]{L^{3/2}_{#1}}
\newcommand{\soninemn}[2]{L^{#1}_{#2}}
\newcommand{\sarg}{\hat{x}}
\newcommand{\gamfn}{\Gamma}
\newcommand{\genfun}{G}
\newcommand{\zarg}{z}
\newcommand{\darg}{s}
\newcommand{\erf}{{\rm erf}}
\newcommand{\gflux}{\Gamma}
\newcommand{\gfluxperp}{\Gamma_{\perp}}
\newcommand{\gfluxsh}{\Gamma_{\rm SH}}
\newcommand{\gfluxd}{\Gamma_{\rm B}}

\newcommand{\ptlouter}{\ptlsub{{\scriptsize\rm{outer}}}}
\newcommand{\ptlzouter}{\ptlsub{{\scriptsize\rm{outer}}}^{(0)}}
\newcommand{\ptlinner}{\ptlsub{{\scriptsize\rm{inner}}}}
\newcommand{\ptlzinner}{\ptlsub{{\scriptsize\rm{inner}}}^{(0)}}
\newcommand{\HHeinner}{\HHesub{{\scriptsize\rm{inner}}}}
\newcommand{\HHezinner}{\HHesub{{\scriptsize\rm{inner}}}^{(0)}}
\newcommand{\HHeouter}{\HHesub{{\scriptsize\rm{outer}}}}
\newcommand{\HHezouter}{\HHesub{{\scriptsize\rm{outer}}}^{(0)}}
\newcommand{\HHeoouter}{\HHesub{{\scriptsize\rm{outer}}}^{(1)}}
\newcommand{\HHezoutertrapped}{\HHesub{{\scriptsize\rm{outer-trapped}}}^{(0)}}
\newcommand{\hhizouter}{\hhisub{{\scriptsize\rm{outer}}}^{(0)}}
\newcommand{\hhizinner}{\hhisub{{\scriptsize\rm{inner}}}^{(0)}}

\newcommand{\nustarcollisional}{ 1.22 }
\newcommand{\nustarcollisionless}{4.70 \times 10^{-3} }
\newcommand{\nustarcollisionlessitg}{5.40 \times 10^{-5} }
\newcommand{\nustarcollisionalitg}{1.22}

\newcommand{\cfreqeecollisionless}{ 5.21 \times 10^{-3} }
\newcommand{\cfreqeecollisionlessitg}{ 5.98 \times 10^{-5} }

\newcommand{\jintref}{j_{\|}^{{\rm{ref}}}}
\newcommand{\phiref}{\phi_{}^{{\rm{ref}}}}
\newcommand{\tempref}{T_{}^{{\rm{ref}}}}
\newcommand{\bmagref}{B_{{\rm{ref}}}}

\newcommand{\trapfrac}{f_{\rm trap}}
\newcommand{\ellipticE}{E}

\newcommand{\solvcnstn}{\mathcal{K}_n}
\newcommand{\solvcnstt}{\mathcal{K}_T}

\newcommand{\source}{S}
\newcommand{\phase}{P}
\newcommand{\dbprime}{\prime\prime}
\newcommand{\intzero}{I_0}
\newcommand{\intdelta}{I_\delta}

\newcommand{\delt}{\Delta \ttime}
\newcommand{\nref}{n^{\scriptsize{\rm{ref}}}}
\newcommand{\bref}{B_{\scriptsize{\rm{ref}}}}

\newcommand{\shat}{\hat{s}}
\newcommand{\radial}{r}
\newcommand{\radialx}{x}
\newcommand{\binormal}{y}
\newcommand{\kky}{k_y}
\newcommand{\kkx}{k_x}
\newcommand{\kkr}{k_r}
\newcommand {\thetaz} {\theta_0}
 \newcommand{\negrid}{n_\varepsilon}
\newcommand{\ntwopi}{n_{2\pi}}
\newcommand{\ntheta}{n_\theta}
\newcommand{\npitch}{n_\lambda}
\newcommand{\rminor}{r}
\newcommand{\rmajor}{R}
\newcommand{\rmajgeo}{R_{\rm geo}}
\newcommand{\rmajorz}{R_0}
\newcommand{\rmajormax}{R_{\rm max}}
\newcommand{\rmajormin}{R_{\rm min}}
\newcommand{\aspect}{\epsilon}
\newcommand{\dpsidx}{d \flxl / d x}
\newcommand{\daldy}{d \fldl / d \binormal}
\newcommand{\kkxnorm}{\rminor \bcur/ \saffac \rmajorz}
\newcommand{\kkynorm}{(\bcur/\rmajorz)\tdrvt{\rminor}{\flxl}}
\newcommand{\kxfac}{\hat{\kappa}}

\newcommand{\growth}{\gamma}
\newcommand{\growthz}{\gamma^{(0)}}
\newcommand{\growthh}{\gamma^{(1/2)}}
\newcommand{\growtho}{\gamma^{(1)}}

\newcommand{\lti}{L_{\tempi}}
\newcommand{\lte}{L_{\tempe}}
\newcommand{\lts}{L_{\temps}}

\newcommand{\rstars}{\rho_{\ast\spe}}
\newcommand{\lln}{L_n}

\newcommand{\lflrs}{\lambda_{\spe}}
\newcommand{\lflri}{\lambda_{{\rm{i}}}}
\newcommand{\lflre}{\lambda_{{\rm{e}}}}
\newcommand{\lflrez}{\lambda^0_{{\rm e }}}
\newcommand{\lflrezth}{\lambda^0_{{\rm{th,e}}}}
\newcommand{\bflrs}{b_{\spe}}
\newcommand{\bflri}{b_{{\rm{i}}}}
\newcommand{\bflre}{b_{{\rm{e}}}}
\newcommand{\wstars}{\omega_{\ast,\spe}}
\newcommand{\wstar}{\omega_{\ast}}
\newcommand{\wstarsn}{\omega^{\dens}_{\ast,\spe}}
\newcommand{\wstaren}{\omega^{\dens}_{\ast,{\rm{e}}}}
\newcommand{\wstare}{\omega_{\ast,{\rm{e}}}}
\newcommand{\wstari}{\omega_{\ast,{\rm{i}}}}
\newcommand{\etas}{\eta_\spe}
\newcommand{\etae}{\eta_{\rm{e}}}

\begin{document}

\title[Extended electron tails in electrostatic microinstabilities]{Extended electron tails in electrostatic microinstabilities and the nonadiabatic response of passing electrons}
\author{M. R. Hardman$^{1}$, F. I. Parra$^{1}$,  C. Chong$^2$, T. Adkins$^{1}$, 
 M. S. Anastopoulos-Tzanis$^{3}$, M. Barnes$^{1}$, D. Dickinson$^{3}$, \\  J. F. Parisi$^{4,1}$, and H. Wilson$^{3}$} 
\address{$^1$ Rudolf Peierls Centre for Theoretical Physics, University of Oxford, \\ Oxford, OX1 3PU, UK}
\address{$^2$ Mathematical  Institute,  University  of  Oxford,  Andrew  Wiles  Building, \\ 
Radcliffe  Observatory  Quarter,  Woodstock  Road,  Oxford,  OX2  6GG,  UK }
\address{$^3$ York Plasma Institute, Department of Physics, University of York, Heslington, York, YO10 5DD, UK}
\address{ $^4$ Culham Centre for Fusion Energy, UKAEA, Abingdon OX14 3DB, UK}
\ead{\texttt{michael.hardman@physics.ox.ac.uk}} 

\begin{abstract}

 Ion-gyroradius-scale microinstabilities typically have a
 frequency comparable to the ion transit frequency.
 Due to the small electron-to-ion mass ratio and the large electron
 transit frequency, it is conventionally assumed
 that passing electrons respond adiabatically in
 ion-gyroradius-scale modes. However, in gyrokinetic simulations
 of ion-gyroradius-scale modes in axisymmetric toroidal magnetic fields, the nonadiabatic response of
 passing electrons can drive the mode, and 
 generate fluctuations with narrow radial layers,
 which may have consequences for turbulent transport in a variety of circumstances.
 In flux tube simulations, in the ballooning representation,
 these instabilities reveal themselves as modes with extended tails. 
 The small electron-to-ion mass ratio limit of linear gyrokinetics
 for electrostatic instabilities is presented, in axisymmetric toroidal magnetic geometry, including the nonadiabatic response of passing
 electrons and associated narrow radial layers.
  This theory reveals the existence of ion-gyroradius-scale modes driven solely by the nonadiabatic passing electron response, 
  and recovers the usual ion-gyroradius-scale modes driven by the response of ions and trapped electrons,
 where the nonadiabatic response of passing electrons is small. 
 The collisionless and collisional limits of the theory are considered, demonstrating 
 parallels in structure and physical processes to neoclassical transport theory. 
 By examining initial-value simulations of fastest-growing eigenmodes,
 the predictions for mass-ratio scaling are tested and verified numerically for a range 
 of collision frequencies. 
 Insights from the small electron-to-ion mass ratio theory may lead to
 a computationally efficient treatment of extended modes.
 
\end{abstract}

\submitto{ \PPCF }
%\maketitle
{\let\newpage\relax\maketitle}
    \section{Introduction}

The leading magnetic confinement fusion experiments achieve single particle
 confinement by exploiting strong magnetic fields
 that have nested toroidal flux surfaces:
 the Lorentz force prevents
 particles crossing the magnetic field in the perpendicular direction,
 but particles are free to stream along magnetic field lines.
 Despite this, there are still particle and heat losses from the confined plasma.
 Neoclassical transport is driven by interparticle Coulomb collisions in
 toroidal magnetic geometry, and turbulent transport is driven by the free energy
 available in the equilibrium temperature and density gradients.
 
 Turbulence forms through the nonlinear saturation of microinstabilities.
 The most important microinstabilities for transport
 have frequencies $\wfreq$ comparable to the transit
 frequency of the constituent particle species, and perpendicular wavenumbers $\kperp$ 
 comparable to the inverse thermal gyroradius of the particles,
 i.e., $\wfreq \sim \vthers / \lscal \sim \rstars \cycfs \ll \cycfs$, and $\kperp \gyrds \sim 1$,
 where $\vthers = \sqrt{2 \temps/\ma}$ is the
 thermal speed of the component species $\spe$, $\lscal$ is a typical equilibrium length scale,
 $\cycfs = \zeds \charge \bmag /\ma \ltsp $
 is the cyclotron frequency of the component species $\spe$, and
 $\rstars = \gyrds / \lscal \ll 1$, with
 $\gyrds = \vthers/\cycfs$ the thermal gyroradius of the species $\spe$.
 In the above definitions, $\temps$ is the species temperature,
 $\ma$ is the species mass, $\zeds$ is the species charge number, $\charge$
 is the proton charge, $\bmag$ is the magnetic field strength, and $\ltsp$ is the speed of light.
 These microinstabilities 
 are extended along magnetic field lines,
 with parallel wave numbers such that $\kpara \saffac \rmajor \sim 1$,
 where $\saffac \rmajor$ is the connection length, $\saffac \sim 1$ 
 is the safety factor and $\rmajor \sim \lscal$ is the major radius. 
 A diffusive random walk estimate for the heat flux $\qfluxs$
 driven by instabilities at the scale $\gyrds$ 
 gives $\qfluxs \sim \qfluxgb = \rstars^2\denss\temps \vthers $,
 with $\denss$ the equilibrium plasma density of species $\spe$.
 To obtain this estimate, we use that the macroscopic profiles have a scale of order $\lscal$,
 and that turbulent eddies transport heat by a step length $\gyrds$ in a timescale $\vthers/\lscal$.
 
 A plasma has multiple particle species: the simplest plasma consists of ions, with charge
 $\zedi \charge$ and mass $\mi$, and electrons, with charge $-\charge$ and mass $\me$. 
 In a fusion plasma with deuterium ions, the separation between the ion and electron masses
 has significant consequences for the nature of the turbulence and the underlying instabilities.
 Since $\sqrt{\mi/\me} \approx 60$, we have that $\gyrdi \gg \gyrde$ and $\vtheri \ll \vthere$,
 i.e., instabilities can be driven over a wide range of space and time scales.
 Historically, research has largely focused on transport and instabilities 
 driven at the larger scale of the ion gyroradius.
 This is for the simple reason that the heat flux estimate $\qfluxgbi$ for $\gyrdi$-scale
 turbulence dominates the heat flux estimate $\qfluxgbe$ for $\gyrde$-scale turbulence by $\massrut \gg 1$.
 However, it is important not to discount the $\gyrde$ scales for several reasons.
 It is known that $\gyrde$-scale turbulence can drive experimentally relevant 
 heat fluxes that exceed
 the $\qfluxgbe$ estimate by a large order-unity factor
 \cite {dorland2000electron,jenko2000electron,jenko2002prediction,STroach2009PPCF}. 
 Recently, expensive direct numerical simulations (DNS) with realistic electron-to-hydrogen-ion-mass ratio \cite{maeyama2015cross,Maeyama_2017_NF}
 and realistic electron-to-deuterium-ion-mass ratio
 \cite {howard2014synergistic,howard2016enhanced,howard2016comparison,maeyama2017supression,Maeyama_2017_NF,Bonanomi_2018_ImpactofES}
 have demonstrated the existence and significance of cross-scale interactions between turbulence at
 the scales of $\gyrdi$ and $\gyrde$.
 Finally, as we will demonstrate in this paper, even familiar 
 long-wavelength modes with binormal wave numbers $\kky \gyrdi \sim 1$ may have narrow radial
 structures near mode-rational surfaces that satisfy $\kkr \gyrde \sim 1$, with $\kkr$ the radial wave number.
 These structures result from the dynamics of passing electrons \cite{HallatschekgiantelPRL2005,DominskinonadPOP015},
 and may be important for understanding cross-scale interactions in multiscale DNS,
 cf. \cite{Maeyama_2017_NF}.
 We will see that there are
 novel $\kky \gyrdi \sim 1$ modes driven by the electron response to the
 electron temperature gradient (ETG) in the $\kkr \gyrde \sim 1$ narrow layer,
 and we will see that even the familiar ion temperature gradient
 (ITG) mode can exhibit $\kkr \gyrde \sim 1 $ features.
 
 Anisotropy between the radial wave number $\kkr $ and the binormal wave number
 $\kky$ arises naturally in linear modes in toroidal magnetic fields because of
 the presence of magnetic shear $\shat$. In the presence of magnetic shear, 
 linear modes
 are conveniently described in terms of \enquote{ballooning} modes that
 follow the magnetic field line many times around the torus \cite{ConnorProRSoc1979ballooning}.
 Ballooning modes have wave fronts that rotate with position along the magnetic field line.
 As we shall describe with more precision later, in a ballooning mode
 the radial wave number $\kkr$ satisfies $\kkr\propto - \kky \shat \lpar$ for large $\lpar$,
 where $\lpar$ is the extended poloidal angle that is used to describe
 the position along the field line as it winds around the torus. 
 Therefore, it is possible for $\kky\gyrdi \sim 1$ modes to have extended \enquote{ballooning tails} at $\lpar \gg 1$
 that correspond to $\kkr \gyrdi \gg 1$ components. 
 In the real-space picture, modes with extended ballooning tails are modes
 with significant amplitude in a layer around mode-rational flux surfaces
 -- flux surfaces where the field line winds onto itself after an integer number
 of toroidal and poloidal turns. With this in mind, we can understand the origin
 of electron-driven ballooning tails with a simple physical argument.
 On irrational flux surfaces, where a single field line covers the flux surface,
 rapidly moving passing electrons can sample the entire flux surface and respond adiabatically.
 However, on mode-rational flux surfaces, passing electrons can only sample a subset of
 the flux surface and hence have a nonadiabatic response.  
 
 Linear modes with extended ballooning
 tails have been observed in simulations with a variety of equilibrium conditions, for example,
 in simulations of electrostatic modes in core tokamak conditions
 \cite{HallatschekgiantelPRL2005,DominskinonadPOP015} and in the pedestal \cite{Parisi_2020},
 as well as in electromagnetic simulations of 
 linear micro-tearing modes in spherical and conventional tokamaks 
 \cite{Applegate_2007,Dickinson_MTMPed_2013,Moradi_2013}.
  Although simulations of linear modes are inexpensive compared to nonlinear simulations of turbulence,
 simulations of modes with extended ballooning tails can be remarkably costly.
 In implicit codes, the computational expense
 arises from the need to resolve variation of geometric quantities
 on the scales of $2\pi$ in ballooning angle,
 combined with the need to simulate very large ballooning angles with scales $\lpar \gg 2 \pi$:
 this leads to an expensive matrix problem. In explicit codes,
 in addition to the size of the problem in $\lpar$,
 small time steps are required to resolve kinetic electron physics
 in modes with frequencies comparable to the ion transit frequency.
 The results presented in this paper are intended as a
 step towards efficient reduced models of extended electron-driven modes.

 In this paper we obtain an asymptotic theory, valid in the limit of $\massrt \rightarrow 0$,
 for electrostatic modes that exist at the long wavelengths
  of the ion gyroradius scale, i.e., $\kky \gyrdi \sim 1$. 
  Reduced models of these modes must provide a reduced treatment of the electron response.
  In the simplest case, for example, the classical ITG
  mode calculation \cite{cowleyPoFB91,RomanelliITG}, the electron response is taken to be adiabatic. 
  More advanced calculations retain the bounce-averaged response
  due to trapped electrons, needed to capture trapped-electron modes (TEMs),
  see, e.g., \cite{Adam1976TEM,beerbouncePoP1996}.
  The nonadiabatic response of 
  passing electrons is traditionally neglected, despite evidence from
  DNS that indicates that the nonadiabatic
  passing electron response can play a significant role in electrostatic modes:
  pioneering work showed that the passing electron response can alter transport in linear modes
  and fully nonlinear turbulence \cite{HallatschekgiantelPRL2005}. This observation has subsequently
  been reinforced by a variety of investigations, see, for example
  \cite{DominskinonadPOP015,Maeyama_2017_NF,BelliIsotope2019,BelliIsotope2020PRL,ajay_brunner_mcmillan_ball_dominski_merlo_2020,ball_brunner_ajay_2020,ajay_brunner_ball_2020}.
  In this paper, we will show that, in the $\massrt \rightarrow 0$ limit, there
  are in fact two classes of modes existing at $\kky\gyrdi \sim 1$: first, the familiar
  ion or trapped-electron response-driven modes (e.g., the ITG mode or the TEM)
  that rely on a potential localised at $\lpar \sim 1$; and second, ETG
  modes that are driven by the passing electron response
  in the large $\lpar$ tail of the ballooning mode. We find the large $\lpar$ equations
  that govern the electron response in the tail of the ballooning mode, 
  and we provide the matching conditions necessary to connect them to the $\lpar \sim 1$ region. We use
  simulations performed with the gyrokinetic code \gstwo~\cite{KOTSCHENREUTHER1995CPC} to show
  that the orderings used to derive these equations are satisfied by numerical
  examples of both classes of mode. 
  
  The clearest physical ordering for the novel passing-electron-response-driven modes is
  $\kky\gyrde \sim \saffac\rmajor\wfreq/\vthere \ll 1$, i.e., they are
  ETG modes at long wavelengths, that feature a radial layer with $\kkr\gyrde \sim 1$,
  with an asymptotic separation between the transit frequency $\vthere/\saffac\rmajor$
  and the frequency of the drive
  $\wstar \sim \wfreq$.
  We can treat the familiar ITG modes and TEMs in the same formalism as the novel passing-electron-driven modes 
  by simply making the maximal ordering $\kky\gyrde \sim \massrt \ll 1$. 
  Mathematically, the two classes of mode are
  distinguished in the formalism by the matching condition for the passing electrons at $\lpar\sim 1$.
    This is true in both the \enquote{collisionless} limit 
  \beqn \frac{\vthere}{\saffac\rmajor} \gg \wfreq \sim \wstar \sim \cfreqee \sim \cfreqei,
  \label{order:nustar-collisionless} \eeqn 
  where $\cfreqee$ and $\cfreqei$ are the electron self-collision
    and the electron-ion collision frequencies, respectively, and in the \enquote{collisional} limit
    where 
    \beqn \frac{\vthere}{\saffac\rmajor} \sim \cfreqee \sim \cfreqei
    \gg \wfreq \sim \wstar . \label{order:nustar-collisional}\eeqn
  
  The remainder of this paper is structured as follows.
  In section \ref{sec:linearelectrostaticgyrokinetics}, we briefly review the electrostatic
  gyrokinetic model that is the starting point for this work.
  Those familiar with gyrokinetics may skip to
  section \ref{section:gyrokineticsrepackage}, where we identify
  a convenient form of the gyrokinetic equation that we use to describe
  electron dynamics. We obtain the asymptotic theory of
  collisionless modes in section \ref{section:modes-in-memi-zero-limit},
  and we obtain the asymptotic  theory of collisional modes in
  section \ref{section:modes-in-collisional-memi-zero-limit}. We compare the results
  of sections \ref{section:modes-in-memi-zero-limit} and
  \ref{section:modes-in-collisional-memi-zero-limit} to numerical simulations
  in section \ref {section:numericalresults}. Finally, in section \ref{section:discussion}, we discuss
  the implications of these results and possible extensions of the theory.
 Included in this paper are appendices with results pertaining 
 to the plasma response at large $\lpar$. 
 First, in \ref{sec:detailed-analysis-of-the-ion-response},
 we give a detailed analysis of the ion nonadiabatic response at large $\lpar$.
 In \ref{sec:electron-inner-collisional-appendix} we obtain the equations governing 
 the electron response in the collisional limit.
  In \ref {sec:parallel-collisional-contributions},
 we solve the Spitzer problem necessary to obtain the neoclassical 
 parallel and perpendicular flux contributions
 to the electron mode equations.
  In \ref {sec:perpendicular-collisional-contributions},
 we obtain the classical perpendicular
 flux contributions to the electron mode equations.
 In \ref{sec:pfirsh-schluter-parallel-collisional-contributions}, we obtain the 
 parallel and perpendicular fluxes for the electron mode equations
 in the highly collisional (Pfirsh-Schl\"{u}ter) limit.
 In \ref {sec:banana-parallel-collisional-contributions}, we obtain
 the parallel and perpendicular fluxes for the electron mode equations
 in the banana regime of collisionality in a small inverse aspect ratio device.
 Finally, in \ref{sec:electron-outer-collisional-small-tail-appendix} we obtain the electron matching 
 conditions in the collisional limit.

  \section{Electrostatic gyrokinetic equations}\label{sec:linearelectrostaticgyrokinetics}
In this section, we briefly review the linear, electrostatic, $\dlf$ gyrokinetic
 model \cite{cattoPP78} that is the starting point for the analysis in this paper.
In gyrokinetic theory, the microinstability mode frequency $\wfreq$ is taken to be much smaller than
 the cyclotron frequency $\cycfs$
 at which particles gyrate around the magnetic field direction $\bu = \bvec / \bmag$,
 where $\bvec$ is the magnetic field.
 The mode frequency $\wfreq$ is taken to be of order the transit frequency
 $\vthers/\lscal$. 
 The spatial scale of the fluctuations perpendicular to the magnetic field line
 is of order the thermal gyroradius
 $\gyrds = \vthers/\cycfs$, and the fundamental gyrokinetic
 expansion parameter is $\rstars = \gyrds/\lscal$.
 In $\dlf$ gyrokinetics, the fluctuating distribution function $\dlfs$
 for each species $\spe$ is a sum
of the nonadiabatic response $\hhs$, and the adiabatic response $- \zeds \charge \ptl \eqlba/\temps$, i.e.,
\beqn \dlfs(\rr,\pvel,\ttime) = \hhs (\gc,\energy,\pitch,\ttime)
 - \frac{\zeds \charge \ptl(\rr,\ttime)}{\temps}  \eqlba \label{eq:dlf}, \eeqn
where $\ptl$ is the fluctuating electrostatic potential, $\eqlba$ is the equilibrium Maxwellian distribution,
 $\rr$ is the particle position, $\pvel$ is the particle velocity,
 and we have indicated that $\hhs$
 is a function of guiding centre position $\gc = \rr - \gyrdvecs$ (with $\gyrdvecs =\bu \xp \pvel / \cycfs$), energy $\energy = \ma \vmag^2 /2$ (with $\vmag = |\pvel|$),
 and pitch angle $\pitch = \vperp^2/\vmag^2 \bmag$ (with $\vperp = |\pvel - \bu \bu \cdot \pvel|$), whereas $\ptl$ is a function of $\rr$ but not of $\pvel$.
 In this paper we consider linear theory, and so we make the eikonal ansatz 
 $ \hhs(\gc,\ttime) = \sum_{\kperpvec} \hhssub{\kperpvec}\expo{\imag (\kperpvec \cdot \gc - \wfreq \ttime)}$, and 
 $\ptl(\rr,\ttime) = \sum_{\kperpvec} \ptlsub{\kperpvec}\expo{\imag (\kperpvec \cdot \rr - \wfreq \ttime)}$,
 where $\kperpvec$ is the perpendicular-to-the-field wave vector.
 Henceforth, we drop the $\kperpvec$ subscripts on the Fourier coefficients.

 \subsection{The gyrokinetic equation and quasineutrality}
 The linear, electrostatic gyrokinetic equation is
\beqn \fl  \vpar \kpar \drv{\hhs{}}{\lpar}
 + \imag (\kperpvec \cdot \vms - \wfreq) \hhs{} - \cops[\hhs{}] 
 = \imag \left( \wstars - \wfreq \right)\bessn{0}\eqlba\frac{\zeds \charge\ptl{}}{\temps}
 , \label{eq:lineargyrokinetic}\eeqn 
  where $\vpar = \bu \cdot \pvel$, $\lpar$ is the poloidal angle coordinate that measures distance along the magnetic field line,
  $\vms =  (\bu/\cycfs) \xp \left( \vpar^2 \bu \cdot \nbl \bu + \fract{\vperp^2 \nbl\bmag}{2 \bmag}\right)$
  is the magnetic drift, 
  and the finite Larmor radius effects are modelled by the 0\textsuperscript{th} Bessel function of the first kind
  $\bessn{0}= \bes{\bflrs}$, with $\bflrs = \kperp \vperp / \cycfs$, and $\kperp = |\kperpvec|$.
  Note that $\cycfi = \zedi \charge \bmag/\mi\ltsp > 0$, whereas $\cycfe = - \charge \bmag/\me\ltsp <0$. 
  The frequency $\wstars$ contains the equilibrium drives of instability:
 $\wstars/ \wstarsn = 1 + \etas\left(\fract{\energy}{\temps}- \fract{3}{2}\right)$,
  with 
  $ \wstarsn =  -(\fract{\ltsp \kkfldl \temps}{ \zeds \charge })\tdrvt{\ln \denss}{\flxl}$, 
  and $\denss$ the equilibrium number density of species $\spe$, $\fldl$ the dimensionless binormal coordinate, $\kkfldl$ the binormal wavenumber with respect to $\fldl$, 
  $\flxl$ the poloidal magnetic flux ($\fldl$ and $\flxl$ are defined in section \ref{sec:magnetic-coordinates}),
  and 
  $ \etas = \tdrvt{\ln \temps}{\ln \denss} $.
  Finally, the collision operator $\cops[\cdot]$ is shorthand for the linearised gyrokinetic
  collision operator of the species $\spe$.

  For ions, the linearised  gyrokinetic collision operator is defined by
   \beqn \copi\left[\hhi\right] = 
   \gav{\expo{\imag \kperpvec\! \cdot\! \gyrdveci}\coplandaui\left[\expo{-\imag \kperpvec\! \cdot \!\gyrdveci}\hhi \right]}{}
   \label{eq:copgk-ion}, \eeqn
    with $\coplandaui[\cdot]$ the linearised self-collision operator of the ion species,
 and $\gav{\cdot}{}$
 the gyrophase average at fixed
 $\energy$ and $\pitch$. 
   The self-collision operator of the species $\spe$, $\coplandaus[\cdot]$, is defined by 
   \beqn \fl \coplandaus\left[ \fulldistf \right] = \frac{2 \pi \zeds^4 \charge^4 \coloumblog}{\ma^2}
   \drv{}{\pvel} \cdot \intvprim{\eqlba\eqlbaprim
    \Ufunc(\pvel - \pvelprim) \cdot \left(\drv{}{\pvel}\left(\frac{\fulldistf}{\eqlba}\right)-
    \drv{}{\pvelprim}\left(\frac{\fulldistfprim}{\eqlbaprim}\right)\right)} , \label{eq:coplandaus} \eeqn
 where $\fulldistf$ is a distribution function, 
    and we have used the shorthand notation
    $\fulldistf = \fulldistf(\pvel)$, $\fulldistfprim = \fulldistf(\pvelprim)$,
    $\eqlba = \eqlba(\pvel)$, $\eqlbaprim = \eqlba(\pvelprim)$, and 
    \beqn\Ufunc(\pvel - \pvelprim) =
    \frac{\eye |\pvel - \pvelprim|^2 - (\pvel - \pvelprim)(\pvel - \pvelprim)}{|\pvel - \pvelprim|^3},
    \label{eq:ufuncdef}
    \eeqn
    with $\eye$ the identity matrix.
    We note that the Coloumb logarithm $\coloumblog \approx 17$ \cite{HazeltineMeiss}.
 We define 
the ion self-collision frequency 
   $ \cfreqii =  \fract{4 \sqrt{\pi}\zedi^4 \densi\charge^4 \coloumblog}{3\mi^{1/2} \tempi^{3/2}}$
    and the electron self-collision frequency
    $ \cfreqee = \fract{4 \sqrt{2\pi}\dense\charge^4 \coloumblog}{3\me^{1/2} \tempe^{3/2}}$ 
    following Braginskii \cite{BraginskiiTransport}, noting the factor of $\sqrt{2}$
    difference in the definitions of $\cfreqee$ and $\cfreqii$.
 
 For electrons, the linearised  gyrokinetic collision operator is defined by
   \beqn \fl \cope\left[\hhe\right] = 
   \gav{\expo{\imag \kperpvec \!\cdot \!\gyrdvece}\coplandaue\left[\expo{-\imag \kperpvec \!\cdot \!\gyrdvece}\hhe \right]}{}
  \nonumber\eeqn \beqn \fl + \gav{\expo{\imag \kperpvec \!\cdot \!\gyrdvece}\coplorentze\left[\expo{-\imag \kperpvec \!\cdot \!\gyrdvece}\hhe - \frac{\me \pvel \cdot \duveci}{\tempe}\eqlbe \right]}{}
    \label{eq:copgk-electron}, \eeqn
     where $\coplandaue[\cdot]$ is the linearised self-collision operator of the electron species,
     defined by equation \refeq{eq:coplandaus}, and 
    \beqn {\coplorentze\left[\fulldistf \right] = \frac{3 \sqrt{\pi} }{8}\cfreqei \vthere^3 
    \drv{}{\pvel} \cdot \left(\frac{\vmag^2 \eye - \pvel \pvel}{\vmag^3}
    \cdot \drv{\fulldistf}{\pvel}\right),} \label{eq:coplorentze}\eeqn 
 is the Lorentz collision operator resulting from electron-ion collisions,
 with the electron-ion collision frequency 
    $\cfreqei = \fract{4 \sqrt{2\pi}\zedi^2 \densi\charge^4 \coloumblog}{3\me^{1/2} \tempe^{3/2}}$
    defined following Braginskii \cite{BraginskiiTransport}. 
  In equation \refeq{eq:copgk-electron},   
  \beqn \duveci = \frac{1}{\densi}\intv{ \left(\besin{0}\vpar \bu
     + \imag \besin{1}\frac{\vperp }{\kperp}\kperpvec\xp \bu\right) \; \hhi} \label{eq:duveci},\eeqn
 where $\bessn{1} = \besfirst{\bflrs}$ is the 1\textsuperscript{st} Bessel function of the first kind.      
  
  For a simple two-species plasma of ions and electrons, quasineutrality
  implies that the equilibrium densities satisfy $\zedi \densi = \dense$. 
  In the electrostatic limit, the system of gyrokinetic equations for the fluctuations
  is closed by the quasineutrality relation.
  The quasineutrality relation has the form 
  \beqn \left(\frac{\zedi \tempe}{\tempi} + 1 \right)\frac{\charge \ptl{}}{\tempe}
  = \frac{\ddensi}{\densi} - \frac{\ddense}{\dense} \label{eq:qn},\eeqn
  where the fluctuating \textit{nonadiabatic} densities $\ddenss$ are defined by
 \beqn \ddenss = \intv{ \bessn{0} \hhs{}} \label{eq:deltans}.\eeqn 
 
  \subsection{Magnetic coordinates and boundary conditions} \label{sec:magnetic-coordinates}
  To describe the plane perpendicular to the magnetic field line, we use the dimensionless binormal 
  field-line-label coordinate $\fldl$, and the flux label $\flxl$,
  defined such that the magnetic field may be written in the Clebsch form 
  \beqn \bvec = \nbl \fldl \xp \nbl \flxl \label{eq:bvec}. \eeqn
  We restrict our attention to axisymmetric magnetic fields of the form
  \beqn \bvec = \bcur \nbl \toragl + \nbl \toragl \xp \nbl \flxl \label{eq:eqilibriumbvec}, \eeqn
  where $\toragl$ is the toroidal angle, and $\bcur(\flxl)$ is the toroidal current function.
  An explicit formula for $\fldl$, in terms of $\flxl$, $\toragl$, and the poloidal angle $\lpar$,
  may be obtained by equating expressions \refeq{eq:bvec} and \refeq{eq:eqilibriumbvec}: 
  \beqn \fldl(\flxl,\toragl,\lpar) = \toragl - \saffac(\flxl) \lpar - \nufunc(\flxl,\lpar), \label{eq:fldl}\eeqn
  with the safety factor \beqn \saffac(\flxl) = \frac{1}{2 \pi} \int^{2 \pi}_0 \frac{\bvec \cdot \nbl \toragl}{\bvec \cdot \nbl \lparprim} d \lparprim, \label{eq:safetyfactor} \eeqn
  and \beqn \nufunc(\flxl,\lpar) = \int^{\lpar}_0 \frac{\bvec \cdot \nbl \toragl}{\bvec \cdot \nbl \lparprim} d \lparprim - \saffac \lpar \label{eq:nufunc}. \eeqn
  Note that $\nufunc(\flxl,2 \pi) = \nufunc(\flxl,0) = 0 $.
  Using the $(\flxl,\fldl)$ coordinates, we write the perpendicular wave vector
  as \beqn \kperpvec = \kkflxl \nbl \flxl + \kkfldl \nbl \fldl, \label{eq:kperpvec} \eeqn
  with the field-aligned radial and binormal wave numbers $\kkflxl$ and $\kkfldl$, respectively.
 
  In the study of linear modes, it is convenient to consider
  the coordinate $\lpar$ as an extended ballooning angle,
  and to replace $\kkflxl$ with $\thetaz = \kkflxl / \saffacprim \kkfldl$, where $\saffacprim = \tdrvt{\saffac}{\flxl}$.
  In this formulation, $\hhsub{\thetaz, \kkfldl}=\hhsub{\thetaz, \kkfldl} ( \lpar)$, with $-\infty < \lpar < \infty$ and the boundary conditions  
 \beqn \hhsub{\thetaz, \kkfldl}(\lpar) = 0 {\rm{~at~}} \lpar \rightarrow -\infty, {\rm{~for~}} \vpar > 0, {\rm~and} \nonumber \eeqn 
 \beqn \hhsub{\thetaz, \kkfldl}(\lpar) = 0 {\rm{~at~}} \lpar \rightarrow \infty,  {\rm{~for~}} \vpar < 0. \label{eq:incomingbc2} \eeqn
  
  Much of the discussion in the following sections of this paper is focused
  on the behaviour of the solution at large $\lpar$. The large $\lpar$ part
  of the mode corresponds to a narrow radial layer in the real-space representation.
  To see this, consider the (contravariant) radial wave number
  \beqn \kkr = \kperpvec \cdot \nbl \radial = (\thetaz - \lpar)\kkfldl \tdrv{\saffac}{\radial} |\nbl \radial|^2
  - \kkfldl(\saffac \nbl \lpar \cdot \nbl \radial + \nbl\nufunc \cdot \nbl\radial), \eeqn
  where $\radial = \radial(\flxl)$ is a minor radial coordinate that is a function of $\flxl$ only, 
  and has dimensions of length.
  For large $|\thetaz - \lpar|$, we find that $\kkr \simeq (\thetaz - \lpar)\kkfldl (\tdrvt{\saffac}{\radial})|\nbl \radial|^2$, i.e.,
  we may obtain narrow radial structures in a ballooning mode by either
  imposing a large $\thetaz =\kkflxl/\saffacprim \kkfldl$, or by following the field line, as a result of magnetic shear. 
  
  It will be interesting to consider the behaviour of the magnetic drift. 
  The term due to the magnetic drift, $\imag \kperpvec \cdot \vms$, may be written in the following convenient form
  \beqn 
  \imag \kperpvec \cdot \vms = 
    \imag \kkfldl  \vms \cdot(\nbl \fldl +  \lpar \nbl \saffac) 
   + \imag \kkfldl \tdrv{\saffac}{\radial} (\thetaz - \lpar) \vms \cdot \nbl \radial.  \label{equation:magneticdrift} \eeqn 
  We note that the quantity $\nbl \fldl +  \lpar \nbl \saffac = \nbl \toragl - \saffac \nbl \lpar - \nbl \nufunc$ contains no secular variation in $\lpar$.
  Hence, for large $|\thetaz - \lpar|$ the magnetic drift is dominated by the radial component:
  $ \imag \kperpvec \cdot \vms \simeq \imag \kkr \vms \cdot \nbl \radial / |\nbl \radial|^2$ for $|\thetaz - \lpar| \gg 1$.
  Thus, the leading behaviour of a ballooning mode at large $\lpar$ should be expected to involve the radial magnetic drift.  
  We will often make use of the identity for the radial magnetic drift in an axisymmetric magnetic field,
  \beqn \vms \cdot \nbl \flxl = \vpar \kpar \drv{}{\lpar} 
    \left( \frac{\bcur \vpar}{\cycfs} \right) \label{equation:magdrift}.\eeqn
    
  Finally, we complete this discussion of coordinates by defining a 
  field-aligned radial wave number and binormal wave number
  with dimensions of length,  $\kkx$ and $\kky$, respectively.
  First, we define local radial and binormal coordinates with units of length,
  $\radialx = (\flxl - \flxlz)(\dpsidx)^{-1}$ and
  $\binormal =  (\fldl - \fldlz)(\daldy)^{-1} $, respectively, where $(\flxlz,\fldlz)$
  are the coordinates of the field line of interest. Then,
  the field-aligned radial wavenumber $\kkx =  \kkflxl (\dpsidx) $ and the
  binormal wave number $\kky =\kkfldl (\daldy)$. We take   
  the proportionality constants to be 
  $\dpsidx = \kkxnorm$ and $\daldy=\kkynorm$.
  The functions $\bcur(\flxl)$, $\radial(\flxl)$, and $\saffac(\flxl)$
  appearing in the proportionality  constants should be evaluated on the local
  flux surface of interest, and $\rmajorz = (\rmajormax + \rmajormin)/2$ is a reference major radius,
  with $\rmajormax$ and $\rmajormin$ the maximum and minimum major radial positions on the flux surface, respectively.
  Note that for circular concentric flux surfaces $\rmajorz$ is the major radius at the magnetic axis.
   Using these normalisations, we find that the true radial wave number
  $\kkr \simeq (\thetaz - \lpar) \kky \shat \kxfac |\nbl \radial|^2$ for $|\thetaz - \lpar| \gg 1$,
  with the magnetic shear defined by $\shat =  (\rminor/\saffac) d \saffac / d \rminor$ and
  the geometrical factor $\kxfac = (\saffac \rmajorz / \bcur \rminor )\tdrvt{\flxl}{\rminor}$.

  \section{A convenient form of the gyrokinetic equation}\label{section:gyrokineticsrepackage}
It is possible
 to use the identity for the radial magnetic drift in
 equation \refeq{equation:magdrift}  
to rewrite 
 the gyrokinetic equation in
 a novel way that simplifies the asymptotic analysis
 of the electron response. Collecting terms due
 to parallel streaming and radial drifts, we find that we can write
 \beqn \fl \vpar \kpar \drv{\hhs}{\lpar}
 + \imag \kkfldl \saffacprim (\thetaz - \lpar) \vpar \kpar \drv{}{\lpar}\left(\neophases\right) \hhs \nonumber \eeqn \beqn \fl \qquad \qquad
 = \expo{-\imag \lflrs (\thetaz -\lpar)}\vpar \kpar \drv{}{\lpar}\left(\expo{\imag \lflrs (\thetaz- \lpar)} \hhs \right)
  + \imag  \lflrs \vpar \kpar \; \hhs \label{eq:repackage}, \eeqn
  with 
  \beqn \lflrs =  \frac{\kkfldl \saffacprim\bcur \vpar}{ \cycfs}. \label{eq:lflrs} \eeqn
  Note that $\lflrs$ should not be confused with the pitch angle coordinate $\pitch$.
  We define the new function $\HHs$ by
  \beqn \HHs = \expo{\imag \lflrs (\thetaz-\lpar)} \hhs, \label{eq:HHs} \eeqn
  and hence we can rewrite the gyrokinetic equation, equation, \refeq{eq:lineargyrokinetic}, as
  \beqn \fl \vpar \kpar \drv{\HHs}{\lpar}
 + \imag (\wmags - \wfreq) \HHs 
   - \copsh[\HHs] 
 = \imag \left(\wstars - \wfreq\right)\expo{\imag \lflrs (\thetaz-\lpar)}\bessn{0}\eqlba\frac{\zeds \charge\ptl{}}{\temps}, \label{eq:gkHH}\eeqn 
 where
 \beqn \wmags =  \kkfldl \left( \vms \cdot \left(\nbl \fldl + \lpar\nbl\saffac \right)
 + \kpar \frac{\saffacprim\bcur \vpar^2}{\cycfs} \right) \label{eq:precessiondrift}, \eeqn
 and 
 \beqn \copsh[\HHs] = \expo{\imag \lflrs (\thetaz-\lpar)}\cops[\expo{-\imag \lflrs (\thetaz-\lpar)}\HHs{}]. \label{eq:HHcollsions} \eeqn
 
 It is also useful to consider the form of the nonadiabatic density
 appearing in the quasineutrality relation, equation \refeq{eq:qn}.
 In terms of $\HHs$, we can write the nonadiabatic, fluctuating density $\ddenss$ as
 \beqn \ddenss = \intv {\expo{-\imag \lflrs (\thetaz-\lpar)}\bessn{0} \HHs{}}\label{eq:deltanHHs}.\eeqn
 When the gyrokinetic equation is written in terms of $\HHs$, the oscillation in the distribution function due
 to the radial magnetic drift appears explicitly as the phase $\expo{\imag \lflrs (\thetaz-\lpar)}$ --
 this phase may be thought of in analogy to the phase $\expo{\imag \kperpvec \cdot \gyrdvecs}$
 arising from the finite Larmor radius in gyrokinetic theory. In fact, the appearance of 
 $\expo{\imag \lflrs (\thetaz-\lpar)}$ is due to the finite particle drift orbit width.
 This may be noted by writing $\lflrs (\thetaz-\lpar) = \kkfldl \saffacprim (\thetaz-\lpar) \Dflxl $, recalling that $\Dflxl = \bcur \vpar / \cycfs$
 is the excursion in flux label $\flxl$ made by trapped particles in a banana orbit \cite{helander}, and finally, noting that,  
 in the limit of large $|\thetaz-\lpar|$, $\lflrs (\thetaz-\lpar) \simeq \kkr (\tdrvt{\radial}{\flxl}) \Dflxl / |\nbl \radial |^2$.

   \section {Long-wavelength collisionless electrostatic modes in the $\massrt\rightarrow 0$ limit}
 \label {section:modes-in-memi-zero-limit}
  
  In this section, we derive reduced model
  equations for long-wavelength, electrostatic modes
  in the $\massrt \rightarrow 0$ limit, using the
  \enquote{collisionless} ordering \refeq{order:nustar-collisionless},
  with $\kky \gyrdi \sim \thetaz \sim 1$.
   In this ordering, the extent of the ballooning mode is controlled by a 
  balance of free streaming, precessional drifts,
  finite-Larmor-radius and finite-orbit-width phases,
  and the orbit-averaged electron collision operator,
  with the result that the mode
  extends to a ballooning angle $\lpar \sim (\kky \shat\gyrde)^{-1} \sim \massrut \gg 1$.
  For consistency with the electron species, we take
  the ion self-collision frequency 
  $\cfreqii \sim \massrt\vtheri/\saffac \rmajor \ll \vtheri/\saffac \rmajor \sim \wfreq $.
  
  We first examine the $\lpar \sim 1$ region of the ballooning mode. 
  This discussion reveals the existence of passing-electron-response-driven modes,
  in addition to the usual ion-response-driven and trapped-electron-response-driven modes, and motivates
  an examination of the $\lpar \sim \massrut$ region of the collisionless
  ballooning mode in section \ref{section:innersolution-collisionless}.
  To aid comprehension, we summarise the results
  for trapped-electron-response-driven
  and ion-response-driven modes in section \ref{section:smalltailmode}, and
  for passing-electron-response-driven modes in section \ref{section:largetailmode}. 
  Finally, in section \ref{eq:relationship-between-GK-averaged-eqs},
  we comment on the relationship between the derivation of gyrokinetics and
  the derivation of the reduced model equations
  for the electron response: although these theories have fundamental differences,
  they have a similar structure, relying on the finite Lamor radius and the finite magnetic
  drift orbit width of particles, respectively.

   \subsection {Outer solution -- $\kkr \gyrdi \sim 1$}\label {section:outersolution}
  We define the outer region of the mode to be the region where $\kkr\gyrdi \sim \lpar \sim 1$.
  In real space, the outer region is the large-scale region 
  far from the rational flux surface. In the collisionless ordering, 
  it is natural to expand the electrostatic potential $\ptl$, distribution functions $\hhs$, and frequency $\wfreq$ in $\massrt$:
  for the potential $\ptl$, we expand 
  \beqn \ptl{} = \ptlz +\ptlo + \Or\left(\massrp{} \ptl\right), \label{eq:collisionless-phi-expansion} \eeqn   
  with $\ptln \sim \massrtn \ptlz$. We make expansions of the same form as \refeq{eq:collisionless-phi-expansion}
  for $\hhs$ and $\wfreq$, with 
$\hhsn \sim \massrtn (\fract{\charge \ptl{}}{\tempe}) \eqlba$ and 
$\wfreqn \sim \massrtn \wfreq$.

    \subsubsection{Ion response in the outer region.} \label{sec:ion_outer}
  For the ion species, we start with the usual form of the gyrokinetic equation,
  equation \refeq{eq:lineargyrokinetic}. 
  In the collisionless limit, 
 the leading order equation for the ion response is
\beqn \fl \vpar \kpar \drv{\hhiz{}}{\lpar} + \imag \left(\kperpvec \cdot \vmi - \wfreqz\right) \hhiz{}
  = \imag \left( \wstari - \wfreqz \right)\besin{0}\eqlbi\frac{\zedi \charge \ptlz }{\tempi}. \label{eq:ion0}\eeqn 
 Using equation \refeq{eq:ion0}, and the estimates $\wfreq \sim \vtheri/\lscal \sim \wstari$ and $\bflri \sim 1$,
 we find that $\hhiz /\eqlbi \sim \charge \ptlz/\tempi$. 
 The equation for the nonadiabatic ion density is
 \beqn  \frac{\ddensns{0}{i}}{\densi} = \intv {\besin{0}\frac{\hhiz}{\densi} }
  \sim \frac{\charge\ptlz}{\tempe} , \label{eq:ion-density-outer}\eeqn
  where we have assumed that $\tempi \sim \tempe$ 
  in the final estimate of equation \refeq{eq:ion-density-outer}.
  As expected, the ion nonadiabatic response contributes at
  leading order to $\ptl{}$ in the outer region.
   \subsubsection{Electron response in the outer region.} \label{sec:electron_outer}
      For the electron species, we use the modified form of the gyrokinetic equation,
      equation \refeq{eq:gkHH}.
      We do this to avoid integrating the radial magnetic drift $\vme\cdot \nbl \flxl$
      by parts in $\lpar$ at every order when applying transit or bounce averages.
      The leading order equation for the electron response is
     \beqn \vpar \kpar \drv{\HHez{}}{\lpar} = 0
     \label{eq:electron0},\eeqn
     where we have used that the electron parallel streaming term is larger
     than every other term in equation \refeq{eq:gkHH}
     by the ordering \refeq{order:nustar-collisionless}. 
     We note that for electrons $\lflre \sim \massrt$, and hence, for $\thetaz \sim \lpar \sim 1$
     the phase $\expo{\imag \lflre (\thetaz-\lpar)} $ may be expanded as
     \beqn \expo{\imag \lflre(\thetaz-\lpar)} =
      1 + \imag \lflre(\thetaz-\lpar) - \frac{\lflre^2}{2}(\thetaz-\lpar)^2 + \order{\massrp{3/2}}. \eeqn    
     As a consequence of equation \refeq{eq:electron0}, the leading-order
     nonadiabatic electron response is independent of $\lpar$ for $\lpar \sim 1$.  
     The remainder of the expansion must be carried out separately for passing and trapped particles.
     
    Trapped particles occupy the range of pitch angles $1/\bmagmax < \pitch  \leq 1/\bmag(\lpar)$,
    with $\bmagmax$ the maximum value of $\bmag(\lpar)$.
    In each well, trapped particles bounce at the upper and lower bounce points, $\lparp$ and $\lparm$, respectively.
    Equation \refeq{eq:electron0} for trapped particles
    states that $\HHez$ is constant in $\lpar$ within each magnetic well.
    Imposing the trapped particle boundary conditions 
    \beqn \hhe(\lparpm,\sign= 1) = \hhe(\lparpm,\sign= -1), \label{eq:trappedbc}\eeqn
    where $\sign = \vpar/|\vpar|$, and using equation \refeq{eq:HHs},
    we find that $\HHez(\sign =  1) = \HHez(\sign =  -1)$, and so
    $\HHez$ is independent of $(\lpar,\sign)$.
    The trapped electron piece of the distribution function $\HHez$ is determined by 
 the equation for the next order in the electron response
 \beqn \fl \vpar \kpar \drv{\HHeo}{\lpar}
 + \imag (\wmage - \wfreqz) \HHez 
   - \copbothe[\HHez]  
 = -\imag \left(\wstare - \wfreqz\right)\eqlbe\frac{\charge\ptlz{}}{\tempe}, \label{eq:electrontrappedouter}\eeqn 
  where we have used that for $\bflre = \Or(\massrt)$, $\besen{0} = 1 + \Or (\massrlt)$,
  and $\expo{\imag \lflre(\thetaz-\lpar)} = 1 + \order{\massrt}$,
   and we have employed  $ \kperpvec \cdot \gyrdvece  \sim \bflre\sim \massrt$ to
   reduce the collision operator in equation \refeq{eq:electrontrappedouter} 
   to the drift-kinetic electron collision operator 
   \beqn \copbothe\left[\cdot\right]
    = \coplandaue\left[\cdot\right]+ \coplorentze\left[\cdot\right]. \label{eq:copbothe}\eeqn
   To close equation \refeq{eq:electrontrappedouter}{,} we introduce
   the bounce average for trapped particles
    \beqn \bouncav{\cdot}= \frac{ \sum_\sign \int^{\lparp}_{\lparm}   d \lpar
    \; \fract{(\cdot)}{|\vpar| \kpar} }{2 \int^{\lparp}_{\lparm} d \lpar  \fract{}{|\vpar| \kpar} }  . \label{eq:bounceav}\eeqn
    Applying $\bouncav{\cdot}$ to equation \refeq{eq:electrontrappedouter}, we find the solvability condition
 \beqn \fl  \imag \left(\bouncav{\wmage} - \wfreqz\right) \HHez 
   - \bouncav{\copbothe[\HHez]}
   = -\imag \left(\wstare - \wfreqz\right)\eqlbe\frac{\charge\bouncav{\ptlz}}{\tempe},
   \label{eq:electrontrappedouter2}\eeqn 
  where we have used the property
  \beqn \bouncav{\vpar \kpar \drv{\fulldistf}{\lpar}} =0 \label{eq:bounceav1b}\eeqn 
    of the bounce average, valid for any $\fulldistf=\fulldistf_{\thetaz,\kkfldl}(\energy,\pitch,\sign,\lpar)$ satisfying the bounce condition
   $\fulldistf_{\thetaz,\kkfldl}(\energy,\pitch,\sign=1,\lparpm) = \fulldistf_{\thetaz,\kkfldl}(\energy,\pitch,\sign=-1,\lparpm)$.  

  Passing particles occupy the range
  of pitch angles $0 \leq \pitch  \leq 1/\bmagmax$,
  and hence, passing particles are free to travel between magnetic wells.
  For passing electrons, equation \refeq{eq:electron0}
  determines that, for a given $(\thetaz,\kkfldl)$
  mode, $\HHez$ is a constant in $\lpar$
  for each sign of the parallel velocity $\sign$, i.e., 
  $ \HHez = \HHez(\energy,\pitch,\sign).$ 
  To determine this constant $\HHez$,
  we need to supply an appropriate incoming
  boundary condition to the $\lpar \sim 1$ region. This requires us
  to consider the $\lpar \gg 1$ region.

  In the conventional treatment of passing electrons,
  it is argued that the incoming boundary condition \refeq{eq:incomingbc2}
  implies that $\HHez = 0$ in the passing piece of velocity space, cf. \cite{hardmanpaper1,abelEMOD}.
  This assumption results in modes driven at scales of $\kky\gyrdi \sim 1$
  by the ion response or the trapped electron response.
  Under this assumption, the leading-order nonadiabatic response
  of passing electrons $\HHeo$ is determined by the first-order equation
  \beqn \vpar \kpar \drv{\HHeo}{\lpar} - \coplandaue[\HHez]
 = -\imag \left(\wstare - \wfreqz\right)\eqlbe\frac{\charge\ptlz{}}{\tempe}, \label{eq:electronpassingouter-smalltails}\eeqn 
  where the magnetic drift, frequency, and electron-ion collision terms
  are neglected because $\HHez =0$ in the passing part of velocity space.
  The collision operator term $\coplandaue[\HHez]$ is retained because $\coplandaue[\HHez]$
  is a nonlocal operator representing the drag of trapped particles on passing particles.
  In this ordering, passing electrons coming from the $\lpar \gg 1$ region receive
  a $\massrt$ small impulse from the $\lpar \sim 1$ electrostatic potential:
  \beqn \fl \HHeo(\lpar,\sign = \pm 1) = \HHeo( \mp \infty,\sign= \pm 1)  \label{eq:HHeo_plusminus} \eeqn
  \beqn + \int^{\lpar}_{\mp\infty} \frac{1}{\vpar \kpar}\left( \coplandaue[\HHez] -\imag \left(\wstare - \wfreqz\right)\frac{\charge\ptlz}{\tempe}\eqlbe \right) d \lparprim, \nonumber\eeqn
  where $\HHeo( \mp \infty,\sign=\pm 1)$
  should be determined consistently in the $\lpar \gg 1$ region.
  Because $\HHeo$ contributes only a small correction to  quasineutrality, the nonadiabatic response 
  of passing electrons is conventionally ignored. 

  One of the key contributions of this paper is to notice a flaw in the conventional argument:
  in fact, $\HHez$ need not vanish, but instead $\HHez$ can be determined self-consistently
  in the $\lpar \gg 1$ region. The resulting class of modes are driven by the nonadiabatic
  response of passing electrons, with no leading-order impact from the ion response or trapped electron
  response in the $\lpar \sim 1$ region. 
  We now turn to the $\lpar \gg 1$ region for the collisionless ordering.
  The equations that we obtain there provide the nonadiabatic
  passing electron response, $\HHez(\energy,\pitch,\sign =\pm 1)$, in the case of
  passing-electron-response driven modes, and the boundary conditions $\HHeo(\lpar=\mp \infty,\energy,\pitch,\sign=\pm 1)$
  in the case of modes in the conventional ordering. 
  
  \subsection {Inner solution -- $\kkr \gyrde \sim 1$}\label {section:innersolution-collisionless}
    In real space, the inner region is the radial layer close to the rational flux surface. 
    The inner region
    is characterised by fine radial scales associated with electron physics.
    In order to capture these scales analytically in the ballooning formalism, we introduce an
    additional ballooning angle coordinate $\lchi$ that measures distance
    along the magnetic field line. The coordinate $\lpar$ will capture periodic variation in ballooning angle
    on the scale of $2\pi$ associated with the equilibrium geometry, whereas the coordinate $\lchi$ will measure
    secular variation on scales much larger than $2\pi$. In the inner region,
    distribution functions and fields become
    functions of the independent variables $\lpar$ and $\lchi$, i.e.,
    \beqn \fulldistf(\lpar) \rightarrow \fulldistf(\lpar,\lchi){,} \label{eq:lchi-ansatz1} \eeqn 
    and parallel-to-the-field-line derivatives become
    \beqn \drv{}{\lpar} \rightarrow \drv{}{\lpar} + \drv{}{\lchi}. \label{eq:lchi-ansatz2} \eeqn
    We order $\drvt{}{\lchi} \sim \massrt \drvt{}{\lpar}$, and we order $\kkr \gyrde \sim 1$,
    whilst keeping $\kky \gyrdi \sim 1$. 
    
    The parallel-to-the-field variable $\lpar$ appears in two forms
    in the gyrokinetic equation \refeq{eq:gkHH}: as the argument of periodic functions
    associated with the magnetic geometry;
    and linearly in the combination $ - \kkfldl \saffacprim (\lpar - \thetaz)$.
    To treat the scale separation within the part of the mode where $\lpar \sim \massrut \gg 1$,
    we send $\lpar \rightarrow \lchi$ where $\lpar$ appears in secular terms (e.g. $ - \kkfldl \saffacprim (\lpar - \thetaz)$), 
    and we take $\lchi \sim \massrut \gg 1$.
    This assignment captures the effect of the secular growth of the
    radial wave number $\kkr$ in the $\lpar \gg 1$ region.
    With this procedure, we note that, in the inner region, we can usefully write
    \beqn \kperpvec = \kperpvecz + \kperpveco \label{eq:kperpvecapprox}, \eeqn 
    with \beqn \kperpvecz = - \kkfldl \lchi \nbl \saffac \label{eq:kperpvecapprox0}\eeqn  and
    \beqn \kperpveco = \kkfldl \thetaz \nbl \saffac +
     \kkfldl (\nbl \fldl + \lpar \nbl \saffac)
     = \order{\lchi^{-1} \kperpvec}, \label{eq:kperpvecapprox1} \eeqn
     where we recall that $\nbl \fldl + \lpar \nbl \saffac = \nbl \toragl - \saffac \nbl \lpar - \nbl \nufunc$
     has no secular dependence on $\lpar$.
    
    We also need to consider the argument of the Bessel function
    \beqn \bflrs =\frac{ \kperp\vperp\ma\ltsp}{\zeds\charge\bmag}
    = \frac{ \kperp(\lpar) \ltsp}{\zeds\charge}\sqrt{\frac{2 \ma \energy \pitch}{\bmag(\lpar)}}    
    \label{eq:bflrs}.\eeqn
    In the region  $\lchi \sim \massrut$, we find that 
    \beqn \bflrs = 
    \kkfldl|\nbl\saffac|(\lpar)\frac{\ltsp}{\zeds\charge}
    \sqrt{\frac{2\ma\energy\pitch}{\bmag(\lpar)}}|\lchi| + \order{\lchi^{-1} \bflrs}. \label{eq:bflrsapprox} \eeqn
    Note that $\bflrs$ has a linear dependence on $\lchi$,
    whereas $\lpar$ appears only through the periodic functions $|\nbl\saffac| = |\saffacprim | |\nbl\flxl|(\lpar)$ and $\bmag(\lpar)$. 
    The plasma is magnetised, and hence $\bmag(\lpar)$ is never close to zero: 
    $\bmag(\lpar)$ has an order unity component independent of $\lpar$.
    Likewise, $|\nbl \flxl|$ 
    will be nowhere zero on any given flux surface (except perhaps if there is an X-point on the last closed flux surface).
    Hence, changes in $\lpar$ cause only order unity oscillations in $\bflrs$,
    whereas changes in $\lchi$ can cause arbitrarily large variations in $\bflrs$.
    
    Finally, to solve for the electron distribution function, we need to 
    impose a $2\pi$ periodic boundary condition on $\lpar$,
    and a \enquote{ballooning} boundary condition on $\lchi$, i.e.,   
    \beqn \hhe{}(\lpar=\pi,\lchi) =
   \hhe{}(\lpar=-\pi,\lchi),  \label{eq:twshinner1}\eeqn
    and \beqn \hhe{}(\lchi = -\infty) = 0, {\rm~for~} \vpar > 0, {\rm~and} \nonumber\eeqn  
    \beqn \hhe{}(\lchi = \infty) = 0, {\rm~for~} \vpar < 0. \label{eq:twshinner2}\eeqn
    The results for large $\lchi$ above, equations \refeq{eq:kperpvecapprox}-\refeq{eq:twshinner2},
    are not peculiar to the ordering $\lchi \sim \massrut$.
    We will reuse results \refeq{eq:kperpvecapprox}-\refeq{eq:twshinner2}
    for $\lchi \sim \massrutp{1/4}$
    when we come to discuss the collisional inner region in section \ref{section:innersolution-collisional}. 

    To solve for the electron response, we will again use
    the modified electron distribution function $\HHe$, defined by equation
    \refeq{eq:HHs}, and the modified electron gyrokinetic equation
    (equation \refeq{eq:gkHH} with $\spe={\rm{e}}$). We note that,
    in the inner region of the collisionless ordering,  
    $\lpar \sim \massrut \gg 1 \sim \thetaz \gg \lflre \sim \massrt$,
    and hence the phase in \refeq{eq:HHs} becomes
    \beqn \fl\expo{\imag \lflre (\thetaz-\lpar)} = \expo{-\imag \lflre \lchi}
    \left(1 + \imag \lflre \thetaz - \frac{\lflre^2 \thetaz^2}{2} + \order{\massrp{3/2}} \right).\eeqn

    Consistent with the expansion in the outer region, in the inner region we expand the
    electrostatic potential $\ptl$, the distribution functions $\hhi$ and $\HHe$, and the frequency
    $\wfreq$ in powers of $\massrt$. However, we leave the relative size of the
    fluctuations in the outer and inner regions to be determined. We will return
    to this point in sections \ref{section:smalltailmode} and \ref{section:largetailmode}.  
      \subsubsection{Ion response in the inner region.} \label{sec:ions:inner-collisionless}
    The leading-order equation for the ion response in the inner region is
    \beqn \fl
    \left({\frac{\kkfldl^2 {\saffacprim}^2 |\nbl \flxl|^2 \lchi^2 \vmag^2}{4\cycfi^2}\left(\cfreqiipar \pitch \bmag
    + \frac{\cfreqiiperp}{2} \left(2 - \pitch \bmag \right)\right) }
    -\imag \kkfldl \saffacprim \lchi \vmi \cdot \nbl \flxl \right) \hhiz \nonumber \eeqn \beqn 
     =  \imag \left( \wstari  - \wfreqz\right)
   \besin{0} \eqlbi\frac{\zedi \charge \ptlz}{\tempi}
 \label{eq:inner_ion1},\eeqn
    where we have defined the collision frequencies
    \beqn \cfreqiipar = \sqrt{\frac{\pi }{2}} \cfreqii
    \frac{\Psifunc(\vmag/\vtheri)}{(\vmag/\vtheri)^3}, \label{eq:cfreqiipar}\eeqn
    and 
    \beqn \cfreqiiperp = \sqrt{\frac{\pi }{2}} \cfreqii
    \frac{\erf(\vmag/\vtheri) - \Psifunc(\vmag/\vtheri)}{(\vmag/\vtheri)^3}, \label{eq:cfreqiiperp}\eeqn
    with the functions \beqn \erf(\zarg) =
 \frac{2}{\sqrt{\pi}} \int^\zarg_0 \expo{-\darg^2} \; d \darg, \label{eq:error-function} \eeqn
    and \beqn \Psifunc(\zarg) = \frac{1}{2 \zarg^2}\left(\erf(\zarg)
  - \frac{2 \zarg}{\sqrt{\pi}} \expo{-\zarg^2} \right). \label{eq:Psifunc-function}\eeqn 
    The first term on the left of equation \refeq{eq:inner_ion1}
    is due to the finite-Larmor-radius terms in the ion gyrokinetic
    self-collision operator \refeq{eq:copgk-ion} (cf. \cite{cattotasngPoF1977,abelPoP08,barnesPoP09}).
    The ion response given by equation \refeq{eq:inner_ion1} is local
    in ballooning angle -- a more detailed analysis demonstrating how this response arises
    is given in \ref{sec:detailed-analysis-of-the-ion-response}.
    We note that {$\besin{0} \sim 
    \lchi^{-1/2} \sim \massrtp{1/4}$} for 
    {$ \bflri \sim \kkr\gyrdi \sim \lchi \sim \massrut \gg 1$}.
    Hence, if $\cfreqii/\wfreq \sim \massrt$, we take the ion nonadiabatic response
    {$\hhiz{}/\eqlbi \sim \lchi^{-3/2} (\charge \ptlz{} / \tempe)$} in the inner region.
    Hence, the contribution of $\hhiz{}$ to $\ptl{}$ is small. 
    Estimating the size of the ion nonadiabatic density $\ddensns{0}{i}$ and ion mean velocity $\duvecns{0}{i}$ in the inner region, 
    we find that 
    \beqn \frac{\ddensns{0}{i}}{\densi} \sim \frac{\duvecns{0}{i}}{\vtheri} \sim
        \massrl \frac{\charge\ptlz{}}{\tempe} \ll \frac{\charge\ptlz{}}{\tempe}.\label{eq:inner_ion3a} \eeqn
    
    We have used the conventional distribution function $\hhi$ 
    and conventional form of the gyrokinetic equation to describe the ion species.
    We could obtain the estimate \refeq{eq:inner_ion3a} 
    by using the alternative form of the gyrokinetic equation, equation \refeq{eq:gkHH}.
    However, if we use the distribution function $\HHi$ and equation \refeq{eq:gkHH} for ions,
    we need to be careful with estimates involving integrals of the phase
    $\expo{-\imag \lflri \lchi}$,
    because $\lflri\lchi \gg 1$ in the inner region.

    \subsubsection{ Electron response in the inner region.} \label{eq:collisionless-electron-response-inner}
    The leading order equation for the electron response
    in the inner region takes the form of
    equation \refeq{eq:electron0}.
    Equation \refeq{eq:electron0} appears to be trivially simple because of the choice to use
    the modified electron gyrokinetic equation \refeq{eq:gkHH}, and modified
    distribution function $\HHe$.
    In terms of $\hhe$, and using equation \refeq{eq:HHs}, equation \refeq{eq:electron0}
    tells us that the leading-order
    electron distribution function 
    has the form
    \beqn \hhez(\lpar, \lchi) = \expo{-\imag \lflre \lchi} \HHez(\lchi), \eeqn
    i.e., the $\lpar$ dependence in $\hhez$
    comes entirely from the radial-magnetic-drift phase $\expo{-\imag \lflre \lchi}$,
    and $\HHez(\lchi)$ is the slowly decaying envelope of $\hhez$.
    This observation motivates the choice to present the
    derivation in terms of $\HHe$ rather than $\hhe$.
    
    The distribution function 
    $\HHez$ is determined by the
    first-order equation for the electron response in the inner region
     \beqn \fl \vpar \kpar \left(\drv{\HHeo}{\lpar}+   \drv{\HHez}{\lchi}\right)
 + \imag \left(\wmage - \wfreqz\right) \HHez 
   - \copehz[\HHez] \nonumber \eeqn \beqn 
 = -\imag \left(\wstare - \wfreqz\right)\expo{-\imag \lflre \lchi}\besenz{0}\eqlbe\frac{\charge\ptlz}{\tempe}, \label{eq:electron_firstorder}\eeqn 
  where \beqn \fl \copehz[\HHez] =
  \expo{-\imag \lflre \lchi}\gav{\expo{\imag \kperpvecz \! \cdot \! \gyrdvece}
  \copbothe\left[\expo{-\imag \kperpvecz \! \cdot \! \gyrdvece}
  \expo{\imag \lflre \lchi}\HHez\right]}{} \label{eq:copehz}, \eeqn
    with $\copbothe\left[\cdot\right]$ defined by equation \refeq{eq:copbothe}, and $\besenz{0} = \bes{\bflre^{(0)}}$.
    In order to solve equation \refeq{eq:electron_firstorder} for passing particles,
    we must impose the solvability condition that $\HHeo$ is periodic in $\lpar$.
    This condition can be imposed by using the transit average
    \beqn  \transav{\cdot}=
    \frac{ \int^{\pi}_{-\pi}  d \lpar \;\fract{(\cdot)}{\vpar \kpar}  }{\int^{\pi}_{-\pi} d \lpar  \fract{}{\vpar \kpar} }  . \label{eq:transitav}\eeqn    
    Applying the transit average to equation \refeq{eq:electron_firstorder}
    results in the equation for $\HHez$;
    \beqn \fl  \transav{\vpar \kpar} \drv{\HHez}{\lchi} 
    + \imag \left(\transav{\wmage} - \wfreqz\right) \; \HHez -\transav{\copehz\left[\HHez\right]}   
     \nonumber\eeqn \beqn  
     = - \imag \left(\wstare - \wfreqz\right)\eqlbe
     \transav{\expo{-\imag \lflre\lchi}\besenz{0} \frac{\charge\ptlz}{\tempe}}
       \label{eq:electron_firstorder_passing}.\eeqn 
        
    For trapped electrons, we need to be careful in our interpretation of the
    two scales in equation \refeq{eq:electron_firstorder}. Physically, trapped particles 
    cannot pass between wells in the magnetic field strength. Trapped particles can
    observe only changes of order unity in poloidal angle as they follow trapped orbits.
    This prohibits large variation in the ballooning poloidal angle $\lchi$ for individual particles.
    The trapped particle distribution function should satisfy the  trapped particle boundary conditions,
    equation \refeq{eq:trappedbc}. Noting that $\lflre(\lparpm) = 0$ as $\vpar(\lparpm)=0$, we have
    that for trapped particles $\HHez(\lparpm,\sign=1) = \HHez(\lparpm,\sign=-1)$ and hence 
    $\HHez$ is constant in both $\lpar$ and $\sign$.  
    To go to higher order, we must impose the solvability condition that
    $\HHeo$ satisfies the bounce conditions
    $\HHeo(\lparpm,\sign = 1) =\HHeo(\lparpm,\sign =- 1) $.
    Hence, to obtain the equation for $\HHez$ for trapped electrons,
    we apply the bounce average $\bouncav{\cdot}$, defined in equation
    \refeq{eq:bounceav},
    to equation \refeq{eq:electron_firstorder}. The result is 
    \beqn \fl   
     \imag \left(\bouncav{\wmage} - \wfreqz\right) \; \HHez  -\bouncav{\copehz\left[\HHez\right]}   
     \nonumber \eeqn \beqn 
     = - \imag \left(\wstare - \wfreqz\right)\eqlbe
     \bouncav{\expo{-\imag \lflre\lchi}\besenz{0} \frac{\charge\ptlz}{\tempe}}
       \label{eq:electron_firstorder_trapped},\eeqn 
    where we have used the property 
    \refeq{eq:bounceav1b} of the bounce average
    to eliminate the parallel derivative in $\lpar$ on the left-hand side of 
    equation \refeq{eq:electron_firstorder}, and we have used the property
    \beqn \bouncav{\vpar \fulldistg} =0 \label{eq:bounceav1a},\eeqn
   for any $\sign$-independent function
   $\fulldistg=\fulldistg_{\thetaz,\kkfldl}(\energy,\pitch,\lpar)$,
   to eliminate the term $\bouncav{\vpar\kpar}\drvt{\HHez}{\lchi}$.
    Note that no derivatives in $\lchi$ appear explicitly in equation \refeq{eq:electron_firstorder_trapped},
    and hence, for trapped particles, $\HHez$ is only a parametric function of $\lchi$. 
    This is the manifestation of the physical intuition that
    trapped particles do not move between magnetic wells. 
    
  \subsection {Modes with $\massrt$ small electron tails}\label {section:smalltailmode}
 
 In this section we describe the class of modes in the collisionless ordering that
 have small electron tails. This class of modes includes the conventional ITG
 mode and the 
 TEM, so much of the discussion will be familiar. 
 To obtain the \enquote{small-tail} modes, we assume a priori that $\HHezouter = 0$
 for passing electrons in the outer region of the mode where $\kkr\gyrdi \sim \lpar \sim 1$. 
 Then, the passing electron response has a leading-order
 nonzero component $\HHeoouter$, given by equation \refeq{eq:HHeo_plusminus}.
 We obtain the leading-order trapped electron response $\HHezouter$ from equation \refeq{eq:electrontrappedouter2},
 and the leading-order ion response $\hhizouter$ from equation \refeq{eq:ion0}.
 No parallel boundary condition is required to solve the trapped-electron equation \refeq{eq:electrontrappedouter2}.
 For equation \refeq{eq:ion0} for the ion response, we supply the zero-incoming boundary
 condition \refeq{eq:incomingbc2}, without referring to the inner region where
 $ \kkr \gyrdi \sim \lpar \gg 1$. This is justified by the fact that in the inner region
 $\hhizinner$ is small. 
 We can regard $\HHezouter$ and $\hhizouter$ as functionals of $\ptlzouter$ and functions
 of $\wfreqz$, i.e., $\HHezouter = \HHezouter[\ptlzouter,\wfreqz]$ and
 $\hhizouter = \hhizouter[\ptlzouter,\wfreqz]$.   
 The frequency $\wfreqz$ and potential $\ptlzouter$ are determined through
 the leading-order quasineutrality relation in the outer region 
   \beqn \left(\frac{\zedi \tempe}{\tempi} + 1 \right)\frac{\charge \ptlzouter}{\tempe}
  = \intv{ \besin{0} \frac{\hhizouter}{\densi}} - \intv{  \frac{\HHezouter}{\dense}}  \label{eq:qnouter:smalltails},\eeqn
  where we have used that $\besen{0} = 1 + \order{\massrlt}$ for $\kky\gyrdi\sim \kkr\gyrdi \sim 1$.
  \begin{figure}
\begin{center}
\includegraphics[clip, trim=0cm 0cm 0cm 0cm, width=0.8\textwidth]{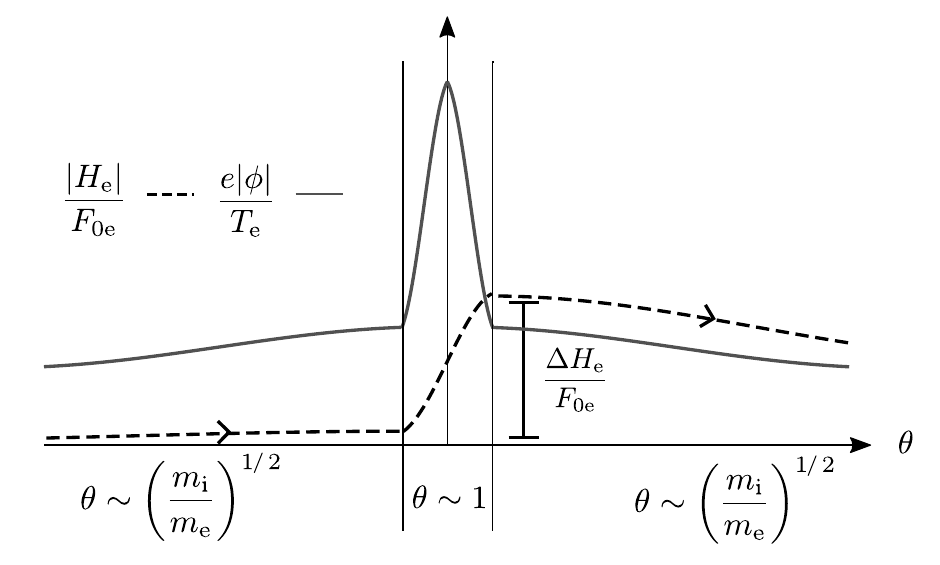}
% trim=left bottom right top, clip
\caption{An illustration showing the
 nonadiabatic passing electron response for forward-going particles in the small-tail limit.
 At leading-order in the $\massrt$ expansion, the mode frequency is determined
 by the response of ions and trapped electrons in the outer region ($\lpar \sim 1$),
 by solving equations \refeq{eq:ion0} and \refeq{eq:electrontrappedouter2}, with quasineutrality \refeq{eq:qnouter:smalltails}.
The passing part of the electron distribution function $\HHe$ is propagated from
 left to right, via equation \refeq{eq:electron_firstorder_passing},
 starting with zero amplitude at $\lpar = - \infty$,
 receiving an impulse $\Delta \HHe$ from the potential $\phi$
 in the outer region (see equation \refeq{eq:jump}),
 and finally, carrying that amplitude into the inner region ($\lpar \sim \massrut$).
 In the inner region, the trapped electron response may be determined with equation \refeq{eq:electron_firstorder_trapped}, 
 and the electron response determines $\phi$, via quasineutrality \refeq{eq:qninner}.  
 } \label{fig:small_tail_modes}
\end{center}
\end{figure}
 
  The small correction $\HHeoouter$ from passing electrons does
  not enter in the leading-order eigenvalue problem, equation \refeq{eq:qnouter:smalltails}.
  As a result, in small-tail modes the nonadiabatic passing electron
  response is a \enquote{cosmetic} feature that does not contribute to
  determining the basic properties of the mode. Nonetheless, observable electron tails can develop
  in $\kkr\gyrde \sim 1$ regions ($\lpar \sim \massrut$).
  We illustrate this in figure \ref{fig:small_tail_modes}. 
  The mode is decomposed into three regions:
  $\lpar \sim 1$, and $ |\lpar |\sim \massrut$ for $\lpar > 0$ and $\lpar < 0$.
  Forward-going passing electrons travel through the $\lpar \sim 1$ region, receiving an impulse
  \beqn {\Delta \HHe =
 \int^{\infty}_{-\infty} \frac{1}{\vpar \kparprim} \left( \coplandaue\left[\HHezouter\right]
 -\imag \left(\wstare - \wfreqz\right)\frac{\charge\ptlzouter}{\tempe}\eqlbe
   \right) d \lparprim }\label{eq:jump} \eeqn 
  from the electrostatic potential $\ptlouter$.  
  The matching condition for the electron nonadiabatic response
  in the $\lpar \sim \massrut$ region
   $\HHeinner$ is obtained from the jump condition \refeq{eq:jump} by demanding that the passing electron
  distribution function is continuous across the boundary between the outer and inner regions, i.e.,
  \beqn \HHeoouter(\lpar = \pm \infty) = \HHezinner(\lchi = 0^{\pm}), {\rm{~for~}} 0\leq\pitch \bmagmax \leq 1. \label{eq:small-tail-continuity}\eeqn
  Combining equations \refeq{eq:HHeo_plusminus}, \refeq{eq:jump} and \refeq{eq:small-tail-continuity},
  we find that the matching condition for solving for the passing electron response in the small-tails limit is
  \beqn \HHezinner(\lchi = 0^+) = \HHezinner(\lchi = 0^-) + \Delta \HHe \label{eq:small-tail-match}, \eeqn
  valid for both $\sign \pm 1$.
  Once $\ptlzouter$ and $\wfreqz$ are determined, we can self-consistently
   obtain the electron tails associated
   with a small-tail mode 
   by solving the inner region equations \refeq{eq:electron_firstorder_passing} and 
   \refeq{eq:electron_firstorder_trapped} for the nonadiabatic
   response of passing electrons and trapped electrons, respectively,
   subject to 
   the jump condition \refeq{eq:small-tail-match} at $\lchi = 0$.
   This obtains the functional $\HHezinner = \HHezinner[\ptlzinner,\ptlzouter,\wfreqz]$.
   Finally, we impose quasineutrality in the inner region
   to obtain a relation for $\ptlzinner$ in terms of the jump over $\ptlzouter$, 
   using the leading-order equation
   \beqn \left(\frac{\zedi \tempe}{\tempi} + 1 \right)\frac{\charge \ptlzinner}{\tempe}
  = - \intv{  \expo{\imag \lflre \lchi} \besenz{0}\frac{\HHezinner}{\dense}}  \label{eq:qninner},\eeqn
    where we have used that the
    ion contribution to quasineutrality is small, 
    cf. equation \refeq{eq:inner_ion3a}.  
  
  We obtain an estimate for the size of $\ptlzinner$ by noting that 
  the impulse \refeq{eq:jump} sets the natural size of $\HHeinner$
  compared to the size of the potential in the outer region: 
  \beqn \HHeinner \sim \Delta \HHe
  \sim \massr \frac{\charge \ptlouter}{\tempe}{\eqlbe}  . 
  \label{eq:small-tail-collisionless-HHe-estimate}\eeqn
    Combining this estimate with equation \refeq{eq:qninner}, we find
    that the electrostatic potential in the inner region $\ptlinner$ is of size 
\beqn \frac{\charge \ptlinner}{\tempe}
  \sim \massr \frac{\charge \ptlouter}{\tempe} \label{eq:small-tail-collisionless-ptlinner-estimate} .\eeqn

  \subsection {Modes with dominant electron tails}\label {section:largetailmode}
 
\begin{figure}
\begin{center}
\includegraphics[clip, trim=0cm 0cm 0cm 0cm, width=0.8\textwidth]{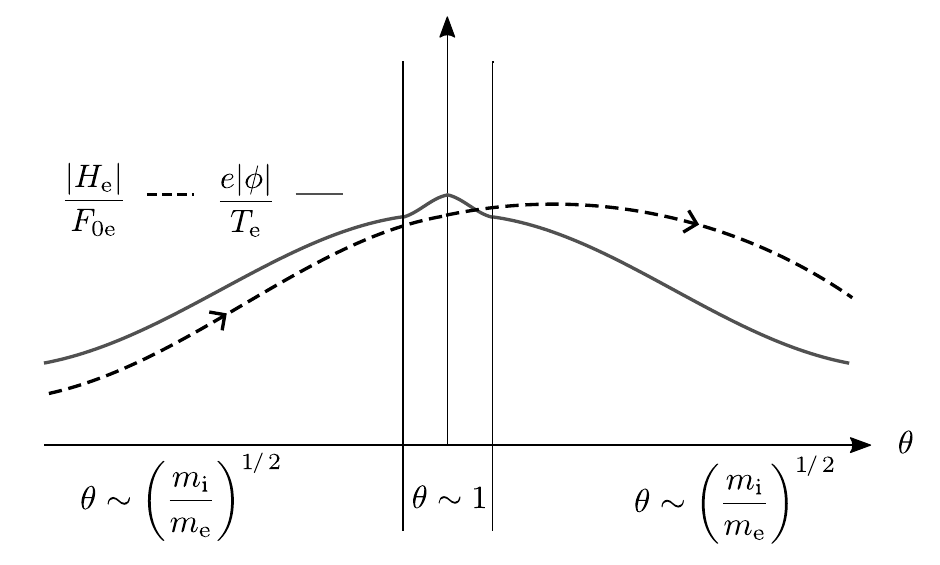}
% trim=left bottom right top, clip
\caption{An illustration showing the
 nonadiabatic passing electron response for forward-going particles in the large-tail limit.
 In this limit, the electron response in the inner region ($\lpar \sim \massrut$)
 determines the mode frequency to leading order in the $\massrt$ expansion: we solve 
 equations \refeq{eq:electron_firstorder_passing} and \refeq{eq:electron_firstorder_trapped}
 for the passing and trapped electron response, respectively, subject to quasineutrality \refeq{eq:qninner}. 
 In the outer region, the electron response $\HHe$ is approximately constant,
 and ions respond passively, without modifying the frequency to leading order. 
 } \label{fig:large_tail_modes}
\end{center}
\end{figure}

 We now turn to the novel class of modes identified in this paper.
 To obtain a \enquote{large-tail} mode in the $\massrt \rightarrow 0$ limit,
 we assume that the leading-order nonadiabatic passing electron response is
 nonzero in the outer region, i.e.,
 \beqn \frac{\HHeouter}{\eqlbe} \sim \frac{\charge \ptlouter}{\tempe}. \label{eq:largetailorder}\eeqn
 We recall from section \ref{sec:electron_outer}
 that $\HHezouter$ is a constant in $\lpar$, and is independent
 of the ion response and the trapped electron response in the outer region.
 As a consequence, in the ordering \refeq{eq:largetailorder} we may solve the leading-order
 equations \refeq{eq:electron_firstorder_passing} and \refeq{eq:electron_firstorder_trapped}
 for $\HHez$ in the inner region with the boundary condition that
 \beqn \HHezinner (\lchi = 0^{-}) = \HHezinner (\lchi = 0^{+}) \label{eq:largetailbc}. \eeqn 
 Imposing quasineutrality via equation \refeq{eq:qninner}
 results in an eigenvalue problem for $\ptlzinner$ and $ \wfreqz$.
We illustrate the mode structure in the large-tail ordering in figure \ref{fig:large_tail_modes}.
  Note in particular that the nonadiabatic passing electron response
  changes by only a small ($\massrt$) amount over the $\lpar \sim 1$ region.
  {As a} consequence of the ordering \refeq{eq:largetailorder}, and 
  the boundary condition \refeq{eq:largetailbc}, we find that the electrostatic
  potential in the inner region has no mass ratio scaling with respect
  to the electrostatic potential in the outer region, i.e.,
\beqn \frac{\charge \ptlinner}{\tempe}
  \sim \frac{\charge \ptlouter}{\tempe} \label{eq:largetailorder2} .\eeqn
  An interesting corollary of these arguments is that
  the leading order complex frequency $\wfreqz$ of a large-tail mode
  should be independent of $\thetaz$. 
  
  Finally, in a large-tail mode the role of the nonadiabatic ion response
  (and nonadiabatic trapped electron response for $\lpar \sim 1$) 
  is to modify the leading-order mode structure at $\lpar \sim 1$
  without modifying the frequency $\wfreqz$. To see this, note that
 equations \refeq{eq:electron_firstorder_passing},
 \refeq{eq:electron_firstorder_trapped}, and \refeq{eq:qninner} determine the frequency $\wfreqz$.
 However, $\ptlzouter$ is not yet determined:
 in the $\lpar \sim 1$ region only the nonadiabatic density
 due to passing electrons $\ddensns{0}{e,passing}$ is fixed by the passing electron tails.
 To obtain $\ptlzouter${,} we solve equation \refeq{eq:ion0}
 for the nonadiabatic ion response $\hhizouter = \hhizouter[\ptlzouter,\wfreqz]$,
 and equation \refeq{eq:electrontrappedouter2}
 for the nonadiabatic trapped electron response $\HHezoutertrapped = \HHezoutertrapped[\ptlzouter,\wfreqz]$,
 where we have indicated that $\hhizouter$ and $\HHezoutertrapped$ are functionals of $\ptlzouter$
 and functions of $\wfreqz$. 
 We then use $\lpar \sim 1$ quasineutrality, equation \refeq{eq:qnouter:smalltails},
 to obtain $\ptlzouter$ as a function of $\ddensns{0}{e,passing}$.
 The role of the nonadiabatic ion response (and the nonadiabatic trapped electron response)
 is to modify the response of the electrostatic potential $\ptlzouter$
 to an input $\wfreqz$ and $\ddensns{0}{e,passing}$.
    
\subsection{{Relating the derivation of gyrokinetics to the derivation of the transit
 and bounce averaged equations for the electron response}} \label{eq:relationship-between-GK-averaged-eqs}
 We conclude this section on collisionless physics by commenting on the relationship between
 the derivation of gyrokinetics and the derivation of the transit
 and bounce averaged equations for the electron response in the inner
 region. We note that in the derivation of the gyrokinetic equation the change of  variables
from $(\rr,\energy,\pitch,\gyrophase)$ to $(\gc,\energy,\pitch,\gyrophase)$
introduces the finite-Larmor-radius phase
 $\expo{\imag \kperpvec \cdot \gyrdvecs}$ into the kinetic equation.
The $\gyrophase$ dependence in the kinetic equation can be removed by a  gyroaverage $\gav{\cdot}{}$
because the field $\ptl(\rr)$ has no dependence on the gyrophase $\gyrophase$,
 and the finite-Larmor-radius phases are converted into a Bessel function $\bes{\bflrs}$
by the gyroaverage $\gav{\cdot}{}$.
In the derivation of the equations for the electron response in the inner region,
 equations \refeq{eq:electron_firstorder_passing} and
 \refeq{eq:electron_firstorder_trapped},
 we find that the leading-order electron distribution function
 $\HHez = \HHez(\lchi,\energy,\pitch,\sign)$ is independent of
 $\lpar$, and the phase $\expo{-\imag \lflre \lchi}$
 keeps track of the electron drift-orbit motion.
 However, the potential $\ptlz = \ptlz(\lpar,\lchi)$ has a
 nontrivial dependence on $\lpar$. This can be observed by inspecting
 the inner region quasineutrality relation, equation \refeq{eq:qninner}, 
 where we see that velocity-space structure in $\HHez$ influences the $\lpar$ structure of $\ptlz$. 
  As a consequence, we may not directly
remove $\lpar$ when solving the system of equations \refeq{eq:electron_firstorder_passing},
 \refeq{eq:electron_firstorder_trapped}, and \refeq{eq:qninner}.

   \section {Long-wavelength collisional electrostatic modes in the $\massrt\rightarrow 0$ limit}
 \label {section:modes-in-collisional-memi-zero-limit}
  
  In this section, we present reduced model
  equations for long-wavelength, collisional, electrostatic modes in the $\massrt \rightarrow 0$ limit.
  We define the collisional limit to be the limit \refeq{order:nustar-collisional}. 
  In the collisional limit, the scale of the mode in extended ballooning angle $\lchi$
  is set by the balance between parallel and perpendicular
  classical and neoclassical diffusion terms appearing
  in the equations for the mode. This means that we expect a balance
  \beqn \frac{\vthere^2}{\saffac^2 \rmajorz^2 \cfreqee}\drvsq{}{\lchi}
   \sim \cfreqee \kky^2 \gyrde^2 \lchi^2 \label{eq:collisional-balance}. \eeqn
     We can rearrange the balance \refeq{eq:collisional-balance}
    to give an estimate for the size of $\lchi$.
    We find that
    \beqn \lchi \sim \left(\frac{\saffac \rmajorz \cfreqee}{  \vthere}\right)^{-1/2} \massrup{1/4}. \label{eq:collisional-lchi}\eeqn
    For the collisional ordering of $\fract{\saffac \rmajorz \cfreqee}{  \vthere}\sim 1$, the scale of the
    electron tail is $\lchi \sim \massrutp{1/4}$. As expected,
    the \enquote{collisionless} ordering of $\fract{\saffac \rmajorz \cfreqee}{  \vthere} \sim \massrt$
    in the estimate \refeq{eq:collisional-lchi} yields the scale $\lchi \sim \massrut$.
    In section \ref{sec:banana-and-collisional-collisionless-matching}, we demonstrate that there is a continuous transition between the collisional
    and the collisionless limits.
    
    We obtain the equations for the
    response of ions and electrons in a $\kky \gyrdi \sim 1$ mode with a $\lpar \sim 1$ outer region,
    and a $\lpar \sim \massrutp{1/4}$ inner region. Although the details of the equations
    obtained here are different to the collisionless case, the final result is
    qualitatively similar: two types of modes exist, large-tail modes
    driven by the nonadiabatic electron response at $\lpar \sim \massrutp{1/4}$ scales,
    and conventional small-tail modes driven by 
    the ion response 
    at $\lpar \sim 1$ scales. 
    Note that in the collisional limit, trapped electrons cannot drive an instability
    because their orbits are disrupted by collisions.
    In order to motivate the $\massrtp{1/4}$ expansion, we first present the
    equations in the $\lpar \gg 1$ region. As in the collisionless case,
    the equations that we obtain in the $\lpar \gg 1$ region are common to both classes of mode.
    The two different classes of mode are distinguished by the boundary matching at $\lpar \sim 1$. 
    Section \ref{sec:electron-outer-collisional-small-tail} 
    provides a description of the boundary matching between the
    outer and inner regions for the small-tail mode,
    in addition to a plenary summary for how to solve the small-tail mode equations.
    Finally, section \ref{sec:electron-outer-collisional-large-tail} 
    provides a description of the boundary matching between the
    outer and inner regions for the large-tail mode, and
    a plenary summary for how to solve the large-tail mode equations.

  \subsection {Collisional inner solution -- $\lpar \sim \massrutp{1/4}$ -- $\kkr \gyrde \sim \massrtp{1/4} $}
\label {section:innersolution-collisional}
 
    To treat the fine radial scales of the collisional inner region, we introduce an
    additional coordinate $\lchi$ measuring distance along the magnetic field line,
    via the substitutions \refeq{eq:lchi-ansatz1} and \refeq{eq:lchi-ansatz2}.
    The coordinate $\lpar$ will measure $2\pi$ periodic variation,
    whereas $\lchi \sim \massrutp{1/4}$ is an extended ballooning
    angle for the envelope of the mode.
    Refer to the discussion in section \ref{section:innersolution-collisionless}
    for the details of the substitution in geometric quantities
    (equations \refeq{eq:kperpvecapprox}-\refeq{eq:bflrsapprox}); and
    the modifications to the boundary conditions on the electron distribution function
    (equations \refeq{eq:twshinner1} and \refeq{eq:twshinner2}). 
    In the collisional limit, we can note the similarity of structure of the derivation
    of the inner region equations to the treatment of resistive ballooning modes
    and semi-collisional tearing modes in toroidal geometry, see,
    for example, \cite{Connor_1985PPCFresistiveballooning} and \cite{ConnorPPCF2008}, respectively.

    To proceed, we expand
    electrostatic potential $\ptl$, distribution functions $\hhs$, and frequency
    $\wfreq$ in powers of $\massrst$, i.e., for the potential we expand
\beqn \ptl{} = \ptlz +\ptlh +\ptlo + \Or\left(\massrp{3/4} \ptl\right), \label{eq:ptl_expand}\eeqn   
  with $\ptln \sim \massrtn \ptl$. Identical expansions are made for $\hhs$ and $\wfreq$, 
  again taking $\hhsn \sim \massrtn (\fract{\charge \ptl{}}{\tempe}) \eqlba$ and 
$\wfreqn \sim \massrtn \wfreq$.
    As in the collisionless case, to solve for the electron response, we use
    the modified electron distribution function $\HHe$, defined by equation
    \refeq{eq:HHs}, rather than the usual distribution function $\hhe$. 
    We leave the relative size of the
    fluctuations in the outer and inner regions to be determined by the matching 
    in sections \ref{sec:electron-outer-collisional-small-tail} and \ref{sec:electron-outer-collisional-large-tail}.

      \subsubsection{Ion response in the collisional inner region.} \label{sec:ions:inner-collisional}
    Before considering the electron response, we first 
    comment on the ion response in the inner region in the collisional limit.
    The analysis proceeds almost identically to the analysis presented
    in section \ref{sec:ions:inner-collisionless} for the ion response in the collisionless limit.
    For $\cfreqii/\wfreq \sim 1$ and $\lchi \sim \massrutp{1/4}$, we find that the leading-order equation for the ion response 
    has the same form as equation \refeq{eq:inner_ion1}, apart
    from the fact that the radial magnetic drift term is neglected. 
    This observation allows us to obtain an estimate for $\hhiz$:
    $\fract{\hhiz}{\eqlbi} \sim \lchi^{-5/2} \; \fract{\charge \ptlz}{\tempe}$,
    where we have employed that $\besin{0} \sim \order{\lchi^{-1/2}}$ for $\lchi \gg 1$.
    This estimate for $\hhiz$ 
    yields estimates for the ion nonadiabatic density $\ddensns{0}{i}$ and the ion mean velocity $\duvecns{0}{i}$, 
    required for the electron-ion piece of the electron collision operator, 
    \beqn \frac{\ddensns{0}{i}}{\densi}\sim \frac{\duvecns{0}{i}}{\vtheri}\sim
        \massrp{3/4} \frac{\charge\ptlz{}}{\tempe} \ll \frac{\charge\ptlz{}}{\tempe},\label{eq:inner_ion2-collisional} \eeqn
    where we have used that $\besin{0}\sim \besin{1} \sim \order{\massrtp{1/8}}$. The estimate \refeq{eq:inner_ion2-collisional} 
    shows that the nonadiabatic ion response has no leading-order contribution to the mode evolution in the inner region.

    \subsubsection{ Electron response in the collisional inner region.} \label{sec:electron-inner-collisional-main-text}
    The calculation of the electron response in the collisional inner region has a structure that is
 reminiscent of neoclassical transport theory. The leading-order equation constrains the
 leading-order electron distribution function to be a perturbed Maxwellian with no flows.
 The first-order equation takes the form of a Spitzer-H\"{a}rm problem \cite{SpitzerHarmPhysRev.89.977,BraginskiiTransport,helander}.
 Physically, the first-order terms control the self-consistent parallel flows that
 result from the leading-order perturbations. The second-order equation
 governs the time evolution of the leading-order fluctuations. 
 A poloidal angle average
    \beqn \lparav{\cdot} = \int^{\pi}_{-\pi} \left(\cdot\right) \; \frac{d \lpar}{\bvec \cdot \nbl \lpar}{\left/\int^{\pi}_{-\pi}  \frac{d \lpar}{\bvec \cdot \nbl \lpar}\right.} \label{eq:lparav} \eeqn 
 of the density and temperature velocity
 moments of the second-order equation yield transport equations for the electron
 density and temperature fluctuations, closing the system of equations. 
  In this section, we give the forms of the transport equations in the $\massrtp{1/4} \rightarrow 0$
  limit, with $\fract{\saffac \rmajorz \cfreqee}{  \vthere} \sim 1$. 
  The full details of the calculation are contained in \ref{sec:electron-inner-collisional-appendix}.

  The leading order equation for the electron response in the inner region
  is a balance between parallel streaming in the periodic coordinate $\lpar$ and collisions:
    \beqn \vpar \kpar \drv{\HHez}{\lpar}  = \coplandaue\left[\HHez\right] + \coplorentze\left[\HHez\right]
 \label{eq:electron_inner0-collisional}.\eeqn
    To solve equation \refeq{eq:electron_inner0-collisional}, in \ref{sec:electron-inner-collisional-appendix}
    we follow the standard H-theorem procedure \cite{hintonRMP76,helander} to prove that $\HHez$
    is a perturbed Maxwellian with no flow, i.e.,
    \beqn \frac{\HHez}{\eqlbe} = \frac{\ddensns{0}{e}}{\dense}
    + \frac{\dtempns{0}{e}}{\tempe}\left(\frac{\energy}{\tempe}
    - \frac{3}{2}\right), \label{eq:HHez-collisional}\eeqn 
    where the nonadiabatic density $\ddensns{0}{e}$ and temperature $ \dtempns{0}{e}$
    are constant in $\lpar$, i.e.,
    \beqn \ddensns{0}{e} = \ddensns{0}{e}(\lchi), 
    \quad {\rm{and}} \quad \dtempns{0}{e} =  \dtempns{0}{e}(\lchi). \label{eq:denstemplchi}\eeqn
    To obtain evolution equations for $\ddensns{0}{e}$
    and $\dtempns{0}{e}$, in \ref{sec:electron-inner-collisional-appendix} we go to second order in the expansion in $\massrtp{1/4}$. 
    
    Before writing down the transport equations, we consider the collisional
    inner-region quasineutrality relation. Using the 
    the ordering \refeq{eq:inner_ion2-collisional},
    and the solution \refeq{eq:HHez-collisional}, with $\besen{0} = 1 + \order{\massrt}$ and $\expo{\imag \lflre(\lchi-\thetaz)} = 1 + \order{\massrtp{1/4}}$
    for $\lchi \sim \massrutp{1/4}$ and $\kky \gyrdi\sim 1$,
    we find that the leading-order quasineutrality relation is
    \beqn \left(\frac{\zedi \tempe}{\tempi} + 1\right) \frac{\charge \ptlz}{\tempe} 
    = - \intv{\frac{\HHez}{\dense}} = - \frac{\ddensns{0}{e}}{\dense}.
    \label{eq:quasineutrality-inner-collisional} \eeqn 
    Equation \refeq{eq:quasineutrality-inner-collisional}
    allows us to note that the electrostatic potential in the inner region
    is not a function of geometric angle $\lpar$,
    i.e., $\ptlz = \ptlz(\lchi)$.
    This is a significant simplification over
    the collisionless case (cf. equation \refeq{eq:qninner}), where 
    $\ptlz = \ptlz(\lpar,\lchi)$. This simplification arises
    in the collisional limit because, first, there is
    no distinction between trapped and passing particles; and second,
    the extent of the mode is shortened to $\lchi \sim \massrutp{1/4}$,
    meaning that finite-Larmor-radius and finite-orbit-width effects do not enter at leading order. 
    
    The final transport equations for $\ddensns{0}{e}$ and $\dtempns{0}{e}$ are 
    are most clearly written in a form where
    the terms admit simple physical interpretations. We have a continuity equation 
       \beqn \fl  \lparav{\kpar}\drv{\dupareeff}{\lchi}
    + \imag \lparav{\wmageprecth}\left(\frac{\ddensns{0}{e}}{\dense} + \frac{\dtempns{0}{e}}{\tempe}\right)
       -\imag \wfreqz \frac{\ddensns{0}{e}}{\dense} 
    - \imag \kky \shat \kxfac \lchi \left(\frac{\dpartc}{\dense} + \frac{\dpartnc}{\dense}\right)
    \label{eq:electron_inner-dens0-collisional} \eeqn \beqn 
     = - \imag (\wstaren - \wfreqz) \frac{\charge \ptlz}{\tempe}
    , \nonumber \eeqn
   and a temperature equation  
 \beqn \fl   \lparav{\kpar}\drv{}{\lchi}\left(\frac{\dqpareeff}{\dense\tempe} + \dupareeff\right)     
    + \imag \lparav{\wmageprecth}\left(\frac{\ddensns{0}{e}}{\dense} + \frac{7}{2}\frac{\dtempns{0}{e}}{\tempe}\right)        
        - \imag  \frac{3}{2}  \wfreqz \frac{ \dtempns{0}{e}}{\tempe} 
       \nonumber  \eeqn \beqn \label{eq:electron_inner-temp0-collisional}
    - \imag \kky \shat \kxfac \lchi \left(\frac{\dpartc}{\dense} + \frac{\dheatc}{\dense \tempe} + \frac{\dpartnc}{\dense} + \frac{\dheatnc}{\dense \tempe}\right) 
     = - \imag \frac{3}{2} \wstaren \etae \frac{\charge \ptlz}{\tempe}
   . \eeqn
   The physical interpretations of the terms
   in equations \refeq{eq:electron_inner-dens0-collisional}   
   and \refeq{eq:electron_inner-temp0-collisional}
   are the following, from left to right: parallel diffusion,
   magnetic (precession) drifts within the flux surface,
   time evolution, classical perpendicular diffusion,
   neoclassical perpendicular diffusion,
   and drives by equilibrium gradients. To write equations \refeq{eq:electron_inner-dens0-collisional}
   and \refeq{eq:electron_inner-temp0-collisional}, we have defined
     the effective parallel velocity and  effective parallel heat flux,
    \beqn \dupareeff = \frac{1}{\lparav{\kpar}}\lparav{ \frac{\kpar}{\dense}\intv{\vpar (\HHeh + \imag \lflre \lchi \HHez)}}
      \label{eq:upareeff},\eeqn     
    and 
    \beqn \fl \dqpareeff = \frac{1}{\lparav{\kpar}}\lparav{\kpar\intv{\vpar 
    \left(\energy -\frac{5\tempe}{2}\right)(\HHeh + \imag \lflre \lchi \HHez)}}
      \label{eq:qpareeff},\eeqn     
    respectively,   
    the thermal magnetic precession drift
    \beqn \wmageprecth = \frac{\kkfldl \vthere^2}{2\cycfe}
    \bu \xp \left(\bu \cdot \nbl \bu + \frac{\nbl \bmag}{\bmag} \right)
    \cdot \left(\nbl \fldl + \lpar\nbl\saffac \right) \label{eq:precessiondrift-thermal-proper},\eeqn
    the fluctuating perpendicular fluxes: the classical particle flux
    \beqn \dpartc = \imag \lparav{\intv{\nbl\radial \cdot \gyrdvece \; 
    \copbothe\left[\kperpvecz\cdot\gyrdvece\; \HHez\right]}},
    \label{eq:classical-particle-flux} \eeqn
    the classical heat flux
    \beqn \dheatc = \imag \lparav{\intv{\left(\energy - \frac{5\tempe}{2}\right)\nbl\radial \cdot \gyrdvece \; 
    \copbothe\left[\kperpvecz\cdot\gyrdvece\; \HHez\right]}},
    \label{eq:classical-heat-flux} \eeqn
    the neoclassical particle flux
    \beqn \dpartnc = -\lparav{\frac{\bcur}{\cycfe}\tdrv{\radial}{\flxl}\intv{\vpar \; 
    \copbothe\left[\HHep{1/2} + \imag \lflre\lchi\HHez - \HHeSH\right]}},
    \label{eq:neoclassical-particle-flux} \eeqn
    and the neoclassical heat flux 
    \beqn \fl \dheatnc = -\lparav{\frac{\bcur}{\cycfe}\tdrv{\radial}{\flxl}\intv{\left(\energy - \frac{5\tempe}{2}\right) \vpar \; 
    \copbothe\left[\HHep{1/2} + \imag \lflre\lchi\HHez - \HHeSH\right]}}
    \label{eq:neoclassical-heat-flux}. \eeqn
    To obtain $\dupareeff$, $\dqpareeff$, $\dpartnc$ and $\dheatnc$ we require the $\massrtp{1/4}$ small distribution functions $\HHeSH$ and 
        $\HHeh$. The distribution function $\HHeSH$  is determined by solving the Spitzer-H\"{a}rm problem \cite{SpitzerHarmPhysRev.89.977,BraginskiiTransport,helander}
    \beqn \vpar \kpar \drv{\HHez}{\lchi} = \copbothe[\HHeSH] 
    \label{eq:electron_inner1a-collisional},     \eeqn
    whereas $\HHeh$ is determined by solving the first-order electron equation \beqn 
     \vpar \kpar \drv{\HHeh}{\lpar}=
     \copbothe\left[\HHeh + \imag \lflre \lchi \HHez - \HHeSH\right]
    . \label{eq:electron_inner1b-collisional}\eeqn
     In general, equation \refeq{eq:electron_inner1b-collisional} is not solvable analytically. 
    To maximise the physical insight from the calculation, we subsequently solve
    equation \refeq{eq:electron_inner1b-collisional} in the subsidiary limits
    of large and small collisionality.

    The classical fluxes $\dpartc$ and $\dheatc$ are due to Larmor orbits being interupted by collisions, 
    and can be evaluated for arbitrary $\fract{\saffac \rmajorz \cfreqee}{  \vthere}$ \cite{helander,BraginskiiTransport}.
    We use the results of \ref{sec:perpendicular-collisional-contributions} to
    write down the classical particle flux $\dpartc$ and classical heat flux $\dheatc$.
    We use result \refeq{eq:lorentz-terms1} to find that
    \beqn \frac{\dpartc}{\dense} =  \imag \kky \shat \kxfac \lchi \frac{\cfreqei \gyrdeovl^2}{2} 
    \lparav{\bmagovl^2\frac{|\nbl \radial|^2}{\bmag^2}}
    \left(\frac{\ddensns{0}{e}}{\dense}- \frac{1}{2}\frac{\dtempns{0}{e}}{\tempe}\right),\eeqn 
    where we have used that $\kperpvecz = - \kkfldl (\tdrvt{\saffac}{\radial}) \lchi \nbl \radial = - \kky \shat \kxfac \lchi \nbl \radial $, with
     $\gyrdeovl = \vthere / \cycfeovl$,  $\cycfeovl = - \charge \bmagovl/\me \ltsp$,
     and $\bmagovl = \lparav{\bmag}$. Similarly, we use the results \refeq{eq:lorentz-terms2}
     and \refeq{eq:landau-terms0final} to find that  
    \beqn \fl \frac{\dheatc}{\dense\tempe} =  \imag \kky \shat \kxfac\lchi \frac{\cfreqei \gyrdeovl^2}{2} 
    \lparav{\bmagovl^2\frac{|\nbl \radial|^2}{\bmag^2}}
    \left(\left(\frac{7}{4} + \frac{\sqrt{2}}{\zedi}\right)\frac{\dtempns{0}{e}}{\tempe} - \frac{3}{2}\frac{\ddensns{0}{e}}{\dense} \right),\eeqn 
    where we have used that $\cfreqee/\cfreqei = 1/ \zedi$.    
    
    The mode evolution equations for the density and the temperature,
    equations \refeq{eq:electron_inner-dens0-collisional} and  \refeq{eq:electron_inner-temp0-collisional},
    respectively, have the promised structure: 
    the envelope of the mode is controlled by the combination of the
    finite-orbit-width and finite-Larmor-radius perpendicular diffusion, and parallel diffusion.
    The perpendicular diffusion terms scale as
    $\cfreqei (\kky \gyrde)^2 \lchi^2$,
    whereas equations \refeq{eq:electron_inner1b-collisional}, \refeq{eq:upareeff} and \refeq{eq:qpareeff}
    show implicitly that the parallel diffusion terms
    scale as $(\vthere^2/\cfreqei \saffac^2 \rmajorz^2)\drvsqt{}{\lchi}$.
    This result justifies the initial ordering \refeq{eq:collisional-lchi} and
    the discussion in section \ref{section:innersolution-collisional}.
    To obtain explicit analytical forms for
    all terms in the transport equations \refeq{eq:electron_inner-dens0-collisional}
    and  \refeq{eq:electron_inner-temp0-collisional}, we consider
    the $\fract{\saffac \rmajorz \cfreqee}{  \vthere} \gg 1$ (Pfirsh-Schl\"{u}ter) regime in the next section. 
    To demonstrate the transition between the collisionless and collisional regimes,
    we consider the $\fract{\saffac \rmajorz \cfreqee}{  \vthere}  \ll 1$ (banana-plateau)
    regime in section \ref{sec:banana-and-collisional-collisionless-matching}.
         
    \subsubsection{Parallel flows and perpendicular diffusion
    in the subsidiary limit of $\fract{\saffac \rmajorz \cfreqee}{  \vthere} \gg 1$
    -- the Pfirsh-Schl\"{u}ter regime.} \label{sec:pfirsh-schluter}
    In order to obtain the analytical form of the transport equations
    in the subsidiary limit $\fract{\saffac \rmajorz \cfreqee}{  \vthere} \gg 1$,
    we must solve equation \refeq{eq:electron_inner1b-collisional}
    to obtain approximate solutions for $\HHeh$. 
    Using the results of \ref{sec:pfirsh-schluter-parallel-collisional-contributions},
    we can write down the effective
    parallel velocity, parallel heat flux, and perpendicular diffusion terms that
    appear in the transport equations \refeq{eq:electron_inner-dens0-collisional} and 
    \refeq{eq:electron_inner-temp0-collisional}
    in the $\fract{\saffac \rmajorz \cfreqee}{  \vthere} \gg 1$ limit.
    We find that 
    \beqn \fl
   \frac{\dupareeff}{\vthere} = -\frac{\vthere}{2\cfreqei}
    \frac{(\lparav{\bvec\cdot\nbl\lpar})^2}{\lparav{\kpar}\lparav{\bmag^2}}
    \left[
    1.97\drv{}{\lchi}\left(\frac{\ddensns{0}{e}}{\dense}\right)
    +3.37\drv{}{\lchi}\left(\frac{\dtempns{0}{e}}{\tempe}\right)\right]
    \label{eq:dupareeff-pfirsh-schluter}
    \eeqn \beqn 
    + \frac{\imag}{2} \frac{   \kky\gyrdeovl \kxfac \shat  \lchi}{\lparav{\kpar}}\bmagovl\bcur \tdrv{\radial}{\flxl}
    \left(\lparav{\frac{\bvec\cdot\nbl\lpar}{\bmag^2}} -
    \frac{\lparav{\bvec\cdot\nbl\lpar}}{\lparav{\bmag^2}}\right)
    \left(\frac{\ddensns{0}{e}}{\dense}+\frac{\dtempns{0}{e}}{\tempe} \right)
    , \nonumber\eeqn
    where we have used equation \refeq{eq:dupareeff-step-last}, with the numerical results 
    \refeq{eq:acoeffn} and \refeq{eq:bcoeffn} for the transport coefficients, assuming $\zedi = 1$.
    Similarly, using  \refeq{eq:dqpareeff-step-last}, we obtain the effective electron parallel heat flux
   \beqn \fl
   \frac{\dqpareeff}{\dense \tempe\vthere}
   = -\frac{5\vthere}{4\cfreqei} 
     \frac{(\lparav{\bvec\cdot\nbl\lpar})^2}{\lparav{\kpar}\lparav{\bmag^2}}
    \left[
    0.56\drv{}{\lchi}\left(\frac{\ddensns{0}{e}}{\dense}\right)
    +2.23\drv{}{\lchi}\left(\frac{\dtempns{0}{e}}{\tempe}\right)\right]
    \label{eq:dqpareeff-pfirsh-schluter}
    \eeqn \beqn 
    + \frac{5\imag}{4}
    \frac{  \kky\gyrdeovl \kxfac \shat  \lchi}{\lparav{\kpar}}\bmagovl\bcur \tdrv{\radial}{\flxl}
    \left(\lparav{\frac{\bvec\cdot\nbl\lpar}{\bmag^2}} -
    \frac{\lparav{\bvec\cdot\nbl\lpar}}{\lparav{\bmag^2}}\right)
   \frac{\dtempns{0}{e}}{\tempe}
     \nonumber.\eeqn
    We note that the terms linear in $\lchi$ in equations \refeq{eq:dupareeff-pfirsh-schluter}
    and \refeq{eq:dqpareeff-pfirsh-schluter} arise from the radial magnetic drift, whereas the terms
    in $\drvt{}{\lchi}$  arise from the effective electric field generated
    by the leading-order electron response (cf. equation \refeq{eq:electron_inner1-collisional}).

    The neoclassical particle flux $\dpartnc$ appearing in the nonadiabatic
    density transport equation, equation \refeq{eq:electron_inner-dens0-collisional},
    can be evaluated using the result
    \refeq{eq:density-neoclassical-perpendicular-diffusion-pfirsh-schluter-step-last}. 
    We find that 
\beqn \fl \frac{\dpartnc}{\dense}\nonumber
   = - \frac{\vthere \gyrdeovl}{2 } \bcur\tdrv{\radial}{\flxl}
 \bmagovl  \left( \frac{\lparav{\bvec\cdot\nbl\lpar}}{\lparav{\bmag^2}}
 - \lparav{\frac{\bvec\cdot\nbl\lpar}{\bmag^2}}\right)
 \left(\drv{}{\lchi}\left(\frac{\ddensns{0}{e}}{\dense}\right)+
   \drv{}{\lchi}\left(\frac{\dtempns{0}{e}}{\tempe}\right)\right)
 \eeqn \beqn \fl 
 +\imag \kky\shat\kxfac \lchi \frac{\cfreqei \gyrdeovl^2}{2 } \left(\bcur\tdrv{\radial}{\flxl}\right)^2
\left( \lparav{\frac{\bmagovl^2}{\bmag^2}}
 - \frac{\bmagovl^2}{\lparav{\bmag^2}}\right)
\left[0.67\frac{\ddensns{0}{e}}{\dense}
 + 0.11\frac{\dtempns{0}{e}}{\tempe}\right] 
 \label{eq:density-neoclassical-perpendicular-diffusion-pfirsh-schluter}
 .\eeqn    
    Similarly, the neoclassical heat flux $\dheatnc$ appearing in the temperature
    transport equation, equation \refeq{eq:electron_inner-temp0-collisional},
    can be evaluated using the result
    \refeq{eq:temperature-neoclassical-perpendicular-diffusion-pfirsh-schluter-step-last}. 
    We find that  
    \beqn \fl \frac{\dheatnc}{\dense \tempe}
     \label{eq:temperature-neoclassical-perpendicular-diffusion-pfirsh-schluter}
  = - \frac{5\vthere \gyrdeovl}{4} \bcur\tdrv{\radial}{\flxl}
 \bmagovl\left( \frac{\lparav{\bvec\cdot\nbl\lpar}}{\lparav{\bmag^2}}
 - \lparav{\frac{\bvec\cdot\nbl\lpar}{\bmag^2}}\right)
\drv{}{\lchi}\left(\frac{\dtempns{0}{e}}{\tempe}\right)
 \eeqn \beqn \fl +\imag \kky\shat\kxfac \lchi \frac{\cfreqei\gyrdeovl^2}{2 } \left(\bcur \tdrv{\radial}{\flxl}\right)^2
\left( \lparav{\frac{\bmagovl^2}{\bmag^2}}
 - \frac{\bmagovl^2}{\lparav{\bmag^2}}\right)
\left[ 1.41\frac{\dtempns{0}{e}}{\tempe} 
 - 0.56 \frac{\ddensns{0}{e}}{\dense}\right] 
  .\nonumber\eeqn 
 Physically, equations \refeq{eq:density-neoclassical-perpendicular-diffusion-pfirsh-schluter}
 and \refeq{eq:temperature-neoclassical-perpendicular-diffusion-pfirsh-schluter}
 indicate that diffusive transport
 arises from the radial magnetic drift (note the terms linear in $\lchi$). 

 We note that the scale of the extended tail, $\lchi$,
 decreases with increasing $\saffac \rmajorz \cfreqee / \vthere$.
 This is explicit in the estimate \refeq{eq:collisional-lchi}.
 Using \refeq{eq:collisional-lchi}, we can see 
 that, for extreme collision frequencies where 
 $\saffac \rmajorz \cfreqee / \vthere \sim \massrut$, there is 
 no separation between the scale of the electron tail
 and the scale of the geometric quantities: for such an extreme collisionality,
 $\lchi \sim 1$.
 The fluid equations for this extreme regime are not examined in this paper.

    \subsubsection{The subsidiary limit of $\fract{\saffac \rmajorz \cfreqee}{  \vthere} \ll 1$ -- the banana-plateau regime.}
    \label{sec:banana-and-collisional-collisionless-matching}
    We now examine equation \refeq{eq:electron_inner1b-collisional} in
    the subsidiary limit $\fract{\saffac \rmajorz \cfreqee}{  \vthere} \ll 1$.
    This discussion will enable us to demonstrate the smooth transition between the
    collisionless and collisional regimes. 
    We will need to go to first-order in the subsidiary
    expansion of $\fract{\saffac \rmajorz \cfreqee}{  \vthere} \ll 1$, and so we expand 
     \beqn {\HHeh = \HHepq{1/2}{0} + \HHepq{1/2}{1} + \order{ \left(\frac{\saffac \rmajorz \cfreqee}{  \vthere}\right)^2\HHepq{1/2}{0}}},
     \label{eq:HHeh-nustar-ll-1}\eeqn
    where $\HHepq{1/2}{0} \sim \imag \lflre\lchi\HHez \sim \HHeSH$ and
    $\HHepq{1/2}{n} \sim \left(\fract{\saffac \rmajorz \cfreqee}{  \vthere}\right)^n \HHepq{1/2}{0}$.
    The leading-order form of equation \refeq{eq:electron_inner1b-collisional} is    
     \beqn
     \vpar \kpar \drv{}{\lpar}\left(\HHepq{1/2}{0}\right) = 0
    , \label{eq:electron-banana-1}\eeqn
    i.e., we learn that $\HHepq{1/2}{0}= \HHepq{1/2}{0}(\lchi,\energy,\lambda,\sign)$.
    Going to first-order terms in the expansion of the drift-kinetic
    equation \refeq{eq:electron_inner1b-collisional}, we find that
         \beqn 
     \vpar \kpar \drv{}{\lpar}\left(\HHepq{1/2}{1}\right)
    = \copbothe\left[\HHepq{1/2}{0} + \imag \lflre \lchi \HHez - \HHeSH \right]
    . \label{eq:electron-banana-2}\eeqn
    We now impose the solvability condition that
    $\HHepq{1/2}{1}(\lpar,\lchi,\energy,\pitch,\sign)$ should be $2\pi$-periodic in $\lpar$. 
    We must treat the passing and trapped part of the velocity space independently.
    For passing particles we apply the transit average $\transav{\cdot}$,
    defined in equation \refeq{eq:transitav}, to obtain
    \beqn \transav{\copbothe\left[\HHepq{1/2}{0}+ \imag \lflre \lchi \HHez - \HHeSH \right]}
    = 0 \label{eq:electron-banana-passing}. \eeqn
    We note that equation \refeq{eq:electron-banana-passing} is 
    a partial differential equation in $(\energy,\pitch)$ at fixed $\lchi$.
    For trapped particles we apply the bounce average $\bouncav{\cdot}$,
    defined in equation \refeq{eq:bounceav}, to obtain
    \beqn \bouncav{\copbothe\left[\HHepq{1/2}{0}\right]}
    = 0, \label{eq:electron-banana-trapped} \eeqn
    where we have used that $\imag \lflre \lchi \HHez$ and $\HHeSH$
    are odd in $\sign = \vpar/|\vpar|$, and therefore vanish under $\bouncav{\cdot}$.
    The trapped particle bounce condition requires that
    \beqn \HHeh(\lparpm,\sign = 1) = \HHeh(\lparpm,\sign = -1), \nonumber \eeqn
    and hence $\HHepq{1/2}{0}$
    is even in $\sign$, by virtue of being constant in $\lpar$.
    In contrast, we can see from equation \refeq{eq:electron-banana-passing} that 
    the passing particle response must be odd in $\sign$.
    A Maxwellian solution to equation \refeq{eq:electron-banana-trapped} is not valid,
    because of the change in the $\sign$ symmetry of $\HHepq{1/2}{0}$ at the trapped-passing boundary,
    and hence we must have that $\HHepq{1/2}{0}=0$ for trapped particles.
    To obtain $\HHepq{1/2}{0}$ for passing particles,
    we must solve equation \refeq{eq:electron-banana-passing}
    subject to continuity in $\HHepq{1/2}{0}$ at the trapped-passing boundary. 
    
    In order to make progress analytically, it is necessary to expand in
    inverse aspect ratio $\aspect = \rminor/\rmajorz \ll 1$, where 
     $\rminor$ is the minor radial coordinate of the flux surface of interest.
    We assume that the normalised collisionality
    \beqn \nustar = \frac{\saffac \rmajorz \cfreqee }{\aspect^{3/2}\vthere } \ll 1, \label{eq:nustar} \eeqn
    and assume that the
    equilibrium can be approximated by the solution with
    circular flux surfaces \cite{HazeltineMeiss,FreidbergMHD}.
    Then, we can use the techniques of neoclassical theory \cite{hintonRMP76,helander}
    to obtain $\HHepq{1/2}{0}$ to leading-order in $\aspect$, and  the velocity $\dupareeff$ and
    flux $\dqpareeff$, and the neoclassical perpendicular diffusion terms
    to order $\aspect^{1/2}$. These calculations are
    performed in \ref{sec:banana-parallel-collisional-contributions}.
    We conclude that for $\nustar \ll 1$
    the electron parallel velocity and electron parallel heat flux have a diffusive character.
    
    Finally, we comment on the smooth transition between the equations for the electron response in the
    collisionless and the collisional regimes.
    To obtain the mode transport equations
    \refeq{eq:electron_inner-dens0-collisional} and \refeq{eq:electron_inner-temp0-collisional} from the equations
    in the collisionless limit for the passing electron response,
    equation \refeq{eq:electron_firstorder_passing},
    and the trapped electron response, equation \refeq{eq:electron_firstorder_trapped},
    we take the following steps: First, in equations \refeq{eq:electron_firstorder_passing}
    and \refeq{eq:electron_firstorder_trapped}, we take the electron
    collision frequency to be large compared to the ion transit frequency, i.e.,
    $ \saffac \rmajorz \cfreqee / \vtheri \gg  1$, 
    and we take the extent of the ballooning mode to be small, with
    \beqn 1 \ll \lchi \sim
    \left(\frac{\saffac \rmajorz \cfreqee}{ \vtheri}\right)^{-1/2} \massru \ll \massru.
    \label{eq:mode-extent-collisional-from-collisionless}\eeqn 
    Then, the leading-order equation for the electron response is 
    \beqn \copbothe\left[\HHez\right] = 0, \label{eq:dfn-collisional-from-collisionless} \eeqn
    i.e., $\HHez$ is a perturbed Maxwellian with no flow, and with no dependence on $\lpar$.
    Second, we collect terms of $\order{(\saffac \rmajorz \cfreqee / \vtheri)^{-1/2}}$
    in the subsidiary expansion, and obtain equations
    for the passing and trapped electron response of the form
    \refeq{eq:electron-banana-passing} and \refeq{eq:electron-banana-trapped}, respectively.
    Finally, we collect terms of $\order{(\saffac \rmajorz \cfreqee / \vtheri)^{-1}}$
    in the subsidiary expansion and obtain the transport equations 
    for the nonadiabatic density and temperature, equations
    \refeq{eq:electron_inner-dens0-collisional}
    and \refeq{eq:electron_inner-temp0-collisional}, respectively. 
    The fact that the extent of the mode shortens when going
    from the collisionless to the collisional limits,
    according to the ordering \refeq{eq:mode-extent-collisional-from-collisionless},
    along with the Maxwellianisation of the distribution function by increasing interparticle collisions,
    cf. equation \refeq{eq:dfn-collisional-from-collisionless},
    ensures that the collisionless inner-region quasineutrality relation \refeq{eq:qninner}
    takes the form of the collisional inner-region
    quasineutrality relation \refeq{eq:quasineutrality-inner-collisional}.
    
  \subsection {Collisional outer solution -- $\lpar \sim 1$ -- $\kkr \gyrde \sim \massrtp{1/2} $}
\label {section:outersolution-collisional}
  Just as in the collisionless case, the class of mode that we obtain in the collisional limit 
  depends on the matching condition that we use to solve for the electron response in the inner 
  region ($\lpar \gg 1$) via equations \refeq{eq:electron_inner-dens0-collisional}
  and \refeq{eq:electron_inner-temp0-collisional}. We obtain the matching conditions by considering the
  outer region ($\lpar \sim 1$), and expanding in powers of $\massrtp{1/4}$, for consistency with the expansion in the inner region.

    For the ions in the collisional ordering, we take $\cfreqii \sim \vtheri / \saffac \rmajor$.
    As the electron mass does not appear in the ion gyrokinetic equation, no approximations
    are possible in this ordering and the gyrokinetic equation for the ions in the outer region is simply
    equation \refeq{eq:lineargyrokinetic} with $\spe = {\rm{i}}$.
    The nonadiabatic response
    of ions $\hhi$ contributes at leading-order to the potential $\ptl$ in the outer region.
    As in the collisionless case (see equation \refeq{eq:ion-density-outer}),
    the estimate for the size of the ion nonadiabatic density
    is $\ddensi /\densi \sim \charge \ptl/\tempe$.

    \subsection{Electron response in the outer region for small-tail modes.} \label{sec:electron-outer-collisional-small-tail}
    In a collisional small-tail mode, the fluctuations must satisfy the ordering
    \beqn \frac{\HHeinner}{\eqlbe} \sim \frac{\charge \ptlinner}{\tempe}
     \sim \frac{\HHeouter}{\eqlbe}\ll \frac{\charge\ptlouter}{\tempe}, \label{eq:small-tail-ordering} \eeqn
    so that the nonadiabatic electron response is subdominant to the nonadiabatic
    ion response in the outer region. This ordering will recover the ITG mode.
    
    In this section, we present the matching condition necessary to
    solve for the electron response in modes that obey the ordering \refeq{eq:small-tail-ordering}.
    The details of the calculation of the matching condition are contained in
    \ref{sec:electron-outer-collisional-small-tail-appendix}.
    At leading order, we find that the electron distribution function
    is a Maxwellian with no flow.
    At first order, we find that the electron parallel flows are determined by the potential 
    generated by ions. To match the solutions in the outer and inner regions, 
    we note that in the inner region,
    the electron flows are $\massrtp{1/4}$ smaller than the density
    and temperature components of the electron response, i.e.,
  \beqn \frac{\dupareeffinner}{\vthere} \sim \frac{\dqpareeffinner}{\vthere \dense \tempe}
  \sim \massrp{1/4} \frac{\ddensns{0}{e,inner}}{\dense}\sim \massrp{1/4} \frac{\dtempns{0}{e,inner}}{\tempe}. \label{eq:upare-qpare-estimate} \eeqn
    This must be true in the outer
    solution for the solutions to be matched.
    The size of $\dupareeffinner$ and $\dqpareeffinner$ are set by the jump in
    the electron parallel flows across the outer region due to the
    presence of the electrostatic potential $\ptlouter$. In terms of estimates, we have that
    \beqn \left[\frac{\dupareouter}{\vthere}\right]^{\lpar = \infty}_{\lpar = -\infty}
    \sim \left[\frac{\dqpareouter}{\vthere \dense \tempe}\right]^{\lpar = \infty}_{\lpar = -\infty}
    \sim \massr  \frac{\charge\ptlouter}{\tempe}.  \label{eq:electron-flow-jump}  \eeqn
    Combining estimates  \refeq{eq:upare-qpare-estimate} and \refeq{eq:electron-flow-jump}
    with a demand that the electron distribution function
   is continuous across the boundary of the outer and inner regions, we find an estimate
  for the size of the fluctuations in the inner region:
  \beqn  \frac{\charge\ptlzinner}{\tempe} \sim \frac{\ddensns{0}{e,inner}}{\dense}
  \sim \frac{\dtempns{0}{e,inner}}{\tempe}
  \sim \massrp{1/4} \frac{\charge\ptlzouter}{\tempe}\label{eq:small-tails-estimate}.\eeqn

In appendix \ref{sec:electron-outer-collisional-small-tail-appendix},
 we show that the above arguments lead to the following matching conditions for
 $\ddensns{0}{e,inner}$, $\dtempns{0}{e,inner}$, $\dupareeffinner$, and $\dqpareeffinner$.
 We have continuity for the density and temperature fluctuations, i.e.,
   \beqn \ddensns{0}{e,inner}   (\lchi =0^+) = \ddensns{0}{e,inner}   (\lchi =0^-), \label{eq:density-matching-collisional} \eeqn
    and 
    \beqn \dtempns{0}{e,inner} (\lchi =0^+)=  \dtempns{0}{e,inner} (\lchi =0^-). \label{eq:temperature-matching-collisional}\eeqn
  The small outer region electron density and temperature are set by
  $\ddensns{1/2}{e,outer} = \ddensns{0}{e,inner}   (\lchi =0)$
  and $\dtempns{1/2}{e,outer} = \dtempns{0}{e,inner}   (\lchi =0)$.
  For the effective parallel velocity and heat flux, we have the jump conditions  
   \beqn \Big[\dupareeffinner\Big]^{\lchi=0^{+}}_{\lchi=0^{-}} = -\imag (\wstaren - \wfreqz)
    \frac{\lparav{\bvec \cdot \nbl \lpar}}{\lparav{\kpar}}\int^{\infty}_{-\infty}
   \frac{\charge \ptlzouter(\lpar)}{\tempe} \frac{d \lpar}{\bvec \cdot \nbl \lpar},
\label{eq:upare-jump-small-tail}   \eeqn 
    and 
    \beqn \fl \left[\frac{\dqpareeffinner}{\dense\tempe}\right]^{\lchi=0^{+}}_{\lchi=0^{-}} = -\imag {\left(\frac{3}{2}\wstaren \etae
        - \wstaren + \wfreqz\right)}
    \frac{\lparav{\bvec \cdot \nbl \lpar}}{\lparav{\kpar}}\int^{\infty}_{-\infty}
   \frac{\charge \ptlzouter(\lpar)}{\tempe} \frac{d \lpar}{\bvec \cdot \nbl \lpar}.
\label{eq:qpare-jump-small-tail}   \eeqn 
   
    \begin{figure}
\begin{center}
\includegraphics[clip, trim=0cm 0cm 0cm 0cm, width=0.8\textwidth]{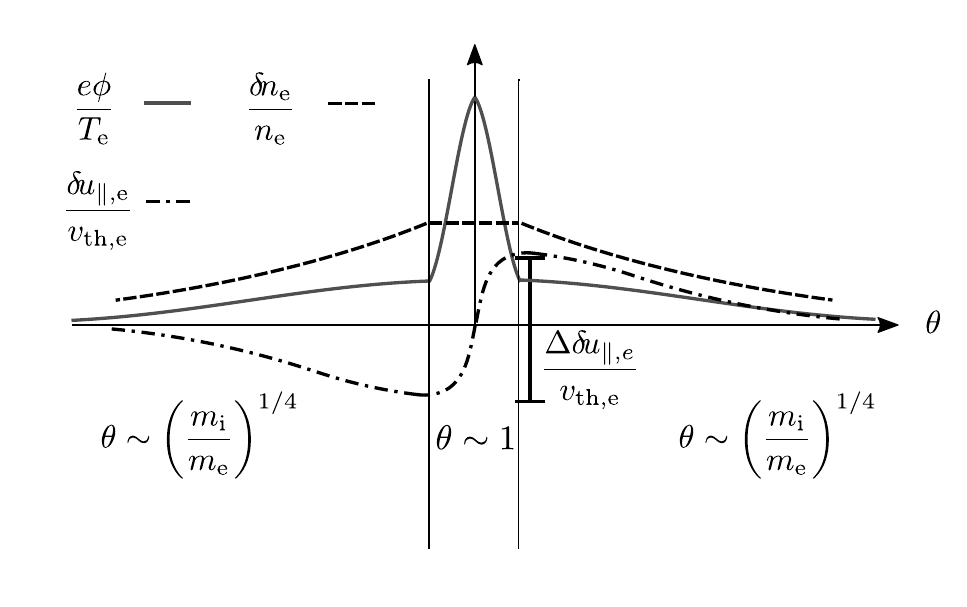}
% trim=left bottom right top, clip
\caption{An illustration showing the
 nonadiabatic electron density $\ddense$ and electron mean velocity $\dupare$
 in the collisional, small-tail limit. 
 The leading-order mode frequency is determined by ions 
 in the $\lpar \sim 1$ (outer) region, by solving equation \refeq{eq:lineargyrokinetic}
  (with $\spe = {\rm{i}}$) subject to quasineutrality, equation \refeq{eq:small-tail-outer-quasineutrality}.
 The electron tails at $\lpar \sim \massrutp{1/4}$ are obtained by solving the transport equations  
 \refeq{eq:electron_inner-dens0-collisional}
    and  \refeq{eq:electron_inner-temp0-collisional}, with inner-region quasineutrality 
 \refeq{eq:quasineutrality-inner-collisional} and the boundary conditions 
 \refeq{eq:density-matching-collisional} - \refeq{eq:qpare-jump-small-tail}.
 From the perspective of the $\lpar \sim \massrutp{1/4} $
 region, the electron density is a cusp, set up by the discontinuity in $\dupare$, $\Delta \dupare$. 
 } \label{fig:collisional_small_tail_modes}
\end{center}
\end{figure}
  
  Finally, we can describe the procedure for solving for the small-tail
  mode in the $\massrt\rightarrow 0$ limit. To determine the frequency $\wfreqz$
  and the potential $\ptlzouter$, we solve the ion gyrokinetic equation \refeq{eq:lineargyrokinetic}
  with $\spe = {\rm{i}}$,
  closed {by} the quasineutrality relation (neglecting the electron nonadiabatic response)
  \beqn \left(\frac{\zedi \tempe}{\tempi} + 1\right)\frac{\charge\ptlzouter}{\tempe} 
  = \intv{\besin{0} \frac{\hhizouter}{\densi}}. \label{eq:small-tail-outer-quasineutrality}\eeqn 
  With $\wfreqz$ and $\ptlzouter$ determined, we can solve for
  the electron response using equations  \refeq{eq:electron_inner-dens0-collisional}
    and  \refeq{eq:electron_inner-temp0-collisional}, with the inner-region quasineutrality 
    equation \refeq{eq:quasineutrality-inner-collisional}. The causal link between the solution
    in the outer region and the inner region is provided by boundary matching conditions
    \refeq{eq:density-matching-collisional} - \refeq{eq:qpare-jump-small-tail}.
    An illustration demonstrating the matching in the collisional
    small-tail mode is given in figure \ref{fig:collisional_small_tail_modes}.

  \subsection{Electron response in the outer region for large-tail modes.} \label{sec:electron-outer-collisional-large-tail}
  
  \begin{figure}
\begin{center}
\includegraphics[clip, trim=0cm 0cm 0cm 0cm, width=0.8\textwidth]{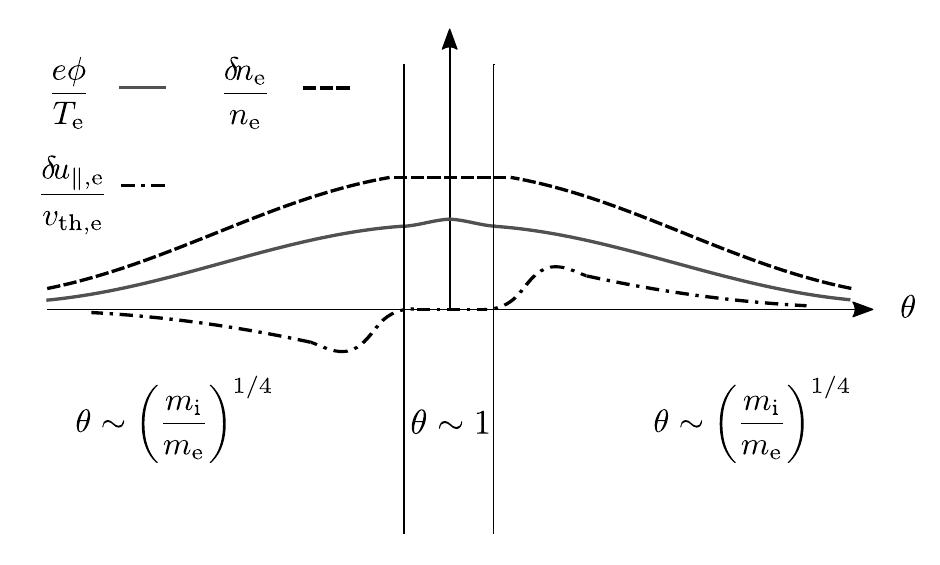}
% trim=left bottom right top, clip
\caption{An illustration showing the
 nonadiabatic electron density $\ddense$ and electron mean velocity $\dupare$
 in the collisional, large-tail limit.
 At leading-order, the mode frequency is determined by the nonadiabatic electron response in the $\lpar \sim \massrutp{1/4}$ (inner)
 region, by solving the transport equations  \refeq{eq:electron_inner-dens0-collisional}
    and  \refeq{eq:electron_inner-temp0-collisional}, with inner-region quasineutrality 
 \refeq{eq:quasineutrality-inner-collisional}, and the boundary conditions
 \refeq{eq:density-matching-collisional}, \refeq{eq:temperature-matching-collisional}, 
 \refeq{eq:upare-matching-collisional}, and \refeq{eq:qpare-matching-collisional}.
 The ion response to the leading-order frequency can be obtained by solving equation \refeq{eq:lineargyrokinetic}
  (with $\spe = {\rm{i}}$) subject to quasineutrality,
  equation \refeq{eq:qnouter:smalltails}.
 In the large-tail mode the leading-order flows are developed in the $\lpar \sim \massrutp{1/4} $
 region, and there is no leading-order electron density cusp near the boundary of the $\lpar \sim 1$ region. 
 } \label{fig:collisional_large_tail_modes}
\end{center}
\end{figure} 
    
    The large-tail ordering is the combination of the orderings \refeq{eq:largetailorder} 
    and \refeq{eq:largetailorder2}.
    As a consequence, the equation for the leading-order electron
    response $\HHezouter$
    takes the form of equation \refeq{eq:electron_inner0-collisional}. 
    Following the same arguments as used in \ref{sec:electron-outer-collisional-small-tail-appendix},
    we can demonstrate that the solution to equation \refeq{eq:electron_inner0-collisional}
    is that the electron distribution is a perturbed Maxwellian
    with a fluctuating nonadiabatic density $\ddensns{0}{e,outer}$ and temperature
    $\dtempns{0}{e,outer}$, 
    no flow, and no dependence on $\lpar$, as required to match to the inner region.
    For the matching conditions on the leading-order distribution function,
    we require that the nonadiabatic density
    and temperature that define the electron distribution function 
    are continuous across $\lchi = 0$ after equations \refeq{eq:density-matching-collisional}
    and \refeq{eq:temperature-matching-collisional}, with 
    $\ddensns{0}{e,outer} = \ddensns{0}{e,inner}   (\lchi =0)$
  and $\dtempns{0}{e,outer} = \dtempns{0}{e,inner}   (\lchi =0)$.
    
    For this class of modes, the frequency is determined by the
    eigenmode equations \refeq{eq:electron_inner-dens0-collisional}
    and  \refeq{eq:electron_inner-temp0-collisional}, with the inner-region quasineutrality 
    equation \refeq{eq:quasineutrality-inner-collisional} and the matching conditions
    \refeq{eq:density-matching-collisional} and \refeq{eq:temperature-matching-collisional}.
    Because the eigenmode equations are second order differential equations in $\lchi$,
    two further matching conditions are required. These conditions are that
    the electron flows $\dupareeff$ and $\dqpareeff$ are continuous across $\lchi =0$, i.e., 
    \beqn \dupareeff(\lchi=0^+) = \dupareeff(\lchi=0^-), \label{eq:upare-matching-collisional}\eeqn
    and     
    \beqn  \dqpareeff(\lchi=0^+) = \dqpareeff(\lchi=0^-). \label{eq:qpare-matching-collisional}\eeqn
    Equations \refeq{eq:upare-matching-collisional}
    and \refeq{eq:qpare-matching-collisional} can be derived by noting
    that the jump in the electron parallel flows across the outer region
    have a fixed size, given by the estimate \refeq{eq:electron-flow-jump}.
    In a large tail mode, we have that 
    \beqn \fl \frac{\dupareinner}{\vthere} \sim \frac{\dqpareinner}{\vthere\dense \tempe} \sim \massrp{1/4} \frac{\charge\ptlinner}{\tempe}
    \sim \massrp{1/4} \frac{\charge\ptlouter}{\tempe} \gg \massrp{1/2} \frac{\charge\ptlouter}{\tempe}. \label{eq:neglect-jump}\eeqn
    and hence the flows are continuous across the outer region to leading order.
    This result can be obtained explicitly by inspecting equations \refeq{eq:upare-jump-small-tail}
    and \refeq{eq:qpare-jump-small-tail}, with the ordering \refeq{eq:neglect-jump}.
    An illustration of the structure of the collisional, large-tail mode
    is presented in figure \ref{fig:collisional_large_tail_modes}.
    
    Finally, we note that 
    the nonadiabatic ion response has no role in determining the leading-order frequency $\wfreqz$.
    Instead, the ions respond passively, serving only to self-consistently determine
    the electrostatic potential $\ptlzouter$ through the quasineutrality equation
    \refeq{eq:qnouter:smalltails} (noting that here the velocity space dependence
    of $\HHez$ is given by equation \refeq{eq:HHez-collisional}).
    Note that $\ptlzouter$ has not entered into the equations that determine the electron response
    in the large-tail mode.

   \section {Numerical results}\label {section:numericalresults}
  
  In this section we present numerical results that support the
  analytical theory of the previous sections.
  We use the gyrokinetic code \gstwo~\cite{KOTSCHENREUTHER1995CPC} to calculate the fastest-growing
  linear modes for parameters where we observe extended electron-driven
  tails in the ballooning eigenfunction.

  We propose a novel method to test the analytical theory:
  for a given mode with extended tails, we scan in our expansion
  parameter $\massrt$ and test the $\massrt$ dependence of the
  eigenmodes and complex frequencies. When we perform the scan in $\massrt$,
  we hold fixed $\kky \gyrdi$, so that we scan in the separation between
  the electron parallel streaming frequency $\vthere/\saffac\rmajorz$, 
  and frequency of the drive $\wstar \sim \wfreq$. We must also choose
  how to treat the collision frequency $\cfreqee$: the normalised electron collision frequency
  $\nustar = \saffac \rmajorz\cfreqee / \vthere \aspect^{3/2}$ is independent of $\me/\mi$,
  and so in a physical mass scan we would vary $\vthere/\saffac\rmajorz$ and $\cfreqee$, holding $\nustar$ fixed.
  This physical mass scan is appropriate for the collisional limit \refeq{order:nustar-collisional}.
  However, to test the \enquote{collisionless} limit \refeq{order:nustar-collisionless}, 
  if $\cfreqee$ is comparable to $\wfreq$,
  we need to enforce $\cfreqee \sim \wfreq \sim \wstar $ as $\massrt\rightarrow 0$,
  meaning that we would vary $\vthere/\saffac\rmajorz$ but hold $\lscal\cfreqee /\vtheri$ fixed.

  In the analytical theory that we have developed here,
  geometrical factors from the magnetic geometry 
  enter into the equations for the inner region only through
  the poloidal angle average $\lparav{\cdot}$. Hence, modes that are driven by the
  electron response in the inner region are unlikely to be sensitive to the details
  of any given magnetic geometry. We therefore choose the 
  simple Cyclone Base Case (CBC) \cite{DimitsCBCPoP2000} magnetic geometry to
  illustrate our theory: we study modes on a circular flux surface centred on the
  magnetic axis. 
  To specify the magnetic geometry, we use the Miller equilibrium parameterisation \cite{Miller98}.
 We take the reference major radius $\rmajorz = 3.0 \lscal$, with the normalising
 length $\lscal$ the half-diameter of the last closed flux surface. We
 examine microstability on the flux surface with minor radius $\rminor = 0.54 \lscal$.
 We take the safety factor to be $\saffac = 1.4$, the magnetic shear to be
 $\shat = (\saffac/\radial)\tdrvt{\saffac}{\radial} = 0.8$, the plasma beta $\pbeta = 0$,
 the Shafranov shift derivative $\tdrvt{\Delta}{\radial} = 0$, the elongation $\kappa = 1.0$,
 the elongation derivative $\tdrvt{\kappa}{\radial} = 0.0$, the triangularity $\delta = 0.0$, and
 the triangularity derivative $\tdrvt{\delta}{\radial} =0.0$. The reference magnetic field
 is given by $\bref = \bcur(\flxl)/\rmajgeo$, i.e., toroidal magnetic field at
 the major radial position $\rmajgeo$. We take $\rmajgeo = \rmajorz$. 
 In section \ref{sec:magnetic-coordinates},
 we define local radial and binormal coordinates with units of length
 $\radialx$ and $\binormal$, respectively,
 and associated radial and binormal wavenumbers $\kkx$ and $\kky$, respectively.
 We parameterise the radial wave number $\kkx$ with $\thetaz = \kkx / \shat \kky$.
 
 We consider a two-species plasma of ions and electrons, with $\zedi = 1$, 
 equal temperatures $\tempi = \tempe$, and an  
 equilibrium density gradient $\lscal/\lln = 0.733$, where the
 length scale $\lln = -\tdrvt{\radial}{\ln \dense}$. We take the equilibrium ion 
 temperature gradient to be $\lscal/\lti = 2.3$, where the equilibrium
 temperature gradient length scale of species $\spe$ is defined by $\lts = -\tdrvt{\radial}{\ln \temps}$.
 These parameters have been chosen to be close to the CBC benchmark equilibrium profiles. 
 In order to examine different instabilities with these CBC-like parameters, we vary $\thetaz$, the equilibrium
 electron temperature gradient length scale $\lte$,
 and the normalised electron collisionality $\nustar = \saffac \rmajorz\cfreqee / \vthere \aspect^{3/2} $,
 where $\aspect = \rminor/\rmajorz = 0.18$. 
 In section \ref{section:numericalresults-collisionless-large-tails},
 we take $\thetaz = 1.57$ and a large electron temperature gradient $\lscal/\lte = 3 \lscal/\lti = 6.9$,
 resulting in novel modes that conform to the large-tail mode ordering.
 In section \ref{section:numericalresults-small-tails}, we take
 $\thetaz = 0.1$ and equal electron and ion temperature gradients $\lscal/\lte = \lscal/\lti = 2.3$. This allows us to
 consider a familiar ITG mode, where we demonstrate that the passing electron response satisfies the small-tail mode ordering.
 Finally, in section \ref{section:theta0-dependence}, we  
 briefly discuss the transition between large-tail and small-tail modes
 as a function of $\thetaz$ in a scenario with
 $\lscal/\lte = 3.45$ and $\lscal/\lti = 2.3$.

 For the simulations presented here, we use the following numerical resolutions:
 $\ntheta = 33$ points per $2\pi$ element in the ballooning angle grid;
 $\npitch = 27$ points in the pitch angle grid;
 and $\negrid =24$ points in the energy grid.
 The energy grid is constructed from a spectral speed grid \cite{barnesPoP10a},
 and the pitch angle grid is constructed from a Radau-Gauss grid
 for passing particles and an unevenly spaced grid for trapped particles.
 We give the number of $2\pi$ elements in the ballooning grid $\ntwopi$
 and timestep size $\delt$ in each of the following subsections.
 The convergence of these resolutions was tested by doubling (or halving) each parameter.

   \subsection {Large tail modes}\label {section:numericalresults-collisionless-large-tails}
  
  In this section we present numerical results that
  are consistent with the asymptotic theory
  of linear modes with large electron tails, summarised
  in sections \ref{section:largetailmode} (the collisionless case)
  and \ref{sec:electron-outer-collisional-large-tail} (the collisional case).
  In order to make the passing-electron-response-driven modes the fastest growing instability
  in the system, it is necessary to increase the electron drive with respect to
  the ion drive: we take 
  $\lscal /\lte = 3 \lscal /\lti = 6.9$, and we focus on modes at 
  $\kky \gyrdi = 0.5$ with $\thetaz = 1.57$.  
  We vary $\nustar$ 
  in order to see the effect of electron collisionality on the mode. 
  The geometry  and physical parameters of the simulations
  are otherwise as described at the start of this section. We use the full $\texttt{GS2}$
  model collision operator \cite{abelPoP08,barnesPoP09},
  including pitch angle scattering, energy diffusion,
  and momentum and energy conserving terms. 
  For the parameters that we consider here, we
  find that the inclusion of pitch-angle scattering collisions
  is crucial for making the large-tail mode the fastest-growing instability.
  
   \begin{figure}
\begin{center}
\includegraphics[clip, trim=0cm 1cm 0cm 0cm, width=0.6\textwidth]{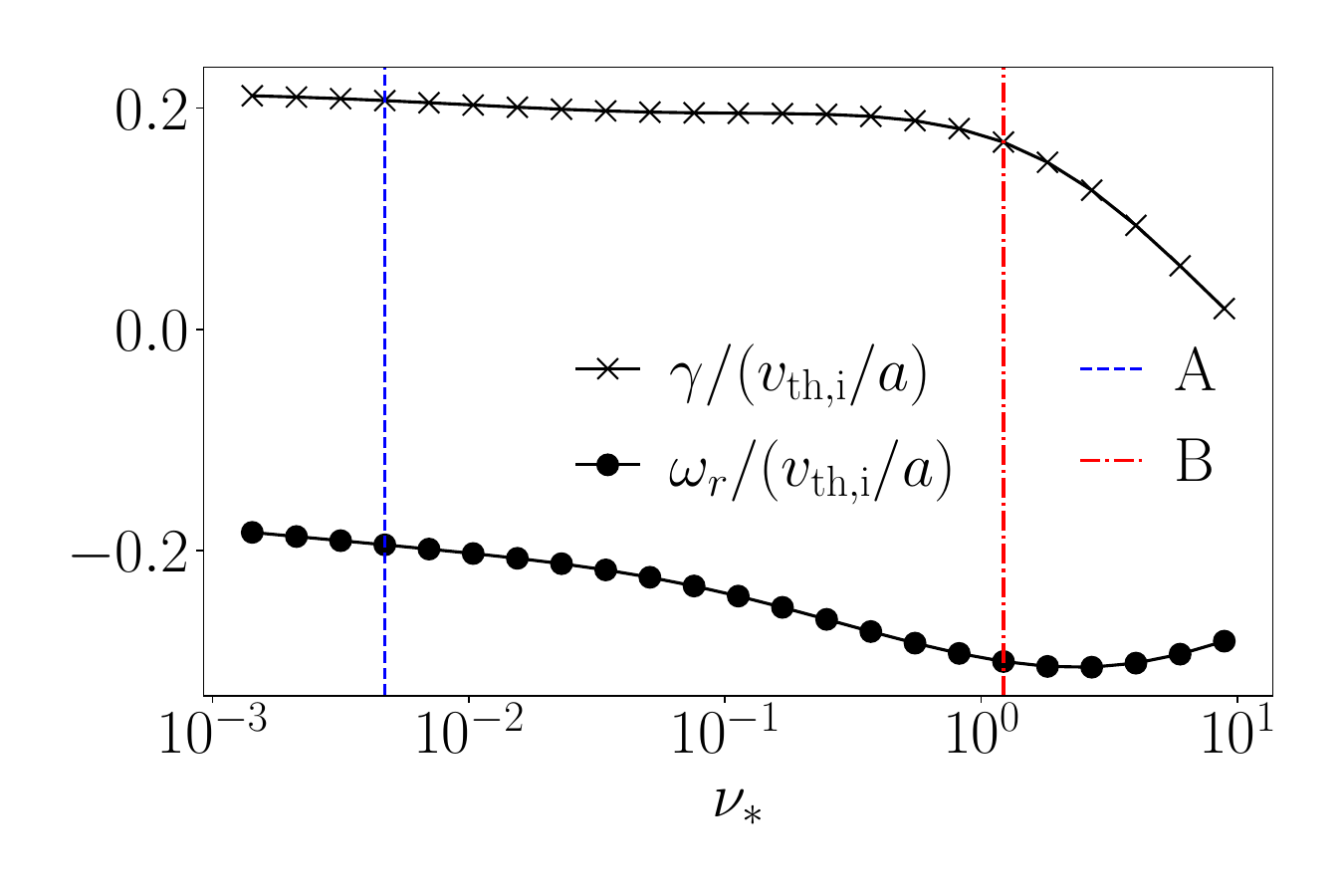}
% trim=left bottom right top, clip
\caption{ The growth rate $\growth$ and frequency $\wfreqr$ for the large-tail mode
 with $\massrut = 61$, $\kky \gyrdi = 0.5$ and $\thetaz = 1.57$, as a function of $\nustar$.
 For $\nustar < 10^{-3}$ the large-tail mode is no longer the
 fastest-growing mode at this $(\kky,\thetaz)$. The vertical dashed lines A and B indicate
 the $\nustar$ of the collisionless and collisional examples
 of {large-tail modes} that are predicted by the theory
 in sections \ref{section:largetailmode} and \ref{sec:electron-outer-collisional-large-tail}, respectively.   } \label{fig:large_tail_modes_nustar_scan}
\end{center}
\end{figure}
 
  In figure \ref{fig:large_tail_modes_nustar_scan}, we show the result of calculating the linear
  growth rate $\growth$ and frequency $\wfreqr$ for the deuterium mass ratio $\massrt \approx 1/61$ and a range of $\nustar$.
  We vary the ion collision frequency $\cfreqii$ consistently
 with $\nustar$, i.e., $\cfreqii = \aspect^{3/2} \nustar\vtheri / \sqrt{2} \saffac \rmajorz$.
  We take $\delt = 0.025 \lscal/\vtheri$ and $\ntwopi = 65$.
  For the range of $\nustar$  shown in figure \ref{fig:large_tail_modes_nustar_scan},
  we identify that the modes are large-tail modes, by a method which we now describe. 
  We recall the cartoons given in figures \ref{fig:large_tail_modes} and \ref{fig:collisional_large_tail_modes}.
  In a large-tail mode, the relative amplitude
  of the potential $\ptl$ has the same size in the outer
  and inner regions as $\massrt \rightarrow 0$, and the size of 
  the passing part of the electron distribution function $\HHe$ in the outer and inner regions remains
  fixed as $\massrt \rightarrow 0$. To demonstrate that the mode is a large-tail mode, we use a procedure with three steps. 
  First, we perform a scan in the value of $\massrt$.
  Second, we identify an integral measure of $\HHe$ and use it to demonstrate that $\HHeouter \sim \HHeinner$, independent of the value of $\massrt$.
  Third, we plot $\ptl$ normalised to the chosen integral measure of $\HHe$
  and demonstrate that $\ptlouter\sim\ptlinner$, independent of the value of $\massrt$.
  We can determine the scale of the inner region $\lchi$ by using the scan in $\massrt$ to 
  determine $\alpha$ such that we can rescale $\HHe(\lpar) \rightarrow \HHe(\lpar \massrtp{\alpha})$
  and so overlay the integral measure of $\HHe$ for the modes with different values of $\massrt$.
  Collisionless modes are expected to have $\alpha = 1/2$, whereas collisional modes are expected to have $\alpha = 1/4$.
    To show how this procedure works in practice, we consider the clean examples of
    collisionless and collisional large-tail modes indicated by
     \enquote{A} and \enquote{B} on figure \ref{fig:large_tail_modes_nustar_scan}, respectively.

  \subsubsection{Case A -- a collisionless large-tail mode.} \label{sec:caseA}
    In the first step in our procedure, we scan in the electron
    mass ratio from $\me / \mi = 5.4 \times 10^{-4}$ to $5.4 \times 10^{-5}$,
    whilst holding fixed $\lscal\cfreqee/\vtheri = \cfreqeecollisionless$.
    The value of $\cfreqee$ is chosen so that $\nustar = \nustarcollisionless$ 
    for the approximate deuterium mass ratio $(\md/\me)^{1/2} = 61$, 
    and $\cfreqii$ is set by $\cfreqii/\cfreqee = (\me/\md)^{1/2}/\sqrt{2} $.
    In this example, we note that $\nustar$ takes a similar value to $\lscal\cfreqee/\vtheri$. This is
    a result of the large value of $\saffac\rmajorz /\lscal\aspect^{3/2} = 55.0 \sim (\md/\me)^{1/2}$.     
    To set $\ntwopi$ for different ion masses $\mi$, we scale the number of $2\pi$ elements
    appropriately with mass ratio, i.e., $\ntwopi = 65 \sqrt{\mi/\md}$. 
    The timestep size is taken to be $\delt = 0.025 \lscal/\vtheri$.
    
\begin{figure}
\begin{center}
\includegraphics[clip, trim=1cm 1cm 1cm 0cm, width=0.5\textwidth]{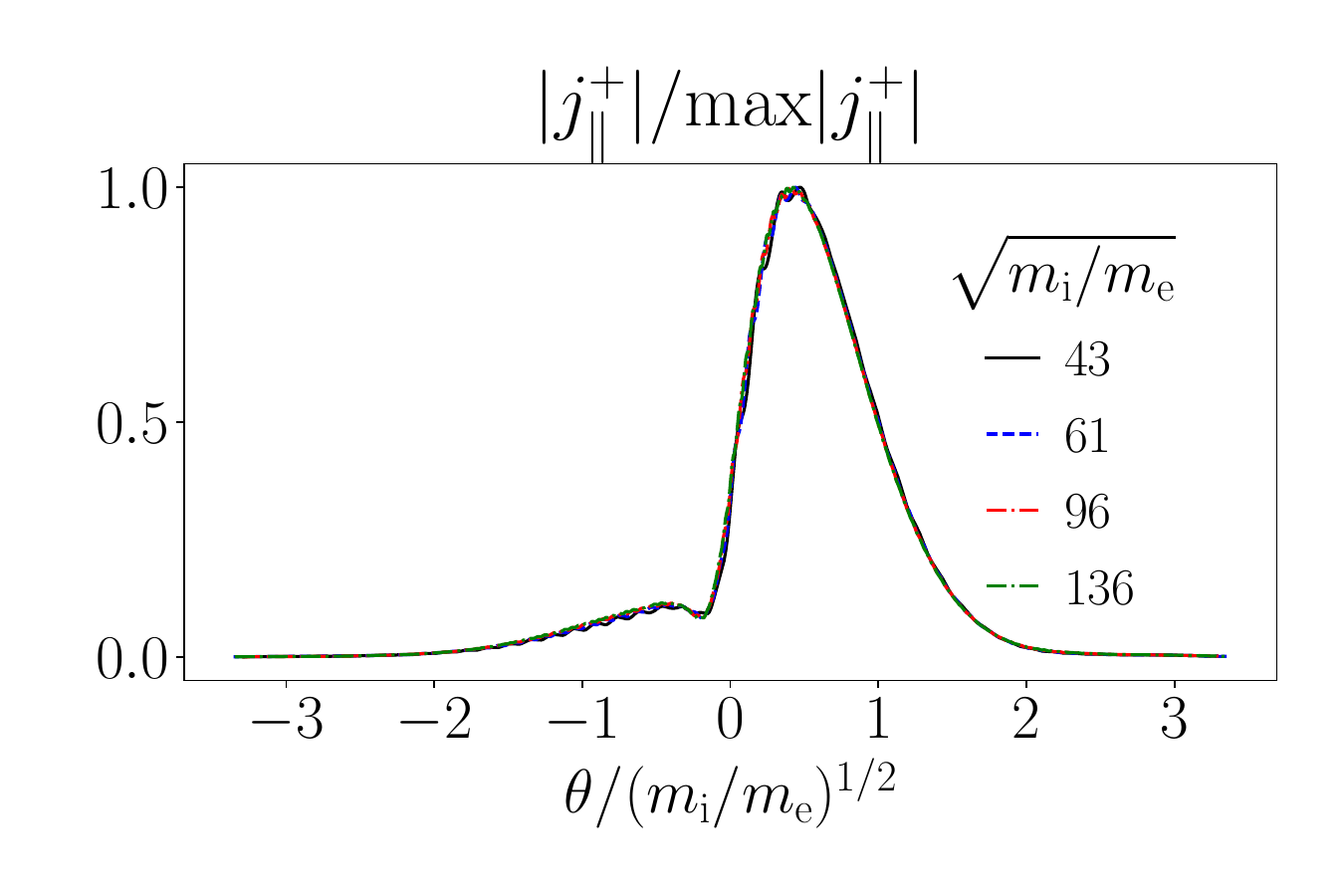}
% trim=left bottom right top, clip
\caption{The field $\jintplus$, calculated for 
  different mass ratios holding fixed $\lscal\cfreqee/\vtheri = \cfreqeecollisionless$ 
 ($\nustar = \nustarcollisionless$ for $\massrut=61$ --  
 case A of figure \ref{fig:large_tail_modes_nustar_scan}). Note that the curves overlay on the
  $\lpar/\massrut$ axis.} 
  \label{fig:jintplus-collisionless-large-tail}
\end{center}
 \end{figure}

    In the second step of the procedure, we define a useful integral measure of the electron distribution function
    \beqn  \jint = \jintplus - \jintminus \label{eq:littlejpar}
    \eeqn
    with
    \beqn  \jintpm = - \charge \int^{\infty}_0 \int^{1/\bmagmax}_0  
     \frac{|\vpar|}{\bmag}\HHe(\sign = \pm 1)
     \; \frac{2\pi \bmag \energy}{\me^2 |\vpar|}  \; d \pitch \;  d \energy
     \label{eq:littlejpar_plusminus}.\eeqn
    The field $\jint$ has dimensions of current over magnetic field strength,
    and the quantities $\jintplus$ and $\jintminus$are the contributions from
    the forward going ($\sign = 1$) and backward going ($\sign = -1$) particles, respectively.   
    The prime usefulness of $\jintpm$ stems from the fact that $\HHez$ is independent
    of the $2\pi$-periodic poloidal angle $\lpar$ in the asymptotic theory, and hence,
    we expect that $\jintpm$ are smoothly varying functions of ballooning angle,
    with minimal geometric $2 \pi$-periodic oscillation. We can use $\jintplus$ as a proxy
    to visualise the distribution of forward-going particles. In figure
    \ref{fig:jintplus-collisionless-large-tail}, we plot
    $|\jintplus|$, normalised to its maximum value, for different values of $\massrut$.
    Figure \ref{fig:jintplus-collisionless-large-tail}
    shows that $\jintplus$ is self-similar for modes with different $\massrt$, provided
    that the ballooning angle $\lpar$ is rescaled to $\lpar/\massrut$. This confirms that
    $\HHeouter \sim \HHeinner$, and that $\lchi \sim \massrut$.
    
\begin{figure}[hbt]
\begin {minipage} {0.495\textwidth}
\begin{center}
\includegraphics[clip, trim=1cm 1cm 1cm 0cm, width=1.0\textwidth]{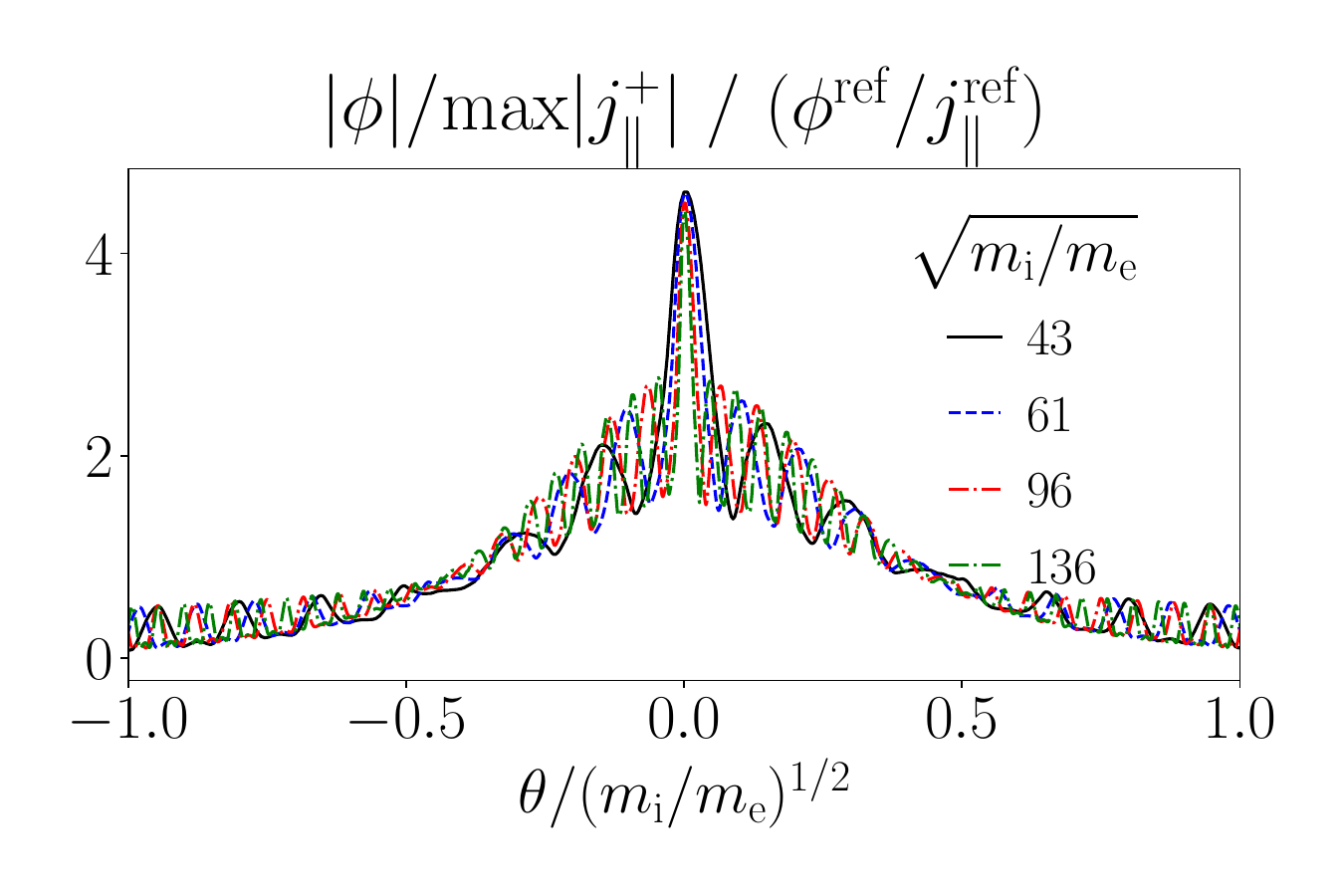}
% trim=left bottom right top, clip
\end{center}
\end{minipage}
\begin {minipage} {0.495\textwidth}
\begin{center}
\includegraphics[clip, trim=1cm 1cm 1cm 0cm, width=1.0\textwidth]{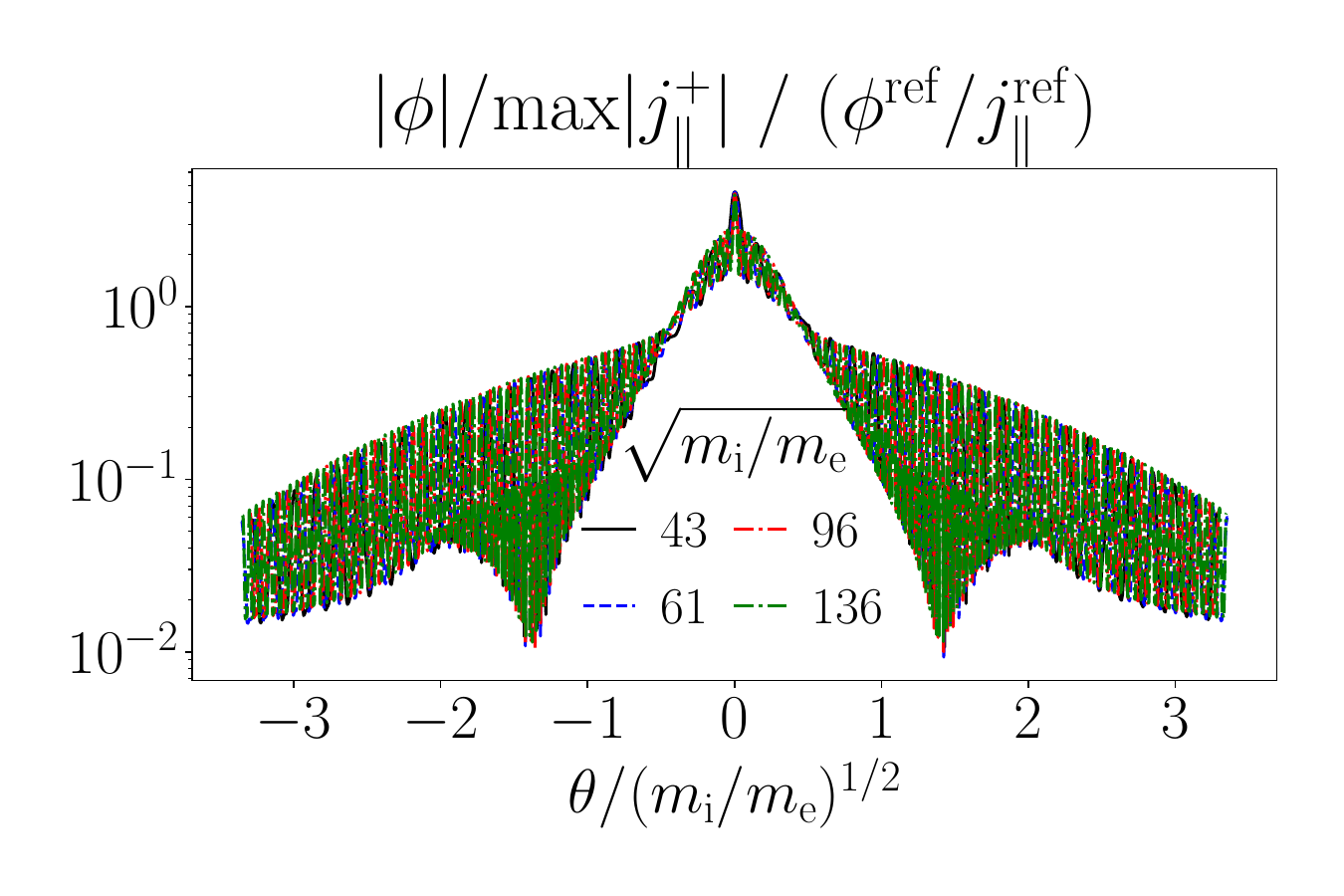}
\end{center}
\end{minipage}
\caption{Two views of the electrostatic potential $\ptl$, for 
  different mass ratios holding fixed $\lscal\cfreqee/\vtheri = \cfreqeecollisionless$
 (corresponding to $\nustar = \nustarcollisionless$ when $\massrut =61$,
 case A of figure \ref{fig:large_tail_modes_nustar_scan}).
 The potential is plotted against the scaled ballooning angle $\lpar / \massrut$,
 and normalised to the maximum value of $\jintplus$ (see equation \refeq{eq:littlejpar_plusminus}
 and figure \ref{fig:jintplus-collisionless-large-tail}).
 The fact that the curves overlay on the
  $\lpar/\massrut$ axis confirms that the mode is a collisionless large-tail mode.
 Note the geometric $2\pi$-periodic oscillation in $\phi$ due to geometric factors
 in the velocity integral over the electron distribution function
 (cf. equation \refeq{eq:qninner}). The dimensions {are} $\phiref = \tempe / \charge$ and 
  $\jintref = \charge \dense \vthere / \bref$.} \label{fig:phi-collisionless}
 \end{figure}

    Finally, in the last step of the procedure, we visualise the electrostatic potential in
    figure \ref{fig:phi-collisionless}.
    We normalise the potential to the maximum value of $|\jintplus|$, and give the result
    $|\ptl |/{\rm max}|\jintplus|$ in the units of $\phiref/\jintref$, where 
    $\phiref = \tempe / \charge$ and $\jintref =  \charge \dense \vthere / \bmagref$.
    In figure \ref{fig:phi-collisionless}, we can see  that $\phi$ 
    has an envelope on the scale of $\lpar \sim \massrut$, and an irreducible geometric structure on
    the scale of $\lpar \sim 2\pi$. The envelope of $|\ptl|$ overlays well in figure \ref{fig:phi-collisionless},
    confirming that $\ptlouter\sim\ptlinner$ and, hence, that the mode is a large-tail mode.
    The geometric $2\pi$-oscillations in $\ptl$ appear because of the geometric poloidal angle dependence 
    in the Jacobian $2\pi\bmag \energy/\me^2|\vpar|$ of the velocity integral, equation 
    \refeq{eq:intv-definition}; because of the inclusion of trapped particles in 
    the velocity integral;  and because of the appearance of the Bessel function 
    $\besenz{0}$ and phase $\expo{\imag \lflre \lchi}$ in the quasineutrality relation
    in the large-$\lpar$ region, equation \refeq{eq:qninner}.

 \begin{figure}[hbt]
\begin {minipage} {0.495\textwidth}
\begin{center}
\includegraphics[clip, trim=1cm 1cm 1cm 0cm, width=1.0\textwidth]{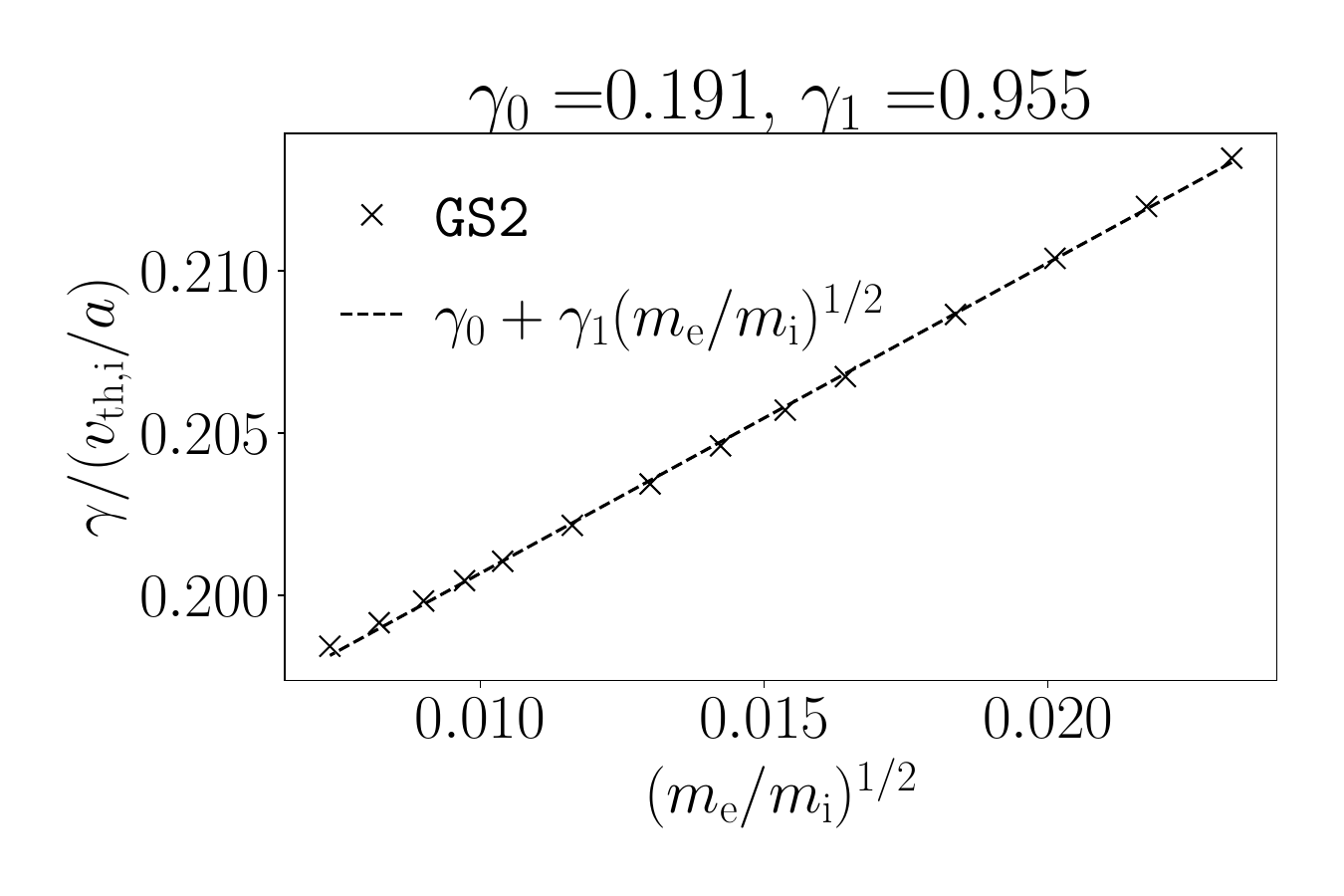}
% trim=left bottom right top, clip
\end{center}
\end{minipage}
\begin {minipage} {0.495\textwidth}
\begin{center}
\includegraphics[clip, trim=1cm 1cm 1cm 0cm, width=1.0\textwidth]{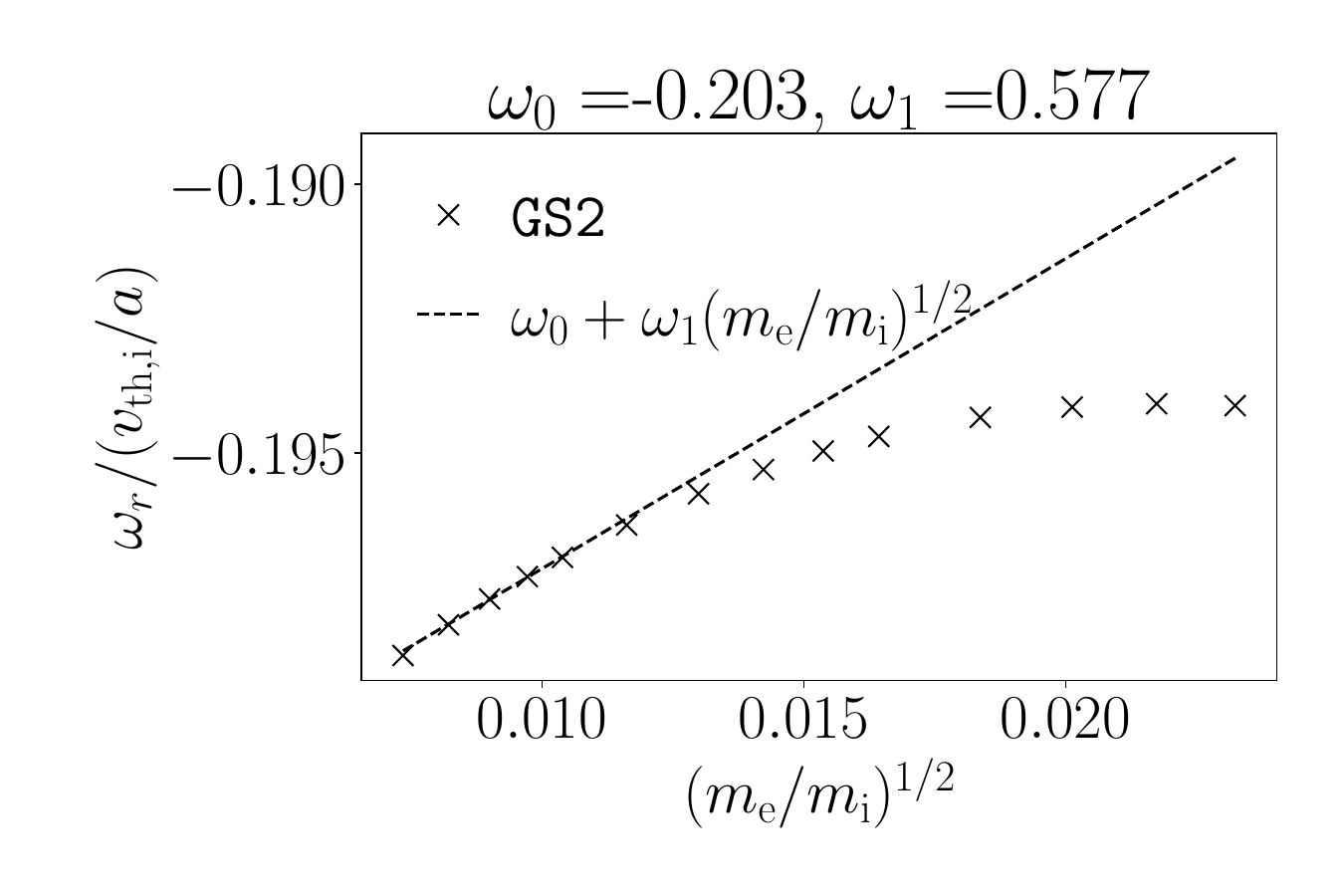}
\end{center}
\end{minipage}
 \caption{Plots of the growth rate $\growth$ (left) and real frequency $\wfreqr$ (right)
 as a function of $\massrt$, for fixed $\lscal \cfreqee/\vtheri = \cfreqeecollisionless$
 (corresponding to case A of figure \ref{fig:large_tail_modes_nustar_scan} when
 $\massrut = 61$ and $\nustar = \nustarcollisionless$).
 We give a linear fit to demonstrate that the dependence of $\growth$ and $\wfreqr$ on $\massrt$
 is consistent with a $\massrt$ expansion
 of the form given by equation \refeq{eq:collisionless-phi-expansion}.} \label{fig:growthwfreqr-collisionless}
 \end{figure}

    From the asymptotic theory in section \ref {section:modes-in-memi-zero-limit},
    we would expect to see that the
    growth rate had a leading order piece $\growthz$, and an $\order{\massrt}$ small correction $\growtho$.
    In figure \ref{fig:growthwfreqr-collisionless}, we plot $\growth$
    (figure \ref{fig:growthwfreqr-collisionless}, left) and $\wfreqr$
    (figure \ref{fig:growthwfreqr-collisionless}, right) as a function of $\massrt$.
    The linear fits are provided to show that the changes with $\massrt$ in $\growth$ and $\wfreqr$
    are consistent with an expansion in $\massrt$: the fit parameters are of order unity,
    and the overall variation in $\growth$ and $\wfreqr$ is small. 
    We note that the linear fit is particularly good for $\growth$ for 
    the whole range of $\massrt$. In the case of $\wfreqr$,
    we see a linear trend arise only at small $\massrt$. 
    In general, we would expect to recover a linear trend for sufficiently small $\massrt$.
    We note that, because $\cfreqee/\wfreq \ll 1$, the qualitative results of figures
    \ref{fig:jintplus-collisionless-large-tail}-\ref{fig:growthwfreqr-collisionless}
    can be reproduced by a scan holding fixed $\nustar$
    rather than $\lscal\cfreqee/\vtheri$, providing that 
    $\cfreqee/\wfreq$ does not approach values of order unity.

  \begin{figure}
\begin {minipage} {0.495\textwidth}
\begin{center}
\includegraphics[clip, trim=1cm 1cm 1cm 0cm, width=1.0\textwidth]{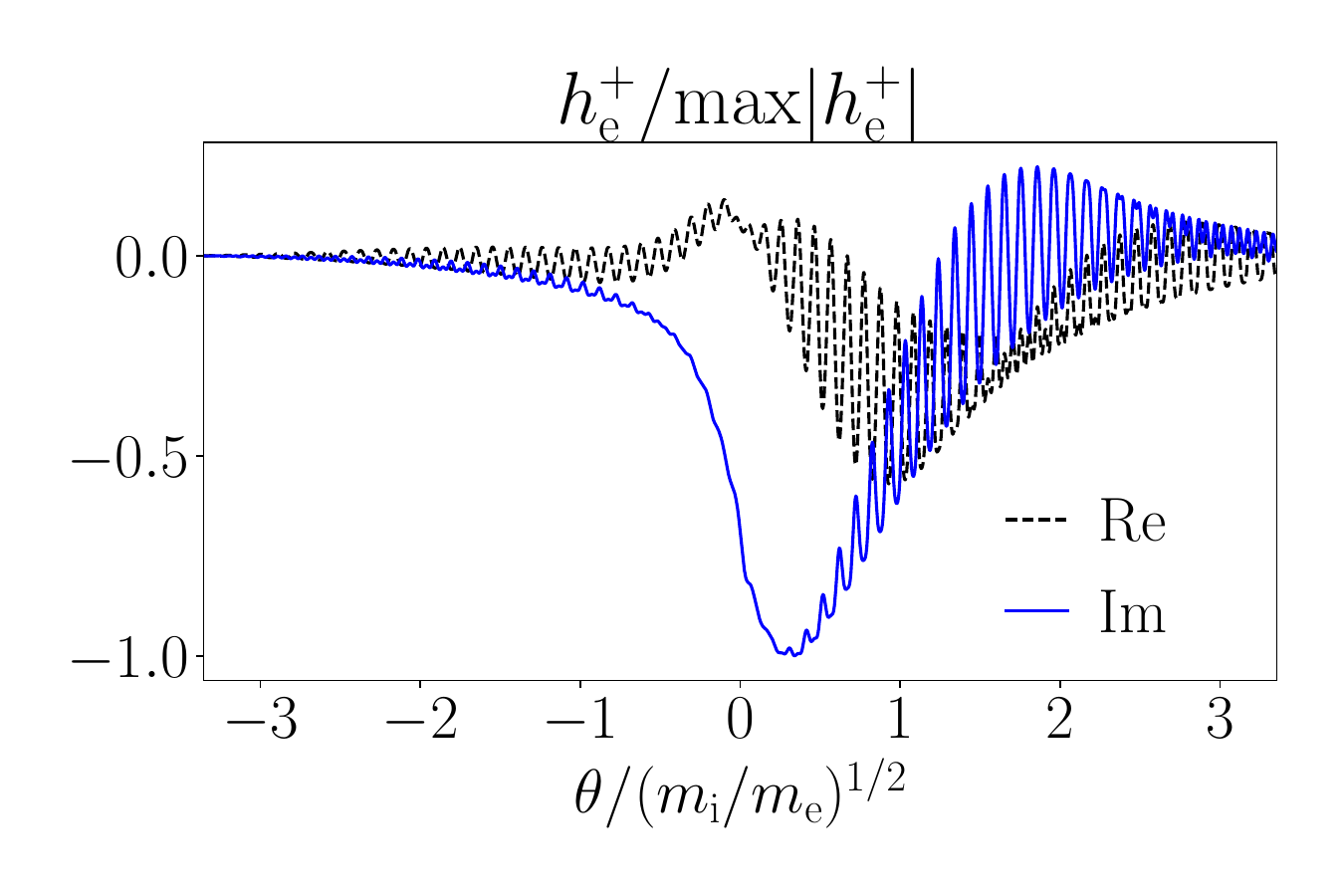}
% trim=left bottom right top, clip
\end{center}
\end{minipage}
\begin {minipage} {0.495\textwidth}
\begin{center}
\includegraphics[clip, trim=1cm 1cm 1cm 0cm, width=1.0\textwidth]{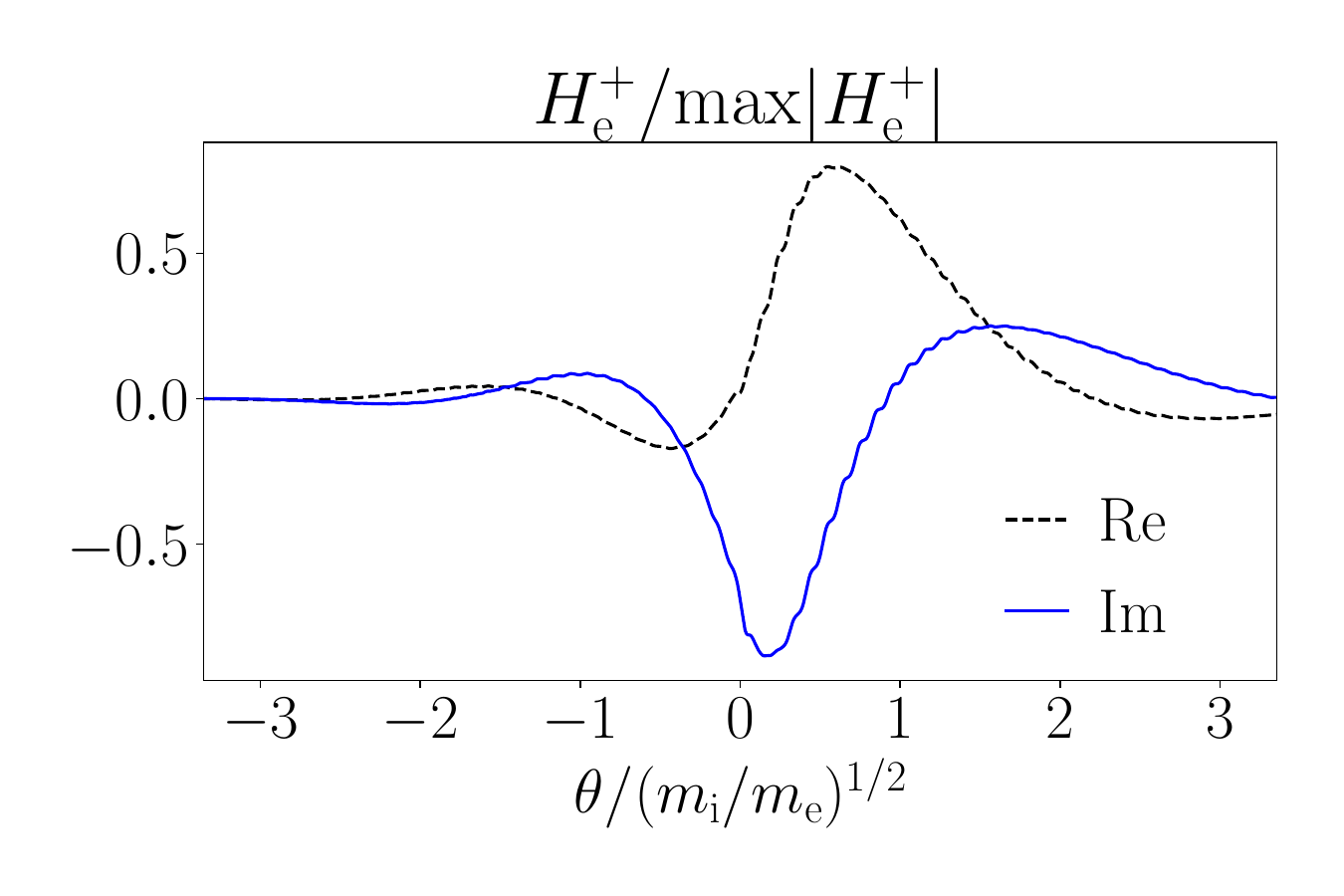}
\end{center}
\end{minipage}
\caption{The distribution function for forward going particles, for the mode
 with $\massrut=61$ in figure \ref{fig:jintplus-collisionless-large-tail}, for 
 $\energy / \tempe = 0.79$ and $\pitch \bmagref = 0.22$. Left, we plot 
 $\hhe$ for forward going particles. Note the rapid oscillation in $\lpar$
  for $\lpar \gg 1$. Right, we plot $\HHe$ for forward going particles. Note the smoothness of $\HHe$
 compared to $\hhe$ for $\lpar \gg 1$.} \label{fig:dfne}
 \end{figure}
    
    Finally, we comment on the use of the modified distribution function
    $\HHe$ in place of the usual nonadiabatic response $\hhe$ in the asymptotic theory.
    In figure \ref{fig:dfne}, we plot the distribution functions
    $\hhe$ and $\HHe$, as a function of $\lpar$, for the velocity space element
    $\energy / \tempe = 0.79$, $\pitch \bmagref = 0.22$ and $\sign = 1$.
    We show the distribution functions for the $\massrut = 61$
    mode featured in figure \ref{fig:jintplus-collisionless-large-tail}.
    We observe that the distribution function $\hhe$ shows large $2\pi$-scale
    oscillations in phase, whereas $\HHe$ is a smoothly varying function.
    In general, $\HHe$ appears to be a smoother variable than $\hhe$ for the parts
    of the electron distribution function where $\vpar \sim \vthere \gg \vtheri$.    
    These observations justify the choice to use the modified distribution
    function $\HHe$ in the asymptotic theory.

%%%%%%%%%%%%%%%%%%%%%%%%%%%%%%%%%%%%%%%%%%%%%%%%%%%%%%%%%%%%%%%%%%
%%%%%%%%%%%%%%%%%%%%%%%%%%%%%%%%%%%%%%%%%%%%%%%%%%%%%%%%%%%%%%%%%%

   \subsubsection{Case B -- a collisional large-tail mode.}
 We now consider an example of a collisional large-tail mode.
 We must demonstrate that $\HHeouter\sim\HHeinner$ and
 $\ptlouter\sim\ptlinner$ as $\massrtp{1/4} \rightarrow 0$, 
 and show that the envelope of the eigenmode scales like $\lchi \sim \massrutp{1/4}$.
 The physics parameters are identical to those 
 of the collisionless large-tail mode described in section \ref{sec:caseA},
 except that the electron collisionality is increased to $\nustar = \nustarcollisional$.
 We scan in the electron mass ratio from $\me / \mi = 5.4 \times 10^{-4}$ to $5.0 \times 10^{-6}$
 while holding $\nustar$ fixed. We take
 $\cfreqii = \aspect^{3/2} \nustar\vtheri / \sqrt{2} \saffac \rmajorz$. 
 Anticipating that the scale of
 the ballooning envelope should go with $\massrutp{1/4}$ in the collisional limit,
 in this case we take $\ntwopi = 39 (\mi/\md)^{1/4}$.
 Perhaps due to the larger collision frequency, we find that convergence is reached only
 with relatively small timesteps for the largest $\massrutp{1/4}$ considered in the scan.
 In the simulations presented here $\delt =  0.00625 \lscal/\vtheri$.
 Obtaining convergence in $\delt$ is more challenging for smaller $\massrtp{1/4}$.

  \begin{figure}[htb]
\begin {minipage} {0.495\textwidth}
\begin{center}
\includegraphics[clip, trim=1cm 1cm 1cm 0cm, width=1.0\textwidth]{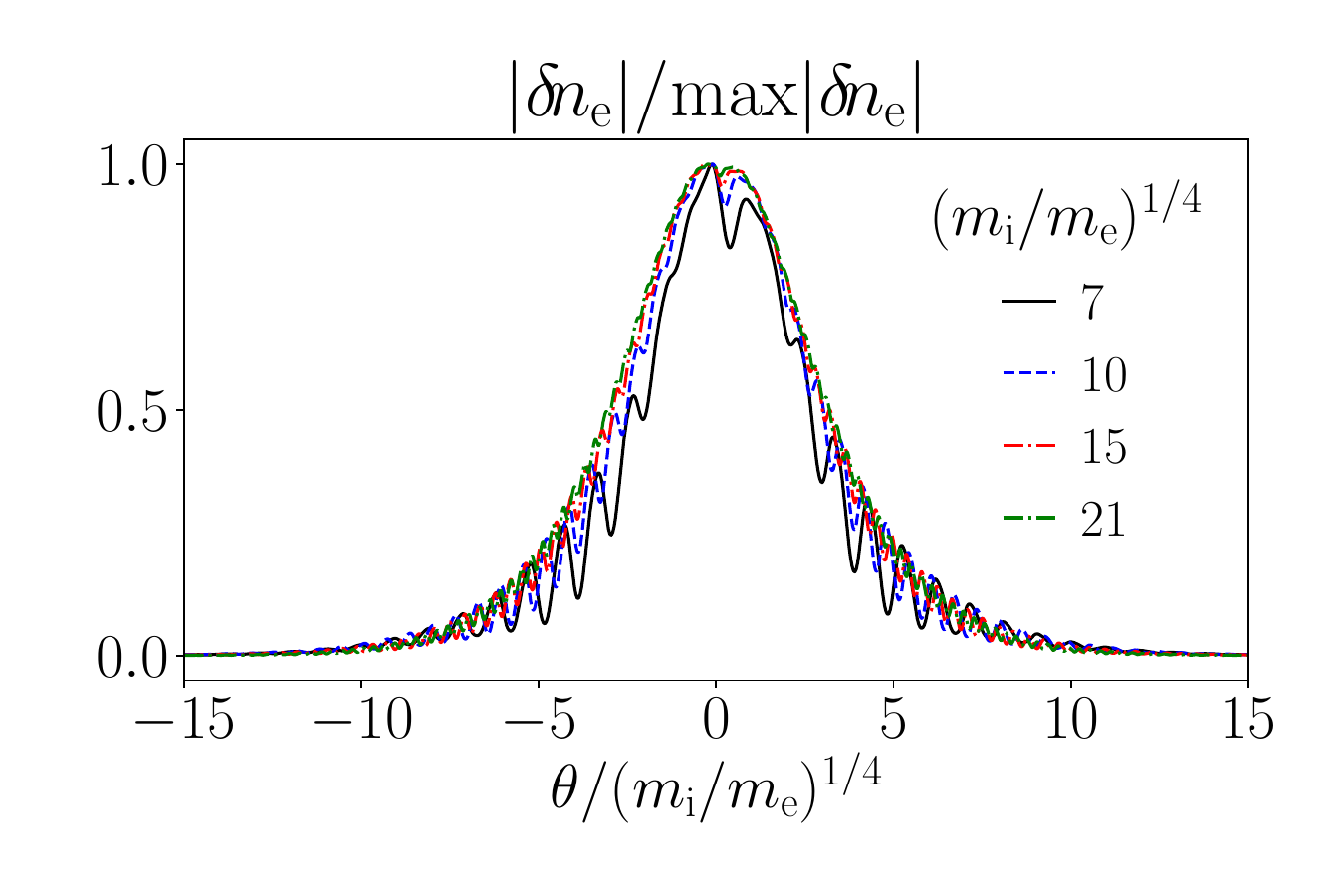}
% trim=left bottom right top, clip
\end{center}
\end{minipage}
\begin {minipage} {0.495\textwidth}
\begin{center}
\includegraphics[clip, trim=1cm 1cm 1cm 0cm, width=1.0\textwidth]{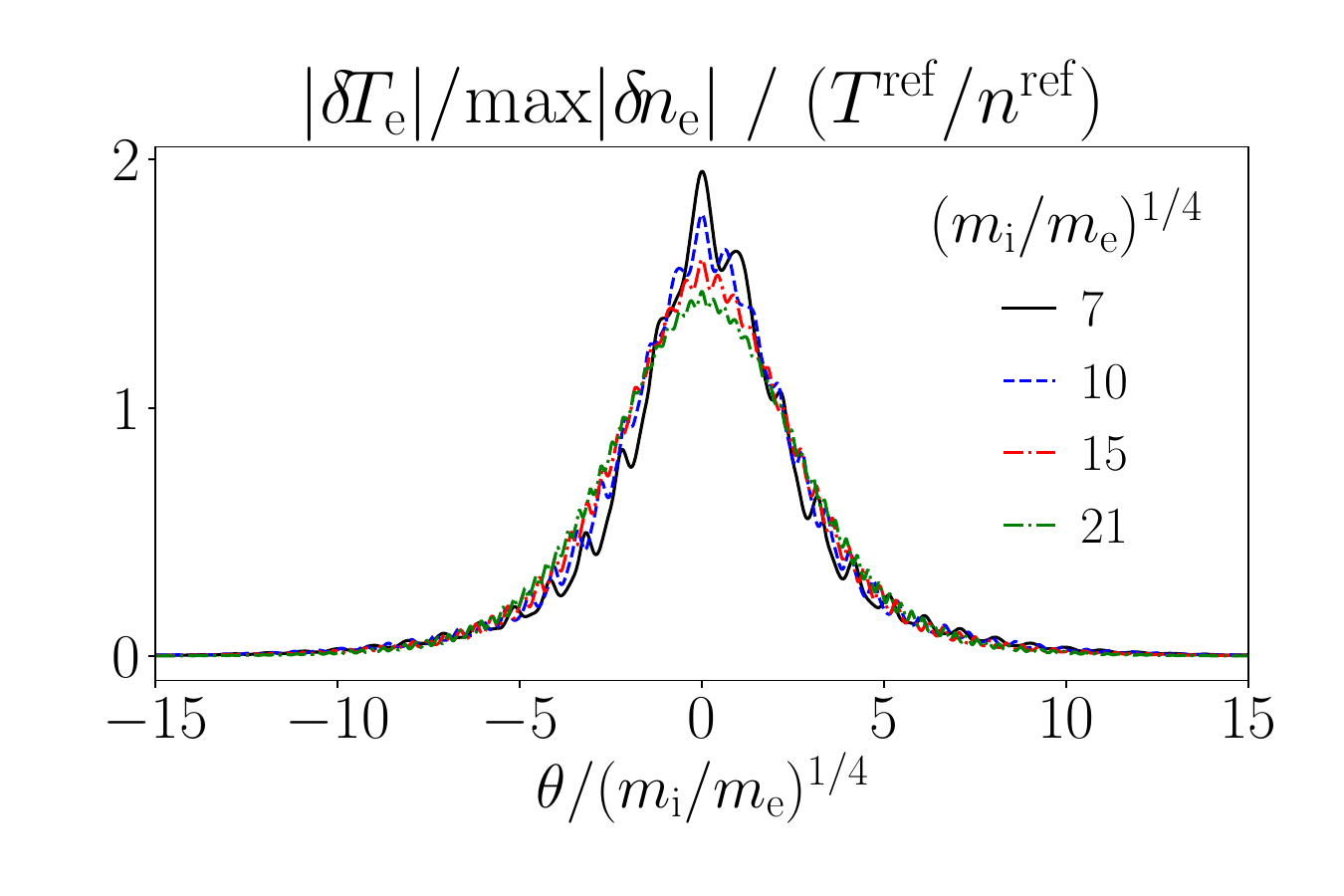}
\end{center}
\end{minipage}
\caption{ (left) The electron nonadiabatic density $\ddense$,
 calculated for $\nustar = \nustarcollisional$
 (case B of figure \ref{fig:large_tail_modes_nustar_scan}) for different mass ratios.
 The density is  normalised to its maximum value,
 and plotted against the scaled ballooning angle $\lpar / \massrutp{1/4}$.
 (right) The electron temperature, normalised to the maximum value of the electron nonadiabatic density.
 Here $\nref = \dense$, $\tempref = \tempe$.}
 \label{fig:ptm+bk+dt16+npd.vnewk.1.27.0.densitye}

 \end{figure}
    \begin{figure}
\begin{center}
\includegraphics[clip, trim=0.5cm 1cm 1cm 0cm, width=0.495\textwidth]{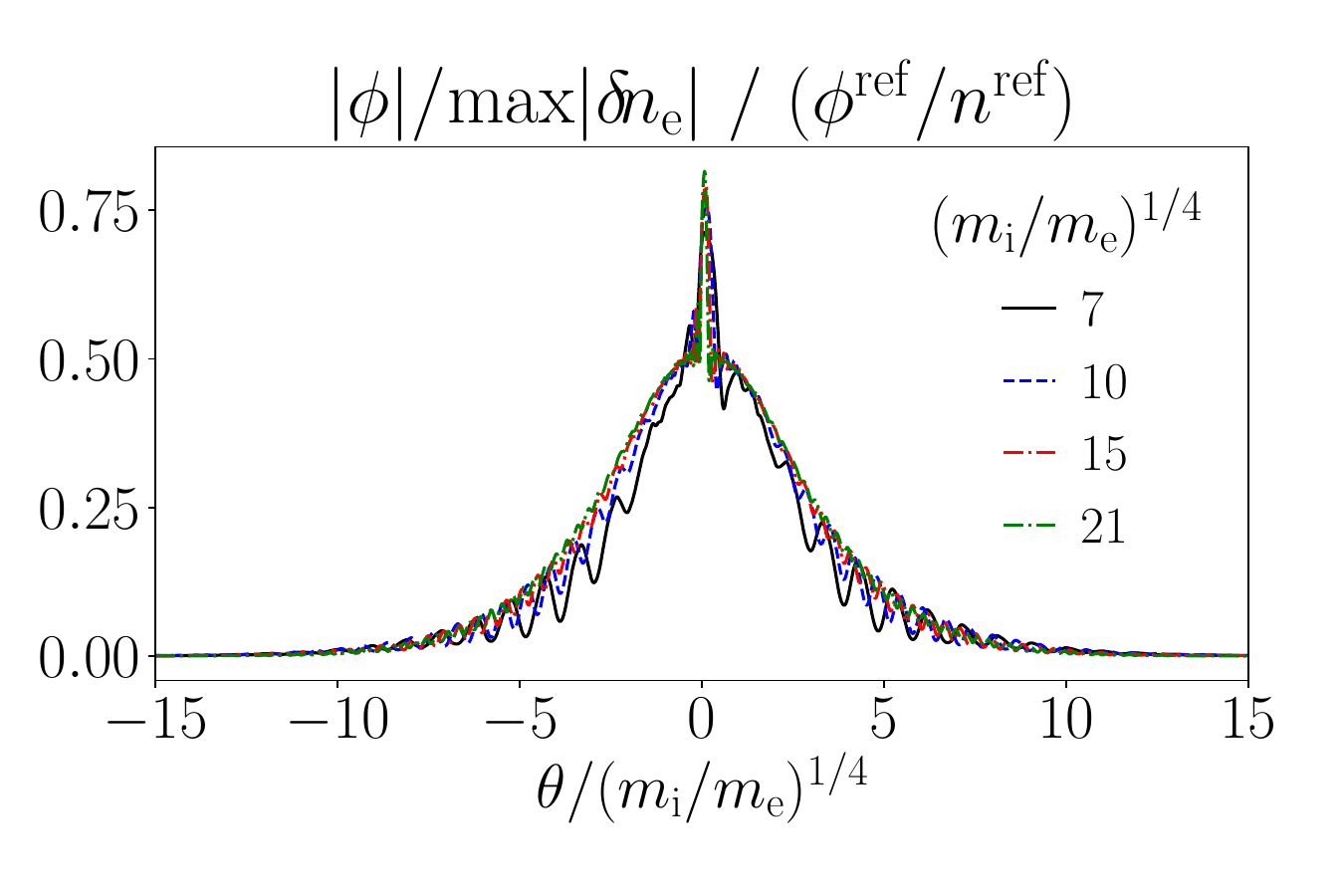}
% trim=left bottom right top, clip
\caption{The electrostatic $\ptl$, calculated for $\nustar = \nustarcollisional$
 (case B of figure \ref{fig:large_tail_modes_nustar_scan}) for 
  different mass ratios. The potential is normalised to the maximum value
  of $\ddense$, ${\rm{max}} |\ddense|$. The good agreement for the rescaled potential normalised to ${\rm{max}} |\ddense|$ suggests that
 this mode can be regarded as a collisional large-tail mode.
 Here $\phiref = \tempe/\charge$ and $\nref = \dense$.  } \label{fig:ptm+bk+dt16+npd.vnewk.1.27.0.phi}
\end{center}
\end{figure}

\begin{figure}
\begin{center}
\includegraphics[clip, trim=1cm 1cm 1cm 0cm, width=0.495\textwidth]{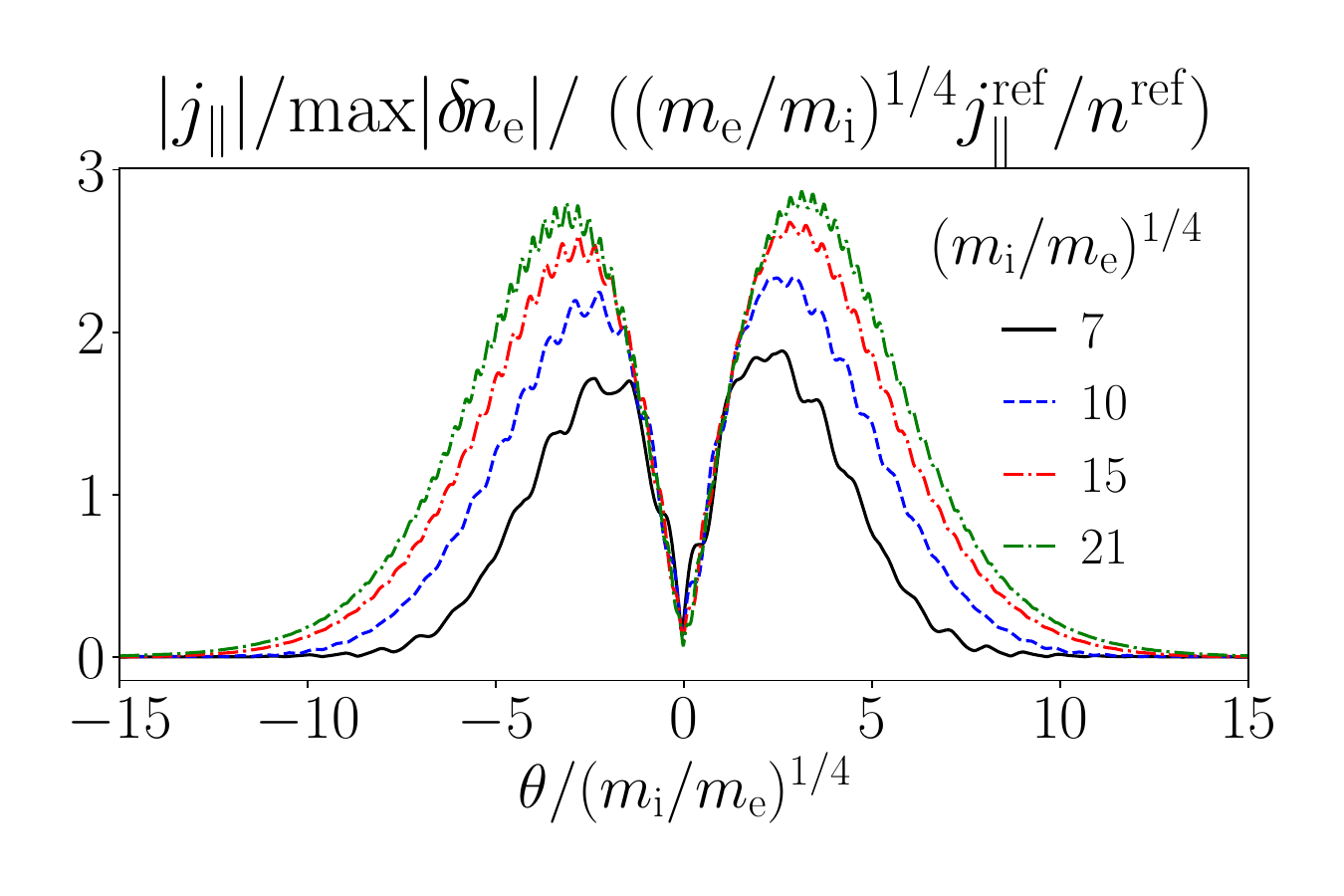}
\caption{The current-like field $\jint$, defined in equation \refeq{eq:littlejpar},
 normalised to the maximum value of $\ddense$, for $\nustar = \nustarcollisional$
 (case B of figure \ref{fig:large_tail_modes_nustar_scan}). Note that $\jint$ is a $\massrtp{1/4}$ small quantity.
 If the numerics perfectly reproduced the asymptotic
 theory of section \ref{section:modes-in-collisional-memi-zero-limit}, the plotted curves would overlay.
 The $\massrutp{1/4}$ rescaling produces better agreement
 than a $\massrutp{1/2}$ rescaling, 
 and we note that the curves appear
 to be converging for the largest $\massrutp{1/4}$. Here,
 $\jintref = \charge \dense \vthere / \bref $ and $\nref = \dense$.} \label{fig:vnewk.4.18.0.littlejpar}
\end{center}
 \end{figure}
 
 In the collisional ordering,
 the asymptotic theory of large-tail modes
 in section \ref{section:modes-in-collisional-memi-zero-limit} indicates
 that there are three leading-order quantities that are free from
 geometric $2 \pi$ poloidal angle oscillations at large $\lpar$: the electron nonadiabatic density $\ddense$,
 the electron temperature $\dtempe$, and the electrostatic potential $\ptl$.
 The electron nonadiabatic density
 and temperature are plotted in figure \ref{fig:ptm+bk+dt16+npd.vnewk.1.27.0.densitye},
 with the ballooning angle $\lpar$ rescaled by $\massrutp{1/4}$.
 We observe good agreement for different $\massrutp{1/4}$ in the mass ratio scan. 
 In figure \ref{fig:ptm+bk+dt16+npd.vnewk.1.27.0.phi}, we plot the electrostatic potential $\ptl$, normalised to the maximum value of $\ddense$.  
  Figure \ref{fig:ptm+bk+dt16+npd.vnewk.1.27.0.phi} shows good agreement for different $\massrutp{1/4}$.
 Together, figures \ref{fig:ptm+bk+dt16+npd.vnewk.1.27.0.densitye} and \ref{fig:ptm+bk+dt16+npd.vnewk.1.27.0.phi} demonstrate that 
 the mode featured is a collisional large-tail mode. 
 
 In the asymptotic theory of the collisional large-tail mode, the parallel-to-the-field flows
 play an important diffusive role, despite being small by $\massrtp{1/4}$.
 In figure \ref{fig:vnewk.4.18.0.littlejpar},
 we plot the current-like field $\jint$, defined in equation \refeq{eq:littlejpar}, with the ballooning angle $\lpar$ rescaled
 by $\massrutp{1/4}$, and the amplitude rescaled by $\massrtp{1/4}$. 
 Although the curves do not overlay perfectly
 in the $\massrutp{1/4}$ ballooning angle rescaling, 
 the curves appear to be converging for the largest $\massrutp{1/4}$ in the scan.
 We note that
 the $\massrutp{1/4}$ ballooning angle rescaling shown in
 figure \ref{fig:vnewk.4.18.0.littlejpar} gives
 better agreement than a $\massrut$ ballooning angle rescaling.
  
 \begin{figure}[hbt]
\begin {minipage} {0.495\textwidth}
\begin{center}
\includegraphics[clip, trim=1cm 1cm 1cm 0cm, width=1.0\textwidth]{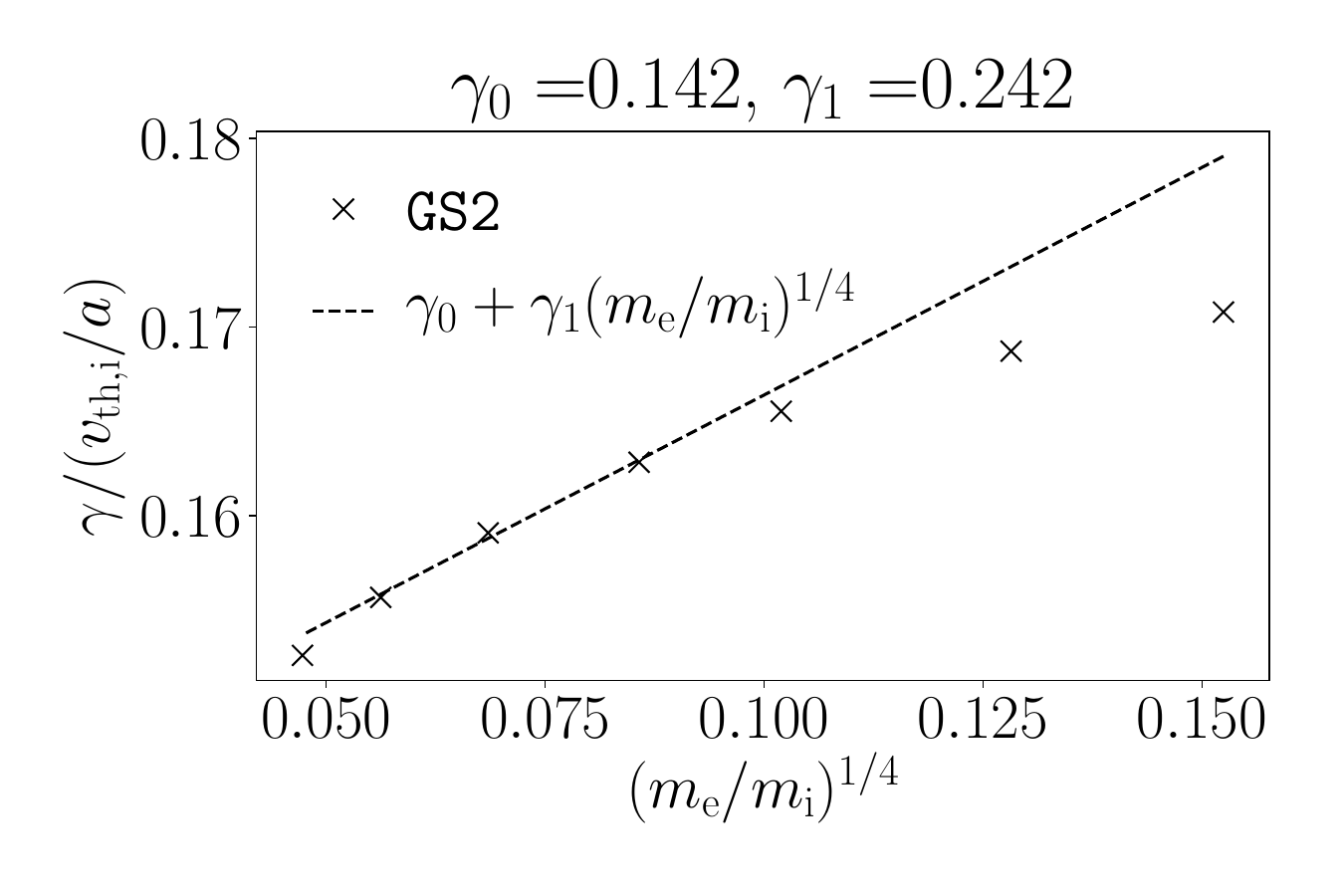}
% trim=left bottom right top, clip
\end{center}
\end{minipage}
\begin {minipage} {0.495\textwidth}
\begin{center}
\includegraphics[clip, trim=1cm 1cm 1cm 0cm, width=1.0\textwidth]{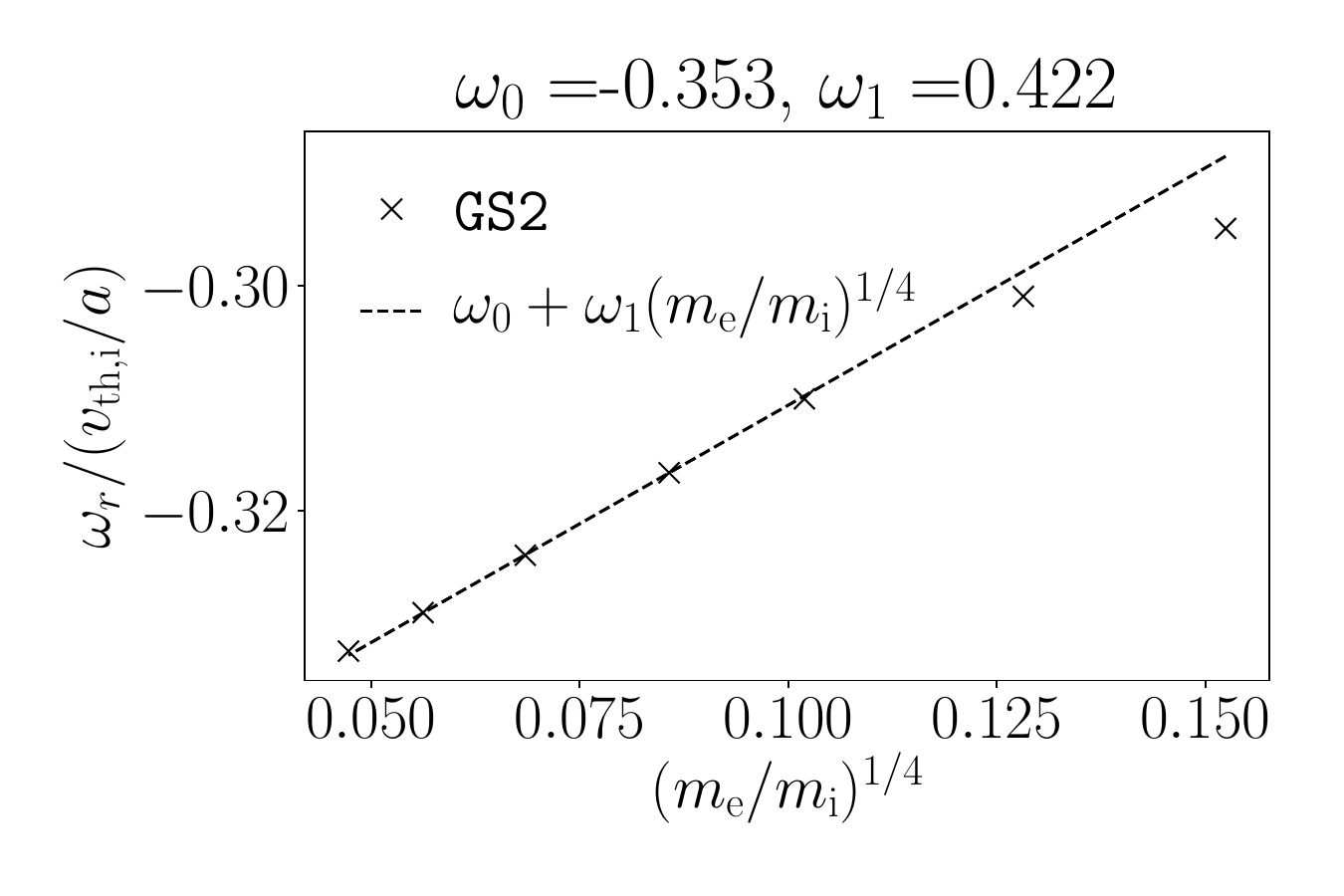}
\end{center}
\end{minipage}
 \caption{Plots of the growth rate $\growth$ (left) and real frequency $\wfreqr$ (right)
 as a function of $\massrtp{1/4}$, for  $\nustar = \nustarcollisional$
 (case B of figure \ref{fig:large_tail_modes_nustar_scan}).
 We give a linear fit to {illustrate} the dependence
 of $\growth$ and $\wfreqr$ on $\massrtp{1/4}$.
 } \label{fig:growthwfreqr-collisional}
 \end{figure}
 
 The asymptotic expansion for the collisional
 large-tail mode is carried out in powers of $\massrtp{1/4}$.
In consequence, we would expect that $\growth$
 and $\wfreqr$ would have leading order components $\growthz$ and $\wfreqrz$, respectively,
 that are independent of mass ratio, and sub-leading components $\growthh$ and $\wfreqrh$,
 respectively, that scale linearly with $\massrtp{1/4}$. 
 In figure \ref{fig:growthwfreqr-collisional} we plot $\growth$ and $\wfreqr$ versus $\massrtp{1/4}$,
 with linear fits given to indicate the order of magnitude of the variation with $\massrtp{1/4}$.
 The fit coefficients are of order unity consistent with a $\massrtp{1/4}$ expansion.
 We note that a nonlinear trend is observed in $\growth$ for the smallest $\massrtp{1/4}$.  
 This may be a result of numerical difficulties, in light of the very small $\delt$ required
 to converge the smallest $\massrtp{1/4}$ points in figure \ref{fig:growthwfreqr-collisional}.

 \subsection {Small tail modes}\label {section:numericalresults-small-tails}
 \begin{figure}[b]
\begin{center}
\includegraphics[clip, trim=0cm 1cm 0cm 0cm, width=0.6\textwidth]{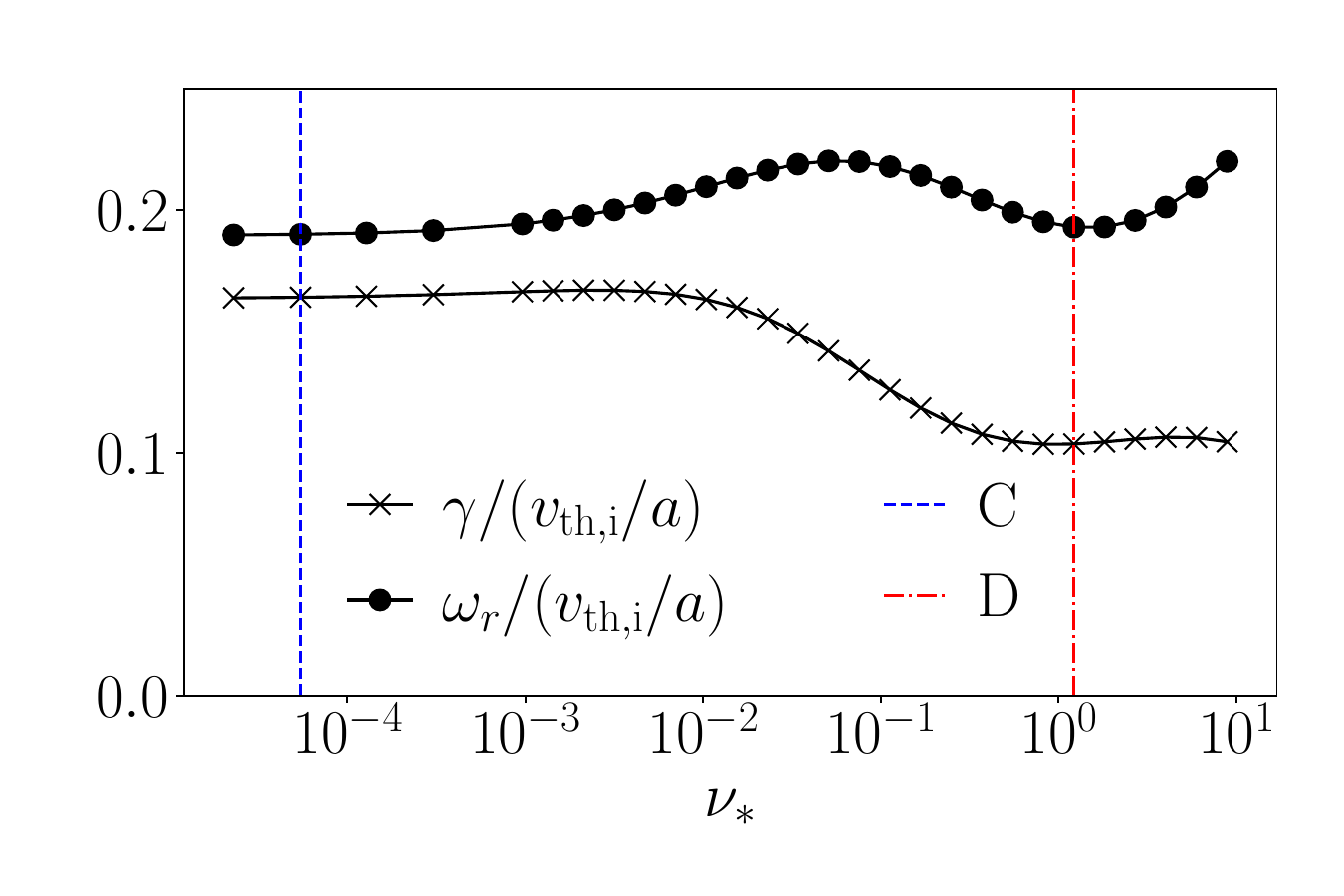}
% trim=left bottom right top, clip
\caption{ The growth rate $\growth$ and real frequency $\wfreqr$ of the (small-tail),
 ITG mode with $\kky\gyrdi = 0.5$, $\thetaz = 0.1$, and $\massrut = 61$, as a function of normalised electron collisionality $\nustar$.
 Note that $\growth$ and $\wfreqr$ experience $\order{1}$
 changes for $\nustar \gtrsim 10^{-2}$.
 We explain this feature in figure \ref{fig:nustar-dependence-small-tails}.
 The vertical dashed lines C and D indicate
 the $\nustar$ of the collisionless and collisional examples
 of the small-tail mode that are predicted by the theory in sections \ref{section:smalltailmode} and
 \ref{sec:electron-outer-collisional-small-tail}, respectively.   } \label{fig:small_tail_modes_nustar_scan}
\end{center}
\end{figure}

 \begin{figure}[htb]
\begin {minipage} {0.495\textwidth}
\begin{center}
\includegraphics[clip, trim=1cm 1cm 1cm 0cm, width=1.0\textwidth]{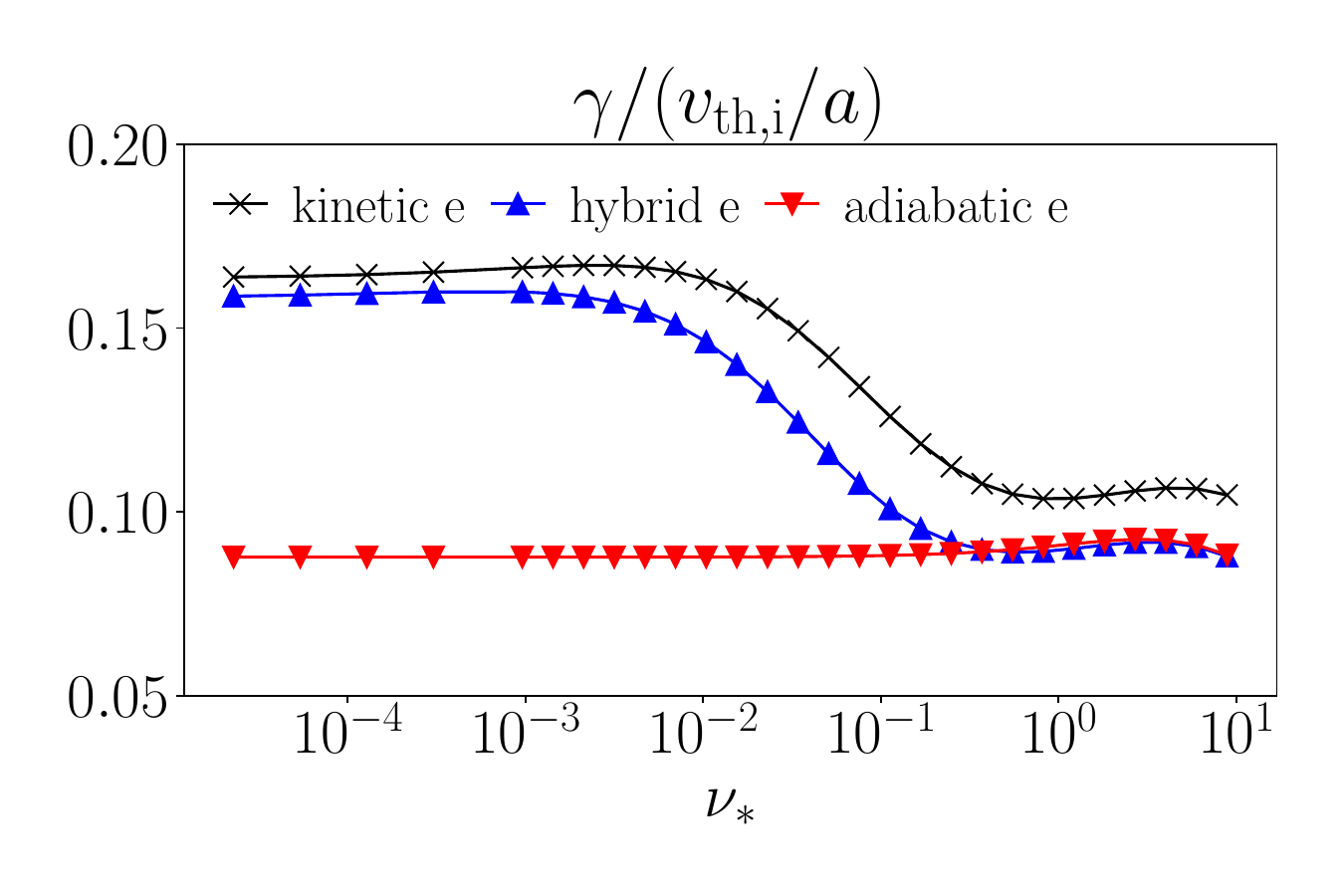}
% trim=left bottom right top, clip
\end{center}
\end{minipage}
\begin {minipage} {0.495\textwidth}
\begin{center}
\includegraphics[clip, trim=0cm 1cm 1cm 0cm, width=1.0\textwidth]{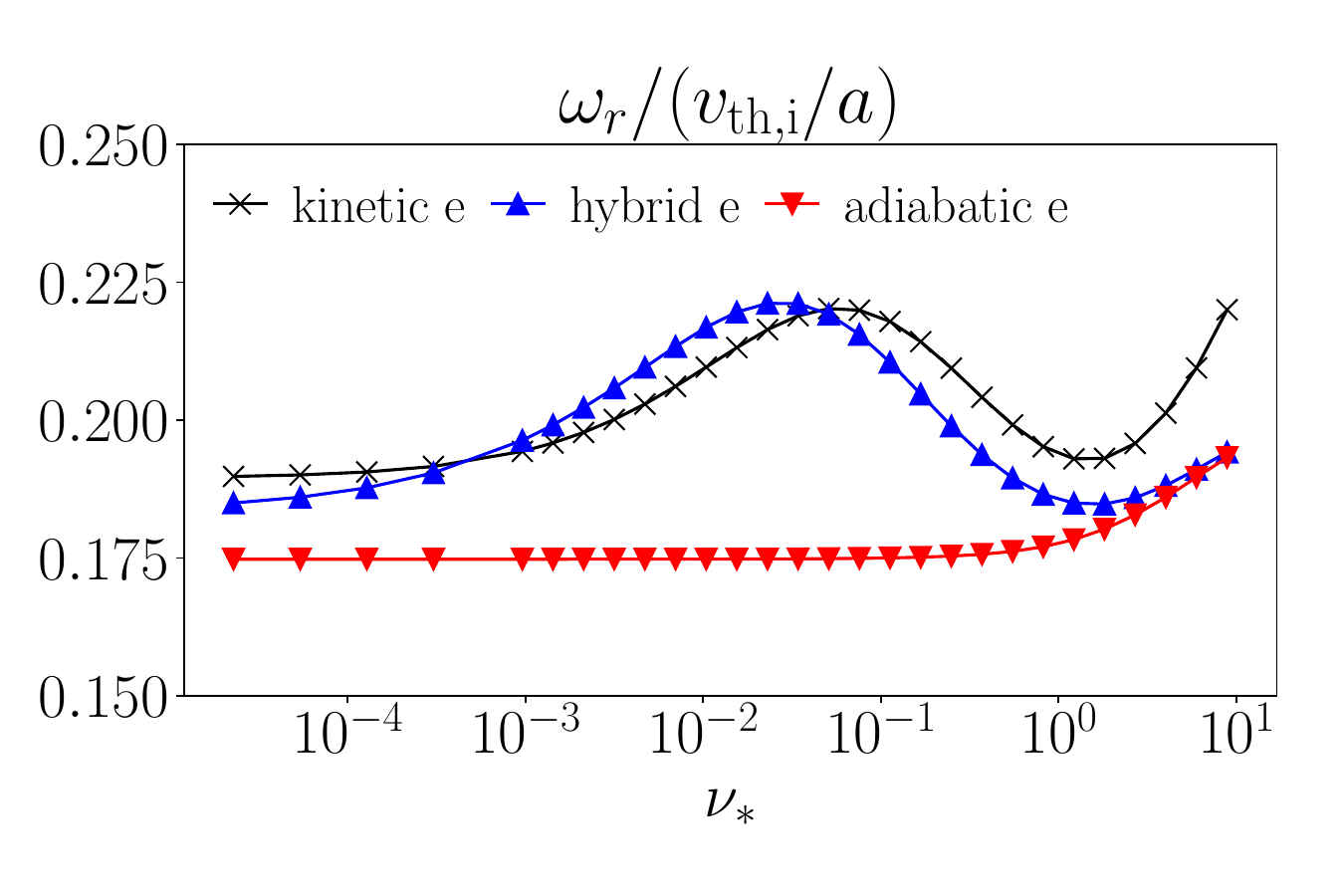}
\end{center}
\end{minipage}
\caption{The ITG mode growth rate $\growth$ (left) and real frequency $\wfreqr$ (right),
 calculated for the physical parameters given in figure \ref{fig:small_tail_modes_nustar_scan},
 for three different electron response models:
 the fully kinetic electron response (black crosses);
 a hybrid electron response, where the passing electrons are forced to respond adiabatically,
 but trapped electrons respond kinetically (blue upward triangles);
 and a fully adiabatic electron response (red downward triangles).
 Note that the fully kinetic electron response results in a similar
 growth rate to the case with a hybrid electron response --
 this implies that trapped electrons are responsible for the $\order{1}$
 variation in $\growth$ and $\wfreqr$.} \label{fig:nustar-dependence-small-tails}
 \end{figure}

 In this section we verify the mass ratio scalings for collisionless small-tail modes,
 described in section \ref{section:smalltailmode}, and collisional small-tail modes,
 described in section \ref{sec:electron-outer-collisional-small-tail}.
 We focus on the example of the ITG mode. We perform simulations using the magnetic
 geometry described at the start of section \ref{section:numericalresults}. We
 we take the temperature and density scale
 lengths to be $\lscal/\lti = \lscal/\lte = 2.3$ and $\lscal/\lln = 0.733$, respectively.
 We examine the mode with $\kky \gyrdi = 0.5$ and $\thetaz = 0.1$.
 
 In figure \ref{fig:small_tail_modes_nustar_scan}, we plot the growth rate $\growth$
 and real frequency $\wfreqr$ as a function of $\nustar$, for $\massrut= 61$.
 We take $\delt = 0.025 \lscal/\vtheri$ and $\ntwopi = 65$. We set the ion 
 collision frequency consistent with $\nustar$, i.e.,
 $\cfreqii = \aspect^{3/2} \nustar\vtheri / \sqrt{2} \saffac \rmajorz$.
 By scanning in $\massrt$,
 we identify that the modes in figure \ref{fig:small_tail_modes_nustar_scan} are small-tail modes
 by verifying that the $\lpar \gg 1$ part of the eigenmode $\charge \ptlinner/\tempe$
 is bounded by the estimates $\massrt \charge \ptlouter/\tempe$ (the collisionless case, 
 where $\cfreqee \lesssim \wfreq$ and we hold $\lscal \cfreqee/\vtheri$ fixed)
 and $\massrtp{1/4} \charge \ptlouter/\tempe$ (the collisional case, where
 $\cfreqee \gg \wfreq$, $\nustar \sim 1$ and we hold $\nustar$ fixed).
 The vertical dashed lines indicate the $\nustar$ of the clean examples of the
 collisionless and collisional small-tail modes that we describe in detail in the following sections.
 Figure \ref{fig:small_tail_modes_nustar_scan}
 shows that $\growth$ and $\wfreqr$ depend on $\nustar$ for
 $\nustar \gtrsim 10^{-2}$. In figure \ref{fig:nustar-dependence-small-tails}, we demonstrate that
 this $\nustar$ dependence arises from the trapped electron response.
 We compare $\growth$ and $\wfreqr$ of the ITG mode calculated using three different
 electron responses (cf. \cite{DominskinonadPOP015,hardmanpaper1,hardmanpaper2}):
 the fully kinetic electron response (black crosses),
 a hybrid electron response with kinetic trapped electrons and adiabatic passing
 electrons (blue triangles), and adiabatic electrons (red triangles).
 By comparing the hybrid electron case to the adiabatic electron case,
 we see that for very small $\nustar$, trapped electrons are decoupled
 from passing electrons and provide an $\order{1}$ modification to the growth rate.
 When $\nustar$ become sufficiently large, the effect of collisions is to detrap 
 the trapped electrons so that the electron response is essentially adiabatic.
 The difference between the cases with the fully kinetic electron response and the
 hybrid electron response shows that passing electrons make
 a small modification to $\growth$ and $\wfreqr$, consistent with the asymptotic theory.
 We note that simulations with adiabatic passing electrons can be converged with small
 $\ntwopi$: for the simulations presented here we take $\ntwopi = 9$.

    \subsubsection{Case C -- a collisionless small-tail mode.}\label{section:numericalresults-collisionless-small-tails}

 \begin{figure}[htb]
\begin {minipage} {0.495\textwidth}
\begin{center}
\includegraphics[clip, trim=1cm 1cm 1cm 0cm, width=1.0\textwidth]{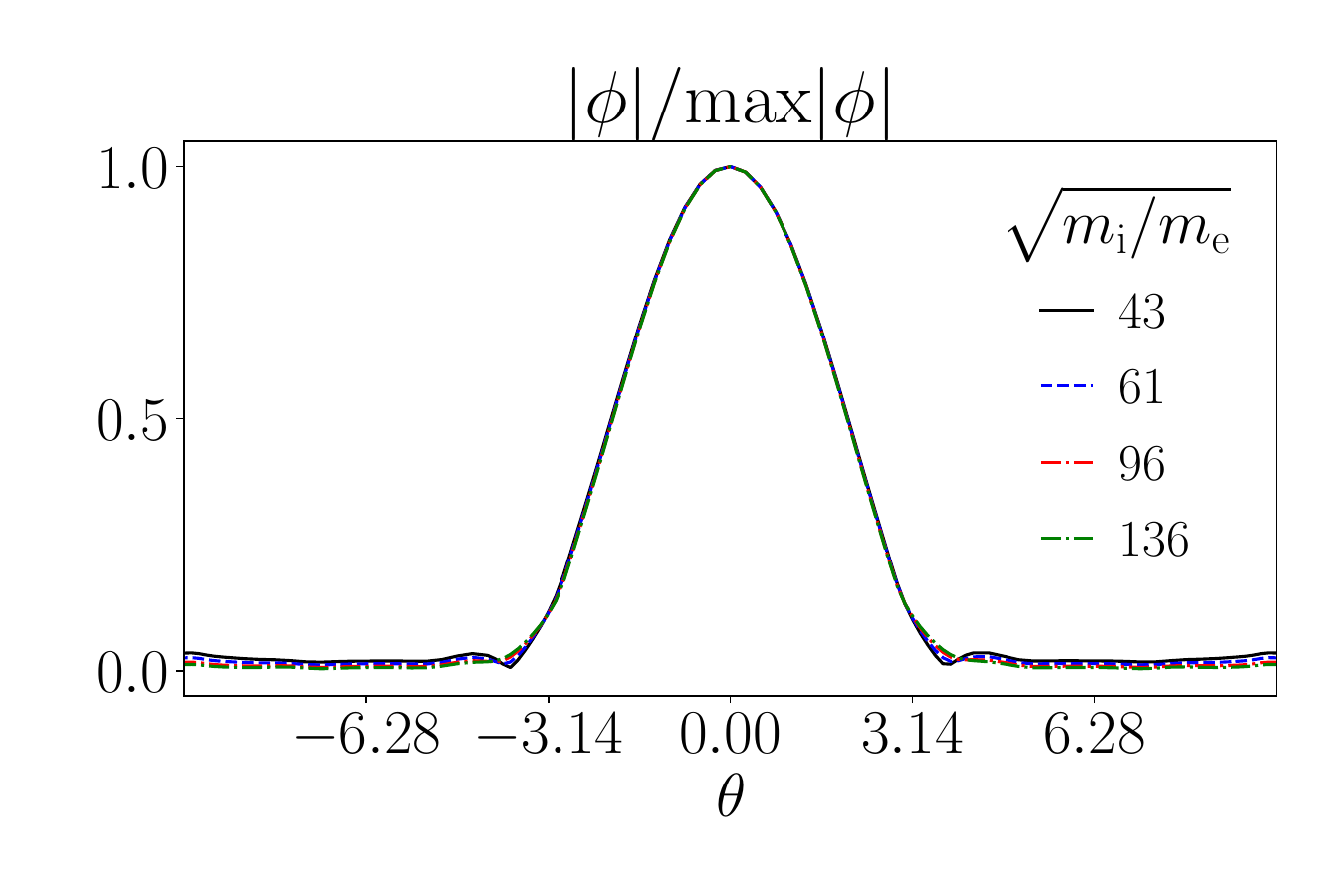}
% trim=left bottom right top, clip
\end{center}
\end{minipage}
\begin {minipage} {0.495\textwidth}
\begin{center}
\includegraphics[clip, trim=1cm 1cm 1cm 0cm, width=1.0\textwidth]{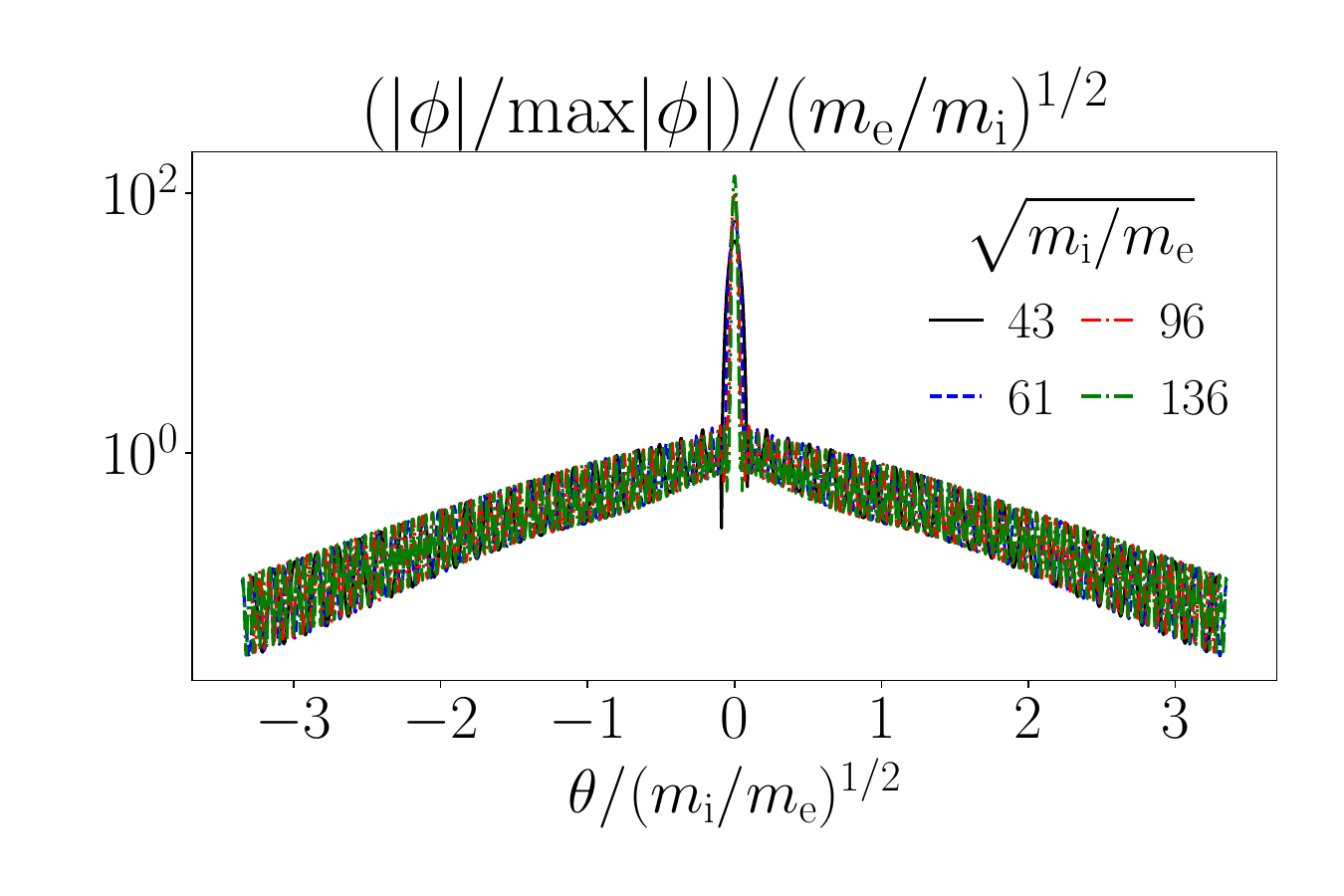}
\end{center}
\end{minipage}
\caption{Two views of the electrostatic potential $\ptl$, 
calculated for different mass ratios
 with fixed $\lscal\cfreqee/\vtheri = \cfreqeecollisionlessitg$,
 corresponding to case C of figure \ref{fig:small_tail_modes_nustar_scan}
 when $\massrut = 61$, and hence, 
 $\nustar = \nustarcollisionlessitg$.
  (left)
 The potential $\ptl$ is plotted against the unscaled ballooning angle $\lpar$,
 and normalised to its maximum value. The fact that the curves overlay for $\lpar \sim 1$
 indicates that the potential eigenmode is independent of $\massrt$ to leading order.
 (right) The potential is plotted against the scaled ballooning angle $\lpar /\massrut$,
 and normalised by a factor of $\massrt {\rm max} |\ptl|$.
 The fact that the curves overlap in the region $\lpar \sim \massrut$
 indicates that the mode is a small-tail mode, satisfying the ordering
 \refeq{eq:small-tail-collisionless-ptlinner-estimate}.}
 \label{fig:phi-collisionless-small-tails}
 \end{figure}

  \begin{figure}
\begin{center}
\includegraphics[clip, trim=1cm 1cm 1cm 0cm, width=0.5\textwidth]{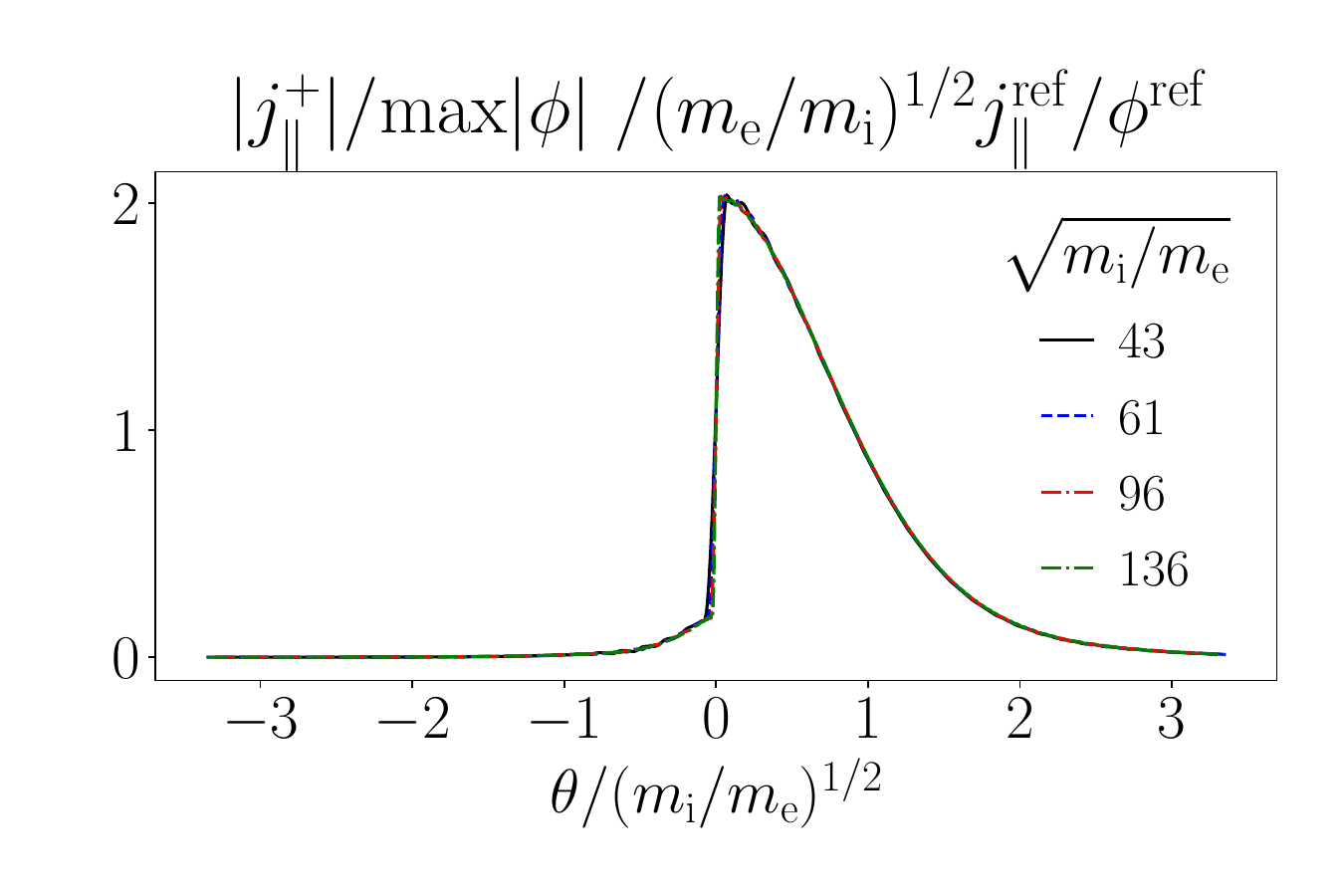}
% trim=left bottom right top, clip
\caption{The field $\jintplus$, calculated for different mass ratios 
  and fixed $\lscal\cfreqee/\vtheri = \cfreqeecollisionlessitg$, 
  corresponding to $\nustar = \nustarcollisionlessitg$ for 
  $\massrut = 61$
 (case C of figure \ref{fig:small_tail_modes_nustar_scan}).
 The fact that the curves overlay on the
  $\lpar/\massrut$ axis confirms that the mode is a collisionless small-tail mode,
  satisfying the ordering \refeq{eq:small-tail-collisionless-HHe-estimate}. }
  \label{fig:jintplus-collisionless-small-tail}
\end{center}
 \end{figure}

    We consider the ITG mode from figure \ref{fig:small_tail_modes_nustar_scan}
    with $\nustar = \nustarcollisionlessitg$. We scan in $\me / \mi = 5.4 \times 10^{-4}$
    to $5.4 \times 10^{-5}$, whilst holding fixed $\lscal\cfreqee/\vtheri = \cfreqeecollisionlessitg$.
    The value of $\cfreqee$ is chosen so that $\nustar = \nustarcollisionlessitg$ 
    for $\massrut = (\md/\me)^{1/2}$, 
    and we take $\cfreqii/\cfreqee = (\me/\md)^{1/2}/\sqrt{2} $. 
    To identify a mode as a collisionless small-tail mode, we must demonstrate several properties.
    First, that there is a $\lpar \sim 1$ region where $\charge\ptl/\tempe$ is independent
    of $\massrt$ at leading order. Second, that the potential in the $\lpar \gg 1$
    region has an amplitude given by estimate \refeq{eq:small-tail-collisionless-ptlinner-estimate},
    and an envelope $\lpar \sim \massrut$. Third, that the electron distribution function
    has a size given by estimate \refeq{eq:small-tail-collisionless-HHe-estimate},
    and an envelope with scale $\lpar \sim \massrut$.
    In figure \ref{fig:phi-collisionless-small-tails},
    we demonstrate that the first and second properties are satisfied.
    In figure \ref{fig:jintplus-collisionless-small-tail}, we use $\jintplus$ as a measure of $\HHe$
    to demonstrate that the third property is satisfied. In these simulations, we take
    $\delt = 0.025 \lscal/\vtheri$ and $\ntwopi = 65\sqrt{\mi/\md}$.

 \begin{figure}
\begin {minipage} {0.495\textwidth}
\begin{center}
\includegraphics[clip, trim=1cm 1cm 1cm 0cm, width=1.0\textwidth]{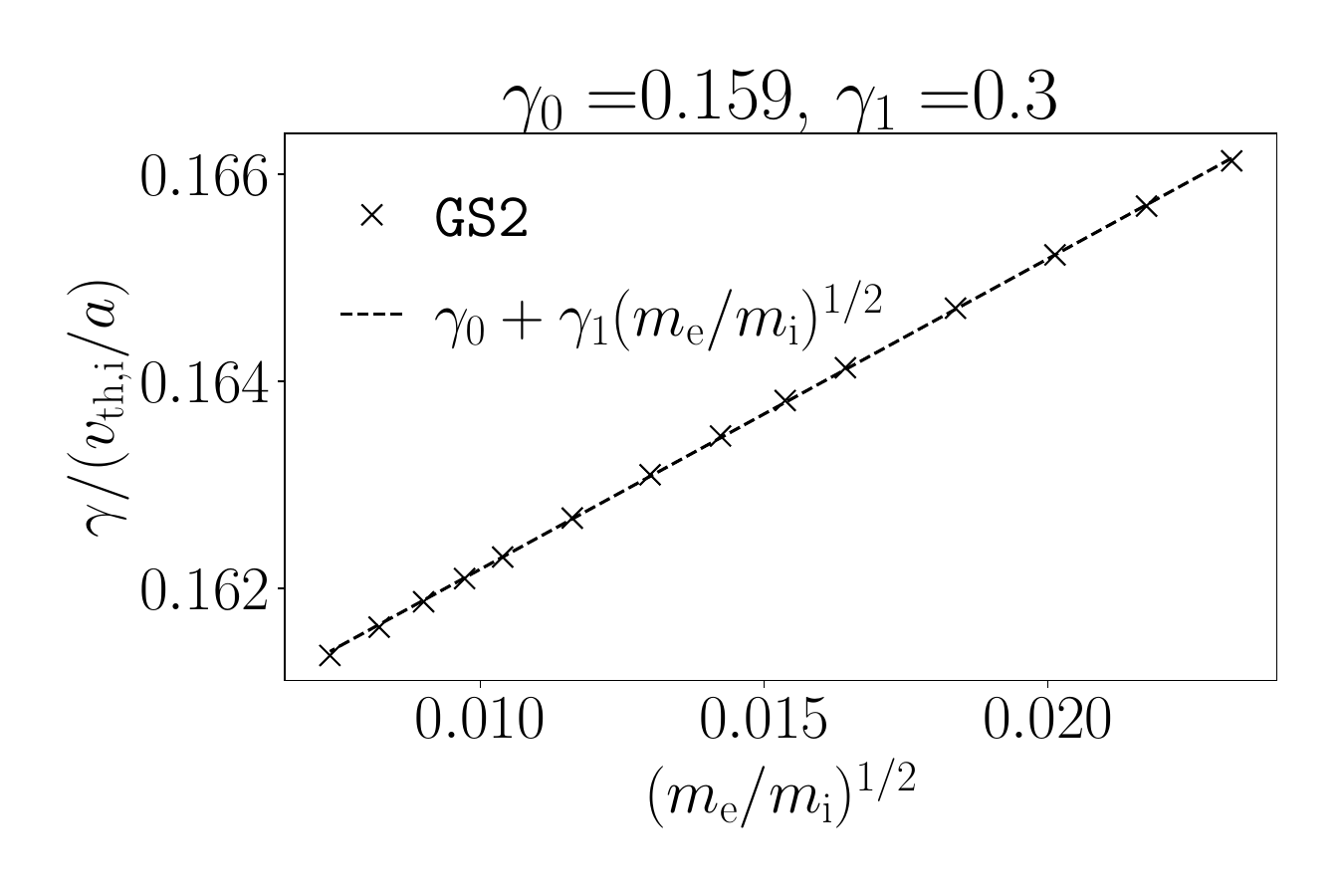}
% trim=left bottom right top, clip
\end{center}
\end{minipage}
\begin {minipage} {0.495\textwidth}
\begin{center}
\includegraphics[clip, trim=1cm 1cm 1cm 0cm, width=1.0\textwidth]{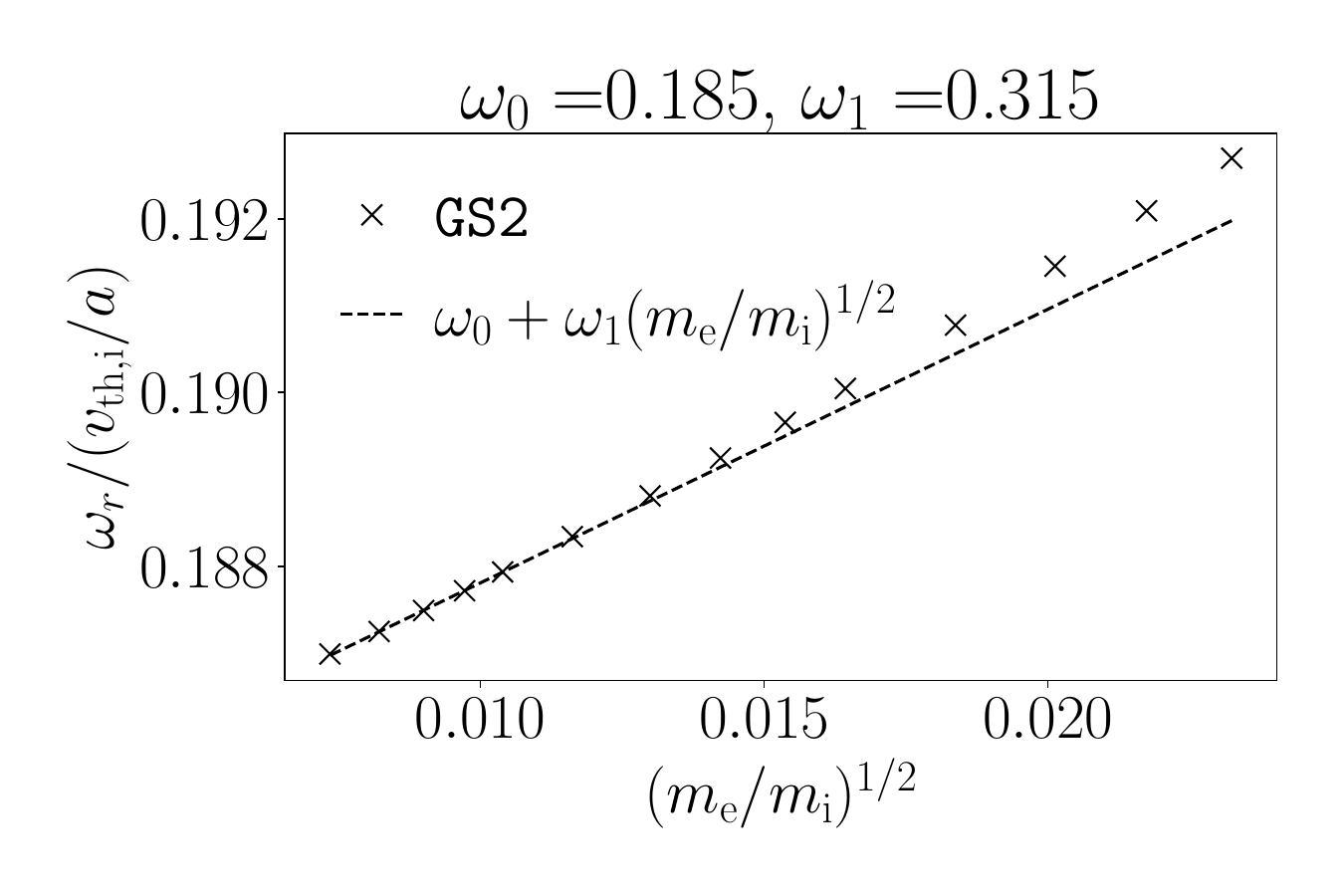}
\end{center}
\end{minipage}
 \caption{Plots of the growth rate $\growth$ (left) and real frequency $\wfreqr$ (right)
 as a function of $\massrt$, for the mode featured in figures \ref{fig:phi-collisionless-small-tails}
 and \ref{fig:jintplus-collisionless-small-tail}.
 We give a linear fit to demonstrate that the dependence of $\growth$ and $\wfreqr$ on $\massrtp{1/2}$
 is consistent with a $\massrtp{1/2}$ expansion.} \label{fig:growthwfreqr-collisionless-small-tails}
 \end{figure}

Finally, we discuss the $\massrt$ dependence of the growth rate $\growth$ and real frequency $\wfreqr$
 in the collisionless example small-tail mode featured in 
 figures \ref{fig:phi-collisionless-small-tails} and \ref{fig:jintplus-collisionless-small-tail}.
 The growth rate and
 frequency are plotted in figure \ref{fig:growthwfreqr-collisionless-small-tails}.
 As the asymptotic expansion is carried out in $\massrt$,
 we expect to see a linear dependence in $\massrt$. This is observed for a wide range of $\massrt$ in $\growth$,
 but for a smaller range of $\massrt$ for $\wfreqr$. We note that the fit coefficients
 are of order unity, confirming that the dependence of $\growth$ and $\wfreqr$ on $\massrt$
 is consistent with the $\massrt$ expansion.

%%%%%%%%%%%%%%%%%%%%%%%%%%%%%%%%%%%%%%%%%%%%%%%%%%%%%%%%%  
%%%%%%%%%%%%%%%%%%%%%%%%%%%%%%%%%%%%%%%%%%%%%%%%%%%%%%%%%  
  \begin{figure}[hbt]
\begin {minipage} {0.495\textwidth}
\begin{center}
\includegraphics[clip, trim=1cm 1cm 1cm 0cm, width=1.0\textwidth]{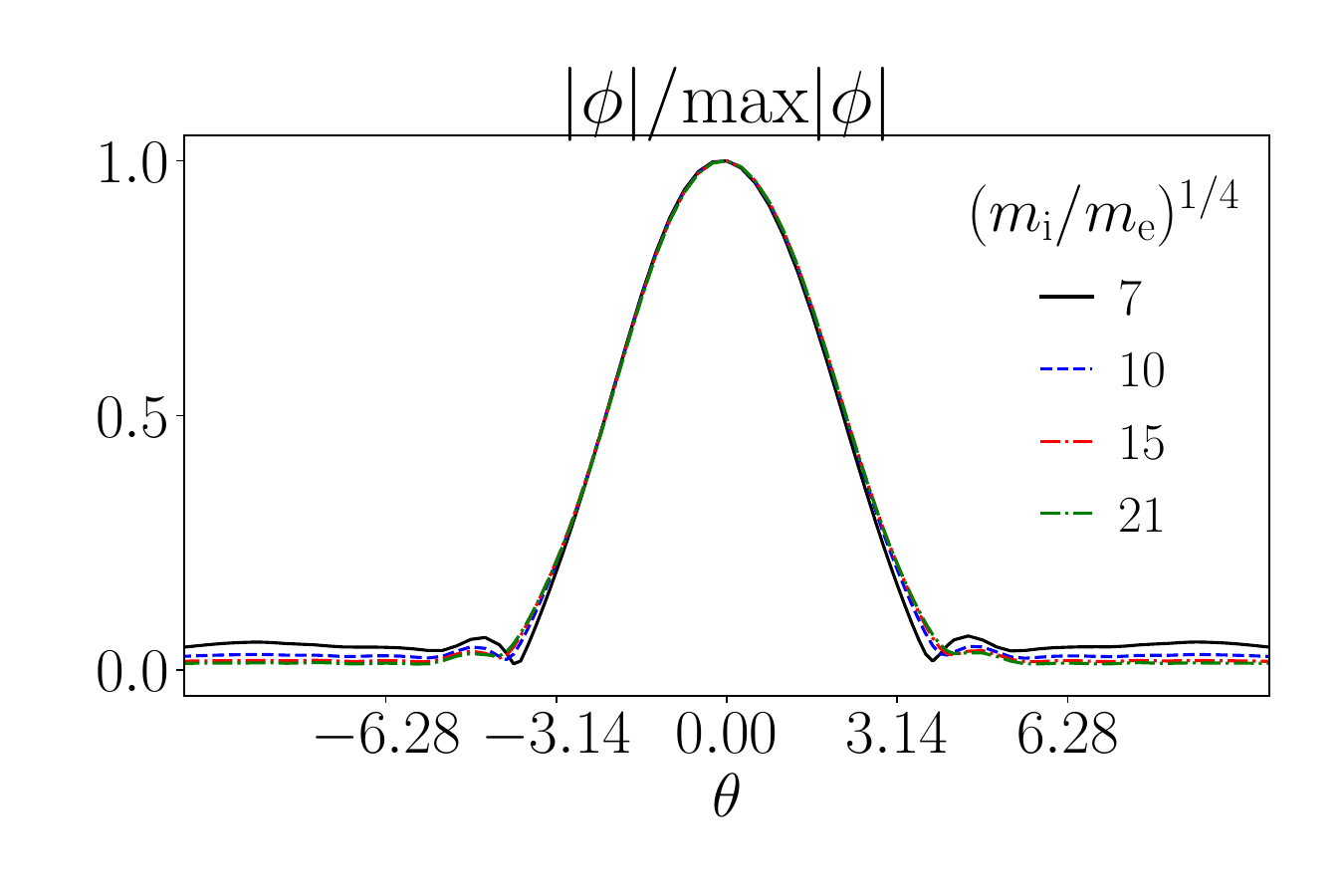}
% trim=left bottom right top, clip
\end{center}
\end{minipage}
\begin {minipage} {0.495\textwidth}
\begin{center}
\includegraphics[clip, trim=1cm 0cm 1cm 0cm, width=1.0\textwidth]{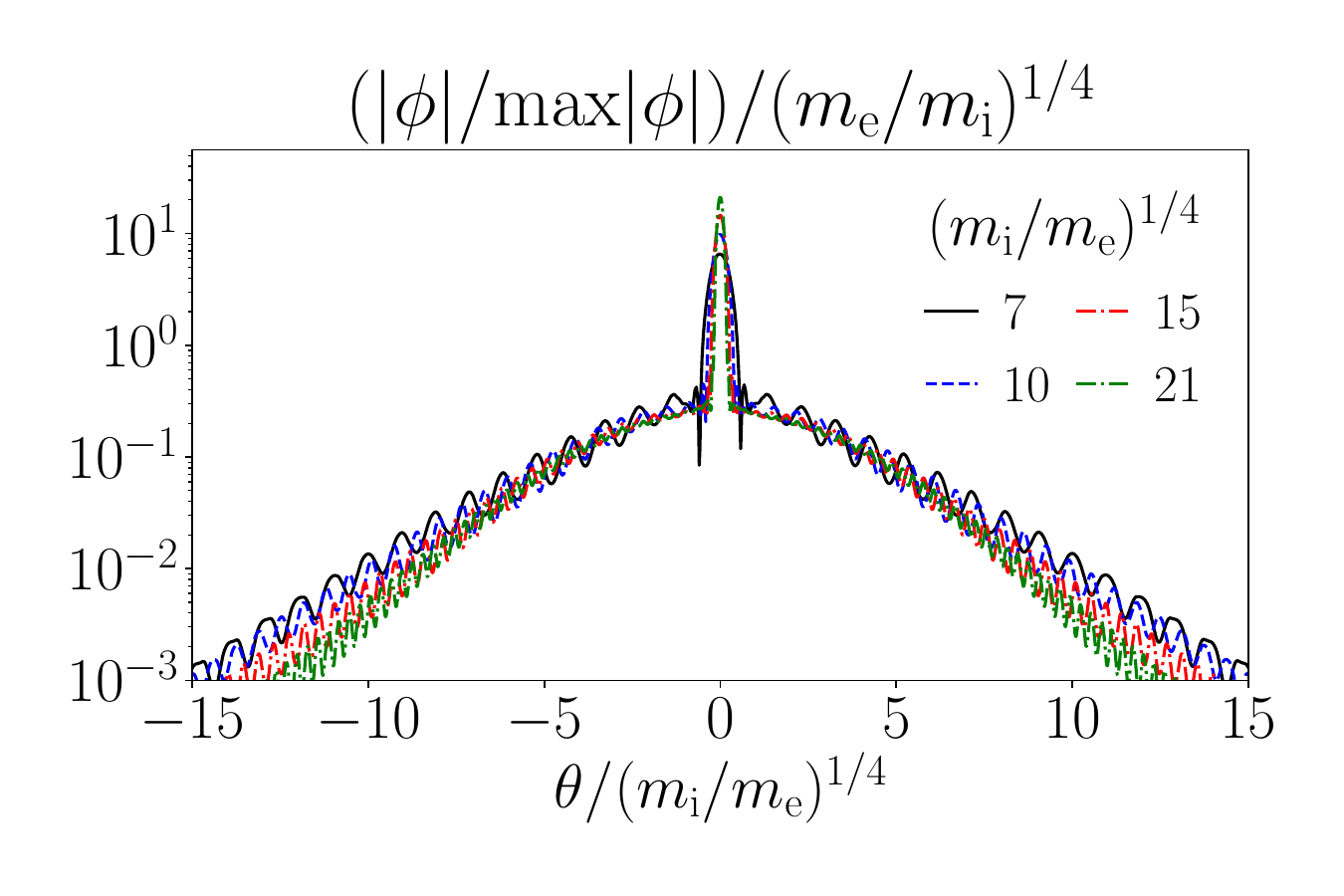}
\end{center}
\end{minipage}
\caption{Two views of the electrostatic potential $\ptl$, calculated for $\nustar = \nustarcollisionalitg$
 (case D of figure \ref{fig:small_tail_modes_nustar_scan}) for 
  different mass ratios.
  (left) The potential $\ptl$ is plotted against the unscaled ballooning angle,
  and normalised to its maximum value. That the curves overlay in the $\lpar \sim 1$
  region indicates that the mode is a small-tail mode. (right)
 The potential $\ptl$ is plotted against the scaled ballooning angle $\lpar / \massrutp{1/4}$,
 and normalised to its maximum value, divided by $\massrtp{1/4}$.
 That the curves overlay for $\lpar \sim \massrutp{1/4}$ indicates that we
 have correctly identified the scaling \refeq{eq:small-tails-estimate}
 for the size of the electron response,
 and the size of the mode envelope.} \label{fig:phi-collisional-small-tails}
 \end{figure}
  
  \subsubsection{Case D -- a collisional small-tail mode.}
In the collisional limit, the electron response of a small-tail mode is
 characterised by a jump in the electron flows across the $\lpar \sim 1$ region.
 This results in the scalings \refeq{eq:small-tails-estimate} for the electrostatic
 potential, electron density, and electron temperature in the $\lpar \gg 1$ region.
 As in the large-tail collisional mode, the size of the envelope
 of the mode is expected to be of scale $\lpar \sim \massrutp{1/4}$.
 To test these scalings, we examine an ITG mode with normalised
 electron collision frequency $\nustar=\nustarcollisionalitg$
 (case D of figure \ref{fig:small_tail_modes_nustar_scan}).
 We scan in $\me / \mi$ from $ 5.4 \times 10^{-4}$ to $5.0 \times 10^{-6}$
 while holding $\nustar$ fixed.
 We plot the electrostatic potential $\ptl$ in figure \ref{fig:phi-collisional-small-tails}.
 We note that $\ptl$ has no mass dependence for $\lpar \sim 1$,
 and that $\ptl$ has the mass scaling given by the estimate 
 \refeq{eq:small-tails-estimate} for $\lpar \sim \massrutp{1/4}$.
 This confirms that the mode is a collisional, small-tail mode.
 In the simulations, we take $\delt = 0.025 \lscal/\vtheri$ and $\ntwopi = 39 (\mi/\md)^{1/4}$.

    \begin{figure}[hbt]
\begin {minipage} {0.495\textwidth}
\begin{center}
\includegraphics[clip, trim=1cm 1cm 1cm 0cm, width=1.0\textwidth]{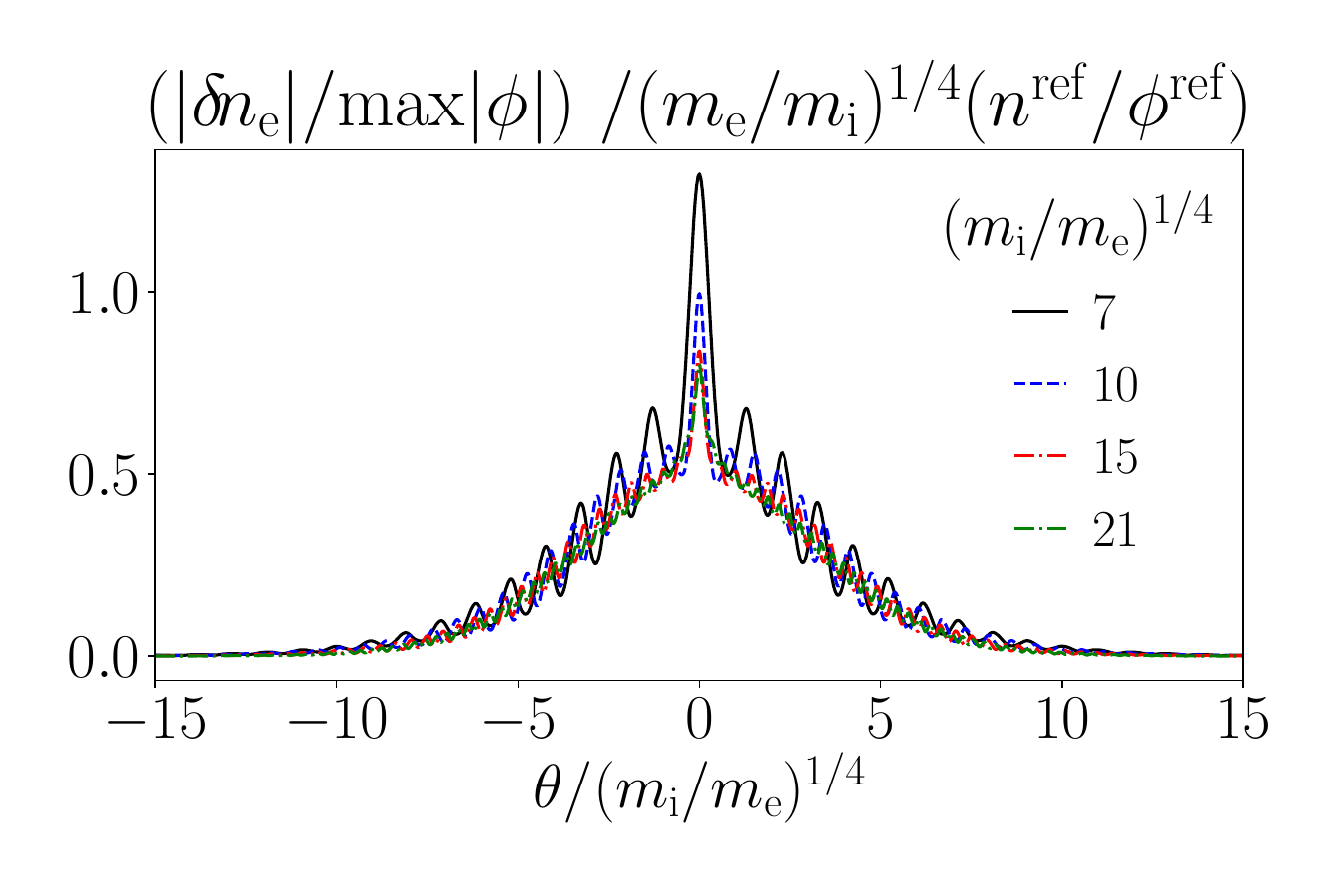}
% trim=left bottom right top, clip
\end{center}
\end{minipage}
\begin {minipage} {0.495\textwidth}
\begin{center}
\includegraphics[clip, trim=1cm 1cm 1cm 0cm, width=1.0\textwidth]{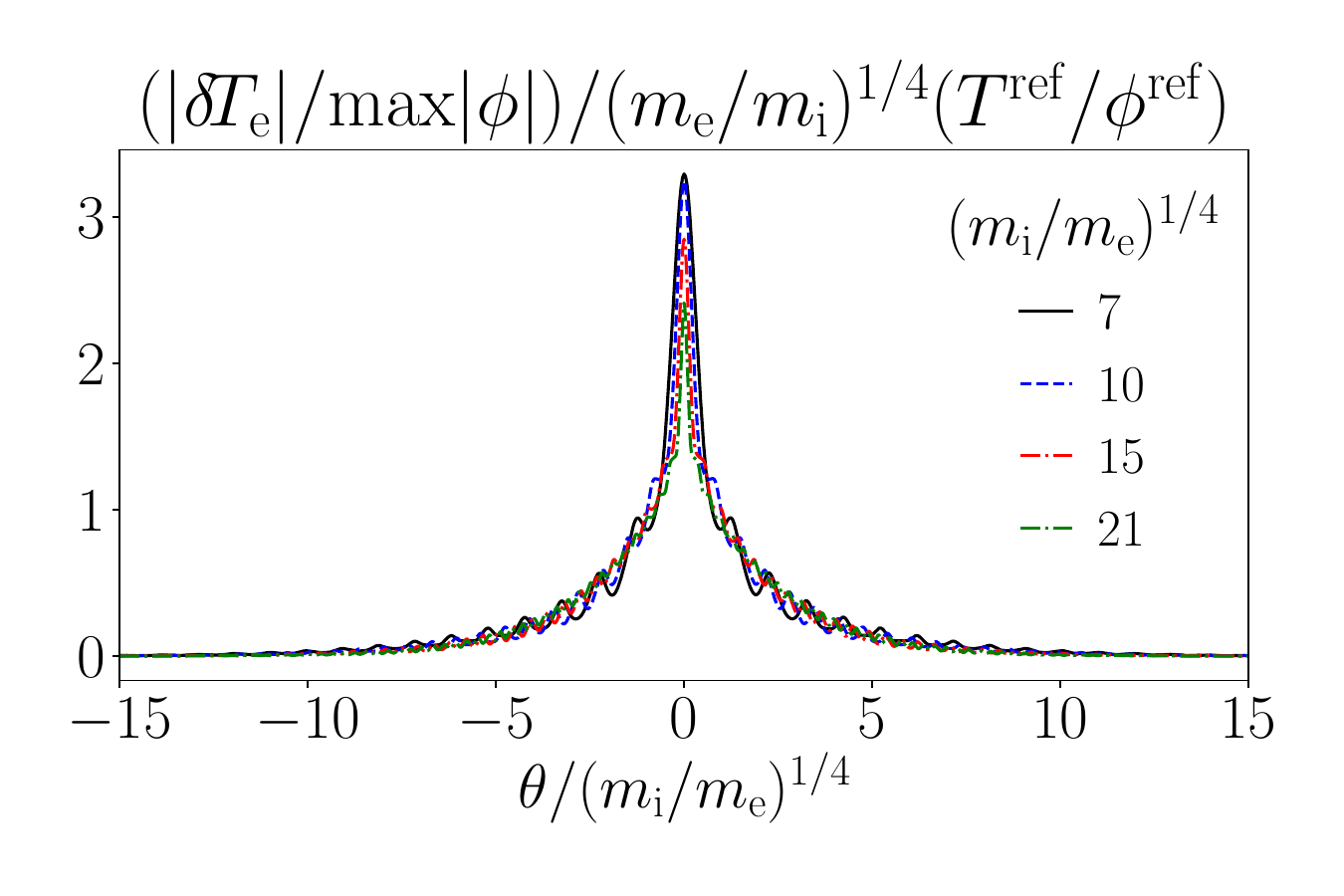}
\end{center}
\end{minipage}
\caption{(left) The electron nonadiabatic density $\ddense$, and (right) 
 the electron temperature $\dtempe$, calculated for $\nustar = \nustarcollisionalitg$
 (case D of figure \ref{fig:small_tail_modes_nustar_scan}) for 
  different mass ratios.} \label{fig:ddense-collisional-small-tails}
 \end{figure}

  \begin{figure}[hbt]
\begin{center}
\includegraphics[clip, trim=1cm 1cm 1cm 0cm, width=0.5\textwidth]{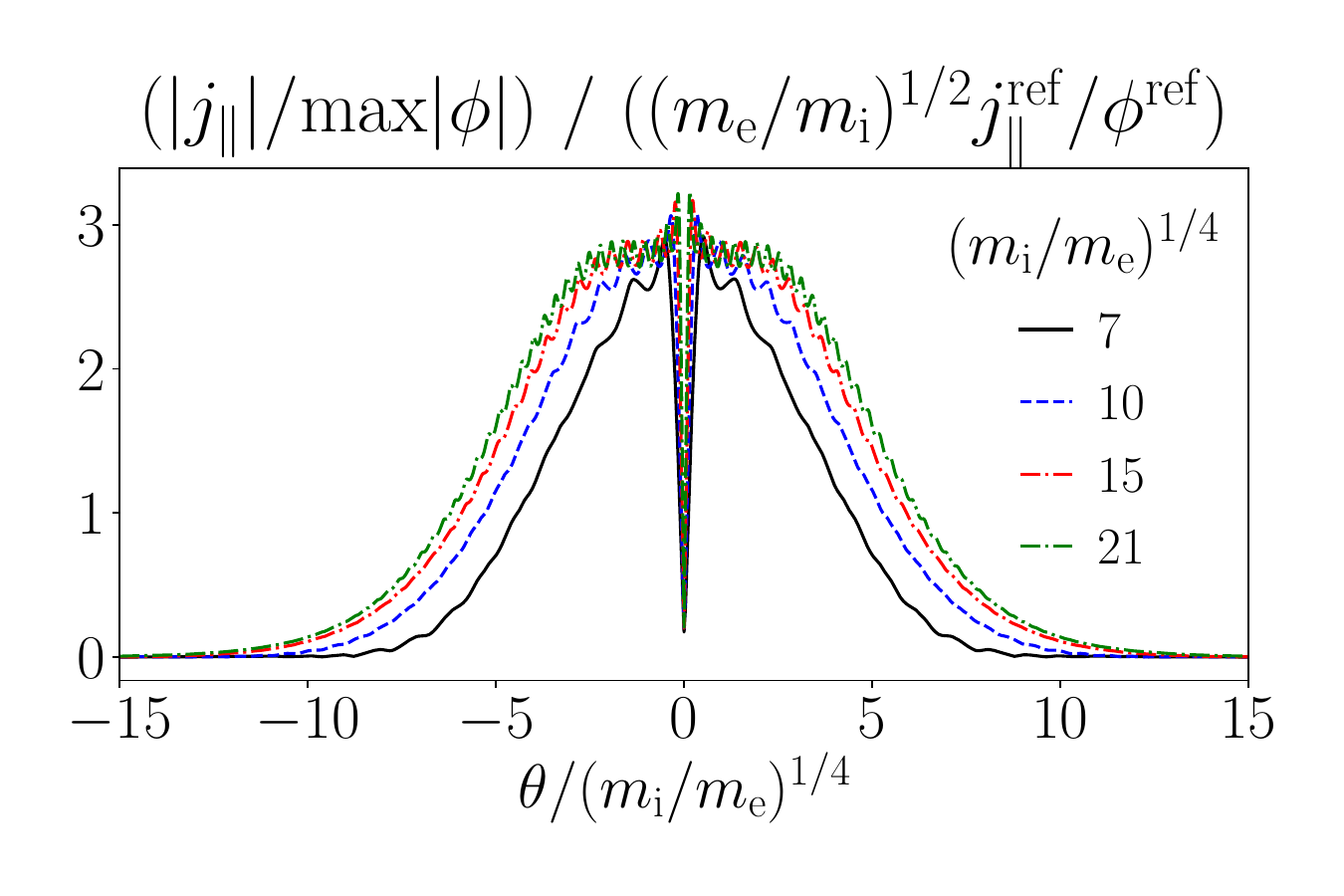}
% trim=left bottom right top, clip
\caption{The field $\jint$, calculated for $\nustar = \nustarcollisionalitg$
 (case D of figure \ref{fig:small_tail_modes_nustar_scan}) for 
  different mass ratios. Whilst the curves do not overlay perfectly,
  we note that the curves appear to be converging for the largest $\massrutp{1/4}$ in the scan.
} \label{fig:jint-collisional}
\end{center}
 \end{figure}
  
To illustrate the electron response further, we plot the nonadiabatic
 electron density $\ddense$ and electron temperature $\dtempe$
 in figure \ref{fig:ddense-collisional-small-tails}.
 The scaling \refeq{eq:small-tails-estimate} is confirmed by the fact that
 the curves overlay with the mass scaling $\massrtp{1/4}$
 in the amplitude, and the mass scaling $\massrutp{1/4}$ in the ballooning angle.
 In figure \ref{fig:jint-collisional},
 we plot the field $\jint$, normalised to the maximum value of $|\ptl|$.
 Consistent with the identification of the mode as a collisional small-tail mode,
 the envelope of $\jint$ appears to scale like $\lpar \sim \massrutp{1/4}$,
 and the amplitude is small by $\massrt$. Although the envelope rescaling is not perfect
 in figure \ref{fig:jint-collisional}, the curves appear to be converging for
 the largest $\massrutp{1/4}$ in the scan. We note that the $\massrutp{1/4}$ rescaling is better than 
 a $\massrut$ rescaling.  
 This figure completes the demonstration of the physical picture for the
 collisional small-tail mode: the ions generate an electrostatic potential
 at $\lpar \sim 1$, the electrons respond with a $\massrt$ small flow, and the
 small electron flow self-consistently sets up a $\massrtp{1/4}$ nonadiabatic
 electron density and temperature response.
\begin{figure}
\begin {minipage} {0.495\textwidth}
\begin{center}
\includegraphics[clip, trim=1cm 1cm 1cm 0cm, width=1.0\textwidth]{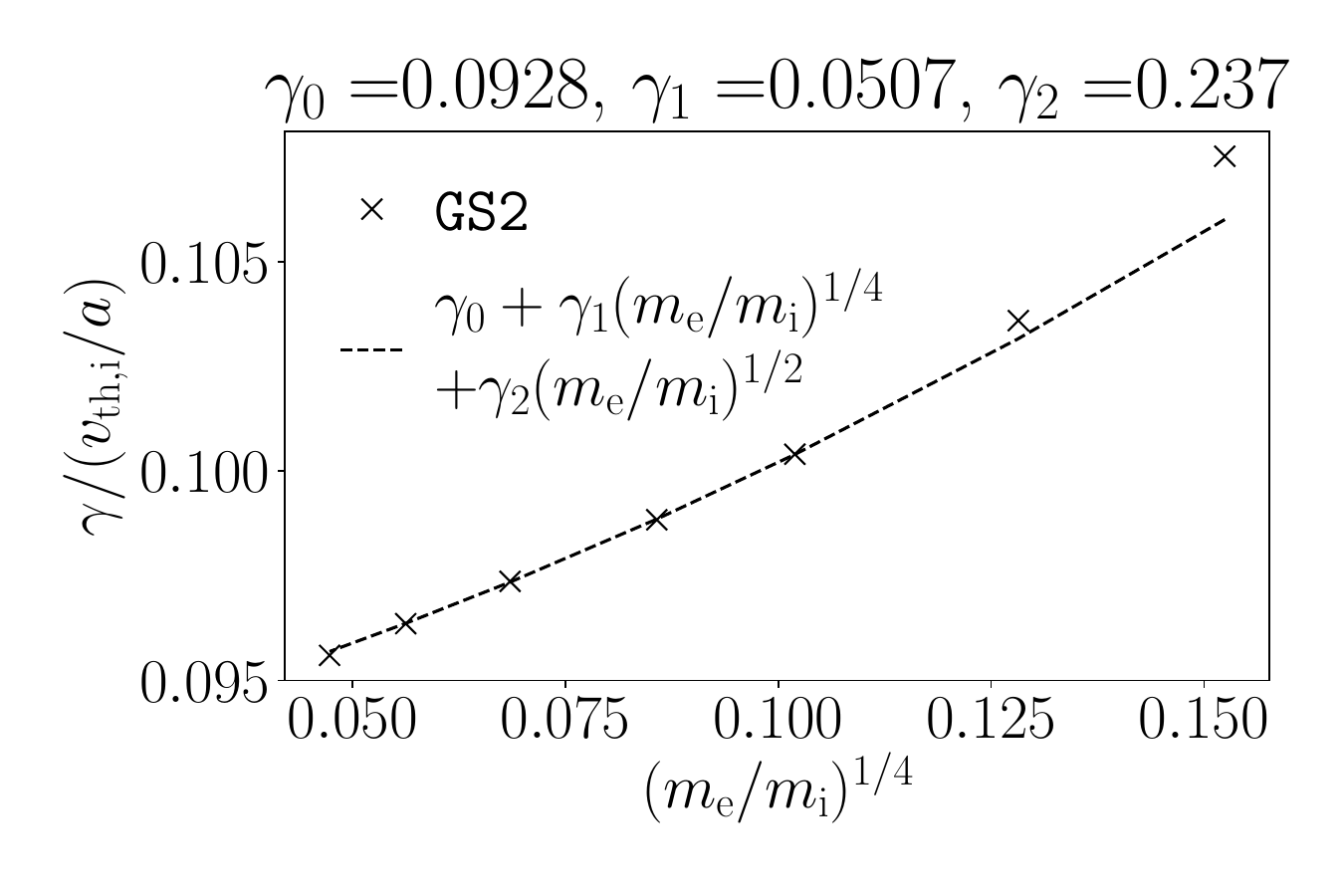}
% trim=left bottom right top, clip
\end{center}
\end{minipage}
\begin {minipage} {0.495\textwidth}
\begin{center}
\includegraphics[clip, trim=1cm 1cm 1cm 0cm, width=1.0\textwidth]{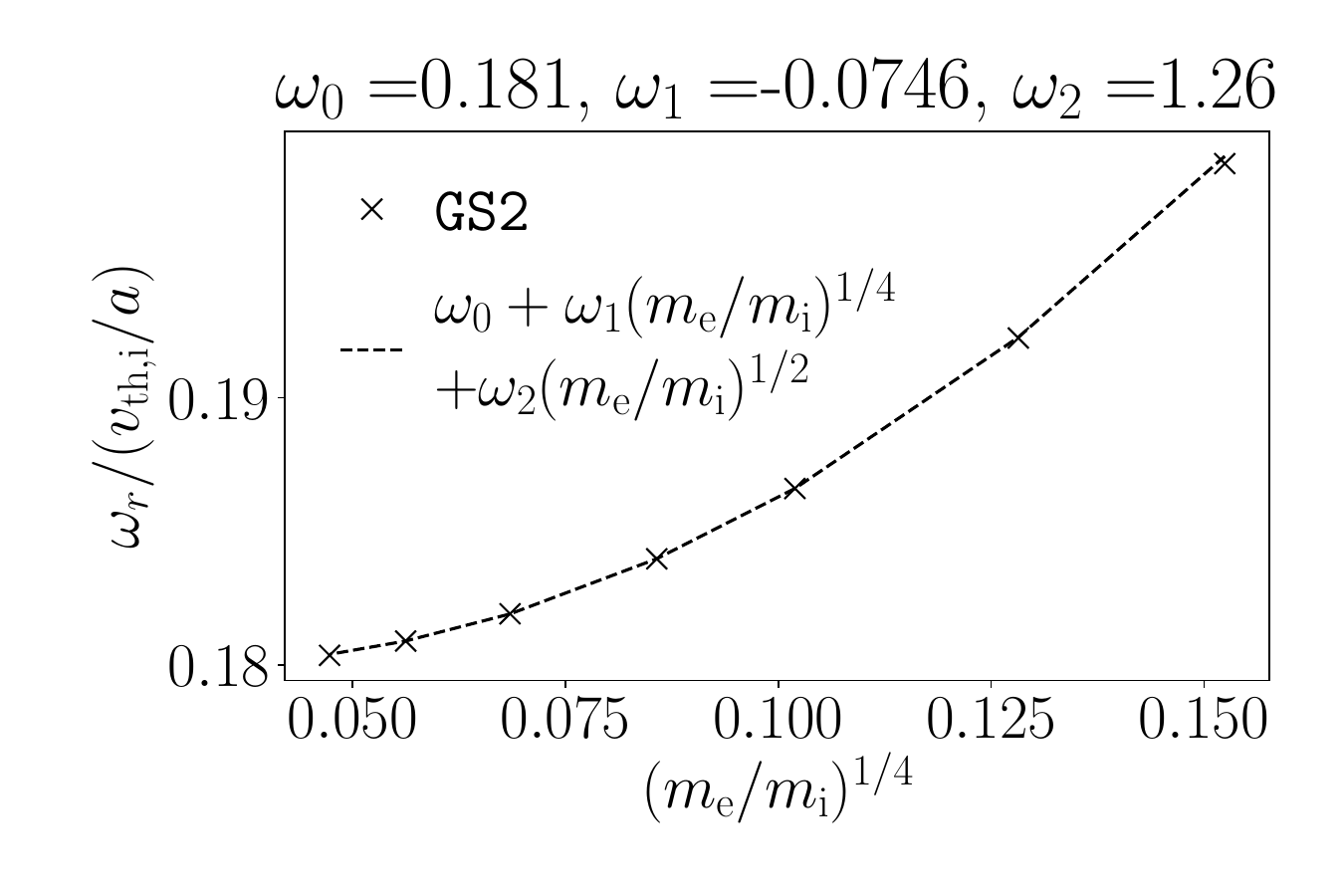}
\end{center}
\end{minipage}
 \caption{Plots of the growth rate $\growth$ (left) and real frequency $\wfreqr$ (right)
 as a function of $\massrtp{1/4}$, for  $\nustar = \nustarcollisionalitg$
 (case D of figure \ref{fig:small_tail_modes_nustar_scan}).
 We give a quadratic fit to demonstrate that the dependence of $\growth$ and $\wfreqr$ on $\massrtp{1/4}$
 is consistent with a $\massrtp{1/4}$ expansion if a $\massrtp{1/2}$ term is included.} \label{fig:growthwfreqr-collisional-small-tails}
 \end{figure}  
 
Finally, in figure \ref{fig:growthwfreqr-collisional-small-tails}
 we plot the growth rate $\growth$ and the real frequency $\wfreqr$ as a function of $\massrtp{1/4}$.
 The asymptotic expansion for the collisional small-tail mode is carried out in powers of $\massrtp{1/4}$.
 Hence, we would expect the leading corrections to the frequency to scale as $\wfreqh/\wfreqz \sim \massrtp{1/4}$,
 with a subleading correction scaling like $\wfreqo/\wfreqz \sim \massrtp{1/2}$.
 Possibly consistent with this, in figure \ref{fig:growthwfreqr-collisional-small-tails}
 we see that the coefficients of plausible quadratic
 fits have parameters $\growth_0$ and $\growth_2$ ($\wfreq_0$ and $\wfreq_2$) of order unity, with 
 $\growth_1$ ($\wfreq_1$) unexpectedly small.
  This may indicate the need for a more complicated asymptotic theory,
  or simply that the $\massrtp{1/4}$ correction is small in this example.
  
  \subsection {The transition between the large-tail and small-tail modes}\label {section:theta0-dependence}
 
 \begin{figure}
\begin {minipage} {0.495\textwidth}
\begin{center}
\includegraphics[clip, trim=1cm 1cm 1cm 0cm, width=1.0\textwidth]{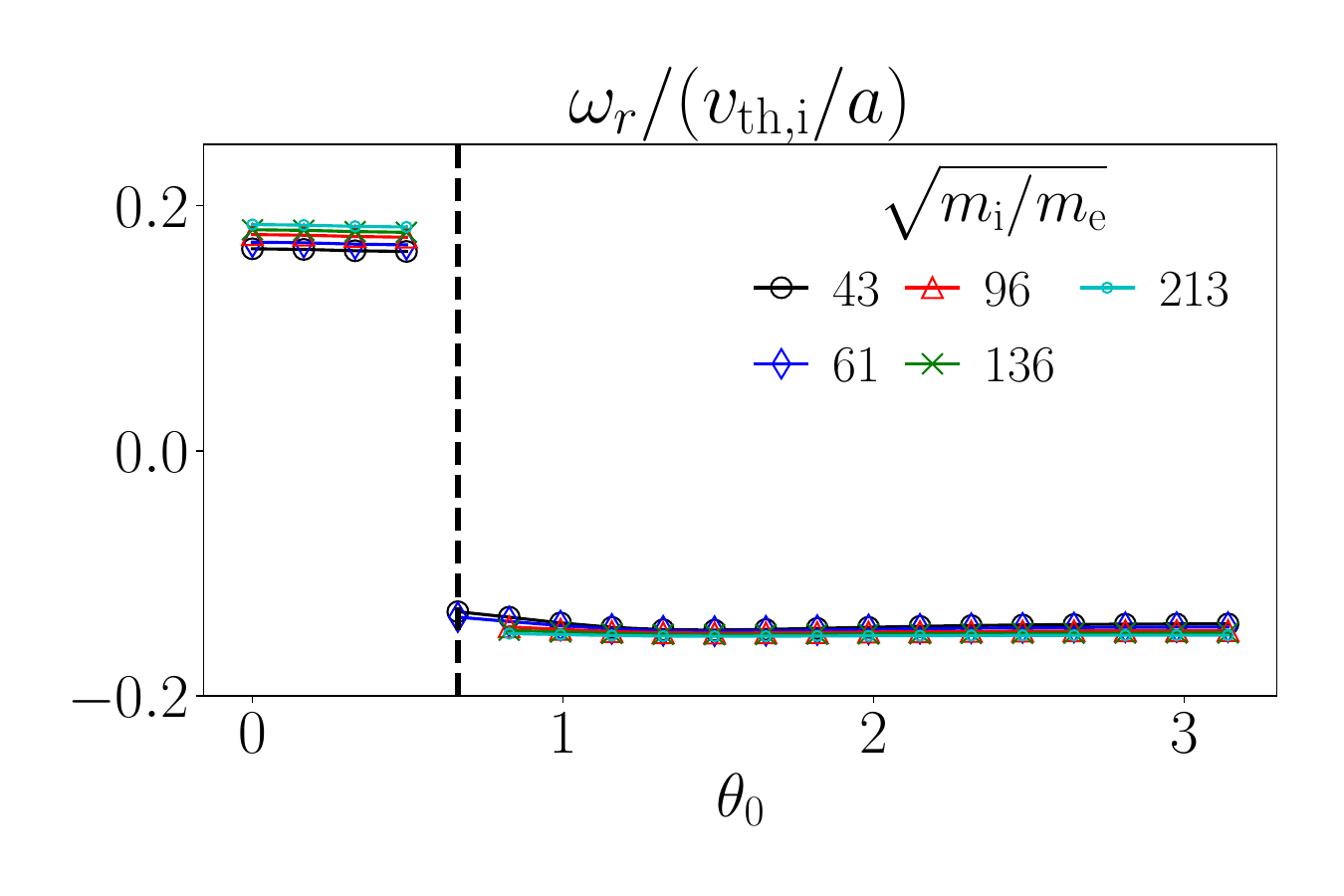}
% trim=left bottom right top, clip
\end{center}
\end{minipage}
\begin {minipage} {0.495\textwidth}
\begin{center}
\includegraphics[clip, trim=1cm 1cm 1cm 0cm, width=1.0\textwidth]{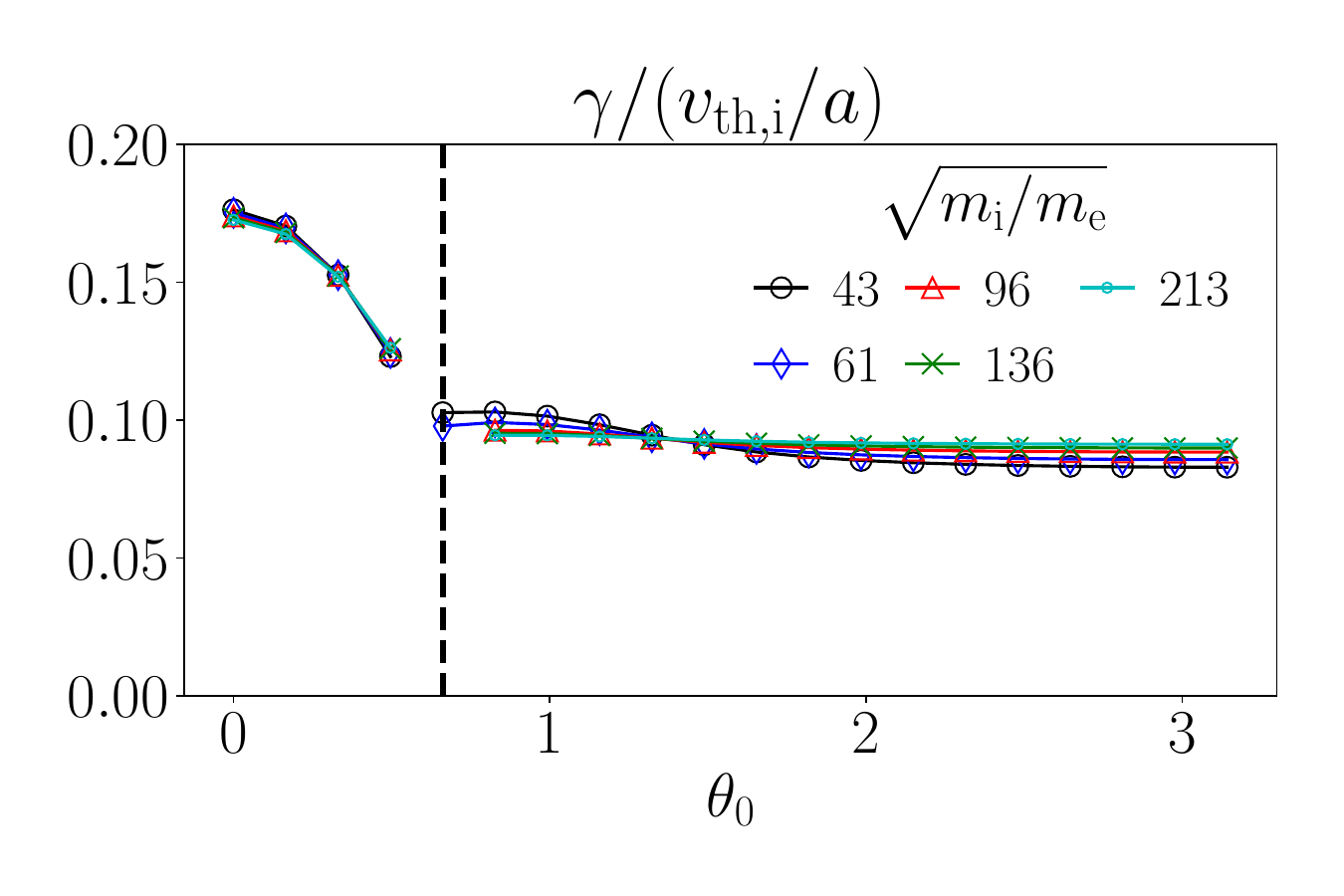}
\end{center}
\end{minipage}
 \caption{The real frequency $\wfreqr$ (left) and growth rate $\growth$  (right)
 as a function of $\thetaz$ and $\massrut$, for  $\lscal\cfreqee/\vtheri = \cfreqeecollisionless$,
 $\lscal/\lte = 3.45$ and $\lscal/\lti = 2.3$.
 The discontinuity in $\wfreqr$ indicates a transition between different instability branches.
 Modes to the left of the dashed line are small-tail modes (driven by trapped electrons and ions),
 whereas modes to the right of the dashed line are large-tail modes (driven by passing electrons).
 Note that $\wfreqr$ and $\growth$ are approximately independent of $\thetaz$ for the large-tail modes
 over a wide range of $\thetaz$. 
 } \label{fig:growthwfreqr-small-large-tail-transition-collisionless}
 \end{figure}

 In the results that we have presented in the previous sections, we have focused on examples
 where either a large-tail mode or a small-tail mode is clearly dominant. However, in practice
 it is possible to find cases where these different asymptotic branches have similar growth rates
 at the same $\kky$ and $\thetaz$,
 meaning that a transition can be observed with the variation of some parameter. To illustrate this,
 we consider modes with $\kky \gyrdi = 0.5$, $\lscal/\lte = 3.45$,
 $\lscal /\lti = 2.3$,
 and a electron collisionality of $\lscal\cfreqee/\vtheri = \cfreqeecollisionless$.
 Figure \ref{fig:growthwfreqr-small-large-tail-transition-collisionless} shows the real frequency
 $\wfreqr$ and growth rate $\growth$ of these modes as a function of 
 $\thetaz$, for different values of $\massrut$ at fixed $\lscal\cfreqee/\vtheri$. 
 There is clear discontinuity in $\wfreqr$, indicating a transition between 
 different mode branches.
 By examining the eigenmodes using the techniques of sections \ref{section:numericalresults-collisionless-large-tails} and \ref{section:numericalresults-small-tails},
 we verify that the modes to the right of the dashed line are collisionless large-tail modes.
 In contrast, we identify that the modes to the
 left of the dashed line are collisionless small-tail modes. 
 We note that the small-tail mode drifts in the ion diamagnetic direction, 
 whereas the large-tail mode drifts in the electron diamagnetic direction.
 As $\massrutp{1/2}$ increases, the $\growth$ and $\wfreqr$ of the large tail mode
 become independent of $\thetaz$,
 in accordance with the predictions of section \ref{section:largetailmode}. 
   
 We note that initial-value simulations of eigenmodes are challenging to converge when distinct
 instabilities exist at the same $(\kky,\thetaz)$ with the same $\growth$.
 The data plotted in figures \ref{fig:growthwfreqr-small-large-tail-transition-collisionless}
 are for modes where the $\wfreqr$ and $\growth$ are converged to
 0.5\% (compared to values averaged over a $5 \lscal/\vtheri$ window) after  $ 500 \lscal/\vtheri$.
 A time step size of $\Delta \ttime = 0.1 \lscal/\vtheri$
 was found to be adequate to resolve the collisionless modes featured in this section.
 The number of $2 \pi$ elements of the ballooning chain was taken to be $\ntwopi = 65 \sqrt{\mi/\md}$.

   \section {Discussion}\label {section:discussion}

 In the conventional treatment of the nonadiabatic electron
 response in modes with binormal wavenumbers on the scale of the
  ion thermal gyroradius, i.e., in modes with $\kky \gyrdi \sim 1$, rapid electron parallel
 streaming is assumed to imply that the nonadiabatic response of passing electrons
 should be small. This assumption leads to the usual ITG-driven modes and TEMs
 where the nonadiabatic passing electron response is subdominant. 
 However, several numerical investigations have revealed the existence of
 long-wavelength modes with extended ballooning tails
 where the nonadiabatic passing electron response appears to play a significant role, see, e.g.,
 \cite{HallatschekgiantelPRL2005,DominskinonadPOP015,Parisi_2020,Applegate_2007,Dickinson_MTMPed_2013,Moradi_2013,ajay_brunner_mcmillan_ball_dominski_merlo_2020,ajay_brunner_ball_2020}.
 In terms of a wavenumber-space description, these electron-driven modes are fluctuations
 with large radial wave numbers, i.e., $\kkr \gyrdi \gg 1$.
 In the real-space description, these modes are fine-scale fluctuations
 with significant amplitudes near mode-rational flux surfaces.
  Examples of these modes may be found in the core of tokamaks
 \cite{HallatschekgiantelPRL2005,DominskinonadPOP015}, and in the pedestal \cite{Parisi_2020}.
 Qualitatively, micro-tearing modes in tokamaks share the same features
 as the extended electrostatic modes in
 \cite{HallatschekgiantelPRL2005,DominskinonadPOP015,Parisi_2020},
 with both extended ballooning tails and an ETG drive
 \cite{Applegate_2007,Dickinson_MTMPed_2013,Moradi_2013}. 
  
 In this paper, 
 we show that it is possible to obtain an asymptotic theory for novel electron-response-driven
 $\kky\gyrdi \sim 1$ modes
 by assuming that the nonadiabatic response
 of passing electrons cannot be neglected, and by carefully considering the
 regions in the mode with large $\kkr$.
 The physics of these novel modes turn out to be dominated by the physics at 
 $\kkr\gyrdi \gg 1$, and surprisingly, the nonadiabatic ion response is unimportant (but not small).
  When the nonadiabatic-electron-response-driven modes are unstable,
  their growth rate is expected to be insensitive to the exact details
  of the magnetic geometry, because the leading-order equations
  for the mode contain only poloidal-angle-averaged geometric quantities.
  As a corollary, the growth rate of the mode is expected to be independent
  of $\thetaz$. Hence, extended, electrostatic electron-driven modes
  may be insensitive to equilibrium flow shear, driving
  turbulence even when the more familiar ITG modes
  and TEMs are flow-shear stabilised.
  This observation may be important for projected tokamaks scenarios
  that rely on equilibrium flow shear to stabilise microinstabilities, 
  cf. \cite{Kotschenreuther_2000_ignited_tokamaks}.

 We identify two limits where there are simple orderings.
 First, we examine the collisionless limit, where
 $\saffac \rmajorz \cfreqee / \vthere \sim \massrt \ll 1$,
 the radial wave number satisfies $\kkr\gyrde \sim 1$,
 and the fundamental expansion parameter is $\massrt \ll 1$.
 In the collisionless ordering, the extent of the mode is set
 by the physics of electron free streaming, and electron finite
 Larmor radius and electron finite orbit width effects.
 Second, we examine the collisional limit where $\saffac \rmajorz \cfreqee / \vthere \sim 1$,
 the radial wave number satisfies $\kkr\gyrde \sim \massrtp{1/4}$,
 and the fundamental expansion parameter is $\massrtp{1/4} \ll 1$.
 In the collisional ordering, the extent of the mode is set by
 parallel and perpendicular classical and neoclassical diffusion. 
 We note that the 
  collisionless and collisional orderings for the extent of the electrostatic modes
  considered in this paper are reminiscient of the
  collisionless and semicollisional orderings for the size of a tearing layer
  in a sheared magnetic field,
  see, for example \cite{DrakeLeePoF1977,Cowley_1986PoFtearing,ConnorPPCF2008,ZoccoPoP2011,ConnorPPCF2012}.
  
 We derive scaling laws for the relative sizes of the electron
 and ion responses in the $\kkr\gyrdi \sim 1$ and $\kkr\gyrdi \gg 1$ regions of the ballooning mode. 
 To confirm our analytically derived scalings, we
 use the gyrokinetic stability code \gstwo~to 
 perform a series of linear simulations for a range of normalised electron
 collisionality $\nustar = \saffac \rmajorz \cfreqee / \aspect^{3/2}\vthere$, and a range of $\me/\mi$. 
 We identify parameters where a novel passing-electron-driven mode is
 the fastest unstable mode. We present two relatively clean examples
 of the passing-electron-driven mode: a collisionless case and a collisional case.
 We perform the same analysis for an ITG mode, and verify the scalings for the
 subdominant nonadiabatic electron response.
 
 Although the theory presented
 here neglects electromagnetic fluctuations, many features of these novel
 electrostatic modes are common to micro-tearing modes. We speculate that
 some classes of micro-tearing modes may be well described by a
 collisionless $\massrt \rightarrow 0$ theory or collisional $\massrtp{1/4} \rightarrow 0$ theory
 similar to the theories presented in this paper.
 Development of these asymptotic theories provides not just physical insight,
 but also the possibility of performing reduced linear simulations of
 nonadiabatic-electron-response-driven modes. Simulations of extended ballooning modes
 can be expensive in comparison to simulations of familiar ITG-driven
 modes: a reduction of the size of the problem by removing the
 geometric $2 \pi$ poloidal-angle scale from the gyrokinetic
 equations may be an advantage. We anticipate that the need
 for computational efficiency in simulating extended ballooning modes
 will become more urgent in light of recent work \cite{bpatelthesis}
 that suggests high-$\pbeta$ spherical tokamak reactor equilibria
 may be unstable to extended micro-tearing modes for a wide range of $\kky$.

 The nonadiabatic response of passing electrons has recently been shown
 to be a significant factor in determining the isotope effect \cite{BelliIsotope2019,BelliIsotope2020PRL}.
 In fact, \cite{BelliIsotope2020PRL} argues that changes in a $\massrt$-small 
 passing electron nonadiabatic response
 can lead to $\order{1}$ changes in the heat fluxes
 as a result of the divergent asymptotic expansion in $\massrt$.
 In this paper we have seen that, in linear modes, the nonadiabatic response
 of passing electrons does not need to be small in $\massrt$.
 Indeed, it is not even obvious that an expansion should be
 carried out in $\massrt$. Instead, $\massrtp{1/4}$ could be the relevant
 expansion parameter for sufficiently large collisionality. This is an important 
 observation, because,  in practice, $\massrtp{1/4} \approx 1/8$ is
 likely to a be worse expansion parameter than $\massrt \approx 1/60$.
 For sufficiently large collisionality, nonasymptotic behaviour may perhaps
 be observed because the physical value of $\massrtp{1/4}$ is not small enough.
 Whilst we do not develop a nonlinear theory in this paper, we speculate
 that the isotope effect may well be the result of 
 the nonadiabatic response of passing electrons in $\kkr\gyrdi \gg 1$
 narrow layers regulating turbulent transport.
 
 The impact of electron-driven narrow radial layers in nonlinearly
 saturated turbulence is the subject of active research.
 Studies of turbulence using DNS have demonstrated that the
 nonadiabatic response of passing electrons in narrow layers near the mode-rational
 flux surfaces can have a significant impact on turbulence saturation levels and fluxes,
  see, e.g., \cite{HallatschekgiantelPRL2005,Waltz2006Corrugations,DominskinonadPOP015,ball_brunner_ajay_2020,ajay_brunner_mcmillan_ball_dominski_merlo_2020,ajay_brunner_ball_2020}.
 We note that electron-driven narrow radial layers
 are not necessarily associated with mode-rational surfaces. In fact, 
  narrow radial structures formed by ion-gyroradius-scale toroidal ETG modes 
 near the top and bottom of the tokamak \cite{Parisi_2020} have recently been observed
 to regulate the fluxes in nonlinear DNS of ETG-driven pedestal turbulence \cite{Parisi_2021draft}.
 The pedestal toroidal ETG modes driving these structures rely on magnetic drift resonances
 and have large radial wave numbers as a
 result of the large gradients of density and temperature, but, in
 contrast to the modes featured in this paper, they rely on favourable local magnetic geometry
 and are relatively localised in the ballooning angle.
 As mentioned in the introduction, electron-tail modes with extended ballooning eigenfunctions may nonetheless
 be observed in the pedestal at very long binormal wavelengths (see Appendix B of \cite{Parisi_2020}):
 extended eigenfunctions arise naturally when there is a separation between
 the frequencies associated with free streaming and the drives of instability.
 
 Further evidence for the importance of the nonadiabatic response of passing
 electrons in narrow layers may be found in DNS
 that bridge $\kky\gyrdi \sim 1$ to $\kky\gyrde \sim 1$ scales: 
  entropy transfer analysis suggests 
  that the nonadiabatic response of passing electrons
  mediates the backreaction of $\kky\gyrde \sim 1$
  eddies on $\kky\gyrdi \sim 1$ turbulence, via $\kkr\gyrdi \gg 1$ narrow
  layers \cite{Maeyama_2017_NF}. These observations suggest that theories
  of turbulence that attempt to capture the $\massrt\rightarrow 0 $
  limit may need to be modified to include the effects of
  electrons in narrow radial layers on saturation:
  this includes theories of turbulence on $\kky\gyrdi \sim 1$
  scales in isolation (cf. \cite{abelEMOD}) 
  and theories of cross-scale interactions between $\kky \gyrdi \sim 1$
  and $\kky\gyrde \sim 1$ scales (cf. \cite{hardmanpaper1,hardmanpaper2}).
  
      Finally, we note that whilst the results presented here are specialised
      to an axisymmetric tokamak by virtue of using the identity \refeq{equation:magdrift},
      we speculate that a similar theory to the one we present
      might be obtained for stellarator geometries.
      Physically, passing-electron-driven resonances at mode rational surfaces can arise
      in tokamaks due to the vanishing average radial magnetic drift -- undisturbed passing
      particles may return many times to the same location on the flux surface. Stellarator
      geometries also guarantee that the average radial magnetic drift of passing particles
      is zero \cite{Helander_2014}, possibly allowing for similar passing-electron-driven modes
      at rational surfaces. 
      In a stellarator, trapped particles only have zero average
      radial magnetic drift if the geometry is omnigeneous 
  \cite{Hall1975PoF,CaryShasharinaOmniPRL,CaryShasharinaOmniPoP,Helander_2014,Parra_2015_Less_constrained_omnigeneous,Calvo_2017_tangential_drifts}.
      As a consequence, in an arbitrary stellarator trapped particles may act
      to disturb passing-electron-drive modes by dragging passing particles off rational surfaces through collisions -- 
      we note that this effect would disappear in the $\nustar \gg 1$ limit  
      in which trapped particles cannot complete an orbit.

 \par \textit{ The authors are grateful for productive discussions with
 N. Christen, O. Beeke, 
 B. Patel, J. Maurino-Alperovich, S. Trinczek, C. M. Roach,
 B. F. McMillan, A. A. Scheckochihin, W. Dorland, J. Ball, S. Brunner,
 I. Calvo, J. M. Garc\'{i}a Rega\~{n}a, H. Thienpondt,
 M. Abazorius, J. Ruiz Ruiz, D. St-Onge, G. Acton and V. Hall-Chen. 
 The simulations were performed
 using the \gstwo~branch} \verb+https://bitbucket.org/gyrokinetics/gs2/branch/ms_pgelres+,\textit{ with the latest
revision at the time of writing being commit ade5780.
 The \gstwo~input files used to perform the gyrokinetic simulations in this study
 are publicly available at} \verb+https://doi.org/10.5287/bodleian:xr5ZApXYm+, \textit{
  alongside scripts used to calculate the neoclassical transport
 coefficients. 
 This work has been carried out within the framework of the EUROfusion Consortium
 and has received funding from the Euratom research and training programme 2014-2018
 and 2019-2020 under grant agreement No 633053, and from EPSRC [Grant Numbers EP/T012250/1 and EP/R034737/1].
 The views and opinions expressed herein do not necessarily reflect those of the European Commission.
 The author acknowledges the use of the EUROfusion High Performance Computer (Marconi-Fusion) under projects MULTEI and OXGK,
 ARCHER through the Plasma HEC Consortium EPSRC Grant Numbers EP/L000237/1 and EP/R029148/1
 under the projects e281-gs2 and e607, the JFRS-1 supercomputer at IFERC-CSC in
 Rokkasho Fusion Institute of QST (Aomori, Japan), and software support 
 through the Plasma-CCP Network under EPSRC Grant Number EP/M022463/1.}
 
\appendix
    \section{ A detailed analysis of the ion response for large $\lpar$
} \label{sec:detailed-analysis-of-the-ion-response}
In this appendix, we give a detailed analysis of the nonadiabatic 
ion response at large ballooning $\lpar$.  
This discussion will illustrate in more detail how the hyperbolic ion gyrokinetic equation,
 equation \refeq{eq:lineargyrokinetic} with $\spe = {\rm i}$,
 reduces to the local algebraic equation \refeq{eq:inner_ion1}
 in the electron-dominated tail of the ballooning mode. 
 
 We note that, for large $\lpar$,
 the leading-order ion gyrokinetic equation has the form
 \beqn \sign \drv{\hhi}{\lpar} + \phase(\lpar) \hhi = \source(\lpar), \label{eq:ion-gk-simple}\eeqn
where the source
$ \source = \imag \left( \wstari - \wfreq \right)\besin{0}\eqlbi \;\fract{\zedi \charge\ptl{}}{\tempi |\vpar|\kpar}$,      
 and the factor $\phase \gg 1$. We argue, in section \ref{sec:ions:inner-collisionless},
 that in the collisionless limit
 \beqn \fl \phase = \frac{1}{|\vpar| \kpar}\left(\frac{\kkfldl^2 {\saffacprim}^2 |\nbl \flxl|^2 \lpar^2 \vmag^2}{4\cycfi^2}\left(\cfreqiipar \pitch \bmag
    + \frac{\cfreqiiperp}{2} \left(2 - \pitch \bmag \right)\right) 
    -\imag \kkfldl \saffacprim \lpar \vmi \cdot \nbl \flxl \right)\label{eq:phase-def-collisionless}, \eeqn
 whereas, in section \ref{sec:ions:inner-collisional}, we argue that in the collisional limit
  \beqn \phase = {\frac{\kkfldl^2 {\saffacprim}^2 |\nbl \flxl|^2 \lpar^2 \vmag^2}{4\cycfi^2 |\vpar| \kpar}\left(\cfreqiipar \pitch \bmag
    + \frac{\cfreqiiperp}{2} \left(2 - \pitch \bmag \right)\right) }
   \label{eq:phase-def-collisional}, \eeqn
 i.e., we can neglect the term due to the radial magnetic drift. 
 In solving for the ion response, 
 we have no need to distinguish between $\lpar$ and $\lchi$
 -- this distinction is only a requirement for solving for the electron response.
 For simplicity, in the subsequent algebra we drop the usage
 of $\lchi$. In writing equation \refeq{eq:ion-gk-simple}, we have emphasised the $\lpar$ dependence of $\source$ and $\phase$,
 and we have neglected the differential terms
 of the ion gyrokinetic collision operator, and the terms due to the time derivative
 of $\hhi$, and the precessional magnetic drift. These terms can be neglected because
 $\lpar \gg 1$, $\kky \gyrdi \sim 1$ and $\wfreq \sim \vtheri/\lscal$, and hence the 
 leading terms are large, i.e., $\phase \gg 1$. Note that the real part of $\phase$ is 
 positive, i.e., $\Re \left[\phase\right] > 0$.
 
 We consider the solution of equation \refeq{eq:ion-gk-simple} for 
 $\lpar > 0$. Integrating equation \refeq{eq:ion-gk-simple} directly,
 we find that, for forward going particles ($\sign = 1$),
 \beqn \fl \hhi = \int^{\lpar}_0  \source(\lpar^{\dbprime})
 \expo{- \int^{\lpar}_{\lpar^{\dbprime}} \phase(\lpar^\prime) d \lpar^{\prime}}d \lpar^{\dbprime} + 
  \hhi(\lpar = 0,\sign=1) \expo{- \int^{\lpar}_{0} \phase(\lpar^\prime) d \lpar^{\prime}}.
  \label{eq:ions-forward-going-1}\eeqn
  For backward-going particles ($\sign=-1$), we find that 
 \beqn  \hhi = \int^{\infty}_\lpar  \source(\lpar^{\dbprime})
 \expo{- \int_{\lpar}^{\lpar^{\dbprime}} \phase(\lpar^\prime) d \lpar^{\prime}}d \lpar^{\dbprime}
 ,\label{eq:ions-backward-going-1} \eeqn
  where we have used $\hhi(\lpar = \infty, \sign = -1) = 0$ as a boundary condition.
  Inspecting the solutions \refeq{eq:ions-forward-going-1} and
  \refeq{eq:ions-backward-going-1}, we note that there are two components:
  a \enquote{local} solution involving $\source$,
  and an exponentially decaying solution (proportional to $\hhi(\lpar=0,\sign =1 )$) due to the
  outgoing particles from $\lpar = 0$. The exponentially decaying part of the
  solution gives rise to the logarithmic boundary layer referred
  to in \ref{sec:electron-outer-collisional-small-tail-appendix}.
  In this discussion, we can neglect the exponentially decaying part of
  the ion response, because the electron tail at large $\lpar$ generates its own
  potential that drives the ions via the \enquote{local} response.
  
  We now consider the form of the local solution when $\phase \gg 1$ and $\Re[\phase] > 0$,
  and the integrals can be treated using the standard Laplace method \cite{BenderOrszag}.
  We treat the case of $\sign = 1$ explicitly.
  First, we note that the dominant contributions to the integrals in equation
  \refeq{eq:ions-forward-going-1} 
  come from where $\lpar \simeq \lpar^{\dbprime}$. In this region, we write
  \beqn \fl \expo{- \int^{\lpar}_{\lpar^{\dbprime}} \phase(\lpar^\prime) d \lpar^{\prime}}
   = \expo{-\phase(\lpar)(\lpar - \lpar^{\dbprime})}
   \left(1 + \order{\frac{1}{\phase} \drv{\phase}{\lpar} (\lpar - \lpar^{\dbprime})}\right), \label{eq:appox-expo}\eeqn
  which is accurate provided that $(\lpar - \lpar^{\dbprime})/\lpar \ll 1$.
  We introduce a parameter $\delta$, so that we can write
 \beqn \hhi = \intzero + \intdelta  \label{eq:ions-forward-going-2}\eeqn
 with \beqn \intzero = 
  \int^{\lpar}_{\lpar - \delta}  \source(\lpar^{\dbprime})
  \expo{- \int^{\lpar}_{\lpar^{\dbprime}} \phase(\lpar^\prime) d \lpar^{\prime}}d \lpar^{\dbprime}
 , \eeqn
and  \beqn \intdelta =    \int^{\lpar- \delta}_0  \source(\lpar^{\dbprime})
 \expo{- \int^{\lpar}_{\lpar^{\dbprime}} \phase(\lpar^\prime) d \lpar^{\prime}}d \lpar^{\dbprime}. \eeqn 
 We can use the appoximation \refeq{eq:appox-expo} in $\intzero$ provided $\delta / \lpar \ll 1$.
 Using \refeq{eq:appox-expo}, the leading contribution to $\intzero$ is given by
\beqn \intzero = \int^{\lpar}_{\lpar - \delta} \source(\lpar)\expo{-\phase(\lpar)(\lpar - \lpar^{\dbprime})} d \lpar^{\dbprime}
 . \label{eq:intzero-intermediate}\eeqn
Evaluating the integral, we find that 
\beqn \intzero = \frac{\source(\lpar)}{\phase(\lpar)}\left( 1 - \expo{-\phase(\lpar) \delta}\right). \eeqn   
Taking $\delta$ such that $\Re[\phase] \delta \gg 1$ (consistent with $\delta / \lpar \ll 1$), we
find that, to leading order,  
\beqn \hhi = \frac{\source(\lpar)}{\phase(\lpar)}, \label{eq:hhi-local-appoximation}\eeqn
where we have used the smallness of $\expo{-\phase(\lpar) \delta}$ to neglect $\intdelta$. 
An analogous calculation can be performed for $\sign = -1$, with the result that $\hhi$ 
satisfies \refeq{eq:hhi-local-appoximation} for both signs of the velocity.    
The result \refeq{eq:hhi-local-appoximation} is identical in form to equations
 \refeq{eq:inner_ion1}. We have demonstrated that,
 although the ion gyrokinetic equation is hyperbolic, the fact that 
 $\phase \gg 1$ means that parallel streaming is unable to effectively
 propagate information at large $\lpar$, and hence, the nonadiabatic response of ions is local in ballooning angle.
 
  \section{ Obtaining the electron transport equations in the collisional inner region} \label{sec:electron-inner-collisional-appendix}
  In this section we calculate the forms of the transport equations
   that describe the electron response in the 
  in the $\massrtp{1/4} \rightarrow 0$
  limit, for $\kky\gyrdi \sim 1$ modes with $\fract{\saffac \rmajorz \cfreqee}{  \vthere} \sim 1$.
        We note that,
        in the ordering for the collisional inner region,
        ${\lchi} \sim \massrutp{1/4} \gg 1 \sim \thetaz \gg \lflre \sim \massrt$, and hence
        the phase in equation \refeq{eq:HHs} for $\HHe$ becomes
        \beqn \expo{\imag \lflre (\thetaz-\lpar)} =
        \left(1 - \imag \lflre \lchi - \frac{\lflre^2 \lchi^2}{2}
        + \imag \lflre \thetaz + \order{\massrp{3/4}} \right). \label{eq:fow-collisional}\eeqn
        In addition, we will need to expand the phase due
        to the finite Larmor radius $\expo{\imag \kperpvec \cdot \gyrdvece}$
        in the collision operator $\cope\left[\cdot\right]$.
        In the inner region, we find
        \beqn \fl \expo{\imag \kperpvec \cdot \gyrdvece} 
        = 1 + \imag \kperpvecz \cdot \gyrdvece - \frac{1}{2} (\kperpvecz \cdot \gyrdvece)^2
        + \imag \kperpveco \cdot \gyrdvece + \order{\massrp{3/4}}, \label{eq:flr-collisional}\eeqn
        where we note that $\kperpvecz \cdot \gyrdvece \sim \massrtp{1/4}$
        and $\kperpveco \cdot \gyrdvece \sim \massrtp{1/2}$.

 The leading order equation for the electron response in the inner region is
  equation \refeq{eq:electron_inner0-collisional}.
  To obtain equation \refeq{eq:electron_inner0-collisional}
    we have used the equations \refeq{eq:fow-collisional} and \refeq{eq:flr-collisional}
    for the finite-orbit-width and finite-Larmor-radius phases, respectively,
    and the estimate \refeq{eq:inner_ion2-collisional} for $\duveci$
    to simplify the collision operators to the drift-kinetic form. We have also
    noted that $\HHez$ is gyrophase independent,
    and $\coplandaue[\cdot]$ and $\coplorentze[\cdot]$ commute with $\gav{\cdot}{}$.
    
    We follow a H-theorem procedure \cite{hintonRMP76,helander}
    to solve equation \refeq{eq:electron_inner0-collisional}:
    first, we multiply equation \refeq{eq:electron_inner0-collisional} by 
    $\HHez/\eqlbe$, with the result
    \beqn \vpar \kpar \drv{}{\lpar}\left(\frac{\left(\HHez\right)^2}{2\eqlbe}\right)
    = \frac{\HHez}{\eqlbe}\coplandaue\left[\HHez\right]
    + \frac{\HHez}{\eqlbe}\coplorentze\left[\HHez\right]
    \label{eq:electron_inner0a-collisional}.    \eeqn
    Second, we integrate over velocity space
    \beqn \fl \bvec \cdot \nbl \lpar \drv{}{\lpar}\left(\intv{\frac{\vpar}{\bmag} \frac{\left(\HHez\right)^2}{2\eqlbe}} \right)
    \nonumber \eeqn \beqn
    = \intv{\frac{\HHez}{\eqlbe}\coplandaue\left[\HHez\right]} 
    + \intv{\frac{\HHez}{\eqlbe}\coplorentze\left[\HHez\right]}
    \label{eq:electron_inner0b-collisional},    \eeqn
    where we have used the form of the velocity integral in $(\energy,\pitch)$ coordinates,
    \beqn \intv{\left(\cdot\right)} = \sum_{\sign}\int^{\infty}_0  \int^{1/\bmag}_{0}  \frac{2 \pi \bmag \energy}{\ma^2|\vpar|} \left(\cdot\right) \; d \pitch \; d\energy, 
      \label{eq:intv-definition}\eeqn 
    and taken the $\drvt{}{\lpar}$ derivative through the integral.
    Finally, we apply the poloidal angle average $\lparav{\cdot}$, defined by equation \refeq{eq:lparav},
    to equation \refeq{eq:electron_inner0b-collisional}, and impose
    periodicity of $\HHez$ in $\lpar$, to obtain
    \beqn \lparav{\intv{\frac{\HHez}{\eqlbe}\;\coplandaue\left[\HHez\right]}}
    + \lparav{\intv{\frac{\HHez}{\eqlbe}\;\coplorentze\left[\HHez\right]}} = 0
    \label{eq:electron_inner0c-collisional}.    \eeqn
    The collision operators $\coplandaue\left[\cdot\right]$ and $\coplorentze\left[\cdot\right]$
    have the properties \cite{helander}
    \beqn \intv{ \frac{\fulldistg}{\eqlbe} \; \coplandaue\left[\fulldistg\right]} \leq 0 \quad
    {\rm{and}} \quad \intv{\frac{\fulldistg}{\eqlbe} \; \coplorentze\left[\fulldistg\right]} \leq 0,
    \label{eq:entropy}\eeqn respectively.
    Collisions always increase the entropy of the system.
    The equality $\intv{(\fulldistg/\eqlbe) \; \coplandaue\left[\fulldistg\right]} = 0$
    is only achieved when $\fulldistg$ is a perturbed Maxwellian so that $\coplandaue\left[\fulldistg\right] = 0$.
    The equality $\intv{(\fulldistg/\eqlbe) \;\coplorentze\left[\fulldistg\right]} = 0$
    is only achieved when $\fulldistg$ is isotropic in $\pvel$ so that $\coplorentze\left[\fulldistg\right] = 0$.
    As a consequence of equation \refeq{eq:electron_inner0c-collisional}, we find that $\HHez$
    satisfies equation \refeq{eq:HHez-collisional}
    with $\ddensns{0}{e}$ and $ \dtempns{0}{e}$ functions of $\lpar$ and $\lchi$ to be determined.
    Returning to equation \refeq{eq:electron_inner0-collisional},
    we now find that $\HHez$ must also satisfy $\vpar \kpar \drvt{\HHez}{\lpar} = 0$.
    For $\drvt{\HHez}{\lpar} = 0$
    to hold for all $\energy$, we must have that $\ddensns{0}{e}$ and $ \dtempns{0}{e}$
    are constant in $\lpar$, i.e., $\ddensns{0}{e}$ and $ \dtempns{0}{e}$
    have the form given by equation \refeq{eq:denstemplchi}.
    
    The first-order equation for the electron response in the inner region takes the form
     \beqn 
     \vpar \kpar \drv{\HHeh}{\lpar}+  \vpar \kpar \drv{\HHez}{\lchi}
    = \copbothe\left[\HHeh + \imag \lflre \lchi \HHez\right]
    , \label{eq:electron_inner1-collisional}\eeqn
    where we have used the definition of the drift-kinetic electron collision operator $\copbothe\left[\cdot\right]$, equation
    \refeq{eq:copbothe}.
    To expand the collision operator \refeq{eq:HHcollsions} for electrons,
    we have used the definition \refeq{eq:copgk-electron},
    equations \refeq{eq:fow-collisional} and \refeq{eq:flr-collisional},
    the estimate \refeq{eq:inner_ion2-collisional}, and the identities
    \beqn \fl \gav{\imag \kperpvecz \cdot \gyrdvece \; \copbothe\left[\HHez\right]}{} = 
    \gav{ \copbothe\left[\imag \kperpvecz \cdot \gyrdvece \; \HHez\right]}{} = 
    \copbothe[\HHez] = 0. \label{eq:copbothe-id1}\eeqn
        
    Equation \refeq{eq:electron_inner1-collisional} bears a resemblance to the
    neoclassical drift-kinetic equation in the banana collisionality regime \cite{hintonRMP76,helander}.
    We note that the term $\vpar \kpar \drvt{\HHez}{\lchi}$
    plays the role of the equilibrium inductive electric field
    in the corresponding neoclassical equation. 
    The resemblance can be made explicit by absorbing the $\vpar \kpar \drvt{\HHez}{\lchi}$
    term into the collision operators by solving
    the Spitzer-H\"{a}rm problem, 
    equation \refeq{eq:electron_inner1a-collisional}.
    It is useful to note that because the collision operator $\copbothe[\cdot]$ is isotropic
    \cite{helander},
    $\HHeSH$ must have the form \beqn \HHeSH = \vpar \KKeSH (\energy,\lchi) \eqlbe, \label{eq:KKeSH} \eeqn
    where $\KKeSH$ is a function of $\energy$ and $\lchi$.
    We determine $\KKeSH$ in \ref{sec:parallel-collisional-contributions}. 
    Using $\HHeSH$,
    we can rewrite equation \refeq{eq:electron_inner1-collisional} in the form
    of equation \refeq{eq:electron_inner1b-collisional}.
    
    To determine the time evolution of $\ddensns{0}{e}$
    and $\dtempns{0}{e}$, we continue to the $\order{\massrt}$ equation. 
    After collecting terms of $\order{\massrt}$,
    we find that the equation that determines $\HHez$ is 
    \beqn \fl \vpar \kpar \drv{\HHeo}{\lpar}+  \vpar \kpar \drv{\HHeh}{\lchi}
    + \imag (\wmage - \wfreqz) \HHez
    \nonumber\eeqn \beqn \fl 
     - \copbothe\left[\HHeo + \imag \lflre \lchi \HHeh
    {-} \left(\frac{1}{2}\left(\lflre^2\lchi^2 + \gav{(\kperpvecz \cdot\gyrdvece)^2}{}\right) {+} \imag \lflre \thetaz \right)\HHez\right]
    \label{eq:electron_inner2-collisional}\eeqn \beqn \fl
    + {\imag\lflre\lchi\copbothe\left[\HHeh + \imag \lflre\lchi\HHez \right]}
    - \gav{\kperpvecz\cdot\gyrdvece \;\copbothe\left[\kperpvecz\cdot\gyrdvece \;\HHez\right]}{}
     = - \imag (\wstare - \wfreqz) \eqlbe \frac{\charge \ptlz}{\tempe}
    , \nonumber\eeqn
    where, to obtain equation \refeq{eq:electron_inner2-collisional}, we have used 
    equations \refeq{eq:fow-collisional} and \refeq{eq:flr-collisional},
    estimate \refeq{eq:inner_ion2-collisional}, identities
    \refeq{eq:copbothe-id1}, that $\besen{0} = 1 + \order{\massrtp{1/2}}$ for
    $\lchi \sim \massrutp{1/4}$,
    and that $\copbothe[\cdot]$ and $\gav{\cdot}{}$ commute.
    
    We can convert equation \refeq{eq:electron_inner2-collisional} into equations for
    $\ddensns{0}{e}(\lchi)$ and $\dtempns{0}{e}(\lchi)$
    by multiplying equation \refeq{eq:electron_inner2-collisional}
    by the appropriate velocity space function ($1$ or $\energy/\tempe - 3/2$),
    integrating over velocity space,
    integrating over $\lpar$, 
    and finally imposing on $\HHeo$ the condition of $2\pi$-periodicity in $\lpar$. 
    After performing these operations, and dividing by $\dense$,
    the equation for the density moment is
   \beqn \fl  \drv{}{\lchi}\left(\lparav{\kpar\; \dUparns{1/2}{e}}\right) 
    + \imag \lparav{\wmageth}\left(\frac{\ddensns{0}{e}}{\dense} + \frac{\dtempns{0}{e}}{\tempe}\right)
       -\imag \wfreqz \frac{\ddensns{0}{e}}{\dense} 
    \nonumber\eeqn \beqn \fl
    + \lparav{{ \frac{1}{\dense}\intv{\imag\lflre\lchi\;\copbothe\left[\HHeh +\imag\lflre\lchi\HHez\right]}}}
    - \lparav{\frac{1}{\dense}\intv{\gav{\kperpvecz\cdot\gyrdvece\;\copbothe\left[\kperpvecz\cdot\gyrdvece\;\HHez\right]}{}}}
    \nonumber\eeqn \beqn 
     = - \imag (\wstaren - \wfreqz) \frac{\charge \ptlz}{\tempe}
    , \label{eq:electron_inner-dens0proto-collisional} \eeqn
    where we have defined the effective thermal magnetic drift frequency
     $\wmageth = \wmageprecth + \fract{\kkfldl \vthere^2 \kpar \saffacprim\bcur }{2\cycfe} $, 
    the $n^{\rm th}$-order component of the $\vpar$ moment of $\HHez$ 
    \beqn \dUparns{n}{e} = \frac{1}{\dense}\intv{\vpar \HHen} \label{eq:upare-nth}, \eeqn
    and used that the collision operator $\copbothe\left[\cdot\right]$ satisfies
    \beqn \intv{ \copbothe\left[\fulldistf\right]} = 0 \eeqn
    for $\fulldistf$ an arbitrary function of $\pvel$.
    Similarly, the equation for the electron temperature is
 \beqn \fl  \drv{}{\lchi}\left(\lparav{\kpar\left(\frac{\dQparns{1/2}{e}}{\dense\tempe} + \dUparns{1/2}{e}\right)} \right)     
    + \imag \lparav{\wmageth}\left(\frac{\ddensns{0}{e}}{\dense} + \frac{7}{2}\frac{\dtempns{0}{e}}{\tempe}\right)        
        - \imag  \frac{3}{2}  \wfreqz \frac{ \dtempns{0}{e}}{\tempe} 
       \nonumber  \eeqn \beqn \fl 
    + \lparav{\frac{1}{\dense}{\intv{ \left(\frac{\energy}{\tempe}-\frac{3}{2}\right)\imag\lflre\lchi\;\copbothe\left[\HHeh+\imag\lflre\lchi\HHez\right]}}}
    \label{eq:electron_inner-temp0proto-collisional}
     \eeqn \beqn \fl 
- \lparav{\frac{1}{\dense}\intv{ \left(\frac{\energy}{\tempe} -\frac{3}{2}\right)\gav{\kperpvecz\cdot\gyrdvece\; \copbothe\left[\kperpvecz\cdot\gyrdvece\; \HHez\right]}{}}}
     = - \imag \frac{3}{2} \wstaren \etae \frac{\charge \ptlz}{\tempe}
   ,\nonumber \eeqn
   where we have defined the $n^{\rm th}$-order component of the $\vpar (\energy/\tempe - 5/2)$ moment of $\HHez$ 
    \beqn \dQparns{n}{e} =\intv{\vpar \left(\energy-\frac{5\tempe}{2}\right) \HHen} \label{eq:qpare-nth}, \eeqn
    and used that the collision operator $\copbothe\left[\cdot\right]$ satisfies
    \beqn  \intv{ \left(\frac{\energy}{\tempe}-\frac{3}{2}\right)\copbothe\left[\fulldistf\right]} = 0.   \eeqn
    
    Equations \refeq{eq:electron_inner-dens0proto-collisional}
    and \refeq{eq:electron_inner-temp0proto-collisional} 
    have simple physical interpretations when written in terms of 
    the leading-order nonzero components of the electron parallel velocity,
     \beqn \duparns{1/2}{e} = \frac{1}{\dense}\intv{\vpar (\HHeh + \imag \lflre \lchi \HHez)},
    \label{eq:true-upare-halfth}\eeqn
    and electron parallel heat flux  \beqn \dqparns{1/2}{e} = \intv{\vpar 
    \left(\energy -\frac{5\tempe}{2}\right)(\HHeh + \imag \lflre \lchi \HHez)},
    \label{eq:true-qpare-halfth}.\eeqn
    After defining the effective parallel velocity and  effective parallel heat flux,
    $ \dupareeff = \fract{\lparav{\kpar \; \duparns{1/2}{e}}}{\lparav{\kpar}} $
      and $ \dqpareeff = \fract{\lparav{\kpar \; \dqparns{1/2}{e}}}{\lparav{\kpar}}$,
    respectively, we obtain equations \refeq{eq:electron_inner-dens0-collisional} and 
   \refeq{eq:electron_inner-temp0-collisional}.

%%%%%%%%%%%%%%%%%%%%%%%%%%%%%%%%%%%%
%% Spitzer calculation
%%%%%%%%%%%%%%%%%%%%%%%%%%%%%%%%%%%%
      
    \section{ Spitzer-H\"{a}rm component of the parallel diffusion collisional terms
} \label{sec:parallel-collisional-contributions}
  In order to evaluate the parallel flow and neoclassical perpendicular
  diffusion terms in equations \refeq{eq:electron_inner-dens0-collisional}
  and \refeq{eq:electron_inner-temp0-collisional},
  we need to solve equation \refeq{eq:electron_inner1a-collisional}
  for $\HHeSH$. To solve for $\HHeSH$, we first note that
  the collision operators $\coplandaue[\cdot]$ and $\coplorentze[\cdot]$, 
  defined in equations \refeq{eq:coplandaus} and \refeq{eq:coplorentze}, respectively,
  are isotropic operators \cite{helander}, and hence $\HHeSH$ may be assumed to have the form
  given in equation \refeq{eq:KKeSH}. Second, we note that
  equation \refeq{eq:electron_inner1a-collisional} is linear in $\ddensns{0}{e}$ and 
  $\dtempns{0}{e}$. Third, we may express the $\energy$ dependence of $\KKeSH$ in a convenient basis
  of polynomials. Hence, $\KKeSH$ is given by
  \beqn \KKeSH =\kpar \drv{}{\lchi}\left(\frac{\ddensns{0}{e}}{\dense}\right) \ffSH (\sarg)
   + \kpar \drv{}{\lchi}\left(\frac{\dtempns{0}{e}}{\tempe} \right) \ggSH (\sarg), \label{eq:trial-solution}\eeqn
  where 
  \beqn   \ffSH (\sarg) = \sum_{\aindex = 0} \acoeffn{\aindex} \soninen{\aindex}(\sarg), \label{eq:ffSH}\eeqn
  and 
  \beqn   \ggSH (\sarg) = \sum_{\aindex = 0} \bcoeffn{\aindex} \soninen{\aindex}(\sarg), \label{eq:ggSH} \eeqn
  with
  $\sarg = \energy/\tempe = \vmag^2/\vthere^2$, $\soninen{\aindex}(\sarg)$
  the $\aindex^{\rm{th}}$ generalised Laguerre polynomial $\soninemn{\mindex}{\aindex}(\sarg)$
  of index $\mindex = 3/2$,
  and $\acoeffn{\aindex}$ and $\bcoeffn{\aindex}$ coefficients to be determined.
  The generalised Laguerre polynomials of index $\mindex=3/2$ are particularly
  convenient for this problem because we will be able to exploit the orthogonality relation \cite{helander}
  \beqn \int^{\infty}_0 \soninemn{\mindex}{\aindex}(\sarg)\soninemn{\mindex}{\bindex}(\sarg)
  \expo{-\sarg} \sarg^\mindex d \sarg = \frac{\gamfn(\aindex + \mindex + 1)}{\aindex !}
  \kron{\aindex,\bindex} \label{eq:Orthogonality},  \eeqn
  where $\gamfn(\mindex) = \int^{\infty}_0 \sarg^{\mindex - 1}\expo{-\sarg} d \sarg$ is the Gamma function,
  and $\kron{\aindex,\bindex}$ is the Kronecker delta.
  The polynomial $\soninemn{\mindex}{\aindex}(\sarg)$
  may be obtained from the generating function \cite{helander}
  \beqn \genfun(\sarg,\zarg) = \frac{\expo{- \sarg \zarg / (1-\zarg)}}{(1 - \zarg)^{\mindex + 1}}
  = \sum_{\aindex= 0} \zarg^\aindex \soninemn{\mindex}{\aindex}(\sarg). \eeqn
  
  With the form of $\KKeSH$ given by equation \refeq{eq:trial-solution}, 
  the problem \refeq{eq:electron_inner1a-collisional} may be cast into two separate Spitzer problems
  for $\ffSH$ and $\ggSH$,
  \beqn \vpar \soninen{0}\eqlbe
  = \copbothe\left[\vpar \ffSH \eqlbe\right], \label{eq:spitzer-density}\eeqn
  and 
  \beqn \vpar (\soninen{0} - \soninen{1}) \eqlbe
  = \copbothe\left[\vpar \ggSH \eqlbe\right], \label{eq:spitzer-temperature}\eeqn
  where  we have used the first two generalised Laguerre polynomials, $\soninen{0}(\sarg) =1$ and 
  $\soninen{1}(\sarg) = 5/2 - \sarg$. We solve equations
  \refeq{eq:spitzer-density} and \refeq{eq:spitzer-temperature}
  by converting them into matrix equations for the coefficients $\acoeffn{\aindex}$
  and $\bcoeffn{\aindex}$, respectively.
  To do this, we define an inner product $\vprod{\cdot}{\cdot}$
  acting on velocity space functions  $\fulldistf = \fulldistf(\pvel)$ and 
  $\fulldistg = \fulldistg(\pvel)$ by
  \beqn \vprod{\fulldistf}{\fulldistg} = \intv{\frac{\fulldistf(\pvel)\fulldistg(\pvel)}{\eqlbe}}
   \label{eq:innerprod},\eeqn and
  we take the inner product of equations \refeq{eq:spitzer-density} and \refeq{eq:spitzer-temperature}
  with the function $\vpar \soninen{\bindex} \eqlbe$.  
  To perform the velocity
  integrals, we use the velocity coordinates
  $(\sarg,\pangle,\gyrophase)$, where $\pangle = \vpar / \vmag$, and we 
  recall that $\gyrophase$ is the gyrophase.
  The velocity integral in these coordinates becomes  
  \beqn \intv{\left(\cdot\right)} =
  \int^{2\pi}_0 \int^{1}_{-1} \int^\infty_0 \left(\cdot\right) \; \frac{\vthere^3}{2}\sqrt{\sarg}
  \; d \sarg \; d \pangle \; d \gyrophase \label{eq:sarg-pangle-int}. \eeqn 
  Using the orthogonality relation \refeq{eq:Orthogonality}, we find that the matrix form of equation
  \refeq{eq:spitzer-density} is 
  \beqn \sum_{\bindex}
   \left(\cfreqee \matrixlandau{\aindex,\bindex}+ \cfreqei \matrixlorentz{\aindex,\bindex} \right)
   \acoeffn{\bindex} = - \kron{0,\aindex} ,
   \label{eq:spitzer-density-matrix} \eeqn 
   where the matrix elements $\matrixlandau{\aindex,\bindex}$ and 
   $\matrixlorentz{\aindex,\bindex}$ are defined by
   \beqn \matrixlandau{\aindex,\bindex}= - \frac{2}{\dense\cfreqee}
   \vprod{\sarg^{1/2}\pangle \eqlbe \soninen{\aindex}}{\coplandaue\left[\sarg^{1/2}\pangle \eqlbe \soninen{\bindex}\right]}
   \label{eq:matrixlandau}\eeqn
   and
   \beqn \matrixlorentz{\aindex,\bindex}= - \frac{2}{\dense\cfreqei}
   \vprod{\sarg^{1/2}\pangle \eqlbe \soninen{\aindex}}{\coplorentze\left[\sarg^{1/2}\pangle \eqlbe \soninen{\bindex}\right]}
   \label{eq:matrixlorentz},\eeqn
   respectively. Similarly, we find that the matrix form of equation
  \refeq{eq:spitzer-temperature} is
\beqn  \sum_{\bindex}
   \left(\cfreqee \matrixlandau{\aindex,\bindex}
   + \cfreqei \matrixlorentz{\aindex,\bindex} \right)\bcoeffn{\bindex}
   = \frac{5}{2}\kron{1,\aindex} -\kron{0,\aindex} 
   \label{eq:spitzer-temperature-matrix}.  \eeqn 
   To solve the problem, we invert the matrix equations   \refeq{eq:spitzer-density-matrix}
   and \refeq{eq:spitzer-temperature-matrix}.
   In practice, we must include a finite number of polynomials, with the series truncated
  at a finite $\aindex=\aindexmax$.
  Velocity moments of $\HHeSH$ will depend only on low-order coefficients $\acoeffn{\aindex}$
   and $\bcoeffn{\aindex}$, and so only a few polynomials are required before convergence is reached.
   This same solution may be obtained using a variational method \cite{helander}.
  
  Although the calculation is tedious, it is relatively straightforward to calculate the matrix
   elements $\matrixlandau{\aindex,\bindex}$ and $\matrixlorentz{\aindex,\bindex}$
   using the generating function $\genfun(\sarg,\zarg)$.
   Truncating the series at polynomial order $\aindexmax = 4$, we find the coefficient matrices (cf. \cite{BraginskiiTransport,helander})
   \beqn \fl \boldmatrixlandau = \sqrt{2}\left(\begin{tabular}{ccccc}
    0 & 0 & 0 & 0 & 0   \\ 
    0 & 1 & 3/4 & 15/32 & 35/128 \\ 
    0 & 3/4 & 45/16 & 309/128 & 885/512 \\
    0 & 15/32 & 309/128 & 5657/1024 & 20349/4096 \\ 
    0 & 35/128 & 885/512 & 20349/4096 & 149749/16384  \\
   \end{tabular}\right),\eeqn 
   and 
   \beqn \fl \boldmatrixlorentz = \left(\begin{tabular}{ccccc}
    1 & 3/2 & 15/8 & 35/16 & 315/128 \\ 
    3/2 & 13/4 & 69/16 & 165/32 & 1505/256  \\ 
    15/8 & 69/16 & 433/64 & 1077/128 & 10005/1024\\ 
    35/16 & 165/32 & 1077/128 & 2957/256 & 28257/2048 \\ 
    315/128 & 1505/256 & 10005/1024 & 28257/2048 & 288473/16384 \\ 
   \end{tabular}\right).\eeqn 
   
   To illustrate the final result of the calculation for a simple case, we
   solve  equations \refeq{eq:spitzer-density-matrix}
   and \refeq{eq:spitzer-temperature-matrix}
   for a hydrogenic plasma with $\zedi =1$, i.e., $\cfreqei = \cfreqee$. 
   To three decimal places,
   we find that the coefficients $\{\acoeffn{n}\}$ and $\{\bcoeffn{n}\}$ are
     \beqn \left(\begin{tabular}{c} 
   $ \acoeffn{0} $ \\
     $ \acoeffn{1} $ \\
     $ \acoeffn{2} $ \\
     $ \acoeffn{3} $ \\
     $ \acoeffn{4} $ \\
        \end{tabular}\right) 
    =  \frac{1}{\cfreqei}\left(\begin{tabular}{c}
     -1.969 \\ 
     0.559 \\ 
     0.017 \\ 
     0.016 \\
     0.027 \\ 
        \end{tabular}\right),  \label{eq:acoeffn}\eeqn and 
        \beqn \left(\begin{tabular}{c} 
      $ \bcoeffn{0} $ \\
     $ \bcoeffn{1} $ \\
     $ \bcoeffn{2} $ \\
     $ \bcoeffn{3} $ \\
     $ \bcoeffn{4} $ \\
        \end{tabular}\right) 
    =\frac{1}{\cfreqei}\left(\begin{tabular}{c}
     -3.366 \\ 
     2.226 \\ 
    -0.635 \\ 
     0.095 \\ 
     0.003 \\
        \end{tabular}\right),  \label{eq:bcoeffn}\eeqn respectively.

   To calculate the parallel flow and the neoclassical perpendicular diffusion
   terms, we need to evaluate velocity integrals of the form 
    \beqn \fl \intv{\vpar \HHeSH} = \label{eq:dens-int-basic} 
   \frac{\dense\vthere^2}{2}\kpar
   \left(\acoeffn{0}\drv{}{\lchi}\left(\frac{\ddensns{0}{e}}{\dense}\right)
   +\bcoeffn{0}\drv{}{\lchi}\left(\frac{\dtempns{0}{e}}{\tempe} \right)\right) ,
   \eeqn
   and 
   \beqn \fl 
   \intv{\vpar \left(\frac{\vmag^2}{\vthere^2}-\frac{5}{2}\right) \HHeSH}  \label{eq:temp-int-basic} \eeqn \beqn 
     =\frac{5\dense\vthere^2}{4}\kpar\left(
     \acoeffn{1}\drv{}{\lchi}\left(\frac{\ddensns{0}{e}}{\dense}\right)
    +  \bcoeffn{1}\drv{}{\lchi}\left(\frac{\dtempns{0}{e}}{\tempe}\right)    
    \right)
     .\nonumber
     \eeqn   

%%%%%%%%%%%%%%%%%%%%%%%%%%
%% Classical diffusion
%%%%%%%%%%%%%%%%%%%%%%%%%%

    \section{ Classical perpendicular diffusion collisional terms} \label{sec:perpendicular-collisional-contributions}
    
    The collision integrals appearing in the definitions of the classical fluxes $\dpartc$ and $\dheatc$,
    equations \refeq{eq:classical-particle-flux}
    and  \refeq{eq:classical-heat-flux}, respectively, have the structural form
    \beqn \intv{\fulldistg(\vmag) \pvel \cdot \tauvec \;
    \copbothe\left[\HHez(\vmag) \pvel \cdot \sigvec\right]},\label{eq:collisional-perpendicular-diffusion}\eeqn 
    with the isotropic function of $\pvel$, $\fulldistg$, satisfying either  
    $\fulldistg = 1$ or $\fulldistg = \vmag^2/\vthere^2 - 5/2$,
    and $\sigvec = \kperpvecz \xp \bu / \cycfe$ and $\tauvec  = \imag \nbl \radial \xp \bu / \cycfe$ velocity independent vectors.
    Note that $\kperpvecz \propto \nbl \radial$, and hence, $\sigvec \propto \tauvec$. 
    We now proceed to evaluate the integral defined by equation \refeq{eq:collisional-perpendicular-diffusion}.
    We first evaluate contributions from the electron Lorentz collision operator $\coplorentze[\cdot]$.
        
    \subsection{Lorentz collision operator contributions.}
    
    The Lorentz collision operator is given by equation \refeq{eq:coplorentze}.
    Inserting the definition \refeq{eq:coplorentze} into
    equation \refeq{eq:collisional-perpendicular-diffusion}, using the form of $\HHez$,
    equation \refeq{eq:HHez-collisional}, and integrating by parts, we find that
    \beqn \fl \intv{\fulldistg(\vmag) \pvel \cdot \tauvec \;
    \coplorentze\left[\HHez(\vmag) \pvel \cdot \sigvec\right]} = \nonumber \eeqn
    \beqn \fl  \frac{3 \sqrt{\pi}}{8}
    \cfreqei \vthere^3\tauvec \sigvec \codot  \intv{\left\{\fulldistg(\vmag)\left(\frac{\ddensns{0}{e}}{\dense}
    + \frac{\dtempns{0}{e}}{\tempe}\left(\frac{\vmag^2}{\vthere^2}-\frac{3}{2}\right) \right)
    \frac{\vmag^2 \eye - \pvel \pvel}{\vmag^3}\eqlbe\right\}}. \label{eq:lorentz-terms0} \eeqn 
    To evaluate equation \refeq{eq:lorentz-terms0} for the appropriate functions $\fulldistg$,
    we use the normalised velocity $\wvel = \pvel/\vthere$, and the identities
    \beqn \intw{ \left(1, \wmag^2, \wmag^4\right)
    \expo{-\wmag^2}  \frac{\wmag^2 \eye - \wvel \wvel}{\wmag^3} } 
    = \frac{4 \pi}{3} \eye \left(1,1,2\right), 
    \label{eq:lorentzUints}\eeqn  with $\wmag = |\wvel|$.
    For the case of $\fulldistg = 1$, we find that
    \beqn \fl \intv{\pvel \cdot \tauvec \;
    \coplorentze\left[\HHez(\vmag) \pvel \cdot \sigvec\right]} =  - \dense \cfreqei \vthere^2 \frac{\tauvec \cdot \sigvec }{2} 
    \left(\frac{\ddensns{0}{e}}{\dense}- \frac{1}{2}\frac{\dtempns{0}{e}}{\tempe}\right).
    \label{eq:lorentz-terms1}    \eeqn
    For the case of $\fulldistg = \vmag^2/\vthere^2 - 5/2$, we find that
    \beqn \fl \intv{\left(\frac{\vmag^2}{\vthere^2}-\frac{5}{2}\right)\pvel \cdot \tauvec \;
    \coplorentze\left[\HHez(\vmag) \pvel \cdot \sigvec\right]} = \nonumber \eeqn
\beqn - \dense \cfreqei \vthere^2 \frac{\tauvec \cdot \sigvec }{2} 
    \left(\frac{7}{4}\frac{\dtempns{0}{e}}{\tempe}- \frac{3}{2}\frac{\ddensns{0}{e}}{\dense}\right).
    \label{eq:lorentz-terms2}    \eeqn
    
    \subsection{Electron self-collision operator contributions.}
    In this section we evaluate the perpendicular-diffusion contributions
    from the electron self-collision operator $\coplandaue\left[\cdot\right]$,
    following \cite{Newton_2010}.
    The electron self-collision operator is defined by equation \refeq{eq:coplandaus}
    with $\spe = {\rm e}$.
    To perform the calculation, first, we substitute the definition \refeq{eq:coplandaus} into the form
    \refeq{eq:collisional-perpendicular-diffusion} of the perpendicular-diffusion integral,
    noting that $2 \pi e^4 \coloumblog / \me^2 = 3\sqrt{\pi} \cfreqee \vthere^3/8\dense$. 
    Second, we integrate by parts, and symmetrise the resulting integral 
    by relabelling the dummy variables $\pvel$ and $\pvelprim$.
    Writing $\fulldistf(\vmag) = \HHez(\vmag)/\eqlbe$, we obtain the result
    \beqn \fl \intv{\fulldistg(\vmag) \pvel \cdot \tauvec \;
    \coplandaue\left[\fulldistf(\vmag) \pvel \cdot \sigvec \eqlbe \right]}= \nonumber\eeqn
    \beqn  -\frac{3\sqrt{\pi}}{16}\frac{\cfreqee \vthere^3}{\dense}  \intv{\intvprim{\eqlbe \eqlbeprim
    \Phifunca \cdot \Ufunc
    \cdot \Phifuncb}},\label{eq:landau-terms0}\eeqn
    where the vectors 
    \beqn \Phifunca(\pvel,\pvelprim) = 
    \tauvec(\fulldistg - \fulldistgprim) + (\pvel\cdot \tauvec)\drv{\fulldistg}{\pvel}
    - (\pvelprim\cdot \tauvec)\drv{\fulldistgprim}{\pvelprim},\label{eq:Phifunca}\eeqn
    and 
    \beqn \Phifuncb(\pvel,\pvelprim) = \sigvec (\fulldistf - \fulldistfprim) + (\pvel\cdot \sigvec)\drv{\fulldistf}{\pvel}
    - (\pvelprim\cdot \sigvec)\drv{\fulldistfprim}{\pvelprim}, \label{eq:Phifuncb}\eeqn
    with $\fulldistgprim = \fulldistg(\vmagprim)$, 
    and $\fulldistfprim = \fulldistf(\vmagprim)$.
    The form of the vector $\Phifunca$, defined in equation 
    \refeq{eq:Phifunca}, shows that there is no self-collision operator contribution
    to the perpendicular-diffusion terms in the density equation, for which $\fulldistg = 1$.
    To evaluate the self-collision operator contribution to the temperature equation, we take
    $\fulldistg(\vmag) = \vmag^2/\vthere^2 - 3/2$ and use that 
    \beqn \fulldistf(\vmag) = \frac{\HHez}{\eqlbe} = \frac{\ddensns{0}{e}}{\dense}
    + \frac{\dtempns{0}{e}}{\dense}\left(\frac{\vmag^2}{\vthere^2}- \frac{3}{2}\right). \eeqn
    After substituting for $\fulldistg$ and $\fulldistf$ in equations \refeq{eq:Phifunca} and 
    \refeq{eq:Phifuncb}, respectively, we find that 
    \beqn \Phifunca(\pvel,\pvelprim) 
    = \tauvec\frac{\vmag^2 - {\vmagprim}^2}{\vthere^2}+ \frac{2(\pvel(\pvel\cdot \tauvec) - \pvelprim(\pvelprim\cdot \tauvec))}{\vthere^2},\eeqn
    and
    \beqn \Phifuncb(\pvel,\pvelprim)  = 
\left(\sigvec\frac{\vmag^2 - {\vmagprim}^2}{\vthere^2}+ \frac{2(\pvel(\pvel\cdot \sigvec) - \pvelprim(\pvelprim\cdot \sigvec))}{\vthere^2}\right)\frac{\dtempns{0}{e}}{\tempe}.
\eeqn     
    To compute the velocity integrals in equation \refeq{eq:landau-terms0}, 
    we convert to the center-of-mass variables
    \beqn  \svel = \frac{\pvel + \pvelprim}{\sqrt{2}\vthere}, \quad {\rm{and}} \quad 
     \wvel = \frac{\pvel - \pvelprim}{\sqrt{2}\vthere}, \eeqn
     with the result 
     \beqn \fl \intv{\intvprim{\eqlbe \eqlbeprim
    \Phifunca \cdot \Ufunc
    \cdot \Phifuncb}} = \frac{4\dense^2}{\sqrt{2}\pi^3 \vthere}\frac{\dtempns{0}{e}}{\tempe} \times \label{eq:landau-terms0RHSb} \eeqn 
   \beqn \fl 
   \intw{\ints{\expo{-\wmag^2 - \smag^2}
    \left( \tauvec(\svel\cdot \wvel)+  (\wvel\cdot\tauvec)\svel \right)\cdot \Ufunchat
    \cdot \left( \sigvec(\svel\cdot \wvel)+  \svel(\wvel\cdot\sigvec)\right)}},
    \nonumber\eeqn
    where we have used that the Jacobian 
    $d^3 \pvelprim d^3 \pvel = \vthere^6 d^3 \svel d^3 \wvel$, 
    the functions 
    $\vmag^2 - {\vmagprim}^2 = 2 \vthere^2 \wvel \cdot \svel$,
     $\pvel \pvel - \pvelprim \pvelprim = \vthere^2 (\wvel \svel + \svel \wvel)$,  
    $\vmag^2 + {\vmagprim}^2 = \vthere^2(\wmag^2+ \smag^2)$, and     
    $\Ufunc(\pvel - \pvelprim) = \Ufunchat(\wvel)/(\sqrt{2} \vthere)$,
    with \beqn \Ufunchat(\wvel) = \frac{\wmag^2 \eye - \wvel \wvel}{\wmag^3} .\label{eq:Ufunchat}  \eeqn    
    Note as well that $\wvel \cdot \Ufunchat(\wvel) =0$. 
    We first evaluate the integral in $\svel$ using the identity
    \beqn \ints{\expo{-\smag^2} \svel \svel} = \frac{\pi^{3/2}}{2}\eye. \label{eq:sint}\eeqn 
    The result is
    \beqn \fl \intv{\intvprim{\eqlbe \eqlbeprim
    \Phifunca \cdot \Ufunc
    \cdot \Phifuncb}} = \label{eq:landau-terms0RHSc} \eeqn 
 \beqn 
 \frac{\sqrt{2}\dense^2}{\pi^{3/2} \vthere}\frac{\dtempns{0}{e}}{\tempe}  \tauvec \sigvec \codot
 \intw{\expo{-\wmag^2 } \left(\wmag^2\Ufunchat +  \wvel \wvel \; \trace{\Ufunchat}\right)},\eeqn
    where $\trace{\Ufunchat} = 2 / \wmag$ is the trace of the tensor $\Ufunchat(\wvel)$.
    Finally, using equations \refeq{eq:landau-terms0} and \refeq{eq:landau-terms0RHSc},
    with the identities
    \refeq{eq:lorentzUints} and 
    \beqn \intw{ \wvel \wvel \trace{\Ufunchat} \expo{-\wmag^2}} = \frac{4\pi}{3} \eye, \label{eq:wint} \eeqn
    we can write down the result of the perpendicular-diffusion collision integral 
    \beqn \fl  \intv{\left(\frac{\vmag^2}{\vthere^2}-\frac{3}{2}\right)\pvel \cdot \tauvec \;
    \coplandaue\left[\HHez(\vmag) \pvel \cdot \sigvec\right]}= 
    - \frac{1}{\sqrt{2}} \cfreqee \dense \vthere^2 \tauvec\cdot \sigvec 
\frac{\dtempns{0}{e}}{\tempe} . \label{eq:landau-terms0final}\eeqn

   %%%%%%%%%%%%%%%%%%%%%%%%%%%%%%%%%
   %%%% Pfirsh-Schl\"{u}ter fluxes
   %%%%%%%%%%%%%%%%%%%%%%%%%%%%%%%%%
   
   \section{Pfirsh-Schl\"{u}ter parallel and perpendicular fluxes}
   \label{sec:pfirsh-schluter-parallel-collisional-contributions}

  In this section, we compute the parallel flows and the perpendicular diffusion terms
    in the subsidiary limit of $\saffac \rmajorz \cfreqee / \vthere \gg 1 $.   
    We must solve equation \refeq{eq:electron_inner1b-collisional}
    to obtain approximate solutions for $\HHeh$.
    We expand
    \beqn \HHeh = \HHepq{1/2}{-1} + \HHepq{1/2}{0}
    + \HHepq{1/2}{1} + \order{\left(\frac{\saffac \rmajorz \cfreqee}{  \vthere}\right)^{-2} \imag\lflre\lchi \HHez}, \label{eq:HHeh-nustar-gg-1}\eeqn
    with \beqn \HHepq{1/2}{n} \sim \left(\frac{\saffac \rmajorz \cfreqee}{  \vthere}\right)^{-n}\HHepq{1/2}{0} \label{eq:pfirsh-schluter-nth-order-size}\eeqn and
    $\HHepq{1/2}{0} \sim  \imag\lflre\lchi \HHez \sim \HHeSH$.    
    The ordering $ \imag\lflre\lchi \HHez \sim \HHeSH$ is
    a manifestation of the ordering \refeq{eq:collisional-lchi} for $\lchi$.
    
    With the expansion \refeq{eq:HHeh-nustar-gg-1}, 
    the leading-order form of equation \refeq{eq:electron_inner1b-collisional} is
     \beqn  \copbothe\left[\HHepq{1/2}{-1}\right]
      = 0, \label{eq:electron_inner1b-collisional-nustar-gg-1-leading-order}\eeqn
    i.e., \beqn \HHepq{1/2}{-1} = \left(\frac{\ddensnsq{1/2}{e}{-1}}{\dense} 
    + \frac{\dtempnsq{1/2}{e}{-1}}{\tempe}\left(\frac{\energy}{\tempe}
    - \frac{3}{2}\right)\right)\eqlbe \label{eq:HHeh-1}\eeqn
    is a perturbed Maxwellian distribution function with no flow. Note that
    $\ddensnsq{1/2}{e}{-1} = \ddensnsq{1/2}{e}{-1}(\lpar,\lchi)$ and 
    $\dtempnsq{1/2}{e}{-1} = \dtempnsq{1/2}{e}{-1}(\lpar,\lchi)$ are
    functions of both the geometric poloidal angle $\lpar$ and the ballooning angle $\lchi$.
    
    To obtain equations for $\ddensnsq{1/2}{e}{-1}$ and $\dtempnsq{1/2}{e}{-1}$, 
    we must go to the second-order equation in the subsidiary expansion.
    We proceed to the first-order equation in the subsidiary expansion, which is
     \beqn 
     \vpar \kpar \drv{}{\lpar}\left(\HHepq{1/2}{-1}\right)=
     \copbothe\left[\HHepq{1/2}{0}+ \imag \lflre \lchi \HHez - \HHeSH\right]
    . \label{eq:electron_inner1b-collisional-nustar-gg-1-first-order}\eeqn
    Equation \refeq{eq:electron_inner1b-collisional-nustar-gg-1-first-order}
    can be solved by inverting an additional Spitzer-H\"{a}rm problem
    \beqn \vpar \kpar \drv{}{\lpar}\left(\HHepq{1/2}{-1}\right)= \copbothe\left[\HHeSHp{1/2}\right].
    \label{eq:nustar-gg-1-first-order-SH}\eeqn 
    With the Spitzer-H\"{a}rm distribution $\HHeSHp{1/2}$
    defined by equation \refeq{eq:nustar-gg-1-first-order-SH},
    we may write equation \refeq{eq:electron_inner1b-collisional-nustar-gg-1-first-order} in the form
     \beqn 
     \copbothe\left[\HHepq{1/2}{0}+ \imag \lflre \lchi \HHez - \HHeSH  - \HHeSHp{1/2}\right] = 0
    . \label{eq:electron_inner1b-collisional-nustar-gg-1-first-order-rewrite}\eeqn
    Hence, we find that
    \beqn \fl  \HHepq{1/2}{0} = \left(\frac{\ddensnsq{1/2}{e}{0}}{\dense} 
    + \frac{\dtempnsq{1/2}{e}{0}}{\tempe}\left(\frac{\energy}{\tempe}
    - \frac{3}{2}\right)\right)\eqlbe - \imag \lflre\lchi\HHez
    + \HHeSH + \HHeSHp{1/2}, \label{eq:HHeh-0}\eeqn
    where  $\ddensnsq{1/2}{e}{0} = \ddensnsq{1/2}{e}{0}(\lpar,\lchi)$ and 
    $\dtempnsq{1/2}{e}{0} = \dtempnsq{1/2}{e}{0}(\lpar,\lchi)$.
    The second-order equation in the subsidiary expansion
    of equation \refeq{eq:electron_inner1b-collisional} is
    \beqn \vpar \kpar \drv{}{\lpar}\left(\HHepq{1/2}{0}\right)=
     \copbothe\left[\HHepq{1/2}{1}\right].
     \label{eq:electron_inner1b-collisional-nustar-gg-1-second-order}\eeqn
    The equations for $\ddensnsq{1/2}{e}{-1}$ and $\dtempnsq{1/2}{e}{-1}$
    are obtained from the solvability conditions of
    equation \refeq{eq:electron_inner1b-collisional-nustar-gg-1-second-order}. These are
    \beqn \bvec\cdot\nbl\lpar \drv{}{\lpar}
    \left(\intv{\frac{\vpar}{\bmag}\HHepq{1/2}{0}}\right)= 0
    \label{eq:collisional-solvability-one}\eeqn 
    and 
    \beqn \bvec\cdot\nbl\lpar \drv{}{\lpar}
    \left(\intv{\frac{\vpar}{\bmag}\left(\frac{\energy}{\tempe}-\frac{5}{2}\right)
    \HHepq{1/2}{0}}\right)= 0 
    \label{eq:collisional-solvability-two}.\eeqn 
    The conditions \refeq{eq:collisional-solvability-one} and \refeq{eq:collisional-solvability-two} are
    obtained by multiplying equation \refeq{eq:electron_inner1b-collisional-nustar-gg-1-second-order} 
    by $1$ and $\energy/\tempe - 5/2$, respectively, and integrating over velocity space.
    Equations \refeq{eq:collisional-solvability-one} and \refeq{eq:collisional-solvability-two}
    indicate that 
    \beqn \intv{\frac{\vpar}{\bmag}\HHepq{1/2}{0}}= \solvcnstn(\lchi)
        \label{eq:pfirsh-schluter-solvcnstn-proto}\eeqn
    and 
    \beqn \intv{\frac{\vpar}{\bmag}\left(\frac{\energy}{\tempe}-\frac{5}{2}\right)
    \HHepq{1/2}{0}} = \solvcnstt(\lchi),  \label{eq:pfirsh-schluter-solvcnstt-proto} \eeqn
    where $\solvcnstn(\lchi)$
    and $\solvcnstt(\lchi)$ are functions
    of the ballooning angle $\lchi$ only. 
    We explicitly evaluate $\solvcnstn$ and $\solvcnstt$
    using the results \refeq{eq:dens-int-basic} and \refeq{eq:temp-int-basic} of
    \ref{sec:parallel-collisional-contributions}. We find that 
    \beqn \fl \intv{\frac{\vpar}{\bmag}\HHepq{1/2}{0}} = \frac{\dense\vthere^2}{2 } \frac{\bvec \cdot \nbl \lpar}{\bmag^2}
   \left[\acoeffn{0}\left(\drv{}{\lchi}\left(\frac{\ddensns{0}{e}}{\dense}\right)
   + \drv{}{\lpar}\left(\frac{\ddensnsq{1/2}{e}{-1}}{\dense}\right)\right) \right. \nonumber
   \eeqn \beqn \fl \left. 
   +\bcoeffn{0}\left(\drv{}{\lchi}\left(\frac{\dtempns{0}{e}}{\tempe}   \right)
   +\drv{}{\lpar}\left(\frac{\dtempnsq{1/2}{e}{-1}}{\tempe}\right)\right)\right]
   - \frac{\dense\vthere^2}{2 } \frac{\imag \kkfldl \saffacprim \lchi \bcur}{\cycfe \bmag } 
   \left(\frac{\ddensns{0}{e}}{\dense} + \frac{\dtempns{0}{e}}{\tempe}\right)  
   , \label{eq:pfirsh-schluter-parallel-particle-flux-proto}\eeqn 
    where $\acoeffn{0} =\fract{ -1.969}{\cfreqei} $, $\bcoeffn{0} =\fract{ -3.366}{\cfreqei}$,
    and we have assumed $\zedi =1$; and 
    \beqn \fl \intv{\frac{\vpar}{\bmag}\left(\frac{\energy}{\tempe}-\frac{5}{2}\right)\HHepq{1/2}{0}} =
    -\frac{5\dense\vthere^2}{4 } \frac{\bvec \cdot \nbl \lpar}{\bmag^2}
   \left[\acoeffn{1}\left(\drv{}{\lchi}\left(\frac{\ddensns{0}{e}}{\dense}\right)
   + \drv{}{\lpar}\left(\frac{\ddensnsq{1/2}{e}{-1}}{\dense}\right)\right) \right. \nonumber
   \eeqn \beqn \fl \left. 
   +\bcoeffn{1}\left(\drv{}{\lchi}\left(\frac{\dtempns{0}{e}}{\tempe}   \right)
   +\drv{}{\lpar}\left(\frac{\dtempnsq{1/2}{e}{-1}}{\tempe}\right)\right)\right]
   - \frac{5\dense\vthere^2}{4 } \frac{\imag \kkfldl \saffacprim \lchi \bcur}{\cycfe \bmag} 
   \frac{\dtempns{0}{e}}{\tempe} 
   ,\label{eq:pfirsh-schluter-parallel-heat-flux-proto}\eeqn 
    where $\acoeffn{1} =\fract{ 0.559}{\cfreqei} $ and  $\bcoeffn{1} =\fract{2.226}{\cfreqei}$.
    We obtain explicit expressions for $\solvcnstn$ and $\solvcnstt$ 
    by multiplying equations \refeq{eq:pfirsh-schluter-parallel-particle-flux-proto} 
    and \refeq{eq:pfirsh-schluter-parallel-heat-flux-proto} by $\bmag^2$,
    and applying poloidal angle average $\lparav{\cdot}$. The results are
    \beqn \fl \solvcnstn(\lchi) = \frac{\dense\vthere^2}{2 } \frac{\lparav{\bvec \cdot \nbl \lpar}}{\lparav{\bmag^2}}
   \left[\acoeffn{0}\drv{}{\lchi}\left(\frac{\ddensns{0}{e}}{\dense}\right)
   +\bcoeffn{0}\drv{}{\lchi}\left(\frac{\dtempns{0}{e}}{\tempe}   \right)
   \right] \nonumber \eeqn \beqn
   - \frac{\dense\vthere^2}{2 } \frac{\imag \kkfldl \saffacprim \lchi \bcur \bmagovl}{\cycfeovl  }\frac{1}{\lparav{\bmag^2}} 
   \left(\frac{\ddensns{0}{e}}{\dense} + \frac{\dtempns{0}{e}}{\tempe}\right)  
   , \label{eq:pfirsh-schluter-solvcnstn}\eeqn 
   and 
    \beqn \fl \solvcnstt(\lchi) = -\frac{5\dense\vthere^2}{4 } \frac{\lparav{\bvec \cdot \nbl \lpar}}{\lparav{\bmag^2}}
   \left[\acoeffn{1}\drv{}{\lchi}\left(\frac{\ddensns{0}{e}}{\dense}\right)
   +\bcoeffn{1}\drv{}{\lchi}\left(\frac{\dtempns{0}{e}}{\tempe}   \right)
   \right] \nonumber \eeqn \beqn
   - \frac{5\dense\vthere^2}{4} \frac{\imag \kkfldl \saffacprim \lchi \bcur \bmagovl}{\cycfeovl  }\frac{1}{\lparav{\bmag^2}} 
   \frac{\dtempns{0}{e}}{\tempe}
   , \label{eq:pfirsh-schluter-solvcnstt}\eeqn 
   respectively.
   Finally, to obtain equations for $\ddensnsq{1/2}{e}{-1}$ and $\dtempnsq{1/2}{e}{-1}$,
   we subtract equation \refeq{eq:pfirsh-schluter-solvcnstn}
   from equation \refeq{eq:pfirsh-schluter-parallel-particle-flux-proto},
   and equation \refeq{eq:pfirsh-schluter-solvcnstt} from
   equation \refeq{eq:pfirsh-schluter-parallel-heat-flux-proto}.
   The result is that
  \beqn \fl  \frac{\bvec \cdot \nbl \lpar}{\bmag^2}
   \left[\acoeffn{0}\drv{}{\lpar}\left(\frac{\ddensnsq{1/2}{e}{-1}}{\dense}\right) 
   +\bcoeffn{0}\drv{}{\lpar}\left(\frac{\dtempnsq{1/2}{e}{-1}}{\tempe}\right)\right]
   = \nonumber \eeqn \beqn 
    \left(\frac{\lparav{\bvec \cdot \nbl \lpar}}{\lparav{\bmag^2}}
    -\frac{\bvec \cdot \nbl \lpar}{\bmag^2}\right)\left[
    \acoeffn{0}\drv{}{\lchi}\left(\frac{\ddensns{0}{e}}{\dense}\right) 
   +\bcoeffn{0}\drv{}{\lchi}\left(\frac{\dtempns{0}{e}}{\tempe}\right)\right]
   \label{eq:pfirsh-schluter-fluctuation-half-eqn-1}\eeqn \beqn 
    +\left(\frac{1}{\bmag^2} - \frac{1}{\lparav{\bmag^2}} \right)
    \frac{\imag \kkfldl \saffacprim \lchi \bcur \bmagovl}{\cycfeovl} 
   \left(\frac{\ddensns{0}{e}}{\dense} + \frac{\dtempns{0}{e}}{\tempe}\right)  
   , \nonumber\eeqn 
    and 
    \beqn \fl  \frac{\bvec \cdot \nbl \lpar}{\bmag^2}
   \left[\acoeffn{1}\drv{}{\lpar}\left(\frac{\ddensnsq{1/2}{e}{-1}}{\dense}\right) 
   +\bcoeffn{1}\drv{}{\lpar}\left(\frac{\dtempnsq{1/2}{e}{-1}}{\tempe}\right)\right]
   = \nonumber \eeqn \beqn 
    \left(\frac{\lparav{\bvec \cdot \nbl \lpar}}{\lparav{\bmag^2}}
    -\frac{\bvec \cdot \nbl \lpar}{\bmag^2}\right)\left[
    \acoeffn{1}\drv{}{\lchi}\left(\frac{\ddensns{0}{e}}{\dense}\right) 
   +\bcoeffn{1}\drv{}{\lchi}\left(\frac{\dtempns{0}{e}}{\tempe}\right)\right]
   \label{eq:pfirsh-schluter-fluctuation-half-eqn-2}\eeqn \beqn 
    -\left(\frac{1}{\bmag^2} - \frac{1}{\lparav{\bmag^2}} \right)
    \frac{\imag \kkfldl \saffacprim \lchi \bcur \bmagovl}{\cycfeovl} 
   \frac{\dtempns{0}{e}}{\tempe}
   . \nonumber\eeqn 
   Inverting equations \refeq{eq:pfirsh-schluter-fluctuation-half-eqn-1} and
   \refeq{eq:pfirsh-schluter-fluctuation-half-eqn-2} for  $\drvt{\left(\ddensnsq{1/2}{e}{-1}\right)}{\lpar}$
   and $\drvt{\left(\dtempnsq{1/2}{e}{-1}\right)}{\lpar}$, we find that
\beqn \fl  \frac{\bvec \cdot \nbl \lpar}{\bmag^2}
   \drv{}{\lpar}\left(\frac{\ddensnsq{1/2}{e}{-1}}{\dense}\right) 
   =  
    \left(\frac{\lparav{\bvec \cdot \nbl \lpar}}{\lparav{\bmag^2}}
    -\frac{\bvec \cdot \nbl \lpar}{\bmag^2}\right)
    \drv{}{\lchi}\left(\frac{\ddensns{0}{e}}{\dense}\right) 
    \label{eq:pfirsh-schluter-density-half}\eeqn \beqn 
    +\left(\frac{1}{\bmag^2} - \frac{1}{\lparav{\bmag^2}} \right)
    \frac{\imag \kkfldl \saffacprim \lchi \bcur \bmagovl}{\cycfeovl} 
   \left(\frac{\bcoeffn{1}}{\acoeffn{0}\bcoeffn{1}-\acoeffn{1}\bcoeffn{0}}
   \frac{\ddensns{0}{e}}{\dense} + 
   \frac{\bcoeffn{1}+\bcoeffn{0}}{\acoeffn{0}\bcoeffn{1}-\acoeffn{1}\bcoeffn{0}}
   \frac{\dtempns{0}{e}}{\tempe}\right)  
   , \nonumber\eeqn 
   and 
\beqn \fl  \frac{\bvec \cdot \nbl \lpar}{\bmag^2}
   \drv{}{\lpar}\left(\frac{\dtempnsq{1/2}{e}{-1}}{\tempe}\right) 
   =  
    \left(\frac{\lparav{\bvec \cdot \nbl \lpar}}{\lparav{\bmag^2}}
    -\frac{\bvec \cdot \nbl \lpar}{\bmag^2}\right)
    \drv{}{\lchi}\left(\frac{\dtempns{0}{e}}{\tempe}\right) 
    \label{eq:pfirsh-schluter-temperature-half}\eeqn \beqn 
    - \left(\frac{1}{\bmag^2} - \frac{1}{\lparav{\bmag^2}} \right)
    \frac{\imag \kkfldl \saffacprim \lchi \bcur \bmagovl}{\cycfeovl} 
   \left(
   \frac{\acoeffn{0}+ \acoeffn{1}}{\acoeffn{0}\bcoeffn{1}-\acoeffn{1}\bcoeffn{0}}
   \frac{\dtempns{0}{e}}{\tempe} + 
   \frac{\acoeffn{1}}{\acoeffn{0}\bcoeffn{1}-\acoeffn{1}\bcoeffn{0}}
   \frac{\ddensns{0}{e}}{\dense} 
   \right)  . \nonumber\eeqn

    To evaluate the effective electron parallel velocity $\dupareeff$,
    defined in equation \refeq{eq:upareeff}, we compute the integral
    \beqn \fl \dupareeff = \frac{1}{\lparav{\kpar}}
    \lparav{\frac{\kpar}{\dense}\intv{\vpar\left(\HHepq{1/2}{-1}
    + \HHepq{1/2}{0} + \imag \lflre \lchi \HHez \right)}}
    ,\label{eq:dupareeff-step-1}\eeqn
    where we have used the definition \refeq{eq:true-upare-halfth}
    and the expansion \refeq{eq:HHeh-nustar-gg-1}.
    With the solutions \refeq{eq:HHeh-1}
    and \refeq{eq:HHeh-0}, and the integral \refeq{eq:dens-int-basic}, we find that
    \beqn \fl \dupareeff = \frac{\vthere^2/2}{\lparav{\kpar}}
    \lparav{\frac{(\bvec\cdot\nbl\lpar)^2}{\bmag^2}}\left[
    \acoeffn{0}\drv{}{\lchi}\left(\frac{\ddensns{0}{e}}{\dense}\right)
    +\bcoeffn{0}\drv{}{\lchi}\left(\frac{\dtempns{0}{e}}{\tempe}\right)\right]
    \label{eq:dupareeff-step-2}\eeqn    \beqn + \frac{\vthere^2/2}{\lparav{\kpar}}
    \lparav{\frac{(\bvec\cdot\nbl\lpar)^2}{\bmag^2}\left[
    \acoeffn{0}\drv{}{\lpar}\left(\frac{\ddensnsq{1/2}{e}{-1}}{\dense}\right)
    +\bcoeffn{0}\drv{}{\lpar}\left(\frac{\dtempnsq{1/2}{e}{-1}}{\tempe}\right)
    \right]}. \nonumber\eeqn
    Finally, using equations \refeq{eq:pfirsh-schluter-density-half} 
    and \refeq{eq:pfirsh-schluter-temperature-half}
    to substitute for $\drvt{\left(\ddensnsq{1/2}{e}{-1}\right)}{\lpar}$
   and $\drvt{\left(\dtempnsq{1/2}{e}{-1}\right)}{\lpar}$, we find that
   \beqn \fl
   \dupareeff = \frac{\vthere^2/2}{\lparav{\kpar}}
    \frac{(\lparav{\bvec\cdot\nbl\lpar})^2}{\lparav{\bmag^2}}
    \left[
    \acoeffn{0}\drv{}{\lchi}\left(\frac{\ddensns{0}{e}}{\dense}\right)
    +\bcoeffn{0}\drv{}{\lchi}\left(\frac{\dtempns{0}{e}}{\tempe}\right)\right]
    \label{eq:dupareeff-step-last}
    \eeqn \beqn 
    +\imag \frac{\vthere}{2} \frac{\kkfldl \saffacprim \bcur \lchi \gyrdeovl \bmagovl }{\lparav{\kpar}}
    \left(\lparav{\frac{\bvec\cdot\nbl\lpar}{\bmag^2}} -
    \frac{\lparav{\bvec\cdot\nbl\lpar}}{\lparav{\bmag^2}}\right)
    \left(\frac{\ddensns{0}{e}}{\dense}+\frac{\dtempns{0}{e}}{\tempe} \right)
    . \nonumber\eeqn
    Using the same techniques, and the
    integral \refeq{eq:temp-int-basic},
    we obtain the effective electron parallel heat flux
   \beqn \fl
   \dqpareeff = -\frac{5}{4}\frac{\dense \tempe\vthere^2}{ \lparav{\kpar}}
     \frac{(\lparav{\bvec\cdot\nbl\lpar})^2}{\lparav{\bmag^2}}
    \left[
    \acoeffn{1}\drv{}{\lchi}\left(\frac{\ddensns{0}{e}}{\dense}\right)
    +\bcoeffn{1}\drv{}{\lchi}\left(\frac{\dtempns{0}{e}}{\tempe}\right)\right]
    \label{eq:dqpareeff-step-last}
    \eeqn \beqn 
    +\imag \frac{5}{4}\dense \tempe \vthere
    \frac{ \kkfldl \saffacprim  \bcur \lchi \gyrdeovl\bmagovl }{\lparav{\kpar}}
    \left(\lparav{\frac{\bvec\cdot\nbl\lpar}{\bmag^2}} -
    \frac{\lparav{\bvec\cdot\nbl\lpar}}{\lparav{\bmag^2}}\right)
   \frac{\dtempns{0}{e}}{\tempe}
    . \nonumber\eeqn
    
    We now turn to the calculation of the
    neoclassical perpendicular diffusion terms appearing in equations
    \refeq{eq:electron_inner-dens0-collisional} and \refeq{eq:electron_inner-temp0-collisional}.
    To evaluate the particle flux $\dpartnc$, 
    defined in equation \refeq{eq:neoclassical-particle-flux},
    we use equations
    \refeq{eq:electron_inner1b-collisional-nustar-gg-1-leading-order} and 
    \refeq{eq:electron_inner1b-collisional-nustar-gg-1-first-order} to show that
    \beqn \dpartnc  
 = - \lparav{ \frac{\bcur}{\cycfe}\tdrv{\radial}{\flxl}\intv{\vpar^2 \kpar \drv{}{\lpar}\left(\HHepq{1/2}{-1}\right)}}. 
\label{eq:density-neoclassical-perpendicular-diffusion-pfirsh-schluter-step-1} \eeqn
    Substituting the solution \refeq{eq:HHeh-1} into equation
    \refeq{eq:density-neoclassical-perpendicular-diffusion-pfirsh-schluter-step-1},
    we find that 
    \beqn \fl \dpartnc  
     \label{eq:density-neoclassical-perpendicular-diffusion-pfirsh-schluter-step-2}
 = - \frac{\dense\vthere^2}{2} \frac{\bcur \bmagovl}{\cycfeovl}\tdrv{\radial}{\flxl}
 \lparav{\frac{\bvec\cdot\nbl\lpar}{\bmag^2}
 \left(\drv{}{\lpar} \left(\frac{\ddensnsq{1/2}{e}{-1}}{\dense}\right)+
   \drv{}{\lpar}\left(\frac{\dtempnsq{1/2}{e}{-1}}{\tempe}\right)\right)}. \eeqn
   Finally, we substitute the results \refeq{eq:pfirsh-schluter-density-half}
   and \refeq{eq:pfirsh-schluter-temperature-half} into
   equation \refeq{eq:density-neoclassical-perpendicular-diffusion-pfirsh-schluter-step-2} to find the neoclassical
   particle flux 
 \beqn \fl \frac{\dpartnc}{\dense}
     \label{eq:density-neoclassical-perpendicular-diffusion-pfirsh-schluter-step-last}
   = - \frac{\vthere^2}{2 } \frac{\bcur\bmagovl }{\cycfeovl} \tdrv{\radial}{\flxl}
 \left( \frac{\lparav{\bvec\cdot\nbl\lpar}}{\lparav{\bmag^2}}
 - \lparav{\frac{\bvec\cdot\nbl\lpar}{\bmag^2}}\right)
 \left(\drv{}{\lchi}\left(\frac{\ddensns{0}{e}}{\dense}\right)+
   \drv{}{\lchi}\left(\frac{\dtempns{0}{e}}{\tempe}\right)\right)
 \eeqn \beqn \fl 
 +\imag \kkfldl \tdrv{\saffac}{\radial}\lchi \frac{\vthere^2}{2 }\!\! \left(\frac{\bcur\bmagovl }{\cycfeovl} \tdrv{\radial}{\flxl}\right)^2
\!\left( \lparav{\frac{1}{\bmag^2}} \!\!
 - \frac{1}{\lparav{\bmag^2}}\right) \!\!
\left[\frac{\acoeffn{1} - \bcoeffn{1} }{\acoeffn{0}\bcoeffn{1} - \acoeffn{1}\bcoeffn{0}}\frac{\ddensns{0}{e}}{\dense}
 + \frac{ \acoeffn{1} +\acoeffn{0} - \bcoeffn{1} - \bcoeffn{0}}{\acoeffn{0}\bcoeffn{1} - \acoeffn{1}\bcoeffn{0}}\frac{\dtempns{0}{e}}{\tempe}\right] 
 .\nonumber\eeqn 
   Following identical steps, we find that the
   neoclassical heat flux $\dheatnc$, 
   defined in equation \refeq{eq:neoclassical-heat-flux},
   is 
    \beqn \fl \frac{\dheatnc}{\dense \tempe}
     \label{eq:temperature-neoclassical-perpendicular-diffusion-pfirsh-schluter-step-last}
  = - \frac{5\vthere^2}{4} \frac{\bcur\bmagovl}{\cycfeovl}\tdrv{\radial}{\flxl}
 \left( \frac{\lparav{\bvec\cdot\nbl\lpar}}{\lparav{\bmag^2}}
 - \lparav{\frac{\bvec\cdot\nbl\lpar}{\bmag^2}}\right)
\drv{}{\lchi}\left(\frac{\dtempns{0}{e}}{\tempe}\right)
 \eeqn \beqn \fl +\imag \kkfldl \tdrv{\saffac}{\radial}\lchi \frac{\vthere^2}{2 } \left(\frac{\bcur\bmagovl }{\cycfeovl} \tdrv{\radial}{\flxl}\right)^2
\left( \lparav{\frac{1}{\bmag^2}}
 - \frac{1}{\lparav{\bmag^2}}\right)
\left[ \frac{ (5/2)(\acoeffn{1} +\acoeffn{0})}{\acoeffn{0}\bcoeffn{1} - \acoeffn{1}\bcoeffn{0}}\frac{\dtempns{0}{e}}{\tempe}
 + \frac{5\acoeffn{1}/2}{  \acoeffn{0}\bcoeffn{1} - \acoeffn{1}\bcoeffn{0}}\frac{\ddensns{0}{e}}{\dense}\right] 
  .\nonumber\eeqn 

%%%%%%%%%%%%%%%%%%%%%%%%%%%%%%%%%%%%
%% Banana Spitzer calculation
%%%%%%%%%%%%%%%%%%%%%%%%%%%%%%%%%%%%
      
    \section{The parallel flows and neoclassical perpendicular diffusion terms
    in the $\nustar \ll 1$, $\aspect \ll 1$ (banana) limit
} \label{sec:banana-parallel-collisional-contributions}
    In this section we calculate the electron distribution function $\HHepq{1/2}{0}$
    by solving equation \refeq{eq:electron-banana-passing} to leading-order
    in the expansion in inverse aspect ratio $\aspect = \radial/\rmajorz \ll 1$. 
    We take the normalised electron collisionality $\nustar = \saffac\rmajorz\cfreqee/\aspect^{3/2}\vthere \ll 1$.
    We assume that the
    equilibrium can be approximated by circular flux surfaces \cite{HazeltineMeiss,FreidbergMHD}.
    We use the fact that
    the $2\pi$-periodic $\lpar$ variation in geometric quantities is small by $\order{\aspect}$.
    For example, the magnetic field strength 
    \beqn \bmag \simeq \bmagz \left(1 - \aspect \cos \lpar \right)
     = \bmagz  + \order{\aspect \bmag}, \eeqn
    where $\bmagz = \bcur/\rmajorz$ is a constant.
    As a consequence, the fraction of velocity space
    occupied by trapped particles
    become small. This can be seen from the definition of 
    \beqn \vpar = \sign \left(\frac{2 \energy}{\me}\right)^{1/2}
    \left(1 - \pitch \bmag(\lpar)\right)^{1/2}: \eeqn 
    passing particles occupy $0 \leq \pitch\bmagz  < \bmagz/ \bmagmax \simeq 1 - \aspect$;
    and trapped particles
    occupy $ \bmagz/\bmagmax \leq \pitch\bmagz \leq \bmagz/\bmag(\lpar) \simeq 1 + \aspect \cos \theta$. 
We can identify two regions in the problem: there is a \enquote{deeply passing} region where
 $\pitch \bmagz \sim 1 \sim 1 - \pitch \bmagz \gg \aspect$, and also the trapped-passing region
 where $\pitch\bmagz = 1 - \order{\aspect}$. In the deeply passing region,
 we can find the leading-order solution by taking $\pitch \bmagz \sim 1 \sim 1 - \pitch \bmagz  $
 and using $\aspect \ll 1$ to approximate the geometric quantities.
 We find that 
 \beqn \HHepq{1/2}{0} = \HHeSHz -\imag \lflrez \lchi \HHez +  \order{\aspect^{1/2} \HHepq{1/2}{0}}, \label{eq:deeplypassing-leading-order}\eeqn
 where \beqn \lflrez = \sign\lflrezth \sqrt{\frac{\energy}{\tempe}}\sqrt{1-\pitch \bmagz}, \eeqn with 
 $\lflrezth = \kkfldl\saffacprim\bcur \vthere / \cycfez$,
 $\cycfez = -\charge \bmagz / \me \ltsp$, and 
 \beqn \HHeSHz = \sign\sqrt{\frac{\energy}{ \tempe}}\sqrt{1-\pitch \bmagz} \vthere\KKeSH \eqlbe.\eeqn
 The $\order{\aspect^{1/2} \HHepq{1/2}{0}}$ correction in equation \refeq{eq:deeplypassing-leading-order}
 arises from the presence of the trapped-passing region.
 To solve for the $\pitch \bmagz = 1 - \order{\aspect}$ region,
 we note that the pitch-angle scattering components of the collision operator, $\pitchscatter[\cdot]$
 are larger than the other test-particle and field-particle terms by $\order{\aspect^{-1}}$. 
 In other words, we need only the pitch-angle scattering collision
 operator (cf. \cite{HazeltineMeiss,helander})
 \beqn \pitchscatter[\cdot] =
 \nueff(\energy) 
 \frac{\sqrt{1- \pitch \bmag}}{\bmag}
 \drv{}{\pitch}\left(\pitch\sqrt{1- \pitch \bmag} \drv{}{\pitch}\left[\cdot\right] \right),
 \label{eq:pitchscatter} \eeqn
 where \beqn \nueff(\energy) = \frac{3\sqrt{\pi}}{2}\left(\frac{\tempe}{\energy}\right)^{3/2}\left(\cfreqei 
 + \cfreqee 
 \left(\erf\left(\sqrt{\energy/\tempe}\right)
 - \Psifunc\left(\sqrt{\energy/\tempe}\right)\right)\right), \eeqn
 with the error function $\erf(\zarg)$ defined by equation \refeq{eq:error-function},
 and the function $\Psifunc(\zarg)$ defined by equation \refeq{eq:Psifunc-function}.
 We note the similarity of the forms of the collision frequencies $\nueff$ and $\cfreqiiperp$,
 defined in equation \refeq{eq:cfreqiiperp}.
 This similarity arises because the terms involving these collision
 frequencies are due to the pitch-angle scattering pieces of the electron
 and ion collision operators, respectively.
 With these considerations, to leading-order in $\aspect$,
 equation \refeq{eq:electron-banana-passing} becomes
 \beqn \fl  
 \frac{2}{\bmagz}\drv{}{\pitch} \left(\pitch \lparav{\sqrt{1- \pitch \bmag(\lpar)}} \drv{}{\pitch}\left(\HHepq{1/2}{0}\right)
 \right) \nonumber \eeqn \beqn 
 + \sign\sqrt{\frac{\energy}{\tempe}}\left( \vthere \KKeSH \eqlbe - \imag  \lflrezth \lchi \HHez\right) 
    =0, \label{eq:boundary-layer-leading-order-c}\eeqn 
 where we have 
 divided by the $\energy$-dependent collision frequency pre-factors,
 we have used the identity
 \beqn \pitchscatter [\vpar \fulldistg(\energy) ] = -\frac{\nueff(\energy)}{2}
 \vpar \fulldistg(\energy)\eeqn
 to simplify the terms proportional to $\vpar$,
 and we have employed the definitions of the transit
 average $\transav{\cdot}$, and poloidal angle average $\lparav{\cdot}$,
 equations \refeq{eq:transitav} and \refeq{eq:lparav}, respectively,
 and finally we have expanded the geometrical factors to leading-order in $\aspect \ll 1$.
 We note that $\sqrt{1-\pitch \bmag(\lpar)}$ may not be usefully expanded in $\aspect$
 because $\pitch \bmagz = 1 + \order{\aspect}$, and the $\lpar$
 dependence of $\bmag(\lpar)$ comes in at $\order{\aspect}$.
 We also note that \beqn \HHepq{1/2}{0} \sim \order{\aspect^{1/2} \massrp{1/4} \HHez} \label{eq:HHep12est}\eeqn
 in the trapped-passing region. We can integrate equation \refeq{eq:boundary-layer-leading-order-c} directly,
 with the boundary conditions $\HHepq{1/2}{0} = 0 $ at $\pitch \bmagz = \bmagz/\bmagmax$, 
 and no divergence at $\pitch = 0$. The result is
 \beqn \fl \HHepq{1/2}{0} = - \sign\sqrt{\frac{\energy}{\tempe}}\left( \vthere \KKeSH \eqlbe - \imag  \lflrezth \lchi \HHez\right)\int^{\pitch}_{1/\bmagmax}
 \frac{ \bmagz d \pitch^\prime /2}{\lparav{\sqrt{1- \pitch^\prime \bmag(\lpar)}}}
 . \label{eq:weaklypassing-leading-order-solution}\eeqn
 We can match to the deeply passing solution \refeq{eq:deeplypassing-leading-order}
 by taking the solution \refeq{eq:weaklypassing-leading-order-solution}, and evaluating the integral
 with $\aspect \rightarrow 0$ and $1 - \pitch \bmagz \gg \order{\aspect}$.
 
 Having evaluated the distribution function using the methods of neoclassical theory,
 we are now able to calculate the electron parallel velocity $\dupareeff$
 and electron parallel heat flux $\dqpareeff$,
 defined in equations \refeq{eq:upareeff} and \refeq{eq:qpareeff}, respectively.
 From the deeply passing solution, equation \refeq{eq:deeplypassing-leading-order},
 we can see that the leading-order contributions {will result from
 the response $\HHeSHz$ to the parallel gradients of density and temperature.}
 As in the neoclassical calculation for the bootstrap current \cite{helander},
 we calculate the additional contribution arising from the interaction between passing
 and trapped electrons in the $\pitch \bmagz = 1 - \order{\aspect}$ region.
 
 To evaluate the electron parallel velocity and the electron parallel heat flux,
 we need to compute an integral of the form 
 \beqn \gflux = \frac{1}{\lparav{\kpar}}\lparav{\kpar \intv{\vpar \fulldistg(\energy)
  \left(\HHepq{1/2}{0} + \imag \lflre\lchi\HHez\right)}},\eeqn
 where for $\fulldistg(\energy) = 1$ we obtain $\dupareeff = \gflux / \dense$, and where
 for $\fulldistg= \energy/\tempe - 5/2$ we obtain $\dqpareeff = \tempe \gflux $.
 First, we use that the result is expected to be close to the Spitzer-H\"{a}rm flows
 obtained from $\HHeSH$. 
 We write \beqn \gflux = \gfluxsh + \gfluxd, \eeqn
 with $\gfluxsh$ defined by 
\beqn \gfluxsh   = \frac{1}{\lparav{\kpar}}\lparav{\kpar \intv{\vpar \fulldistg(\energy)
  \HHeSH}},\label{eq:gfluxsh}\eeqn 
 and 
  \beqn \fl \gfluxd 
 = \frac{1}{\lparav{\kpar}}\lparav{\kpar 
 \intv{\vpar \fulldistg(\energy)\left(\HHepq{1/2}{0} - \HHeSH + \imag \lflre \lchi \HHez \right)}}.
 \label{eq:gfluxd-def1}\eeqn
 We can calculate $\gfluxsh$ using the results of \ref{sec:parallel-collisional-contributions}.
 To evaluate the leading nonzero component of $\gfluxd$ 
 requires that we calculate the sub-leading corrections to $\HHepq{1/2}{0}$ everywhere in $\pitch$.
 To avoid this, we convert the integral \refeq{eq:gfluxd-def1}
 into an integral where the dominant contribution
 comes from the trapped-passing region, where we can use solution \refeq{eq:weaklypassing-leading-order-solution}.
 We localise the integral to the trapped-passing
 region by introducing $\copbothe[\cdot]$ into the integral.
 We do this by using the Spitzer-H\"{a}rm problem \refeq{eq:spitzer-density}
 to replace $\vpar$ in equation \refeq{eq:gfluxd-def1}, with the result
 \beqn \fl \gfluxd =
 \frac{1}{\lparav{\kpar}}\lparav{ \kpar
 \intv{ \fulldistg(\energy) \frac{\copbothe\left[\vpar\ffSH \eqlbe \right]}{\eqlbe}  \left(\HHepq{1/2}{0} - \HHeSH + \imag \lflre \lchi \HHez \right)}}.
 \label{eq:gfluxd-def1a}\eeqn
 Now using the self-adjointness of $\copbothe[\cdot]$ with respect to the inner product \refeq{eq:innerprod} \cite{helander},
 the integral becomes 
 \beqn \fl \gfluxd =
 \frac{1}{\lparav{\kpar}}\lparav{ \kpar
 \intv{\vpar \ffSH \; \copbothe\left[\fulldistg(\energy) (\HHepq{1/2}{0} - \HHeSH + \imag \lflre \lchi \HHez) \right]}}.
 \label{eq:gfluxd-def2}\eeqn
 Finally, we estimate the size of the contributions
 to $\gfluxd$ from the deeply passing region and the trapped-passing region.
 In the deeply passing region, we find that the contribution is of size
 \beqn \gfluxd \sim \aspect \massrp{1/4}\vthere \ddensns{0}{e},
 \label{eq:deeplypassing-contribution-estimate}\eeqn
 since $\copbothe\left[\HHepq{1/2}{0} - \HHeSH + \imag \lflre \lchi \HHez \right]$ is small
 by $\order{\aspect}$ in the deeply passing region.
 In the trapped-passing region, we find that the contribution is of size
 \beqn \gfluxd \sim \aspect^{1/2} \massrp{1/4} \vthere \ddensns{0}{e},
 \label{eq:trappedpassing-contribution-estimate}\eeqn
 since $\HHepq{1/2}{0} - \HHeSH + \imag \lflre \lchi \HHez \sim \aspect^{1/2} \massrtp{1/4}\HHez$
 by the estimates \refeq{eq:HHep12est} and $\vpar \sim \aspect^{1/2}\vthere$,
 and $\pitchscatter[\cdot] d^3 \pvel /\vthere^3 \sim (\cfreqei / \aspect) \aspect^{1/2}  \sim \cfreqei \aspect^{-1/2} $.
 As the contribution from the trapped-passing region is larger than
 the contribution from the deeply passing region,
 we replace $\copbothe\left[\cdot\right]$ by $\pitchscatter\left[\cdot\right]$
 when we evaluate the integral in equation \refeq{eq:gfluxd-def2}.

 To evaluate the integral in equation, we insert the definition of $\pitchscatter\left[\cdot\right]$,
 equation \refeq{eq:pitchscatter},
 with $d^3 \pvel = (\bmag \energy / \me^2 |\vpar|) d \energy d \pitch d \gyrophase$, 
 and integrate by parts once in $\pitch$.
 The integrals in $\energy$ and $\pitch$ are separable, and we find the intermediate result
\beqn \fl \gfluxd = \frac{ \pi \vthere^4}{3}
 \trapfrac 
\int^\infty_0 \fulldistg(\energy)\nueff(\energy) \left(\frac{\energy}{\tempe}\right)^{3/2}
 \ffSH \left(\vthere \KKeSH \eqlbe - \imag \lflrezth \lchi \HHez \right)  \; \frac{d \energy}{\tempe}
  \label{eq:gfluxd-intermediate}, \eeqn where following \cite{helander} we have defined the fraction of trapped particles
\beqn \fl \trapfrac = \frac{3 \bmagz^2}{4} \lparav{\left[\int^{1/\bmag(\lpar)}_0 \frac{\pitch d \pitch}{\sqrt{1- \pitch \bmag(\lpar)}}
 - \int^{1/\bmagmax}_0 \frac{\pitch d \pitch}{\lparav{\sqrt{1- \pitch \bmag(\lpar)}}} \right]},
\label{eq:trapped-fraction} \eeqn
 and taken $\aspect \rightarrow 0 $ in the other geometrical
 quantities appearing in equation \refeq{eq:gfluxd-intermediate}.
 We note that the $\pitch$ limits of the integrals in equation \refeq{eq:trapped-fraction}
 are determined by the fact that $\HHepq{1/2}{0}$ is nonzero for passing particles only,
 whereas $\KKeSH\eqlbe$ and $\HHez$ have both trapped and passing particle components.
 Standard manipulations can be used to simplify $\trapfrac$
 in the limit $\aspect \rightarrow 0 $. To leading order \cite{helander}
 \beqn \trapfrac = \frac{3 \sqrt{2}}{2}\left[1 -
 \int^1_0 \left(\frac{\pi}{2\ellipticE(\zarg)} -1\right) \frac{d \zarg}{\zarg^2} \right]\; \aspect^{1/2}
 = 1.462 \; \aspect^{1/2},
\label{eq:trapped-fraction-value} \eeqn
 where 
\beqn \ellipticE(\zarg) = \frac{1}{2}\int^\pi_0 \sqrt{1 - \zarg^2 \sin^2 \left(\frac{\lpar}{2}\right)} d \lpar
\label{eq:ellipticE}\eeqn
 is the elliptic integral of the second kind.
 Finally, using the result in equation \refeq{eq:gfluxd-intermediate},
 we can calculate the \enquote{bootstrap} corrections to the electron
 parallel velocity and heat flux, $\dupareeffboot$ and $\dqpareeffboot$, respectively. We find that
 \beqn \fl 
\dupareeffboot=\vthere\frac{\trapfrac\cfreqei}{2}
 \left[ \frac{\vthere}{\saffac \rmajorz}\left(
   \sum_{\aindex,\bindex}\acoeffn{\aindex}\dmatrix{\aindex,\bindex}\acoeffn{\bindex}
   \drv{}{\lchi}\left(\frac{\ddensns{0}{e}}{\dense}\right) 
   + \sum_{\aindex,\bindex}\acoeffn{\aindex}\dmatrix{\aindex,\bindex}\bcoeffn{\bindex}
  \drv{}{\lchi}\left(\frac{\dtempns{0}{e}}{\tempe}\right) \right) \right. \nonumber\eeqn
  \beqn \left. - \imag \lflrezth \lchi \left( 
  \sum_{\aindex} \acoeffn{\aindex}\dmatrix{\aindex,0} \frac{\ddensns{0}{e}}{\dense}
  + \sum_{\aindex} \acoeffn{\aindex}(\dmatrix{\aindex,0}- \dmatrix{\aindex,1}) \frac{\dtempns{0}{e}}{\tempe}\right) \right]
  \label{eq:dupareeffboot},\eeqn
  with the matrix element
  \beqn \dmatrix{\aindex,\bindex} = \int^\infty_0 \expo{-\sarg} \soninen{\aindex}(\sarg) \soninen{\bindex}(\sarg) \nueffhat (\sarg)\; d \sarg, \label{eq:dmatrix}\eeqn
  and the function 
  \beqn \nueffhat (\sarg) = 1 + \erf(\sarg^{1/2}) - \Psifunc(\sarg^{1/2}).\eeqn
  Similarly, we find that
 \beqn \fl 
  \dqpareeffboot =\vthere \dense \tempe \frac{\trapfrac\cfreqei}{2}
 \left[ \frac{\vthere}{\saffac \rmajorz}\left(
   \sum_{\aindex,\bindex}\acoeffn{\aindex}\qmatrix{\aindex,\bindex}\acoeffn{\bindex}
   \drv{}{\lchi}\left(\frac{\ddensns{0}{e}}{\dense}\right) 
   + \sum_{\aindex,\bindex}\acoeffn{\aindex}\qmatrix{\aindex,\bindex}\bcoeffn{\bindex}
  \drv{}{\lchi}\left(\frac{\dtempns{0}{e}}{\tempe}\right) \right) \right. \nonumber\eeqn
  \beqn \left. - \imag \lflrezth \lchi \left( 
  \sum_{\aindex} \acoeffn{\aindex}\qmatrix{\aindex,0} \frac{\ddensns{0}{e}}{\dense}
  + \sum_{\aindex} \acoeffn{\aindex}(\qmatrix{\aindex,0}- \qmatrix{\aindex,1}) \frac{\dtempns{0}{e}}{\tempe}\right) \right]
  \label{eq:dqpareeffboot},\eeqn
  with the matrix element
  \beqn \qmatrix{\aindex,\bindex} = 
  \int^\infty_0 \expo{-\sarg} \left(\sarg - \frac{5}{2}\right)\soninen{\aindex}(\sarg)
  \soninen{\bindex}(\sarg) \nueffhat (\sarg)\; d \sarg. \label{eq:qmatrix}\eeqn
  The numerical coefficients appearing in equations \refeq{eq:dupareeffboot} and 
  \refeq{eq:dqpareeffboot} may be evaluated by using the 
  truncated polynomial solution of order $\aindexmax = 4$
  that is obtained in \ref{sec:parallel-collisional-contributions}.
  We use the values of $\{\acoeffn{\aindex}\}$ and $\{\bcoeffn{\aindex}\}$
  given in equations \refeq{eq:acoeffn} and \refeq{eq:bcoeffn}, respectively.
  We compute the matrix elements $\dmatrix{\aindex,\bindex}$ and $\qmatrix{\aindex,\bindex}$,
  with the results (to two decimal places)
  \beqn \dbmatrix = \left(\begin{tabular}{ccccc}
 1.53 & 2.12 & 2.53 & 2.85 & 3.13\\
  2.12 & 4.64 & 5.88 & 6.78 & 7.53\\
  2.53 & 5.88 & 9.25 & 11.15 & 12.61 \\
  2.85 & 6.78 & 11.15 & 15.34 & 17.91 \\
  3.13 & 7.53 & 12.61 & 17.91 & 22.88 \\
  \end{tabular}\right) ,
\eeqn
  and 
      \beqn \qbmatrix = -\left(\begin{tabular}{ccccc}
 2.12 & 4.64 & 5.88 & 6.78 & 7.53 \\ 
  4.64 & 7.79 & 13.07 & 15.87 & 17.98 \\ 
  5.88 & 13.07 & 17.03 & 25.14 & 29.65\\ 
  6.78 & 15.87 & 25.14 & 29.80 & 40.79\\ 
  7.53 & 17.98 & 29.65 & 40.79 & 46.09\\ 
      \end{tabular}\right) . \eeqn
  Combining these results, we find that
\beqn \fl 
\dupareeffboot=\vthere\frac{\trapfrac}{2}
 \left[ \frac{\vthere}{\saffac \rmajorz \cfreqei}\left(
   2.55\drv{}{\lchi}\left(\frac{\ddensns{0}{e}}{\dense}\right) 
   3.51\drv{}{\lchi}\left(\frac{\dtempns{0}{e}}{\tempe}\right) \right) \right. \nonumber\eeqn
  \beqn \left. + \imag \lflrezth \lchi \left( 
   1.66\frac{\ddensns{0}{e}}{\dense}
  0.47\frac{\dtempns{0}{e}}{\tempe}\right) \right]
  \label{eq:dupareeffboot-numerical},\eeqn
and
\beqn \fl 
  \dqpareeffboot =\vthere \dense \tempe \frac{\trapfrac}{2}
 \left[ \frac{\vthere}{\saffac \rmajorz \cfreqei}\left(
    2.98\drv{}{\lchi}\left(\frac{\dtempns{0}{e}}{\tempe}\right) 
   -0.07\drv{}{\lchi}\left(\frac{\ddensns{0}{e}}{\dense}\right) \right)
   \right. \nonumber\eeqn
  \beqn \left. 
  - \imag \lflrezth \lchi \left( 
  1.19\frac{\ddensns{0}{e}}{\dense}
    -2.63 \frac{\dtempns{0}{e}}{\tempe}\right) \right]
  \label{eq:dqpareeffboot-numerical}.\eeqn

  Including both the Spitzer-H\"{a}rm and the bootstrap contributions,
  the results for the flows are (to $\order{\aspect^{1/2}}$)
    \beqn \fl  \frac{\dupareeff}{\vthere} =  \imag \frac{\saffac\shat \kky \gyrdez  \lchi }{2 \aspect^{1/2}}
   \left( 2.43   
   \frac{\ddensns{0}{e}}{\dense}
    + 0.69  \frac{\dtempns{0}{e}}{\tempe}\right)   
   \label{eq:upare-nustar-ll-1} \eeqn \beqn \fl  - \frac{\vthere}{2 \saffac \rmajorz \cfreqei}
   \left[1.97 \left(1 - 1.90 \aspect^{1/2}\right) 
   \drv{}{\lchi}\left(\frac{\ddensns{0}{e}}{\dense}\right)
  + 3.37 \left(1 - 1.52 \aspect^{1/2}\right) 
  \drv{}{\lchi}\left(\frac{\dtempns{0}{e}}{\tempe} \right)\right] 
   ,\nonumber\eeqn
    and  
    \beqn \fl \frac{\dqpareeff}{\vthere \dense \tempe} =  -
    \imag\frac{5\saffac\shat \kky \gyrdez  \lchi }{4 \aspect^{1/2}}\left(
     0.70 
    \frac{\ddensns{0}{e}}{\dense} 
      -1.54 
    \frac{\dtempns{0}{e}}{\tempe}\right) 
    \label{eq:qpare-nustar-ll-1}\eeqn \beqn \fl  - \frac{5\vthere}{4 \saffac \rmajorz \cfreqei }
   \left[0.56 \left(1 + 0.07 \aspect^{1/2}\right)  
   \drv{}{\lchi}\left(\frac{\ddensns{0}{e}}{\dense}\right)
  + 2.23 \left(1 -0.78\aspect^{1/2}\right) 
  \drv{}{\lchi}\left(\frac{\dtempns{0}{e}}{\tempe} \right)\right] 
\nonumber,   \eeqn
    where we have defined $\gyrdez = \vthere/\cycfez$, 
    and used that, for $\aspect \ll 1 $ and circular flux surfaces,
    $\trapfrac = 1.46 \aspect^{1/2}$, $\kxfac \simeq 1$,
    $\kpar \simeq 1/\saffac \rmajorz$ 
    and $\bcur \tdrvt{\radial}{\flxl} \simeq \saffac/\aspect$.
    We note that $\lflrezth = \shat \kky \gyrdez \saffac/ \aspect$.
    
    We now turn to the calculation
    of the transport due to perpendicular diffusion via the neoclassical fluxes 
    $\dpartnc$ and $\dheatnc$, defined in equations \refeq{eq:neoclassical-particle-flux} and \refeq{eq:neoclassical-heat-flux}.
    To evaluate these fluxes, we need to compute integrals of the form
     \beqn \gfluxperp = -\lparav{\frac{\bcur}{\cycfe}\tdrv{\radial}{\flxl} \intv{ \vpar \fulldistg(\energy)
  \copbothe\left[\HHepq{1/2}{0} + \imag \lflre\lchi\HHez - \HHeSH \right]}}.
  \label{eq:banana-perpendicular-collisional-contributions}\eeqn
    We note that the form of the integral \refeq{eq:banana-perpendicular-collisional-contributions} 
    is structurally similar to the integral defined in equation \refeq{eq:gfluxd-def2}.
    To be precise, we can use the estimates \refeq{eq:deeplypassing-contribution-estimate}
    and \refeq{eq:trappedpassing-contribution-estimate} to justify replacing $\copbothe\left[\cdot\right]$
    with $\pitchscatter\left[\cdot\right]$ when evaluating
    \refeq{eq:banana-perpendicular-collisional-contributions}.
    Again, since the integrals in $\energy$ and $\pitch$ are separable in 
    \refeq{eq:banana-perpendicular-collisional-contributions},
    we find the intermediate result 
    \beqn \fl \gfluxperp =  -\frac{\bcur}{\cycfez}\tdrv{\radial}{\flxl}\frac{ \pi \vthere^4}{3}
 \trapfrac 
\int^\infty_0 \fulldistg(\energy)\nueff(\energy) \left(\frac{\energy}{\tempe}\right)^{3/2}
\left(\vthere \KKeSH \eqlbe - \imag \lflrezth \lchi \HHez \right)  \; \frac{d \energy}{\tempe}
  \label{eq:banana-perpendicular-collisional-contributions-intermediate}, \eeqn
    where we note the similarity to \refeq{eq:gfluxd-intermediate}.
    To compute the neoclassical particle flux, we set $\fulldistg(\energy) = 1$
    in equation \refeq{eq:banana-perpendicular-collisional-contributions-intermediate},
    and obtain the result (correct to $\order{\aspect^{1/2}}$)
    \beqn \fl \frac{\dpartnc}{\dense} =
     \label{eq:density-neoclassical-perpendicular-diffusion-banana-particle-last}
 - \gyrdez \frac{\saffac}{\aspect} \frac{\trapfrac\cfreqei}{2}
 \left[ \frac{\vthere}{\saffac \rmajorz}\left(
   \sum_{\bindex}\dmatrix{0,\bindex}\acoeffn{\bindex}
   \drv{}{\lchi}\left(\frac{\ddensns{0}{e}}{\dense}\right) 
   + \sum_{\bindex}\dmatrix{0,\bindex}\bcoeffn{\bindex}
  \drv{}{\lchi}\left(\frac{\dtempns{0}{e}}{\tempe}\right) \right) \right. \nonumber\eeqn
  \beqn \left. - \imag \kky \gyrdez \shat \lchi \frac{\saffac}{\aspect}  \left( 
  \dmatrix{0,0} \frac{\ddensns{0}{e}}{\dense}
  + (\dmatrix{0,0}- \dmatrix{0,1}) \frac{\dtempns{0}{e}}{\tempe}\right) \right]
 .\nonumber\eeqn
  Similarly, to compute the neoclassical heat flux, we set $\fulldistg(\energy) = \energy/\tempe - 5/2$
    in equation \refeq{eq:banana-perpendicular-collisional-contributions-intermediate},
    and obtain the result (correct to $\order{\aspect^{1/2}}$)
    \beqn \fl  \frac{\dheatnc}{\dense \tempe}=
     \label{eq:density-neoclassical-perpendicular-diffusion-banana-heat-last}
- \gyrdez \frac{\saffac}{\aspect} \frac{\trapfrac\cfreqei}{2}
 \left[ \frac{\vthere}{\saffac \rmajorz}\left(
   \sum_{\bindex}\qmatrix{0,\bindex}\acoeffn{\bindex}
   \drv{}{\lchi}\left(\frac{\ddensns{0}{e}}{\dense}\right) 
   + \sum_{\bindex}\qmatrix{0,\bindex}\bcoeffn{\bindex}
  \drv{}{\lchi}\left(\frac{\dtempns{0}{e}}{\tempe}\right) \right) \right. \nonumber\eeqn
  \beqn \left. - \imag \kky \gyrdez \shat \lchi \frac{\saffac}{\aspect}\left( 
  \qmatrix{0,0} \frac{\ddensns{0}{e}}{\dense}
  + \left(\qmatrix{0,0}- \qmatrix{0,1}\right) \frac{\dtempns{0}{e}}{\tempe}\right) \right]
 .\nonumber\eeqn
 Inserting the numerical coefficients, we find that
    \beqn \fl \frac{\dpartnc}{\dense} =
     \label{eq:density-neoclassical-perpendicular-diffusion-banana-last}
  \gyrdez \frac{\saffac}{\aspect} \frac{\trapfrac}{2}
  \frac{\vthere}{\saffac \rmajorz}\left(
   1.66
   \drv{}{\lchi}\left(\frac{\ddensns{0}{e}}{\dense}\right) 
    + 1.75
  \drv{}{\lchi}\left(\frac{\dtempns{0}{e}}{\tempe}\right) \right) \nonumber\eeqn
  \beqn  + \imag \cfreqei \kky \shat \lchi (\gyrdez)^2 \left(\frac{\saffac}{\aspect}\right)^2  \frac{\trapfrac}{2} \left( 
  1.53 \frac{\ddensns{0}{e}}{\dense}
  -0.59 \frac{\dtempns{0}{e}}{\tempe}\right) 
 ,\nonumber\eeqn 
     and
   \beqn \fl  \frac{\dheatnc}{\dense \tempe}=
     \label{eq:density-neoclassical-perpendicular-diffusion-banana-numerical}
  \gyrdez \frac{\saffac}{\aspect} \frac{\trapfrac}{2} \frac{\vthere}{\saffac \rmajorz}\left(
    0.11
  \drv{}{\lchi}\left(\frac{\dtempns{0}{e}}{\tempe}\right)
 -1.19
   \drv{}{\lchi}\left(\frac{\ddensns{0}{e}}{\dense}\right) 
     \right) \nonumber\eeqn
  \beqn  +    \imag \cfreqei \kky \shat \lchi (\gyrdez)^2\left(\frac{\saffac}{\aspect}\right)^2\frac{\trapfrac}{2}\left( 
   2.51 \frac{\dtempns{0}{e}}{\tempe}
 - 2.12 \frac{\ddensns{0}{e}}{\dense}\right)
 .\nonumber\eeqn

    \section{ Obtaining matching conditions for the inner region: 
    the electron response in the outer region} \label{sec:electron-outer-collisional-small-tail-appendix}
  
    In this section we examine the equations to the electron response in the outer region,
    and derive the matching conditions given in sections 
    \ref{sec:electron-outer-collisional-small-tail} and \ref{sec:electron-outer-collisional-large-tail}.
    We consider the case of small electron tails, noting that the large-tail case follows trivially. 
    To satisfy the ordering \refeq{eq:small-tail-ordering}, in the outer region we take $\HHez =0$.
    Expanding in $\massrtp{1/4}$, the next order equation is
    \beqn \vpar \kpar \drv{\HHeh}{\lpar}  = \coplandaue\left[\HHeh\right] + \coplorentze\left[\HHeh\right]
 \label{eq:electron_outer0.5-collisional}.\eeqn
   Superficially, equation \refeq{eq:electron_outer0.5-collisional} has
   an identical form to equation \refeq{eq:electron_inner0-collisional}. 
However, in the outer region, $\HHeh$ cannot be assumed to be periodic in $\lpar$. 
    To solve for $\HHeh$, we multiply equation
    \refeq{eq:electron_outer0.5-collisional} by $\HHeh/\eqlbe$,
    and integrate over velocity and $\lpar$. We obtain
     \beqn \fl \int^\infty_{-\infty} \left[
    \intv{\frac{\HHeh}{\eqlbe} \; \coplandaue\left[\HHeh\right]} 
    + \intv{\frac{\HHeh}{\eqlbe} \;\coplorentze\left[\HHeh\right]}\right]\frac{d \lpar}{\bvec\cdot\nbl\lpar}
 \label{eq:electron_outer0.5-collisional-v2}\eeqn
 \beqn = \left[\intv{\frac{\vpar}{\bmag}\frac{\left(\HHeh\right)^2}{2 \eqlbe}}\right]^{\lpar = \infty}_{\lpar = -\infty}
    = 0, \nonumber \eeqn
    where in the final equality we have assumed
    continuity of the leading-order piece of $\HHe$ in the matching region -- 
    and hence the velocity moment vanishes to leading order at $\lpar = \pm \infty$.
    With the entropy production properties \refeq{eq:entropy},
   equation \refeq{eq:electron_outer0.5-collisional-v2} shows that
   \beqn \frac{\HHeh}{\eqlbe} = \frac{\ddensns{1/2}{e}}{\dense}
    + \frac{\dtempns{1/2}{e}}{\tempe}\left(\frac{\energy}{\tempe}
    - \frac{3}{2}\right)\, \label{eq:HHeh-outer-collisional}\eeqn 
   where $\ddensns{1/2}{e}$ and $\dtempns{1/2}{e}$ are a constant density and temperature, respectively,
   as required to match to the inner region. These results provide the density and temperature matching
   conditions \refeq{eq:density-matching-collisional} and \refeq{eq:temperature-matching-collisional}.
   
   To calculate the electron flow matching conditions, 
   we proceed to the next order equation 
    \beqn \fl \vpar \kpar \drv{\HHeo}{\lpar}  - \coplandaue\left[\HHeo\right]
    - \coplorentze\left[\HHeo - \frac{\me \vpar \duparns{0}{i}}{\tempe}\eqlbe \right]
  = - \imag \left(\wstare - \wfreqz\right)\eqlbe\frac{\charge \ptlz}{\tempe}
 \label{eq:electron_outer1-collisional},\eeqn
  where 
  \beqn \duparns{0}{i} = \frac{1}{\densi}\intv{\vpar \besin{0}\hhiz}. \label{eq:dupari-definition}\eeqn
  We extract equations for the leading-order (nonzero)
  electron mean velocity $\duparns{1}{e}$, and electron heat flux $\dqparns{1}{e}$.
      Noting that $\hheo = \HHeo$ for $\HHez =0$, $\lflre \sim \massrt$ and $\lpar \sim 1$,
      by virtue of expanding the definition \refeq{eq:HHs}, we obtain
      that $\duparns{1}{e} = \dUparns{1}{e}$ and $\dqparns{1}{e} = \dQparns{1}{e}$,
      where $\dUparns{1}{e}$ and $\dQparns{1}{e}$ are the moments of $\HHeo$ defined by equations
      \refeq{eq:upare-nth} and \refeq{eq:qpare-nth}, respectively.
   Taking density and temperature velocity moments, we find that 
  \beqn \bvec \cdot \nbl \lpar \drv{}{\lpar}\left(\frac{\duparns{1}{e}}{\bmag}\right)
  = - \imag (\wstaren - \wfreqz) \frac{\charge \ptlz}{\tempe},\label{eq:dupare-outer-1}\eeqn and 
  \beqn \bvec \cdot \nbl \lpar \drv{}{\lpar}\left(\frac{\dqparns{1}{e}}{\bmag \dense \tempe}
   {+ \frac{\duparns{1}{e}}{\bmag}}\right)
  = - \imag \frac{3}{2} \wstaren \etae \frac{\charge \ptlz}{\tempe}.\label{eq:dqpare-outer-1}\eeqn
  
  Equations \refeq{eq:dupare-outer-1} and \refeq{eq:dqpare-outer-1}
  can be integrated to obtain the leading-order jump in
  $\dupare$ and $\dqpare$ across the outer region. We find that
  \beqn \left[\frac{\duparns{1}{e}}{\bmag}\right]^{\lpar = \infty}_{\lpar = -\infty}
   = - \imag (\wstaren - \wfreqz) \int^{\infty}_{-\infty}
   \frac{\charge \ptlz(\lpar)}{\tempe} \frac{d \lpar}{\bvec \cdot \nbl \lpar}, \label{eq:dupare-outer-2} \eeqn
   and 
  \beqn \left[\frac{\dqparns{1}{e}}{\dense \tempe \bmag}\right]^{\lpar = \infty}_{\lpar = -\infty}
   = - \imag { \left(\frac{3}{2}\wstaren \etae - \wstaren + \wfreqz\right) }\int^{\infty}_{-\infty}
   \frac{\charge \ptlz(\lpar)}{\tempe} \frac{d \lpar}{\bvec \cdot \nbl \lpar}. \label{eq:dqpare-outer-2}\eeqn
  Equations \refeq{eq:dupare-outer-2} and \refeq{eq:dqpare-outer-2}
  give the estimate  \refeq{eq:electron-flow-jump} 
  for the jump in the electron flows across the outer region.
 There is an implicit assumption that the potential due to the nonadiabatic ion response
 decays for large $\lpar$ in the outer region,
 such that the integrals in equations \refeq{eq:dupare-outer-2}
 and \refeq{eq:dqpare-outer-2} exist.
 In fact, it is possible to show that there is a matching
 region of size $(\ln (\mi/\me))^\delta$ between the outer and inner regions where
 the nonadiabatic ion response decays exponentially with $\lpar$,
 and both the nonadiabatic ion and electron responses contribute to a $\massrtp{1/4}$ small potential.
 The quanity $\delta$ is an order unity number that we have not determined.
 Formally, we can neglect this matching region in our analysis because the electron density,
 temperature, and flows remain constant over the matching region, and because no information
 about the ions in this region is propagated into either the outer or inner regions.

    The jump conditions on $\dupareeffinner$
    and $\dqpareeffinner$ can be obtained by taking the following steps.
    First, we note that taking the $|\lpar| \rightarrow \infty$ limit
    in equations \refeq{eq:dupare-outer-1} and \refeq{eq:dqpare-outer-1} leads to the results
    \beqn \bvec \cdot \nbl \lpar \drv{}{\lpar}\left(\frac{\duparns{1}{e,outer}}{\bmag}\right)
  =0,\label{eq:dupare-outer-match}\eeqn and 
  \beqn \bvec \cdot \nbl \lpar \drv{}{\lpar}\left(\frac{\dqparns{1}{e}}{\bmag \dense \tempe}
   {+ \frac{\duparns{1}{e,outer}}{\bmag}}\right)
  = 0,\label{eq:dqpare-outer-match}\eeqn
   where we have used that $\charge \ptlzouter /\tempe$
   becomes exponentially small for large $|\lpar|$,
   due to the decaying nonadiabatic ion response. 
   Equations \refeq{eq:dupare-outer-match} and \refeq{eq:dqpare-outer-match}
   state that $\duparns{1}{e,outer}/\bmag$ and $\dqparns{1}{e,outer}/\bmag$
   are independent of $\lpar$ at large $|\lpar|$.
    Second, we note that, in the inner region, we can show that the $\vpar/\bmag$ moments of $\HHeh$, $\dUparns{1/2}{e,inner}/\bmag$
    and $\dQparns{1/2}{e,inner}/\bmag$, 
    are independent of $\lpar$ for $\lchi \ll \massrutp{1/4}$ by taking the density and temperature
    moments of equation \refeq{eq:electron_inner1b-collisional}. Hence, the flows $\duparns{1/2}{e,inner}/\bmag$
    and $\dqparns{1/2}{e,inner}/\bmag$ are independent of $\lpar$ for $\lchi \ll \massrutp{1/4}$.
    Third, we demand that $\dupare /\bmag$ and $\dqpare/\bmag$ should be continuous over the
    boundaries between the outer and inner regions, i.e.,
    $\dupare$ and $\dqpare$ should satisfy
    \beqn \left.\frac{\duparns{1}{e,outer}}{\bmag}\right|_{\lpar \rightarrow \pm \infty}
    = \left.\frac{\duparns{1/2}{e,inner}}{\bmag}\right|_{\lchi \rightarrow 0^{\pm}}, \label{eq:upare-matching-collisional-small-tail} \eeqn
    and 
    \beqn \left.\frac{\dqparns{1}{e,outer}}{\bmag}\right|_{\lpar \rightarrow \pm \infty}
    = \left.\frac{\dqparns{1/2}{e,inner}}{\bmag}\right|_{\lchi \rightarrow 0^{\pm}}, \label{eq:qpare-matching-collisional-small-tail} \eeqn
    Finally, we combine equations \refeq{eq:dupare-outer-2}, \refeq{eq:dqpare-outer-2},
    \refeq{eq:upare-matching-collisional-small-tail}, and \refeq{eq:qpare-matching-collisional-small-tail}
    to find the boundary conditions \refeq{eq:upare-jump-small-tail} and 
    \refeq{eq:qpare-jump-small-tail} 
    on $\dupareeffinner$ and $\dqpareeffinner$, respectively.

%\bibliography{refs} 

\begin{thebibliography}{10}
\expandafter\ifx\csname url\endcsname\relax
  \def\url#1{{\tt #1}}\fi
\expandafter\ifx\csname urlprefix\endcsname\relax\def\urlprefix{URL }\fi
\providecommand{\eprint}[2][]{\url{#2}}

\bibitem{dorland2000electron}
Dorland W, Jenko F, Kotschenreuther M and Rogers B~N 2000 {\em Phys. Rev.
  Lett.\/} {\bf 85} 5579--5582

\bibitem{jenko2000electron}
Jenko F, Dorland W, Kotschenreuther M and Rogers B~N 2000 {\em Phys. Plasmas\/}
  {\bf 7} 1904--1910

\bibitem{jenko2002prediction}
Jenko F and Dorland W 2002 {\em Phys. Rev. Lett.\/} {\bf 89} 225001

\bibitem{STroach2009PPCF}
Roach C~M, Abel I~G, Akers R~J, Arter W, Barnes M, Camenen Y, Casson F~J,
  Colyer G, Connor J~W, Cowley S~C, Dickinson D, Dorland W, Field A~R,
  Guttenfelder W, Hammett G~W, Hastie R~J, Highcock E, Loureiro N~F, Peeters
  A~G, Reshko M, Saarelma S, Schekochihin A~A, Valovic M and Wilson H~R 2009
  {\em Plasma Phys. Control. Fusion\/} {\bf 51} 124020

\bibitem{maeyama2015cross}
Maeyama S, Idomura Y, Watanabe T~H, Nakata M, Yagi M, Miyato N, Ishizawa A and
  Nunami M 2015 {\em Phys. Rev. Lett.\/} {\bf 114} 255002

\bibitem{Maeyama_2017_NF}
Maeyama S, Watanabe T~H, Idomura Y, Nakata M, Ishizawa A and Nunami M 2017 {\em
  Nucl. Fusion\/} {\bf 57} 066036

\bibitem{howard2014synergistic}
Howard N~T, Holland C, White A~E, Greenwald M and Candy J 2014 {\em Phys.
  Plasmas\/} {\bf 21} 112510

\bibitem{howard2016enhanced}
Howard N~T, Holland C, White A~E, Greenwald M and Candy J 2016 {\em Nucl.
  Fusion\/} {\bf 56} 014004

\bibitem{howard2016comparison}
Howard N~T, Holland C, White A~E, Greenwald M, Candy J and Creely A~J 2016 {\em
  Phys. Plasmas\/} {\bf 23} 056109

\bibitem{maeyama2017supression}
Maeyama S, Watanabe T~H and Ishizawa A 2017 {\em Phys. Rev. Lett.\/} {\bf 119}
  195002

\bibitem{Bonanomi_2018_ImpactofES}
Bonanomi N, Mantica P, Citrin J, Goerler T and and B~T 2018 {\em Nucl.
  Fusion\/} {\bf 58} 124003

\bibitem{HallatschekgiantelPRL2005}
Hallatschek K and Dorland W 2005 {\em Phys. Rev. Lett.\/} {\bf 95} 055002

\bibitem{DominskinonadPOP015}
Dominski J, Brunner S, G\"{o}rler T, Jenko F, Told D and Villard L 2015 {\em
  Phys. Plasmas\/} {\bf 22} 062303

\bibitem{ConnorProRSoc1979ballooning}
Connor J~W, Hastie R~J and Taylor J~B 1979 {\em Proceedings of the Royal
  Society of London. A. Mathematical and Physical Sciences\/} {\bf 365} 1--17

\bibitem{Parisi_2020}
Parisi J~F, Parra F~I, Roach C~M, Giroud C, Dorland W, Hatch D~R, Barnes M,
  Hillesheim J~C, Aiba N, Ball J, Ivanov P~G and contributors J 2020 {\em Nucl.
  Fusion\/} {\bf 60} 126045

\bibitem{Applegate_2007}
Applegate D~J, Roach C~M, Connor J~W, Cowley S~C, Dorland W, Hastie R~J and
  Joiner N 2007 {\em Plasma Phys. Control. Fusion\/} {\bf 49} 1113--1128

\bibitem{Dickinson_MTMPed_2013}
Dickinson D, Roach C~M, Saarelma S, Scannell R, Kirk A and Wilson H~R 2013 {\em
  Plasma Phys. Control. Fusion\/} {\bf 55} 074006

\bibitem{Moradi_2013}
Moradi S, Pusztai I, Guttenfelder W, F\"{u}l\"{o}p T and Moll{\'{e}}n A 2013
  {\em Nucl. Fusion\/} {\bf 53} 063025

\bibitem{cowleyPoFB91}
Cowley S~C, Kulsrud R~M and Sudan R 1991 {\em Phys. Fluids B\/} {\bf 3} 2767

\bibitem{RomanelliITG}
Romanelli F 1989 {\em Phys. Fluids B: Plasma Physics\/} {\bf 1} 1018--1025

\bibitem{Adam1976TEM}
Adam J~C, Tang W~M and Rutherford P~H 1976 {\em Phys. Fluids\/} {\bf 19}
  561--566

\bibitem{beerbouncePoP1996}
Beer M~A and Hammett G~W 1996 {\em Phys. Plasmas\/} {\bf 3} 4018--4022

\bibitem{BelliIsotope2019}
Belli E~A, Candy J and Waltz R~E 2019 {\em Phys. Plasmas\/} {\bf 26} 082305

\bibitem{BelliIsotope2020PRL}
Belli E~A, Candy J and Waltz R~E 2020 {\em Phys. Rev. Lett.\/} {\bf 125} 015001

\bibitem{ajay_brunner_mcmillan_ball_dominski_merlo_2020}
{CJ} A, Brunner S, McMillan B, Ball J, Dominski J and Merlo G 2020 {\em J.
  Plasma Phys.\/} {\bf 86} 905860504

\bibitem{ball_brunner_ajay_2020}
Ball J, Brunner S and {CJ} A 2020 {\em J. Plasma Phys.\/} {\bf 86} 905860207

\bibitem{ajay_brunner_ball_2020}
{CJ} A, Ball J and Brunner S 2020 {\em arXiv:2104.12585\/}

\bibitem{KOTSCHENREUTHER1995CPC}
Kotschenreuther M, Rewoldt G and Tang W 1995 {\em Comput. Phys. Commun.\/} {\bf
  88} 128 -- 140

\bibitem{cattoPP78}
Catto P~J 1978 {\em Plasma Phys.\/} {\bf 20} 719

\bibitem{HazeltineMeiss}
Hazeltine R~D and Meiss J~D 2003 {\em Plasma Confinement\/} (New York: Dover)

\bibitem{BraginskiiTransport}
Braginskii S~I 1957 {\em J. Exptl. Theoret. Phys. (U.S.S.R.)\/} {\bf 33}
  459--472

\bibitem{helander}
Helander P and Sigmar D~J 2002 {\em Collisional transport in magnetized
  plasmas\/} (Cambridge: Cambridge University Press)

\bibitem{hardmanpaper1}
Hardman M~R, Barnes M, Roach C~M and Parra F~I 2019 {\em Plasma Phys. Control.
  Fusion\/} {\bf 61} 065025

\bibitem{abelEMOD}
Abel I~G and Cowley S~C 2013 {\em New J. Phys.\/} {\bf 15} 023041

\bibitem{cattotasngPoF1977}
Catto P~J and Tsang K~T 1977 {\em Phys. Fluids\/} {\bf 20} 396--401

\bibitem{abelPoP08}
Abel I~G, Barnes M, Cowley S~C, Dorland W, Hammett G~W and Schekochihin A~A
  2008 {\em Phys. Plasmas\/} {\bf 15} 122509 arXiv:0806.1069

\bibitem{barnesPoP09}
Barnes M, Abel I~G, Dorland W, Ernst D~R, Hammett G~W, Ricci P, Rogers B~N,
  Schekochihin A~A and Tatsuno T 2009 {\em Phys. Plasmas\/} {\bf 16} 072107

\bibitem{Connor_1985PPCFresistiveballooning}
Connor J~W and Hastie R~J 1985 {\em Plasma Phys. Control. Fusion\/} {\bf 27}
  621--639

\bibitem{ConnorPPCF2008}
Connor J~W, Hastie R~J and Helander P 2008 {\em Plasma Phys. Control. Fusion\/}
  {\bf 51} 015009

\bibitem{hintonRMP76}
Hinton F~L and Hazeltine R~D 1976 {\em Rev. Mod. Phys.\/} {\bf 25} 239--308

\bibitem{SpitzerHarmPhysRev.89.977}
Spitzer L and H\"arm R 1953 {\em Phys. Rev.\/} {\bf 89} 977--981

\bibitem{FreidbergMHD}
Freidberg J 2014 {\em Ideal {MHD}\/} (Cambridge: Cambridge University Press)

\bibitem{DimitsCBCPoP2000}
Dimits A~M, Bateman G, Beer M~A, Cohen B~I, Dorland W, Hammett G~W, Kim C,
  Kinsey J~E, Kotschenreuther M, Kritz A~H, Lao L~L, Mandrekas J, Nevins W~M,
  Parker S~E, Redd A~J, Shumaker D~E, Sydora R and Weiland J 2000 {\em Phys.
  Plasmas\/} {\bf 7} 969--983

\bibitem{Miller98}
Miller R~L, Chu M~S, Greene J~M, Lin-Liu Y~R and Waltz R~E 1998 {\em Phys.
  Plasmas\/} {\bf 5} 973--978

\bibitem{barnesPoP10a}
Barnes M, Dorland W and Tatsuno T 2010 {\em Phys. Plasmas\/} {\bf 17} 032106

\bibitem{hardmanpaper2}
Hardman M~R, Barnes M and Roach C~M 2020 {\em J. Plasma Phys.\/} {\bf 86}
  905860601

\bibitem{Kotschenreuther_2000_ignited_tokamaks}
Kotschenreuther M, Dorland W, Liu Q, Zarnstorff M, Miller R and Lin-Liu Y 2000
  {\em Nucl. Fusion\/} {\bf 40} 677--684

\bibitem{DrakeLeePoF1977}
Drake J~F and Lee Y~C 1977 {\em Phys. Fluids\/} {\bf 20} 1341--1353

\bibitem{Cowley_1986PoFtearing}
Cowley S~C, Kulsrud R~M and Hahm T~S 1986 {\em Phys. Fluids\/} {\bf 29}
  3230--3244

\bibitem{ZoccoPoP2011}
Zocco A and Schekochihin A~A 2011 {\em Phys. Plasmas\/} {\bf 18} 102309

\bibitem{ConnorPPCF2012}
Connor J~W, Hastie R~J and Zocco A 2012 {\em Plasma Phys. Control. Fusion\/}
  {\bf 54} 035003

\bibitem{bpatelthesis}
Patel B 2021 Ph.D. thesis University of York

\bibitem{Waltz2006Corrugations}
Waltz R~E, Austin M~E, Burrell K~H and Candy J 2006 {\em Phys. Plasmas\/} {\bf
  13} 052301

\bibitem{Parisi_2021draft}
Parisi J~F, Parra F~I, Roach C~M, Hardman M~R, Barnes M, Dorland W, St-Onge D,
  Ball J, Hatch D~R, Dickinson D, Abel I~G, Saarelma S, Chapman B, Giroud C,
  Hillesheim J~C, Aiba N and {JET Contributors} 2021 {\em In preparation\/}

\bibitem{Hall1975PoF}
Hall L~S and McNamara B 1975 {\em Phys. Fluids\/} {\bf 18} 552--565

\bibitem{CaryShasharinaOmniPRL}
Cary J~R and Shasharina S~G 1997 {\em Phys. Rev. Lett.\/} {\bf 78} 674--677

\bibitem{CaryShasharinaOmniPoP}
Cary J~R and Shasharina S~G 1997 {\em Phys. Plasmas\/} {\bf 4} 3323--3333

\bibitem{Helander_2014}
Helander P 2014 {\em Rep. Prog. Phys.\/} {\bf 77} 087001

\bibitem{Parra_2015_Less_constrained_omnigeneous}
Parra F~I, Calvo I, Helander P and Landreman M 2015 {\em Nucl. Fusion\/} {\bf
  55} 033005

\bibitem{Calvo_2017_tangential_drifts}
Calvo I, Parra F~I, Velasco J~L and Alonso J~A 2017 {\em Plasma Phys. Control.
  Fusion\/} {\bf 59} 055014

\bibitem{BenderOrszag}
Bender C~M and Orszag S~A 1999 {\em Advanced Mathematical Methods for
  Scientists and Engineers {I}: {A}symptotic Methods and Perturbation
  Theory.\/} (New York: Springer)

\bibitem{Newton_2010}
Newton S~L, Cowley S~C and Loureiro N~F 2010 {\em Plasma Phys. Control.
  Fusion\/} {\bf 52} 125001

\end{thebibliography}
\providecommand{\newblock}{}

\end{document}